\documentclass[a4paper,12pt,english]{memoir}
\usepackage[english]{babel}
\usepackage[T1]{fontenc}
\usepackage{times}
\usepackage[utf8]{inputenc}

\usepackage{geometry}
\usepackage{textcomp}
\usepackage{amsmath}
\usepackage{amssymb}
\usepackage{amsfonts}
\usepackage{upgreek} 
\usepackage{nicefrac}

\usepackage{graphicx}
\usepackage{esint}
\usepackage{epstopdf}

\usepackage[
  style=phys,
  backend=biber,
  sorting=none,
]{biblatex}

\usepackage[
  colorlinks=true,
  linkcolor=blue,
  citecolor=blue,
  urlcolor=blue
]{hyperref}

\DeclareFieldFormat[article]{title}{%
  \href{\thefield{url}}{#1}%
}

\DeclareFieldFormat[thesis]{title}{%
  \href{\thefield{url}}{#1}%
}

\usepackage{pdfpages}

\usepackage{multirow}
\usepackage{dcolumn}
\newcolumntype{d}{D{.}{.}{-1} } 

\makeatletter

\newcommand{\cmi}{cm$^{-1}$}
\newcommand{\lntp}{Ln$^{3+}$}
\newcommand{\narb}{$^{23}$Na$^{87}$Rb}
\newcommand{\nak}{$^{23}$Na$^{39}$K}

\newcommand{\qnpr}[1]
{
  #1^{\phantom{\prime}}
}
\newcommand{\brapr}[1]
{
  \left\langle #1^{\phantom{\prime}} \right|
}

\newcommand{\ninej}[9]
{
  \left\{ \begin{array}{ccc} #1 & #2 & #3 \\
    #4 & #5 & #6 \\ #7 & #8 & #9 \end{array} \right\}
}
\newcommand{\sixj}[6]
{
  \left\{ \begin{array}{ccc} #1 & #2 & #3 \\
    #4 & #5 & #6 \end{array} \right\}
}
\newcommand{\thrj}[6]
{
  \left( \begin{array}{ccc} #1 & #2 & #3 \\
    #4 & #5 & #6 \end{array} \right)
}

\captiondelim {  --  }
\captiontitlefont {\itshape}

\usepackage{babel}
\addto\extrasfrench{\providecommand{\fg}{\ifdim\lastskip>\z@\unskip\fi~\frqq}}
\makeatother

\begin{document}
\thispagestyle{empty}

\vspace*{\fill}

\begin{center}
\begin{sffamily}
\begin{bfseries}
  \noindent{\Large Universit{\'e} Bourgogne Europe}
  \vspace{6mm}

  \noindent{\Huge Habilitation à diriger des Recherches}
  \vspace{15mm}

  \noindent{\large pr{\'e}sent{\'e}e par}
  \vspace{6mm}

  \noindent{\Huge Maxence Lepers}
  \vspace{15mm}

  \noindent{\large Titre~:}
  \vspace{6mm}

  \noindent \Huge Structure et interactions d'atomes et de molécules diatomiques~: des gaz ultra-froids aux solides dopés
  \vspace{15mm}

  \noindent{\large Title~:}
  \vspace{6mm}

  \noindent{\Huge Structure and interactions of atoms and diatomic molecules: from ultracold gases to doped solids}
  \vspace{15mm}

  \large soutenue le 7 octobre 2025, devant le jury compos\'{e} de~:
  \vspace{6mm}

  \begin{tabular}{ll} 
    Mme Lydia Tchang-Brillet & Rapportrice\\
    M. Daniel Comparat & Rapporteur\\
    M. Thierry Stoecklin & Rapporteur\\
    M. Andrea Simoni & Examinateur\\
    M. G{\' e}rard Colas des Francs & Examinateur\\
  \end{tabular}

\end{bfseries}
\end{sffamily}
\end{center}

\vspace*{\fill}

\cleardoublepage

\pagenumbering{roman} \setcounter{page}{1}
\setcounter{tocdepth}{2}
\tableofcontents*

\chapter*{Acknowledgments}  \addcontentsline{toc}{chapter}{Acknowledgments}\markboth{ACKNOWLEDGMENTS}{ }

\pagenumbering{arabic} \setcounter{page}{1}

I would like first to thank each member of my jury for accepting to review my work. I thank all the people -- family, friends, colleagues -- who attended, in person or online, my defense on October 7, 2025, and who helped to prepare it. You made that day an unforgettable moment of sharing.

Research is a very demanding activity. I thank my wife, my sun, as well as my parents and whole family, for their invaluable and constant support, allowing me for achieving that challenging task.

In this manuscript, I mention many people with whom I had the great opportunity to collaborate, learn, discuss. In this respect, I dedicate this manuscript to the memory of Jean-Fran{\c c}ois Wyart, who taught me the semi-empirical method for calculating atomic spectra. I also warmly thank colleagues and friends currently and formerly at Laboratoire Aim{\' e} Cotton, especially Olivier Dulieu, Nadia Bouloufa-Maafa and {\'E}liane Luc, who played a crucial role in my research work and beyond, but also Romain Vexiau, {\'E}tienne Brion, Mireille Aymar, Andrea Orb{\'a}n, Zsolt Mezei, Ioan Schneider, Goulven Qu{\'e}m{\'e}ner, Ting Xie and Viatcheslav Kokoouline.

I also thank my colleagues from Laboratoire Interdisciplinaire Carnot de Bourgogne in Dijon, especially Vincent Boudon, Cyril Richard, Claude Leroy, Pascal Honvault, Gr{\'e}goire Guillon, G{\'e}rard Colas des Francs and Fr{\'e}d{\'e}ric Chaussard. I also thank the administrative and technical staff members of my labs, CNRS, universities and funding agencies, who make research possible in everyday life. I thank the staff of the {}``Centre de Calcul de l'Universit{\'e} de Bourgogne'' (CCUB) and of the {}``Institut du D{\'e}veloppement et des Ressources en Informatique Scientifique'' (IDRIS).  

I also thank the colleagues, mainly experimentalists, with whom I had the chance to interact, compare and improve my results and understanding: Francesca Ferlaino and her team, Raphael Lopes, Sylvain Nascimbene, Jean Dalibard and their team, Igor Ferrier-Barbut and his team, Jean-Fran{\c c}ois Cl{\'e}ment, Matias Velazquez, Richard Moncorg{\'e} and Yannick Guyot.

Finally, I express my special thanks to the young colleagues (PhD and post-docs) whom I was glad to supervise and guide, and to whom I wish a successful career, in particular Gohar Hovhannesyan, Charbel Karam and Hui Li.

\chapter*{Short curriculum vitae}  \addcontentsline{toc}{chapter}{Short curriculum vitae}\markboth{SHORT CURRICULUM VITAE}{ }

\vspace{18pt}

\noindent Web page:
\href{https://icb.cnrs.fr/equipe/maxence-lepers}
     {https://icb.cnrs.fr/equipe/maxence-lepers}

\noindent 46 peer-reviewed articles in international journals, 1 book chapter 

\noindent Guest editor with Ugo Ancarani and Federica Agostini of the special issue \textit{Quantum dynamics in molecular systems}, in the European Physical Journal Special Topics (2023), dedicated to the French research network {}``GDR Thems'' \cite{ancarani2023}.

\vspace{18pt}

\noindent \textbf{POSITIONS HELD} 

\vspace{9pt}

\begin{tabular}{ll}
  \textbf{2005-2009:} & 
    PhD at Laboratoire PhLAM, Université de Lille \\
    & \textit{Title (French): Dynamique d'atomes dans des potentiels optiques~:} \\
    & \qquad \qquad \qquad  
    \textit{du chaos quantique au chaos quasi-classique} \\
  \vspace{6pt}
    & Supervisors: Véronique Zehnlé and Jean-Claude Garreau; \\
  \vspace{6pt}
  \textbf{2009-2011:} & 
    Post-doc at Laboratoire Aimé Cotton (LAC), Université Paris-Saclay; \\
  \vspace{6pt}
  \textbf{2011-2017:} & 
    CNRS Junior Researcher (Chargé de Recherche), LAC; \\
  \textbf{2017-present:} & 
    CNRS Junior Researcher, Laboratoire Interdisciplinaire Carnot de \\
    & Bourgogne (ICB), Université Bourgogne Europe. \\
\end{tabular}

\vspace{18pt}

\noindent \textbf{Funded projects as a PI or co-PI}

\vspace{9pt}

\begin{tabular}{ll}
  \textbf{2015-2016:} & 
    Project {}``InterDy'' of DIM Nano-K (Région {\" I}le de France),  \\
  \vspace{6pt}
   & 1-year post-doctoral fellowship; \\
  \textbf{2018-2021:} & 
    Project {}``ThéCUP'' (Région Bourgogne Franche Comté), including \\
  \vspace{6pt}
    & purchase of nodes for the computing center of Université de Bourgogne Europe; \\
  \vspace{6pt}
  \textbf{2020-2024:} & 
    ANR JC-JC {}``NeoDip'' (French Research Agency), including a PhD fellowship; \\
  \textbf{2025-2027:} & 
    ANR {}``FewBoDyK'', including a 2.5-year post-doctoral fellowship. \\
\end{tabular}

\chapter*{Introduction}  \addcontentsline{toc}{chapter}{Introduction}\markboth{INTRODUCTION}{ }

Since the beginning of my scientific career, I have worked on the modeling of ultracold gases, but with two different points of view. During my PhD, I have studied the center-of-mass motion of isolated ultracold atoms and Bose-Einstein condensates, submitted to various types of laser-generated potentials, in the context of quantum chaos \cite{lepers2008a, lepers2008b, lepers2010a, lepers2011d}. Such potentials are often built from so-called optical lattices that consist in pairs of retro-reflected beams in one, two or three dimensions. For example in one dimension (1D), the potential exerted on the atoms' center of mass (COM) is equal to $V(x) = V_0 \sin^2 (2\pi x/\lambda)$ with $x$ the COM position and $\lambda$ the beam wavelength. The standing wave created by the retro-reflected beam induces a time-independent, space-periodic potential composed of wells with a depth of $V_0$ proportional to the laser intensity.

As an example, when the optical lattice is not turned on continuously, but as a train of very short pulses, one obtains the cold-atom version of the quantum kicked rotor \cite{santhanam2022}. The latter is a paradigmatic system of quantum chaos which can also be implemented on the rotational motion of molecules submitted to trains of laser pulses (but not standing waves) \cite{bitter2016}. The laser-induced potentials acting on ultracold atoms can therefore mimic potentials obtained with very different systems, which is the basic idea of quantum simulation \cite{gross2017}. In order to highlight this general and transverse feature of the systems under study, one often uses scaled units of distances and energies to perform numerical simulations and interpret their results.

In cold-atom physics, this simple description of the atom-field interaction is enabled by the simple structure of the widely used alkali-metal atoms, made up of a single electron orbiting a closed-shell core. It allows in particular for applying the two-level and rotating-wave approximations to model the atom-field interactions. However, in ultracold gases, there is a long-time trend consisting in investigating more complex -- or richer -- systems like molecules or many-electron atoms, that offer additional possibilities of control. They also open the door to new realms of phenomena like collisions or chemical reactions in the quantum regime. But in turn, they require a detailed knowledge of their structure and interactions. Studying ultracold gases composed of particles with a complex structure has been at the heart of my research activities since my post-doctoral stay at Laboratoire Aimé Cotton (LAC).

In this general context, I have followed two main directions. Firstly, I have studied the atomic structure and spectroscopy of lanthanide elements, and their implications on laser-cooling and trapping. This will be the scope of the first part of my manuscript. As an illustration, for lanthanides, the depth $V_0$ of the aforementioned optical-lattice potential depends on the laser wavelength, but also on its polarization. This opens the possibility to nullify the potential for a given atomic states, or equate the potentials felt by two different states, yielding so-called magic trapping. The light polarization that allows for reaching those peculiar situations depend on the atomic spectrum in a rather involved manner. Understanding that dependence can be achieved by measuring or calculating the dynamic dipole polarizability of the atomic levels at the considered wavelength. In Chapter \ref{chap:ddp}, I will present such calculations for different lanthanide neutrals, in particular erbium and dysprosium relevant for ultracold experiments. Those polarizabilities depend on atomic energies and transition intensities, calculated using the semi-empirical method of Robert Cowan's suite of codes and extended by us. Those atomic data can also serve to determine the laser-cooling feasibility for yet unexplored atoms like neodymium, which will be done in Chapter \ref{chap:atoStr}. Those chapters contain a selection of my publications that I comment and enrich with additional figures or tables.

The second main direction, treated in the second part of this manuscript, deals with long-range interactions between atoms and/or molecules. In the ultracold regime, since the kinetic energy of the particles are below 1~mK, their relative motion is strongly sensitive to small variations of the interactions energies between them. Such variations take place where the particles are far away from each other, namely at long range, beyond the region of chemical bonding. The most famous example of those long-range interactions is certainly the van der Walls or dispersion forces between pairs of ground-state molecules. In ultracold matter, various collisional phenomena are determined by long-range interactions, for example photoassociation \cite{jones2006}. Moreover, the physics of dipolar gases stems from the magnetic and/or electric dipole moment of their constituents and from the resulting dipole-dipole interaction (DDI). In that second part, I will discuss several examples of long-range interactions involving atoms and molecules. The aim of this almost chronological presentation is to highlight the intellectual progression which led to a more elaborate account for multipolar terms, atomic fine and hyperfine structure, as well as molecular rotational structure, external electromagnetic fields, and symmetries of the collisional complex. I also describe the crucial importance of the referential frame in which the interaction energy is calculated, either the frame of the complex in Chapter \ref{chap:lriBF}, or the frame of the laboratory in Chapter \ref{chap:lriSF}. As for Chapter \ref{chap:lriPres}, it recalls the essential features and equations of long-range interactions, complementing the book chapter that I wrote with Olivier Dulieu a few years ago \cite{lepers2018}.

Up to now, I have solely mentioned the context of ultracold gases. However the atomic-structure calculations of lanthanides discussed in the first part are also employed to characterize the luminescent properties of solids doped with trivalent lanthanide ions. Such systems are crucial in various domains of current technologies, and characterizing their radiative transition intensities is of prime importance. Such calculations rely on an accurate modeling of the spectrum of the free lanthanide ions, since the latter are only slightly perturbed by their neighboring ligands in the solid. This is the purpose of Chapter \ref{chap:ln3+}. Moreover, in Chapter \ref{chap:lriBF}, I present long range potential energy curves between an oxygen atom and an oxygen molecule, which are relevant for the formation of atmospheric ozone.

\part{Spectral properties of lanthanide atoms and ions}

\chapter{Atomic-structure calculations of lanthanide elements}
\label{chap:atoStr}

In the first part of this manuscript, I present results based on atomic-structure calculations of various properties of lanthanide atoms and ions, such as energy levels or transition intensities. Those quantities are relevant in various fields of research: astrophysics of chemically-peculiar stars or neutron-star mergers \cite{cowley1978, kasen2013}, lasers and optical fibers with doped materials \cite{walsh2006, hehlen2013}, as well as ultracold dipolar gases \cite{norcia2021, chomaz2022}. Here the two first chapters are dedicated to laser-cooling and trapping of neutral lanthanides, and the third one is dedicated to the luminescent properties of solids doped with trivalent lanthanide ions.

Our atomic-structure calculations are performed with the semi-empirical method, combining \textit{ab initio} and least-squares fitting calculations of energy levels and transition probabilities, a method that I learnt from Jean-Fran{\c c}ois Wyart who was one of its internationally recognized experts. The \textit{ab initio} and energy-fitting steps are implemented in Cowan's suite of codes \cite{cowan1981, kramida2019}, while the fitting of transition probabilities is carried out with our home-made code {}``FitAik'', designed to work in interface with Cowan's codes \cite{lepers2023, fitaik-gitlab}.

In Section \ref{sec:atostrUCold}, I describe the principles of those calculations, as well as their motivations in the field of ultracold gases composed of lanthanide neutral atoms. I put the stress on the necessity of having at our disposal an extensive set of reliable atomic data, in order to determine the feasibility of laser-cooling and trapping. I apply those ideas in Sections \ref{sec:atostrEr+} and \ref{sec:atostrNd}. In the former one, I present our article describing the fitting method of transition probabilities and its application to the Er$^+$ ion. The latter section is dedicated to the feasibility of laser-cooling of the yet unexplored neodymium atom.

\section{Atomic-structure calculation and ultracold gases}
\label{sec:atostrUCold}

\subsection{Motivation of the calculations}

Historically, the first laser-cooling and trapping experiments dealt with alkali-metal atoms like rubidium (Rb) or cesium (Cs) which presented many advantages: a simple electronic structure with a single electron surrounding a closed-shell core, broad visible and near-infrared transitions easily accessible by laser \cite{cohen-tannoudji1998, phillips1998, metcalf2003}. Later, alkaline-earth and related atoms containing two valence electrons, such as strontium (Sr) or ytterbium (Yb), were also involved in ultracold experiments \cite{kurosu1992}. They also present broad transitions in the visible range, but also narrow intercombination ones \cite{loftus2004}, especially suitable for optical clocks \cite{letargat2013}. Unlike alkali-metals, they possess stable isotopes without hyperfine structure. A similar evolution was at play for cold-ion experiments, which first dealt with single-valence-electron species like mercury (Hg$^+$), and then with more complex ones \cite{eschner2003}.

During the 2000's decade, paramagnetic atoms with more a complex electronic structure, were also cooled down to ultralow temperatures. The first of these achievements was obtained with chromium \cite{griesmaier2005, chicireanu2006}, a transition metal of atomic number $Z=24$. In 2006 at NIST, the first demonstration of magneto-optical trapping without repumping of the lanthanide (Ln) atom erbium (Er) was really surprising \cite{mcclelland2006}, since one could expect such a complex atom to prevent the existence of closed or quasi-closed absorption-emission cycles, required for laser-cooling. This pioneering work triggered many experiments around the world not only with erbium \cite{aikawa2012, frisch2012, aikawa2014, phelps2020}, but also with other Ln elements like dysprosium (Dy) \cite{lu2010, lu2011, lu2012}, thulium \cite{sukachev2010, davletov2020}, holmium \cite{miao2014, hemmerling2014}, europium \cite{inoue2018, miyazawa2021, miyazawa2022}, and even Er-Dy mixtures \cite{trautmann2018}.

The dense energy spectrum of Cr or Ln atoms is \textit{a priori} unfavorable for laser-cooling, since it is expected to comprise many lossy transitions from cooling cycles. But because this rich structure is due to several unpaired d or f electrons, it also offers great advantages, like a strong magnetic moment ideal for dipolar quantum gases \cite{norcia2021, chomaz2022}, or a wide variety of transition line widths, from Hz to MHz domains suitable for many application in quantum sciences and technologies \cite{chalopin2018a, golovizin2019, golovizin2021, satoor2021, patscheider2021}. Dipolar gases composed of paramagnetic Ln atoms constitute a suitable platform for quantum simulation, allowing for the observation of supersolidity \cite{chomaz2018, natale2019}, Fermi-surface deformation \cite{aikawa2014a}, extended Bose-Hubbard models \cite{baier2016}, quantum magnetism \cite{lepoutre2019}, quantum droplets \cite{chomaz2016, kadau2016}, synthetic gauge fields and topological matter \cite{burdick2016, fabre2024}, or dipolar solids \cite{su2023}.

\begin{figure}
  \begin{center}
  \includegraphics[width=0.7\linewidth]
    {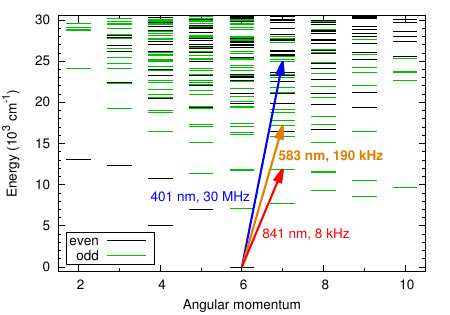}
  \caption{Experimental energy levels of neutral erbium (Er) plotted as functions of the electronic angular momentum and sorted by parity. The arrows represent the main laser-cooling transitions with their wavelengths and linewidths.
  \label{fig:atostrErLev}}
  \end{center}
\end{figure}

Establishing the feasibility of laser-cooling for a given element thus requires a good knowledge of its spectrum, including energy levels and transition strengths. In this respect, a few months before the first article on erbium laser-cooling \cite{mcclelland2006}, a groundwork article was published by the same group \cite{ban2005}, containing characteristics of the possible laser-cooling transitions, some of which were calculated with the Cowan codes. Figure \ref{fig:atostrErLev} presents three of these transitions, drawn as arrows on the energy diagram of Er. The energy levels are plotted as functions of the electronic angular momentum $J$ and sorted by parity.

The ground level of Er belongs to the electronic configuration [Xe]4f$^{12}$6s$^2$, where [Xe] denotes the ground configuration of xenon, omitted in what follows. Its parity is even and its term is $^3$H$_6$, \textit{i.e.}~the orbital, spin and total angular-momentum quantum numbers are respectively $L=5$, $S=1$, $J=6$. Because of electric-dipole (ED or E1) selection rules, the upper level of the cooling transition must be odd, and with electronic angular momentum $J=5$--7. Since all the even energy levels below 15000~\cmi{} have $J<6$, taking an upper level with $J=7$ suppresses the risk of leakage toward those low-lying even levels. The transition at 841~nm is thus totally closed, meaning that from this level the atoms have 100~\% probability to decay back to the ground level by spontaneous emission. However, owing to its narrowness, this transition can only be used as a secondary cooling transition \cite{berglund2008}.

By contrast, the broad 401-nm transition is inescapable. Its configuration and term is 4f$^{12}(^3$H$_6)$6s6p$(^1$P$_1^\circ) \, (6,1)_7^\circ$, corresponding to the following coupling scheme:
\begin{equation}
  \left. \begin{array}{rrl}
    \mathrm{4f}^{12}:& \mathbf{L}_c + \mathbf{S}_c &= \mathbf{J}_c
  \\ \mathrm{6s6p}  :& \mathbf{L}_v + \mathbf{S}_v &= \mathbf{J}_v
  \end{array} \right\} \; 
  \mathbf{J}_c + \mathbf{J}_v = \mathbf{J} \,.
\end{equation}
To the 4f$^{12}$ core subshell are associated the quantum numbers $L_c=5$, $S_c=1$ and $J_c=6$, identical to the ground level. the valence orbitals 6s and 6p are coupled to give the quantum numbers $L_v=1$, $S_v=0$ and $J_v=1$ with odd parity. The angular momenta of the core and valence orbitals are coupled to give the total angular momentum $J=7$. In this {}``blue'' transition, the 4f electrons look like spectators, which explains why there exist similar transitions with nearby wavelengths in neighboring Ln atoms, in particular Yb where the 4f subshell is filled. Figure \ref{fig:atostrErLev} would tend to indicate many possible leakages toward even levels with $J=6$--8, which is actually not the case due to two main reasons:
\begin{enumerate}
  \item The corresponding transitions have transition energies below 10000~\cmi{}, while the one to the ground level is around 25000~\cmi{}. The cubic energy dependence of the Einstein coefficient for spontaneous emission thus disadvantages the former.
  \item The configuration and term to which those even levels belong, namely 4f$^{11}$6s$^2$6p and 4f$^{12}(^3$H$_6)$5d6s$(^3$D$)$ do not favor strong transitions due to E1 selection rules.
\end{enumerate}
In consequence, the blue cooling cycle do not give rise to significant losses: it is used as the first step of cooling. Indeed, the lowest attainable temperature, called the Doppler limit, is proportional to the transition line width, which has the largest value for 421~nm, see Fig.~\ref{fig:atostrErLev}.

The upper level of the orange transition can be labeled as 4f$^{12}(^3$H$_6)$6s6p$(^3$P$_1^\circ) \, (6,1)_7^\circ$. The only difference with the blue one is the triplet nature ($S_v=1$) of the valence subshells. The 583-nm transition can thus be viewed as an intercombination transition, as observed in alkaline earths and ytterbium. Due to the same argument as above, the leakages from that cooling cycle are negligible, and it is often used as a second step of cooling, in order to reach temperatures in the microkelvin ($\mu$K) range. Let us mention finally the lowest $J=7$ odd level at 7696.956~\cmi{} (1299~nm), which has a lifetime of 111~ms (line width of 2.1~Hz). The transition is too narrow for laser-cooling, but the metastable nature of that level, as well as its location in the telecom band, is interesting for various applications \cite{patscheider2021}.

Laser-cooling of other Ln atoms works in a similar manner. For instance, the ground level of Dy is 4f$^{10}$6s$^2\,{}^5$I$_8$. In a first step, it is cooled using the broad transition toward the odd level 4f$^{10}(^5$I$_8)$6s6p$(^1$P$_1^\circ) \, (8,1)_9^\circ$ at 23736.610~\cmi{} (421~nm), and then using the narrow transition toward the level 4f$^{10}(^5$I$_8)$6s6p$(^3$P$_1^\circ) \, (8,1)_9^\circ$ at 15972.35~\cmi{} (626~nm). There also exists a long-lived $J=9$ odd-parity level at 9990.974~\cmi{} \cite{petersen2020}.

Apart from cooling transitions, it is also crucial to determine to which extent the atoms can be held in laser beams via optical trapping \cite{grimm2000}. At the atomic scale, this is determined by the dynamic dipole polarizability (DDP), see Chapter \ref{chap:ddp}, which depends on the laser wavelength and polarization, and on the atomic level. It is calculated with a sum-over-state formula stemming from second-order perturbation theory. Therefore, it requires the knowledge of transition energies and transition dipole moments towards a large number of excited levels. In Ln atoms, many of those quantities have been measured in spectroscopic laboratories like the one of Jim Lawler and Elizabeth Den Hartog at the University of Wisconsin, see for instance Refs.~\cite{denhartog1999, stockett2007, lawler2008, wyart2009, stockett2011} among many others. But in view of the rich spectrum of Ln atoms, there are significantly more transitions that come into play in the DDP calculations, hence the need to compute transition energies and dipole moments.

\subsection{The semi-empirical method}

To do so, we use throughout this manuscript the semi-empirical or Racah-Slater method implemented in Robert Cowan's suite of codes \cite{cowan1981}, either the version by Cornac McGuiness at Trinity College Dublin \cite{mcguinness-cowan}, or the one by Alexander Kramida at NIST \cite{kramida2019}. This method has been successfully used with various Ln atoms and ions by Jean-Fran{\c c}ois Wyart at Laboratoire Aim{\' e} Cotton \cite{wyart2011}. Here we summarize the different steps of the method, taking the example of neutral erbium.

\paragraph{HFR method and parameters' calculation \: --}

First of all, in the RCN code, one chooses the electronic configurations that one wants to characterize: for example 4f$^{12}$6s$^2$ + 4f$^{12}$5d6s for the even parity, and 4f$^{11}$5d6s$^2$ + 4f$^{12}$6s6p for the odd parity of Er. For each subshell of those configurations, RCN calculates the one-electron radial wave functions $P_{n\ell}(r)$ using the Hartree-Fock + relativistic (HFR) method, which assumes that each electron is submitted to the mean and central field induced by the others, to which one-electron relativistic corrections are added. The trial wave function of the $N$-electron atom is taken as a Slater determinant, an antisymmetrized product of $P_{n\ell}(r)$ wave functions which satisfies the Pauli principle. In addition to the $P_{n\ell}(r)$ wave functions, this self-consistent field (SCF) calculation results in the center-of-gravity energy $E_\mathrm{av}$ of each configuration.

Using the one-electron wave functions, the RCN2 code aims to calculate radial parameters which are the building blocks of the atomic Hamiltonian (see below): spin-orbit integrals $\zeta_{n\ell}$ for non-s electrons, and Coulombic integrals describing the electron-electron repulsion. To that end, the inverse distance $1/r_{ij} = 1/|\mathbf{r}_i - \mathbf{r}_j|$ between two electrons is expanded as
\begin{equation}
  \frac{1}{r_{ij}} = \sum_k \frac{r_<^k}{r_>^{k+1}}
    \left( \mathrm{C}_k (\theta_i,\phi_i)
    \cdot  \mathrm{C}_k (\theta_j,\phi_j) \right)
  \label{eq:atostr1oRij}
\end{equation}
where $r_<$ ($r_>$) is the smaller (larger) distance among $r_i$ and $r_j$, $(r_{i,j}, \theta_{i,j}, \phi_{i,j})$ are the spherical coordinates of electrons $i,j$ with respect to the nucleus, $\mathrm{C}_{kq}$ are Racah spherical harmonics, related to the usual ones by $\mathrm{C}_{kq} = \sqrt{4\pi / 2k+1} \times \mathrm{Y}_{kq}$, and $(\cdot)$ their scalar product,
\begin{equation}
  \left( \mathrm{C}_k (\theta_i,\phi_i)
    \cdot  \mathrm{C}_k (\theta_j,\phi_j) \right)
    = \sum_{q=-k}^{+k} (-1)^q \mathrm{C}_{k,-q} (\theta_i,\phi_i)
    \mathrm{C}_{kq} (\theta_j,\phi_j) \,.
\end{equation}
Their matrix elements in the one-electron basis $\{| n\ell m_\ell m_s \rangle \}$ are equal to
\begin{equation}
  \langle n\ell m_\ell m_s | \mathrm{C}_{kq} (\theta,\phi) 
    | n'\ell' m'_\ell m'_s \rangle
    = \delta_{\qnpr{m_s}m'_s} \sqrt{\frac{2\ell'+1}{2\ell+1}}
    C_{\ell' 0k0}^{\ell 0} C_{\ell' m'_\ell kq}^{\ell m_\ell} \,,
\end{equation}
where $C_{a\alpha b\beta}^{c\gamma} = \langle a\alpha b\beta | abc\gamma \rangle$ are Clebsch-Gordan (CG) coefficients in the notation of Varshalovitch \cite{varshalovich1988}.
They impose strong restrictions on the possible values of $k$, when calculating the matrix elements $\langle tu| r^{-1}_{ij} | t'u' \rangle$ between pair states of electrons.

For equivalent electrons, \textit{i.e.}~belonging to the same subshell $t=t'=u=u'=(n\ell)$, one has $k=0, 2, \cdots, 2\ell$, and the radial integral is 
\begin{equation}
  \int_0^{+\infty} dr_i \int_0^{+\infty} dr_j
    \frac{r_<^k}{r_>^{k+1}}
    [P_{n\ell}(r_i) P_{n\ell}(r_j)]^2
    \equiv F^k(n\ell,n\ell) .
\end{equation}
For 4f equivalent electrons, one has the parameters $F^0$, $F^2$, $F^4$ and $F^6$. Non-equivalent electrons in the same configuration give rise to the direct $F^k(n_1\ell_1,n_2\ell_2)$ and exchange $G^k(n_1\ell_1,n_2\ell_2)$ integrals, respectively corresponding to $t=t'=(n_1\ell_1)$, $u=u'=(n_2\ell_2)$ and $t=u'=(n_1\ell_1)$, $u=t'=(n_2\ell_2)$. The conditions on $k$ are $0 \le k \le \min(2\ell_1, 2\ell_2)$ for $F^k$ and $|\ell_1 - \ell_2| \le k \le \ell_1 + \ell_2$ for $G^k$, by steps of 2 in both cases. For instance in the 4f$^{11}$5d6s$^2$ configuration, the relevant radial parameters are $F^{2,4}($4f,5d$)$ and $G^{1,3,5}($4f,5d$)$, in addition to $F^{0,2,4,6}($4f,4f$)$.

When the bra and the ket belong to different configurations, the radial configuration-interaction (CI) parameters are denoted $R^k (n_1\ell_1 n_2\ell_2, n'_1\ell'_1 n'_2\ell'_2)$, with $|\ell_1 - \ell'_1| \le k \le \ell_1 + \ell'_1$, $|\ell_2 - \ell'_2| \le k \le \ell_2 + \ell'_2$ and steps of 2. All the other subshells of the configurations must be identical in the bra and in the ket. For instance, in the pair 4f$^{11}$5d6s$^2$ + 4f$^{12}$6s6p, since eleven 4f and one 6s electrons are present in both configurations, the CI parameters involve one 4f, one 6s, the 5d and the 6p electrons. Namely, the relevant parameters are $R^{1}($5d6s,4f6p$)$ and $R^{3}($5d6s,6p4f$)$,

\paragraph{Setting up and diagonalizing the Hamiltonian \: --}

Once all the radial parameters $E_\mathrm{av}$, $\zeta$, $F^k$, $G^k$, $R^k$ are computed, the full atomic Hamiltonian is built and diagonalized by the program RCG. In this purpose, for each total electronic angular momentum and both parities, the program builds the coupled angular-momentum basis sets (sometimes called configuration state functions, CSFs \cite{froese-fischer2019}) in, say the Russel-Sanders (LS) coupling scheme. In the 4f$^{12}$6s$^2$, the possible LS terms are $^1$S$_0$, $^1$D$_2$, $^1$G$_4$, $^1$I$_6$, $^3$P$_{0,1,2}$, $^3$F$_{2,3,4}$, $^3$H$_{4,5,6}$. Therefore, the matrix elements of the atomic Hamiltonian can be written as the linear combination $H_{ij} = \sum_p A_{ij,p} \, X_{p}$, where $X_{p}$ are the radial (Slater) parameters discussed above, and $A_{ij,p}$ are angular coefficients that are calculated exactly using Racah algebra, namely Wigner 3-j, 6-j, 9-j symbols and coefficients of fractional parentage (CFPs), appearing in some well-suited operators, see Ref.~\cite{cowan1981}, Ch.~11. The diagonalization yields eigenvalues that are the level energies and eigenvectors that allow for labeling the levels and for calculating various properties like Landé g-factors, transition line strengths $S_{ik}$, oscillator strengths $f_{ik}$ and transition probabilities of spontaneous emission $A_{ik}$.

\paragraph{Least-squares fitting of energies \: --}

At this point which is purely \textit{ab initio}, the computed energies have a limited accuracy, which stems from the HFR approximation following which the radial parameters are calculated. To improve the accuracy of calculated energies compared to experimental ones, the RCE program enables to adjust the radial parameters using a least-squares fitting procedure between calculated and experimental energies. The combination of \textit{ab initio} and fitting methods justifies the adjective {}``semi-empirical''. Moreover, to further increase the precision of the results, the program offers the possibility to add some {}``effective parameters'', that is to say radial quantities designed to account for CI mixing with configurations absent from the calculation. Such parameters cannot be calculated \textit{ab initio}, and their initial values are taken from similar spectra. The quality of the fit is determined by the root-mean-square deviation (or standard deviation in RCE)
\begin{equation}
  \sigma = \left[ \frac{1}{N_\mathrm{lev} - N_\mathrm{par}}
    \sum_{i=1}^{N_\mathrm{lev}} \left( E_{\mathrm{cal},i}
     - E_{\mathrm{exp},i} \right)^2 \right]^{1/2} 
\end{equation}
where $E_{\mathrm{cal/exp},i}$ are calculated/experimental energies, $N_\mathrm{lev}$ is the number of experimental levels included in the fit, and $N_\mathrm{par}$ the number of free-parameter groups. Indeed, RCE offers the possibility to fix some parameters and to constrain some groups to vary with the same ratio between their initial and final values. In this respect, one often defines the ratio or scaling factors between the final (fitted) parameter value and the initial (HFR) one, $f_X = X_\mathrm{fitted} / X_\mathrm{HFR}$. Those ratios usually range between 0.6 and 1.2, and a given parameter, \textit{e.g.}~$F^2($4f,4f$)$ presents similarities along the Ln series.

\section{Least-squares fitting of transition probabilities}
\label{sec:atostrEr+}

\begin{figure}[h!]
  \begin{center}
  \includegraphics[width=0.6\linewidth]
    {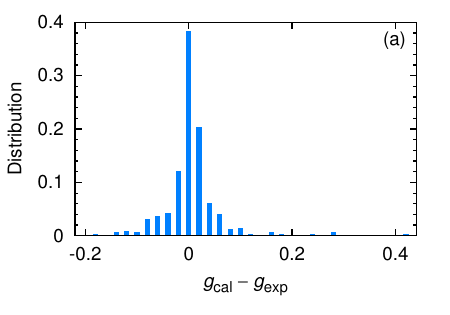}
  \caption{Er$^+$ : normalized level distribution as functions as the difference between calculated and experimental Landé g-factors. Levels of both parities are mixed.
  \label{fig:atostrEr+DistrLande}}
  \end{center}
\end{figure}

When the set of fitted parameters is obtained by RCE, a last run of RCG can be made, in order to obtain final eigenvectors, Landé g-factors and transition probabilities. One thus expects the fitting of energies to improve the accuracy on those properties. Even if the agreement on Landé factors is often very satisfactory as Figure \ref{fig:atostrEr+DistrLande} shows, significant discrepancies, on the order of 20-30~\%, can remain on transition probabilities or radiative lifetimes. Indeed, those quantities depend on one-electron transition integrals, see Eqs.~(1)--(4) of Ref.~\cite{lepers2016},
\begin{equation}
  r_{n\ell,n'\ell'} = \langle n\ell |r| n'\ell' \rangle 
    = \int_0^{+\infty} dr P_{n\ell}(r) r P_{n'\ell'}(r)
  \label{eq:atostrRIntgr}
\end{equation}
which, alongside Slater parameters, are calculated by the Cowan code RCN2 using HFR wave functions. Consequently, the transition integrals induce similar inaccuracies to those created on energies by Slater parameters. So it seems sensible to adjust the $r_{n\ell,n'\ell'}$ quantities by least-squares fitting of experimental and calculated transition probabilities. It requires to have at our disposal extensive sets of experimental $A_{ik}$ coefficients, which is the case for many Ln elements.

Together with Jean-Fran{\c c}ois Wyart and Olivier Dulieu, I have written a program called {}``FitAik'' that performs this least-squares fitting procedure on transition probabilities. This code works in interface with Cowan's code RCG, whose input file contains the one-electron transition integrals \eqref{eq:atostrRIntgr} in a human-readable (and writable) form. Our methodology is described in Ref.~\cite{lepers2023}, hereafter denoted as Paper I. Even if it was published in 2023, we had used that methodology for several years, in our articles dealing with Ln atom DDPs \cite{li2016, li2017, becher2018, chalopin2018}, and also in the one dedicated to the laser-cooling of the Er$^+$ ion \cite{lepers2016}.

Paper I, which illustrates the methodology on Er$^+$, appears as the continuation of Ref.~\cite{lepers2016}. At the time, the choice of Er$^+$ was motivated by the exploration of a new family of systems to laser-cool -- the Ln$^+$ ions -- whose advantages are described in the introduction of Ref.~\cite{lepers2016}, by the detailed interpretation on energy levels by Wyart and Lawler who provided fitted energy parameters a few years before \cite{wyart2009}, and by the existence of an extensive set of 418 experimental transition probabilities from Lawler's group \cite{stockett2007, lawler2008}. Ln$^+$ ions turn out to be a good playground for those calculations, since they present many transitions in the optical region, unlike more charged ions, and the CI effect, although sizable, is less pronounced than in neutrals.

For both parities, the electronic configurations included in the modeling, number of experimental levels and free-parameter groups, and standard deviation are \cite{wyart2009}
\begin{itemize}
  \item even parity: 4f$^{12}$6s, 4f$^{12}$5d, 4f$^{11}$6s6p, 4f$^{11}$5d6p; $N_\mathrm{lev} = 130$; $N_\mathrm{par} = 25$; $\sigma = 55$~\cmi{};
  \item odd parity: 4f$^{11}$6s$^2$, 4f$^{11}$5d6s, 4f$^{11}$5d$^2$, 4f$^{12}$6p, 4f$^{13}$; $N_\mathrm{lev} = 233$; $N_\mathrm{par} = 21$; $\sigma = 63$~\cmi{}.
\end{itemize}
The 4f$^{13}$ configuration is included for technical purpose, but no experimental levels were observed in it. The results of the fits are very good: since the experimental levels cover an energy range of approximately 45000~\cmi{}, the standard deviations represent at most 0.14~\% of that range. A good test of the eigenvectors' quality is made by comparing experimental and calculated Landé g-factors. On Figure \ref{fig:atostrEr+DistrLande}, the normalized distribution of the differences between experimental and calculated ones is plotted as histograms of width 0.02, showing a very good agreement. It indicates that 38~\% of the Landé factors have differences between -0.01 and 0.01. As for the standard deviation, it is equal to 0.055.

The considered configurations give rise to 10 $r_{n\ell,n'\ell'}$ transition integrals: 3 with $(n\ell,n'\ell') = $ (6s,6p), 3 with (6p,5d) and 4 with (5d,4f). Their HFR values are given in Table 2 of Paper I. During the fit, each 6s-6p integral evolves freely, while 6p-5d integrals on the one hand, and 5d-4f integrals on the other hand are constrained to evolve with an identical scaling factor. In analogy with energy parameters, the latter can be defined as the ratio of fitted $r_{n\ell,n'\ell'}$ integral over its HFR value, see Eq.~(7) of Paper I.

\includepdf[pages=-]{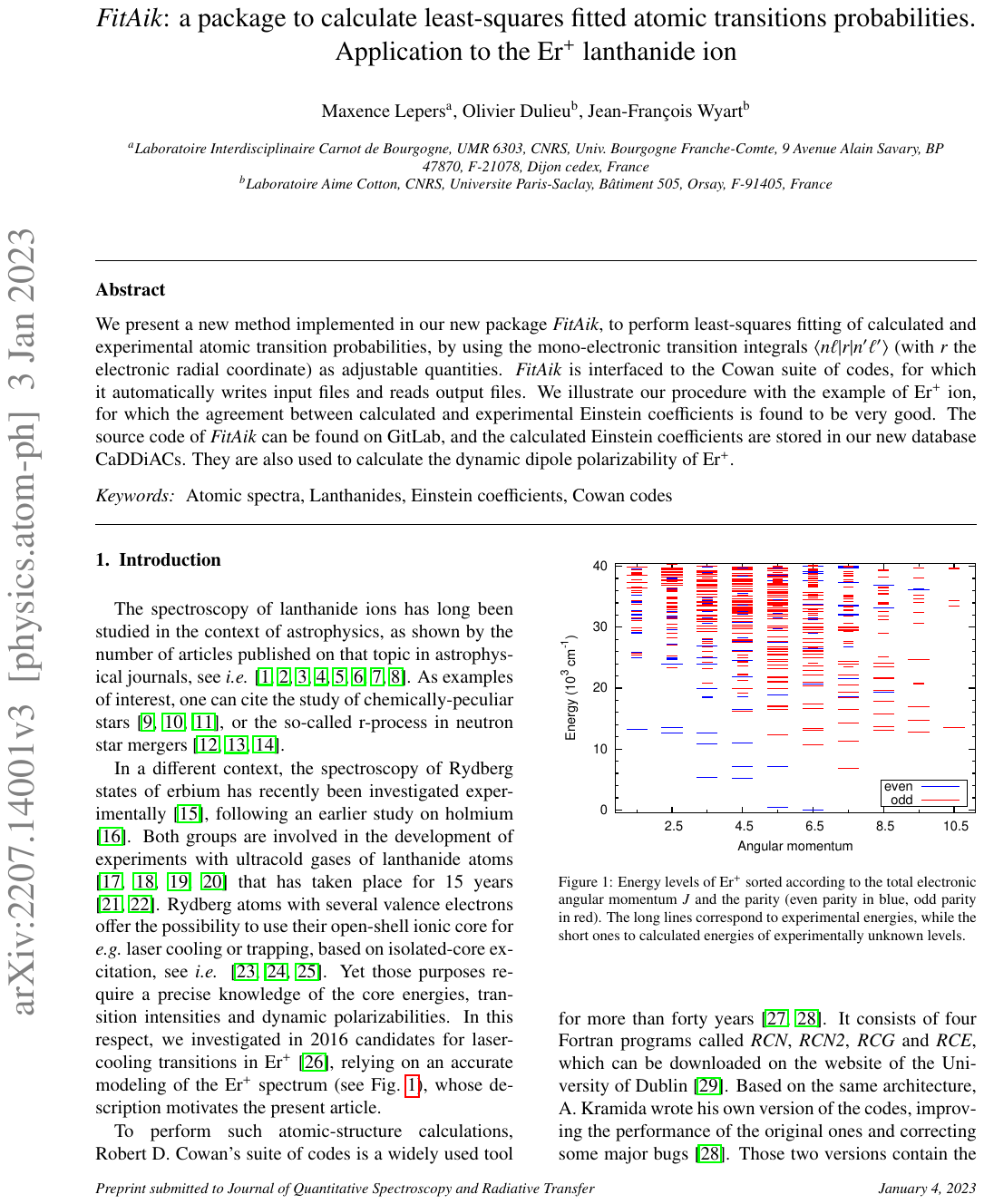}

\begin{figure}
  \begin{center}
  \includegraphics[width=0.6\linewidth]
    {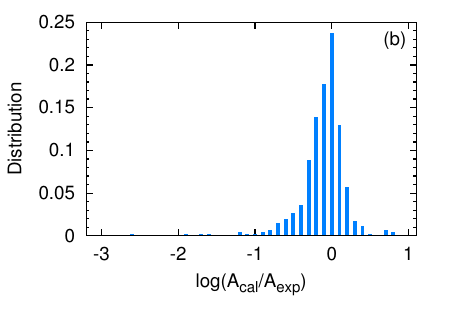}
  \caption{Er$^+$ : normalized distributions of transitions as functions of the 10-base logarithmic ratio between calculated and experimental transition probabilities.
  \label{fig:atostrEr+DistrAik}}
  \end{center}
\end{figure}

The results of the fitting procedure is satisfactory. The best standard deviation on Einstein coefficients, see Eq.~(5) of Paper I, is equal to $\sigma_A = 4.57 \times 10^6$~s$^{-1}$, representing 2.2~\% of the largest experimental $A_{ik}$ coefficient. The scaling factors on $r_{n\ell,n'\ell'}$'s range from 0.79 to 0.89. This result was obtained after excluding 22 transitions from the fit, those for which the ratio $A_\mathrm{cal}/A_\mathrm{exp}$ is smaller than 0.2 or larger than 5. Removing those outliers has a strong influence on the logarithmic standard deviation $\sigma_{\log A}$ given in Eq.~(6) of Paper I, which decreases from 0.52 to 0.22. The distribution of transitions as a function of $\log(A_\mathrm{cal}/A_\mathrm{exp})$ is presented on Fig.~\ref{fig:atostrEr+DistrAik}, as histograms of width 0.1 and including some outliers. The distribution is rather sharply centered around 0 (corresponding to $A_\mathrm{cal} = A_\mathrm{exp}$) and is assymetric toward the small ratios, meaning that there are more underestimated transition probabilities than overestimated ones.

The idea of modifying the Cowan codes to improve the accuracy of transition probabilities has already been applied. P.~Quinet and coworkers modified the RCN code in order to account for core-polarization effects on the one-electron wave functions. This resulted in a significant improvement of radiative lifetimes and oscillator strengths \cite{quinet1999, quinet2002}. As in the present work, very few transitions with large discrepancies are observed. Following a similar semi-empirical methodology to ours, J.~Ruchkowski and coworkers developed their own numerical code to fit energies, hyperfine constants, and oscillator strengths, which they applied \textit{e.g.}~to the scandium ion Sc$^+$ \cite{ruczkowski2014} and Sr atom \cite{ruczkowski2016}.

The FitAik package can be downloaded on GitLab \cite{fitaik-gitlab}, and the E1 and M1 $A_{ik}$ coefficients of Er$^+$ computed with it can be found on our home-made database CaDDiAcS \cite{caddiacs}. As a prospect, I would like to benchmark FitAik with other atoms and ions, especially lanthanides. Indeed, the least-squares fitting seems appropriate for them, since a large number of transitions depend on a rather small number of transition integrals. However, generating a list of transition intensities worthwile to publish in a database turns out to be a hard task for neutrals (see next section). In the database, I also plan to add an indication about the accuracy of the computed transition probabilities, similar to the letters {}``A+'', {}``A'', {}``B+'', etc. given in the NIST database \cite{NIST-ASD}. Following Ref.~\cite{kramida2013}, we see on Fig.~2 of Paper I that transitions with the smallest $A_\mathrm{cal}/A_\mathrm{exp}$ have very small calculated line strengths $S_\mathrm{cal} < 10^{-2}$. Conversely, transitions with the largest $S_\mathrm{cal}$ are the most precise ones. The quantity $S_\mathrm{cal}$ could therefore serve to evaluate a confidence interval.

The computed $A_{ik}$ coefficients were used in Ref.~\cite{lepers2016} to predict possible laser-cooling transition of Er$^+$, similarly to \cite{ban2005}. Our extensive set has also allowed to predict the branching ratios toward the leaking levels. And because the latter can spontaneously decay to even lower levels, and so on, we have highlighted a recycling phenomenon \cite{mcclelland2006} in which part of the ions end up in the ground levels after a cascade of spontaneous emissions. Furthermore, we used our $A_{ik}$ coefficients to calculate the DDP of Er$^+$ in various levels. These quantities are relevant for experimentalists studying ultracold Rydberg atoms of erbium \cite{trautmann2021} and holmium \cite{hostetter2015}, whose optical trapping depends on the polarizability of their ionic core.

\section{Laser-cooling of neodymium atoms}
\label{sec:atostrNd}

Following the previous section, the motivation of the study  presented in the present one was also to propose a laser-cooling scheme to a yet unexplored neutral atom: neodymium (Nd). The result was obtained by Gohar Hovhannesyan and published in Ref.~\cite{hovhannesyan2023a}, hereafter denoted as Paper II. Up to now, laser-cooled open-shell Ln atoms all belong to the right part of the series, from Eu to Tm. This can be explained as the spectrum of these atoms contain a few strong transitions among a forest of weak ones. Those strong transitions are especially well-suited for laser-cooling or Zeeman slowing. This distinction between weak and strong transitions is not so pronounced for the atoms of the left part of the row, from lanthanum to samarium, but still there exist some strong transitions, that are potential candidates for laser-cooling \cite{stockett2011}.

Investigating the laser-cooling and trapping feasibility for atoms of the left part of the Ln row was the main objective of my ANR JCJC project {}``NeoDip'' supported by the French Research Agency \cite{anr-neodip}. Among all atoms, I have identified Nd as the most promising candidate, since (i) it possesses, like Dy and Er, several stable bosonic and fermionic isotopes, the bosonic ones being free of hyperfine structure; (ii) it possesses pairs of nearby opposite-parity energy levels that could be coupled by an external AC electric field to induce an electric dipole moment, in addition to the naturally present magnetic moment. These ideas are discussed in details for Dy in Section \ref{sec:atDblPol}, where we see that the radiative lifetime of one level, in the $\mu$s-range, limits our proposed scheme. Because in Nd the candidate levels are lower in energy (around 11000~\cmi{} and do not have the same spin multiplicity as most of the levels lower in energy, they are likely to have a larger radiative lifetime.

The energy diagrams of Nd and Dy are plotted on Figure 1 of Paper II. The spectrum of Nd is denser than Dy especially between 10000 and 20000~\cmi{}. Because the ground level of Nd is 4f$^4$6s$^2 \, {}^5$I$_4$ (whereas it is 4f$^{10}$6s$^2 \, {}^5$I$_8$ for Dy), the Nd laser-cooling transitions should preferentially imply odd upper levels with $J=3$, so as to prevent decay to other levels of the lowest manifold $^5$I. As the figure shows, there are several potential candidates below 20000~\cmi{}, and many others between 20000 and 25000~\cmi{}, as well as many potential leaking transitions. To determine the most suitable transition, we aimed to compute an extensive set of energies and transition probabilities using our semi-empirical approach, in the same spirit as Paper I. The first necessary step was to model energy levels in both parities, in order to obtain reliable eigenvectors that can be used afterwards to calculate precise transition probabilities. This first step was the scope of Paper II.

\includepdf[pages=1-13]{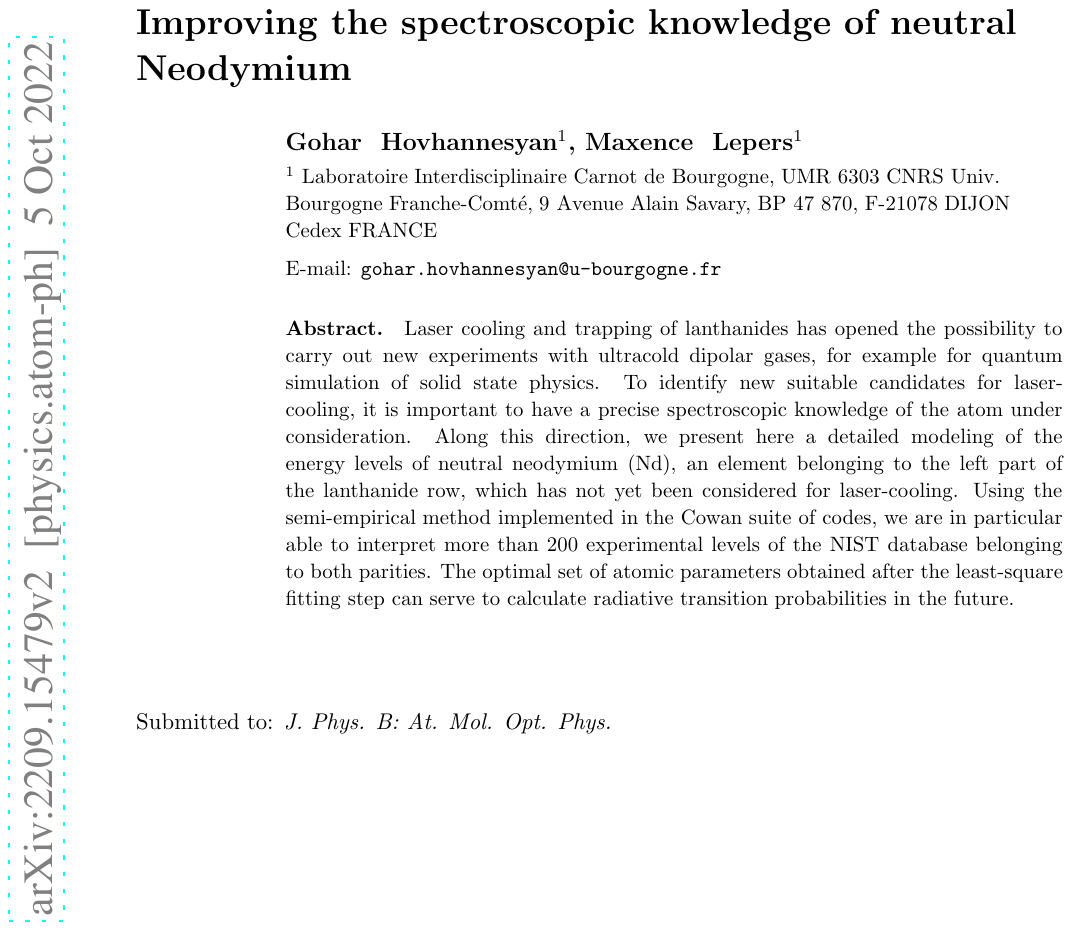}

In our model, we included the following configurations and obtained the following results:
\begin{itemize}
  \item even parity: 4f$^4$6s$^2$, 4f$^4$5d6s, 4f$^3$5d6s6p; $N_\mathrm{lev} = 83$; $N_\mathrm{par} = 12$; $\sigma = 90$~\cmi{};
  \item odd parity: 4f$^3$5d6s$^2$, 4f$^3$5d$^2$6s, 4f$^4$6s6p; $N_\mathrm{lev} = 298$; $N_\mathrm{par} = 15$; $\sigma = 74$~\cmi{}.
\end{itemize}
In both parities, the spectrum is so dense that we first made separate calculations for the first two on the one hand, and the third on the other hand. As results of the final calculations, we have distinguished three types of levels, see Fig.~2 of the paper: those having an interpretation in the NIST database \cite{NIST-ASD}, those having no interpretation in the database and that we interpreted in Paper II, and the unobserved ones that we predicted in the paper. The latter are numerous, especially in the even parity and in the extremal $J$-values of the odd parity.

\begin{figure}
  \begin{center}
  \includegraphics[width=0.6\linewidth]
    {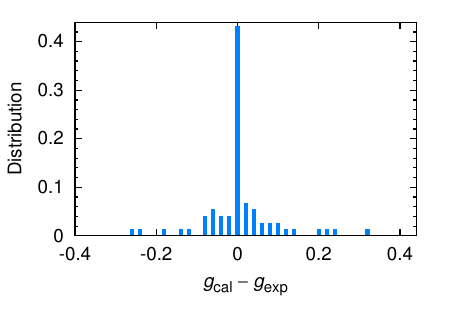}
  \caption{Nd : normalized distribution of levels of both parities as function of the difference between calculated and experimental Landé g-factors.
  \label{fig:atostrNdDistrLande}}
  \end{center}
\end{figure}

The accuracy of Landé g-factors is evaluated on paper II's figure~3, where the difference between calculated and experimental values is plotted on two different scales as function of the experimental energy. The normalized distribution of differences is shown as histograms on Fig.~\ref{fig:atostrNdDistrLande}. The distribution is strongly localized around zero, with 43~\% of the Landé factors with an accuracy better than 0.01. On figure 3 though, we see three outliers with differences, one at -0.45 and two above 0.8, not shown on the distribution. The resulting standard deviation is equal to 0.199 and 0.088, with and without the three outliers.

\paragraph{Transition probabilities \: --}

The natural continuation of the work is to calculate transition probabilities using the FitAik package and the experimental data set of Ref.~\cite{stockett2011}. In such a fitting procedure, the strongest transitions like those of Table \ref{tab:atostrNdTrs} play an important role to stabilize the fit. However the corresponding transitions appearing in our calculations possess transition probabilities several orders of magnitude smaller than the experimental ones. For the ground levels, transitions with an Einstein coefficient $A_{ik} > 10^7$~s$^{-1}$ are only visible for upper levels above 30000~\cmi{}. Transitions given in Ref.~\cite{stockett2011} imply upper levels above 17000~\cmi{}, a region where the levels have strongly mixed eigenvectors, and so where transition dipole moments are very sensitive to those eigenvector compositions.

\begin{table*}
  \begin{center}
  \begin{tabular}{|ccc|ccc|ccc|}
    \hline
    \multicolumn{3}{|c|}{Upper level} 
    & \multicolumn{3}{c|}{Lower level} & Wave
    & Transition & \multirow{3}{*}{\shortstack[c]
      { Width\vspace{3pt} \\ (MHz)} }
    \\ \cline{1-3} \cline{4-6}
    Energy & \multirow{2}{*}{Parity} 
    & \multirow{2}{*}{$\quad J\quad$} 
    & Energy & \multirow{2}{*}{Parity} 
    & \multirow{2}{*}{$\quad J\quad$} & length & probability & \\
    (\cmi{}) & & & (\cmi{}) & & & (nm) & ($10^6$~s$^{-1}$) & \\
    \hline
    17787.0 & odd & 3 &      0 & even & 4 & 562 & 13.1 &  2.09  \\
    17976.9 & odd & 3 &      0 & even & 4 & 556 &  3.3 &  0.525 \\
    21572.6 & odd & 3 &      0 & even & 4 & 463 & 88.0 & 14.0   \\
    23487.0 & odd & 3 &      0 & even & 4 & 425 & 10.7 &  1.70  \\
    24935.0 & odd & 9 & 5048.6 & even & 8 & 503 & 12.8 &  2.04  \\
    25141.5 & odd & 9 & 5048.6 & even & 8 & 498 & 18.5 &  2.94  \\
    25518.7 & odd & 9 & 5048.6 & even & 8 & 488 &  108 & 17.2   \\
    \hline
  \end{tabular}
  \caption{Nd possible laser-cooling transitions starting from the ground level and from the highest level of the lowest manifold $^5$I$_8$. Transition probabilities are experimental values taken from Ref.~\cite{stockett2011}.}
  \label{tab:atostrNdTrs}
  \end{center}
\end{table*}

In consequence, we select the candidates for laser-cooling from transitions with an experimental Einstein coefficient. We have identified four transitions from the ground level $^5$I$_4$ to odd levels with $J=3$ (in order to eliminate leakages to other $^5$I$_J$ levels) and with $A_{ik} > 10^6$~s$^{-1}$. Their wavelengths are in the visible window, from 425 to 562~nm. Using a $J \to J-1$ transition is not the most straightforward way to achieve laser-cooling, because the stretched sublevels $M = \pm 4$ are so-called dark states, meaning that they are insensitive to $\sigma^\pm$-polarized laser beams. However, this problem has been circumvented in so-called type-II magneto-optical traps (MOTs) \cite{nasyrov2001} used for molecules \cite{fitch2021} or Zeeman slowers \cite{petzold2018}, operating on such transitions.

More usual transition $J \to J+1$ could be employed if laser-cooling were achieved in the metastable level 4f$^4$6s$^2 \, {}^5$I$_8$ at 5048.6~\cmi{}. Upper levels of cooling transitions would be odd with $J=9$, a situation similar to Dy, which would prevent decay toward the $^5$I$_{6,7}$ levels. Table \ref{tab:atostrNdTrs} presents three of those transitions with $A_{ik} > 10^6$~s$^{-1}$, which have wavelengths around 500~nm. Cooling in this level would require to transfer to it the atomic population from the ground level.

In all cases, branching ratios from the selected upper levels are necessary to determine the feasibility of laser-cooling. For the upper levels of Tab.~\ref{tab:atostrNdTrs}, the transition toward the ground or metastable levels are the only ones detected in Ref.~\cite{stockett2011}. This is annoying as it does not allow to draw any conclusion on leaking transitions, but it could also mean that there are no detectable transitions, hence no significant losses, from those upper levels. Our work triggered discussions with the experimental group of Jean-Fran{\c c}ois Cl{\'e}ment and Vincent Jacques from University of Lille.

\,

In this chapter, I have presented the principles of the atomic-structure calculations used throughout this manuscript. They are based on the semi-empirical Racah-Slater method provided by Cowan's suite of codes, and extended by us in order to include least-squares fitting of transition probabilities. This is done by our home-made package FitAik designed to work in interaction with Cowan's code RCG. This methodology is particularly adequate with complex atoms and ions where a large number of energy levels and transition probabilities depend on a limited number of fitted radial parameters. The accuracy of the calculated quantities, especially energies, is better than in purely \textit{ab initio} methods even based on Dirac's equation \cite{radziute2021, bondarev2024}. Of course fitting procedures require the availability of experimental energies and transition probabilities, which is the case for neutral or singly ionized lanthanides, especially coming from the groups at Meudon and Madison.

Those calculations are applied in the framework of ultracold gases, in order to predict the feasibility of laser-cooling. The latter requires the existence of broad, quasi-cycling radiative transitions, in which the atoms or ions absorb and emit a large number of photons (typically $10^5$) to be slowed down. Knowing the probabilities of these potential cooling transitions, as well as leaking transitions, \textit{i.e.}~implying other lower levels, is therefore crucial, in particular for complex atoms like lanthanides. Similar calculations are usually performed with molecules in order to propose new coolable candidates, see \textit{e.g.}~\cite{jiang2021, smialkowski2021, zhang2021, oleynichenko2022, yuan2022} among the most recent articles. 

In the case of Er$^+$, we have obtained a reasonably accurate set of transition probabilities, that enabled us to predict laser-cooling and leaking transitions, and to highlight a recycling mechanism due to cascades of spontaneous emissions. However, if the statistical properties of the fit are satisfactory, it would be appropriate to evaluate a confidence interval for each individual transitions in the same spirit of the NIST ASD database. Our computed transition intensities are published in our home-made database CaDDiAcS.

In the case of Nd, we have improved the interpretation of energy levels in both parities, by assigning several tens of experimental levels of the NIST database. Using experimental Einstein coefficients, we have identified seven promising cooling transitions, from the ground level and from an interesting metastable level. However, it was not possible to determine whether those transitions suffer from losses from their cooling cycle.

Since Cowan's least-squares fitting procedure is nonlinear, one should have a good guess of initial fitting parameters, especially for those on which the computed energies weakly depend. In order to improve the fit stability, Uyling and coworkers have developed a method based on so-called orthogonal operators \cite{uylings2019}. It allows for including more terms in the Hamiltonian compared to Cowan, such as two-body spin-orbit interaction, which significantly reduces the standard deviation between calculated and experimental energies. The method is in progress for d$^w$ and f$^w$ subshells with $w>2$.

Other alternatives to Cowan's codes are purely \textit{ab initio} ones, in which the absence of least-squares fitting is somewhat compensated by the inclusion of many more electronic configurations in the atomic Hamiltonian. This is the case in the Multi-Configuration Hartree-Fock (MCHF) code \cite{froese-fischer2007} that solves the many-body Schr{\"o}dinger equation. The same group has also written the relativistic counterpart GRASP \cite{froese-fischer2019} solving Dirac's equation. Using a similar methodology, the Flexible Atomic Code (FAC) not only computes atomic energies and radiative transition probabilities, but also \textit{e.g.}~ionization by electron impact or photoionization \cite{gu2008}.

\chapter{Interaction of ultracold lanthanide atoms with electromagnetic fields}
\label{chap:ddp}

In the previous chapter, I have discussed the application of atomic-structure calculations, in order to characterize the laser-cooling mechanism. Another crucial point of ultracold gases is the ability to confine them in a small region of space, away from any solid material, for instance the walls of a cell or a bottle. This trapping process can be achieved with electromagnetic fields. But for neutral atoms or molecules, the interaction forces, stemming from polarization effects, are much weaker than for charged ones.

In magneto-optical traps (MOTs), the confinement comes from the magnetic-field gradient that, in interaction with the atomic magnetic moment, induces a position-dependent potential. Once the cooling stage is over, atoms are often loaded in an optical trap which is very shallow, namely in the $\mu$K range. The trap can be designed with a single, usually Gaussian laser beam (optical dipole trap) \cite{grimm2000}, an optical lattice made by the standing wave of a retro-reflected beam \cite{gross2017}, or more recently with a single or an array of optical tweezers, which are tightly focused beams \cite{kaufman2021}. The depths of the resulting potential wells are proportional to the laser intensity and to the atomic dynamic dipole polarizability (DDP) which characterizes the single-particle response to the field. The DDP is function of the quantum state of the particle and of the laser wavelength and polarization. Furthermore, in many situations, it is necessary that the DDP of two different states be equal, corresponding to so-called {}``magic trapping conditions'' \cite{derevianko2010a, ruttley2025}.

In the case of ground-level alkali or alkaline-earth metals, the trapping potential depends on the frequency but not on the polarization, because the outermost s electrons possess isotropic orbitals. This is in stark contrast to lanthanides, whose unfilled 4f orbital is anisotropic. The DDP of a given Zeeman sublevel can then be written as a linear combination of three quantities, the scalar, vector and tensor polarizabilities, which do not depend on the sublevel or the beam polarization. Their knowledge allows for determining the DDP in all sublevels and polarizations. The latter is therefore an efficient knob \textit{e.g.}~to obtain magic trapping conditions \cite{patscheider2021, grun2024}.

To a large extent, the present chapter is dedicated to the DDP calculations of lanthanide atoms that we made throughout the years. Section \ref{sec:ddpDeriv} derives the DDP expressions in the most general case of an arbitrary light polarization and atomic sublevel, and of a multi-level atom. To that end, we use Floquet's formalism together with the flexible sum-over-state formula inherent to second-order perturbation theory on degenerate energy levels. The derived expression are also valid for a molecule, for which the sum runs on electronic, vibrational and rotational levels \cite{vexiau2017}. Using algebra of irreducible tensors, we obtain expressions for the scalar, vector and tensor DDPs that are functions of transition energies and squares of the reduced transition dipole moments (TDMs), in other words line strengths. In consequence, for practical calculations, we need an extensive set of atomic data, which is the purpose of the previous chapter. In section \ref{sec:ddpCompar}, I discuss the various comparisons between our calculations and measurements by experimental colleagues, with an emphasis on a joint publication on erbium with Francesca Ferlaino's group at Innsbruck \cite{becher2018}. I also present the assessment of uncertainties on our DDP calculations as given in Refs.~\cite{lepers2023, bloch2024}.

In addition to lasers, static electric and magnetic fields can also be employed to control the particles' state, for instance to polarize them. Electric fields act on Rydberg atoms or heteronuclear diatomic molecules by inducing an electric dipole moment. In section \ref{sec:atDblPol}, I describe a particular situation in dysprosium where two quasi-degenerate energy levels can be mixed by a static electric field, in order to induce a sizable electric dipole moment in addition to the strong magnetic moment. In such a combination of states, dysprosium atoms could form a so-called doubly dipolar quantum gas.

\section{Derivation of the AC Stark shift}
\label{sec:ddpDeriv}

In this section, I derive the equations to describe the interaction of a non-spherically symmetric atom with an oscillating electric field. The derivations are mostly taken from Ref.~\cite{beloy2009}, but some can also be found in \textit{e.g.}~Refs.~\cite{angel1968, manakov1986, grimm2000}. In a first step, I use the Floquet and perturbation formalisms to express the second-order AC Stark shift. This approach allows to go beyond the widely used two-level and rotating-wave approximations. Then, I use the formalism of irreducible tensor operators \cite{varshalovich1988} to introduce the scalar, vector and tensor polarizabilities of the atom. In this manuscript, I will focus on the real part of the polarizability, because, although I was involved in the calculations of the imaginary part, see Refs.~\cite{lepers2014, li2016, li2017, vexiau2017}, no experimental measurements were done to check those results.

\subsection{AC Stark shift and Floquet formalism}
\label{sub:ddpFloq}

This point is discussed in Appendix D of Ref.~\cite{beloy2009} and originally in Ref.~\cite{sambe1973}. We consider an atom-field system described by a time-periodic Hamiltonian $\mathrm{H}(\xi,t)$ of period $T$, where $\xi$ represents the spatial and spin variables. According to Floquet's theorem \cite{sambe1973}, the solution $\Psi(\xi,t)$ of the time-dependent Schr{\"o}dinger equation can be expanded as $\Psi(\xi,t) = e^{-i\varepsilon t/\hbar} \psi(\xi,t)$ where $\varepsilon$ is called the quasi-energy and $\psi(\xi,t)$ has also the periodicity of $T$. If one expands it as a Fourier series, $\psi(\xi,t) = \sum_q \psi_q(\xi) e^{iq\omega t}$, with $\omega = 2\pi/T$ the angular frequency, one can show that the functions $\psi_q(\xi)$ are solutions of the stationary Schr\"odinger equation 
\begin{equation}
  \left( \varepsilon - q\hbar\omega \right) \psi_q(\xi) 
   = \sum_{q'} \psi_{q'}(\xi) \frac{1}{T} \int_{-T/2}^{+T/2} dt
   \mathrm{H}(\xi,t) e^{i(q'-q)\omega t} .
  \label{eq:ddpFloqH1}
\end{equation}
We go one step further by expanding the $\psi_q(\xi)$ functions on a complete basis $\{f_n(\xi)\}$ of the field-free part of the Hamiltonian $\mathrm{H}_0$ (namely the bare atom), $\psi_q(\xi) = \sum_n c_{nq} f_n(\xi)$. Equation \eqref{eq:ddpFloqH1} becomes
\begin{equation}
  \left( \varepsilon - q\hbar\omega \right) c_{nq} 
   = \sum_{n'q'} c_{n'q'} \int d\xi f_{n'}(\xi) 
   f_n^*(\xi) \frac{1}{T} \int_{-T/2}^{+T/2} dt
   \mathrm{H}(\xi,t) e^{i(q'-q)\omega t} .
  \label{eq:ddpFloqH2}
\end{equation}
Equation \eqref{eq:ddpFloqH2} appears as a time-independent Schr\"odinger equation $\mathrm{H}_\mathrm{eff} |\bar{\Psi} \rangle = \varepsilon |\bar{\Psi} \rangle$ with the effective hamiltonian $\mathrm{H}_\mathrm{eff} = \mathrm{H} + q\hbar\omega \mathrm{I}$, with $\mathrm{I}$ the identity matrix, expressed in the basis $\{|nq\rangle\}$, combining the atomic states $n$ and the so-called Floquet blocks $q$. The right-hand side of Eq.~\eqref{eq:ddpFloqH2} is thus $\sum_{n'q'} c_{n'q'} \mathrm{H}_{nq,n'q'}$. The eigenvalues are the quasi-energy and the eigenvectors are $|\bar{\Psi} \rangle = \sum_{nq} c_{nq} |nq\rangle$.

We decompose the Hamiltonian as $\mathrm{H}(\xi,t) = \mathrm{H}_0(\xi) + \mathrm{V}(\xi,t)$, where $\mathrm{H}_0(\xi)$ describes the bare atom and $\mathrm{V}(\xi,t)$ describes the periodic atom-field interaction such that
\begin{equation}
  \mathrm{V}(\xi,t) = V_0(\xi) \cos \omega t 
   = \frac{1}{2} \left( V_+(\xi) e^{+i\omega t} 
                      + V_-(\xi) e^{-i\omega t}\right) .
  \label{eq:ddpFloqH3}
\end{equation}
The Hamiltonian $\mathrm{H}_0$ is diagonal in the basis $\{|nq\rangle\}$. Its matrix elements, equal to the atomic energies $E_n$, do not depend on $q$. Equation \eqref{eq:ddpFloqH2} indicates that for each $q$, the diagonal matrix elements of $\mathrm{H}_\mathrm{eff}$ are equal to $E_n + q\hbar \omega$, the energies of the so-called {}``dressed" atom. In this respect, $q$ can be viewed as a photon number, but one should keep in mind that the Floquet formalism is valid for any oscillatory problem, regardless of the photonic nature of light.

Plugging Eq.~\eqref{eq:ddpFloqH3} into Eq.~\eqref{eq:ddpFloqH2} gives the matrix elements
\begin{equation}
  \left\langle nq \right| \mathrm{V} \left| n'q' \right\rangle
   = \frac{1}{2} \left( \delta_{q,q'+1}
     \left\langle n \right| \mathrm{V}_+ \left| n' \right\rangle
     + \delta_{q,q'-1}
     \left\langle n \right| \mathrm{V}_- \left| n' \right\rangle
   \right)
  \label{eq:ddpFloqH4}
\end{equation}
which couples neighboring blocks $q' = q \pm 1$, while $\langle n | \mathrm{V}_\pm |n'\rangle$ is characteristic of the atomic structure. Because $\left\langle nq \right| \mathrm{V} \left| n'q' \right\rangle$ does not depend on $q$ itself, the ony $q$ dependence of $\mathrm{H}_{\mathrm{eff}}$ comes from the diagonal terms $q\hbar\omega$. Therefore, the spectrum of quasi-energies obtained for \textit{e.g.}~$q=0$ is duplicated and shifted by $q\hbar\omega$ for each block $q$. In the electric-dipole approximation, one has
\begin{equation}
  \mathrm{V}_-  = -\mathcal{E} \left( \mathbf{d}
                    \cdot \mathbf{e} \right) \\
  \text{\;and\;} \mathrm{V}_+ = \mathrm{V}_-^*
                = -\mathcal{E} \left( \mathbf{d}
                    \cdot \mathbf{e}^* \right)
\end{equation}
where $\mathcal{E}$ is the electric-field amplitude, $\mathbf{e}$ its unit vector of polarization, and the $\mathbf{d}$ the (assumed real) atomic dipole moment operator.

In the perturbative regime $|\langle nq |\mathrm{V}| n'q' \rangle| \ll \hbar\omega, |E_{n+1}-E_n| \, \forall n$, the quasi-energies $\varepsilon_{nq}$ can be expanded in a sum of zeroth, first, second-order, ect.~corrections, $\varepsilon_{nq} \approx \varepsilon_{nq}^{(0)} + \varepsilon_{nq}^{(1)} + \varepsilon_{nq}^{(2)} + \cdots$. The unperturbed ones are the dressed-atom energies, $\varepsilon_{nq}^{(0)} = E_n + q\hbar\omega$. In a ground or moderately excited atomic level, the dipole moment has zero matrix elements, hence $\varepsilon_{nq}^{(1)} = 0$. The leading atom-field interaction results in the second-order ac Stark shift
\begin{align}
  \varepsilon_{nq}^{(2)} & = -\sum_{(n'q')\neq(n,q)}
    \frac{ \left\langle nq \right| \mathrm{V} 
      \left| n'q' \right\rangle
      \left\langle n'q' \right| \mathrm{V} 
      \left| nq \right\rangle }
      {\varepsilon_{n'q'}^{(0)} - \varepsilon_{\qnpr{nq}}^{(0)}}
  \nonumber \\
  & = -\frac{1}{4} \sum_{n'\neq n} \left(
    \frac{ \left\langle n \right| \mathrm{V}_+ 
      \left| n' \right\rangle \left\langle n' \right| 
      \mathrm{V}_- \left| n \right\rangle }
      {E_{n'} - E_n - \hbar\omega}
  + \frac{ \left\langle n \right| \mathrm{V}_- 
      \left| n' \right\rangle \left\langle n' \right| 
      \mathrm{V}_+ \left| n \right\rangle }
      {E_{n'} - E_n + \hbar\omega}
  \right)
  \nonumber \\
  & = -\frac{\mathcal{E}^2}{4} \sum_{n'\neq n} \left(
    \frac{ \left( \left\langle n\right| \mathbf{d} 
      \left| n'\right\rangle \cdot \mathbf{e}^* \right)
      \left( \left\langle n'\right| \mathbf{d} 
      \left| n\right\rangle \cdot \mathbf{e} \right)}
      {E_{n'} - E_n - \hbar\omega}
  + \frac{ \left( \left\langle n\right| \mathbf{d} 
      \left| n'\right\rangle \cdot \mathbf{e} \right)
      \left( \left\langle n'\right| \mathbf{d}
      \left| n\right\rangle \cdot \mathbf{e}^* \right)}
      {E_{n'} - E_n + \hbar\omega}
  \right) .
  \label{eq:ddp2strk1}
\end{align}
The terms $\langle n| \mathrm{V}_\pm |n'\rangle \langle n'|  \mathrm{V}_\pm |n\rangle$ do not contribute since they couple blocks $q$ and $q\pm 2$. The shift $\varepsilon_{nq}^{(2)}$ is $q$-independent, and so $q=0$ can be taken without loss of generality. For each $n'$ term, Equation \eqref{eq:ddp2strk1} contains a contribution in $E_{n'} - E_n - \hbar\omega = -2\pi\hbar \Delta$ which is present in the two-level rotating-wave approximation ($\Delta$ being the frequency detuning). By contrast, the off-resonant contribution in $E_{n'} - E_n + \hbar\omega$ is usually neglected.

Let's finish with two remarks. In the case of an electric field polarized in $z$ direction, Equation \eqref{eq:ddp2strk1} gives the usual relationship $\varepsilon_{nq}^{(2)} = -\alpha_{zz}(\omega) \mathcal{E}^2/4$ with $\alpha_{zz}(\omega)$ the $zz$ component of the polarizability tensor. Moreover, Equation \eqref{eq:ddp2strk1} is valid for a non-degenerate atomic level. In the degenerate case, one defines an effective operator $\mathrm{W} = -\sum_{n'q'} \mathrm{V}| n'q' \rangle \langle n'q'| \mathrm{V} / (\varepsilon_{n'q'}^{(0)} - \varepsilon_{\qnpr{nq}}^{(0)})$ which is diagonalized in the subspace of degeneracy \cite{landau2013}. This point will be discussed in the context of long-range interactions, see Section \ref{sec:lriPert}.

\subsection{AC Stark shift and tensor operators}
\label{sub:ddpTens}

We write the atomic levels as $|n\rangle = |\beta JM\rangle$, where $J$ and $M$ denote the quantum numbers related to the total electronic angular momentum and its $z$-projection, and $\beta$ denotes the other relevant quantum numbers. Following Ref.~\cite{beloy2009}, we introduce the frequency-dependent resolvent operators
\begin{equation}
  \mathrm{R}_{\beta J}^{(\pm)} = \sum_{\beta'J'M'}
    \frac{\left| \beta'J'M' \right\rangle
          \left\langle \beta'J'M' \right|}
         {E_{\beta'J'}-E_{\beta J} \pm \hbar\omega}
\end{equation}
which is a tensor operator of rank 0. Setting $\varepsilon_{nq}^{(2)} = \Delta E_{\beta JM}^{(2)}$, Equation \eqref{eq:ddp2strk1} becomes
\begin{align}
  \Delta E_{\beta JM}^{(2)} = -\frac{\mathcal{E}^2}{4}
    \left\langle \beta JM \right| \left(
      \left( \mathbf{d} \cdot \mathbf{e}^* \right) 
      \mathrm{R}_{\beta J}^{(-)}
      \left( \mathbf{d} \cdot \mathbf{e} \right) 
    + \left( \mathbf{d} \cdot \mathbf{e} \right) 
      \mathrm{R}_{\beta J}^{(+)}
      \left( \mathbf{d} \cdot \mathbf{e}^* \right) 
    \right) \left| \beta JM \right\rangle .
  \label{eq:ddp2strk2}
\end{align}
We work out the first term of Eq.~\eqref{eq:ddp2strk2} by gathering the dipole moments on the one hand, and the field polarizations on the other hand. Recalling that both are rank-1 tensors, we use the recoupling relation of Eq.~\eqref{eq:AppLriRecpl1}, which gives
\begin{align}
  \left( \mathbf{d} \cdot \mathbf{e}^* \right) 
    \mathrm{R}_{\beta J}^{(-)}
    \left( \mathbf{d} \cdot \mathbf{e} \right)
  & = \sum_{k=0}^{2} \left(-1\right)^{k}
    \left( {\{\mathbf{e}^* \otimes \mathbf{e}\}}_{k} \cdot 
    {\{ \mathbf{d} \otimes \mathrm{R}_{\beta J}^{(-)} 
    \mathbf{d} \}}_{k} \right)
\end{align}
where $\{\otimes\}_k$ is the rank-$k$ tensor product of operators, see Ch.~3 of Ref.~\cite{varshalovich1988}, which are for example
\begin{align}
  {\{\mathbf{e}^* \otimes \mathbf{e}\}}_{00} &
    = -\frac{1}{\sqrt{3}} \left( 
      \mathbf{e}^* \cdot \mathbf{e} \right)
    = -\frac{1}{\sqrt{3}} \\
  {\{\mathbf{e}^* \otimes \mathbf{e}\}}_{10} &
    = \frac{i}{\sqrt{2}} \left( 
      \mathbf{e}^* \times \mathbf{e} \right)
      \cdot \mathbf{e}_z \\
  {\{\mathbf{e}^* \otimes \mathbf{e}\}}_{20} &
    = \frac{1}{\sqrt{6}} \left[
      3 \left( \mathbf{e}^* \cdot \mathbf{e}_z \right) 
        \left( \mathbf{e}   \cdot \mathbf{e}_z \right)
      - \left( \mathbf{e}^* \cdot \mathbf{e} \right) \right] 
    = \frac{1}{\sqrt{6}} \left[
      3 \left| \mathbf{e} \cdot \mathbf{e}_z \right|^2
      - 1 \right] .
\end{align}

Expanding the scalar product and making use of the relation ${\{\mathbf{e} \otimes \mathbf{e}^* \}}_{k} = (-1)^k {\{\mathbf{e}^* \otimes \mathbf{e}\}}_{k}$, we can rewrite Eq.~\eqref{eq:ddp2strk2} as
\begin{align}
  \Delta E_{\beta JM}^{(2)} & = -\frac{\mathcal{E}^2}{4}
    \sum_{k=0}^{2} \sum_{q=-k}^{+k} \left(-1\right)^{k+q}
    {\{\mathbf{e}^* \otimes \mathbf{e}\}}_{k,-q}
  \nonumber \\
    & \qquad \times \left\langle \beta JM \right| \left[
      {\{ \mathbf{d} \otimes
      \mathrm{R}_{\beta J}^{(-)} \mathbf{d} \}}_{kq}
    + \left(-1\right)^{k} {\{ \mathbf{d} \otimes
      \mathrm{R}_{\beta J}^{(+)} \mathbf{d} \}}_{kq}
    \right] \left| \beta JM \right\rangle .
  \nonumber \\
    & = -\frac{\mathcal{E}^2}{4}
    \sum_{k=0}^{2} \sum_{q=-k}^{+k} \left(-1\right)^{k+q}
    {\left\{ \mathbf{e}^* \otimes \mathbf{e} \right\}}_{k,-q}
    \left\langle \beta JM \right| \upalpha_{(11)kq}
    (\omega) \left| \beta JM \right\rangle
  \label{eq:ddp2strk3}
\end{align}
where we introduced the tensor operator $\upalpha_{(11)kq}$ associated with the dipole polarizability, of rank $k$ and component $q$. The {}``$(11)$" means that $\upalpha_{(11)kq}$ is constructed by coupling two rank-1 tensors \textit{i.e.}~two dipole moments. As such, they are related to the usual polarizabilities $\upalpha_{1m1m'}$ (for instance $\upalpha_{1010} = \upalpha_{zz}$) by $\upalpha_{(11)kq} = \sum_{kq} C_{1m1m'}^{kq} \upalpha_{1m1m'}$. Moreover, as a tensor operator, $\upalpha_{(11)kq}$ satisfies the Wigner-Eckart theorem
\begin{equation}
  \left\langle \beta JM \right| \upalpha_{(11)kq}(\omega)
    \left| \beta JM \right\rangle = 
    \frac{C_{JMkq}^{JM}}{\sqrt{2J+1}}
    \left\langle \beta J \right\| \upalpha_{(11)k}(\omega)
    \left\| \beta J \right\rangle
  \label{eq:ddpWEPola}
\end{equation}
where $C_{JMkq}^{JM}$ is a Clebsch-Gordan (CG) coefficient, imposing here $q=0$, and $\langle \beta J \| \upalpha_{(11)k}(\omega) \| \beta J \rangle$ is the reduced matrix element given by
\begin{align}
  \left\langle \beta J \right\| \upalpha_{(11)k}(\omega)
    \left\| \beta J \right\rangle \equiv \alpha_{(11)k} 
   & = \sqrt{2k+1} \sum_{\beta'J'}
    \left(-1\right)^{J+J'} \sixj{1}{1}{k}{J}{J}{J'}
    \left| \left\langle \beta'J' \right\| \mathbf{d}
    \left\| \beta J \right\rangle \right|^2
  \nonumber \\
   & \phantom{\sqrt{2k+1}} \times \left[
     \frac{(-1)^{k}} {E_{\beta'J'}-E_{\beta J}-\hbar\omega}
   + \frac{1} {E_{\beta'J'}-E_{\beta J}+\hbar\omega} \right]
  \label{eq:ddpAlphaK}
\end{align}
where we used Eq.~\eqref{eq:AppLriProdTens} and $\langle \beta J \| \mathbf{d} \| \beta'J' \rangle = (-1)^{J'-J} \langle \beta'J' \| \mathbf{d} \| \beta J \rangle$. For $k=0$, 1 and 2, the latter is proportional to the scalar, vector and tensor polarizability respectively
\begin{align}
  \alpha_\mathrm{scal}(\omega) & = -\frac{\alpha_{(11)0}(\omega)}
                                         {\sqrt{3(2J+1)}}
    = \frac{2}{3(2J+1)} \sum_{\beta'J'}
    \frac { \left( E_{\beta'J'}-E_{\beta J} \right) \, 
      \left| \left\langle \beta'J' \right\| \mathbf{d}
      \left\| \beta J \right\rangle \right|^2 }
    { \left( E_{\beta'J'}-E_{\beta J} \right)^2 - \hbar^2\omega^2 } 
  \label{eq:ddpScal}
  \\
  \alpha_\mathrm{vect}(\omega) & = \sqrt{\frac{2J}{(J+1)(2J+1)}}
    \, \alpha_{(11)1}(\omega) 
  \nonumber \\ 
    & = - 2 \sqrt{ \frac{6J}{(J+1)(2J+1)} }
    \sum_{\beta'J'} (-1)^{J+J'} \sixj{1}{1}{1}{J}{J}{J'}
    \frac { \hbar\omega \, 
      \left| \left\langle \beta'J' \right\| \mathbf{d}
      \left\| \beta J \right\rangle \right|^2 }
    { \left( E_{\beta'J'}-E_{\beta J} \right)^2 - \hbar^2\omega^2 }  
  \label{eq:ddpvect}
  \\
  \alpha_\mathrm{tens}(\omega) & =
    \sqrt{\frac{2J(2J-1)}{3(J+1)(2J+1)(2J+1)}}
    \, \alpha_{(11)2}(\omega)
  \nonumber \\ 
    & = 2 \sqrt{ \frac{10J(2J-1)} {3(J+1)(2J+1)(2J+3)} }
    \sum_{\beta'J'} (-1)^{J+J'} \sixj{1}{1}{2}{J}{J}{J'}
  \nonumber \\
    & \qquad \qquad \qquad \qquad \qquad \qquad \qquad 
    \times \frac { \left( E_{\beta'J'}-E_{\beta J} \right) 
    \, \left| \left\langle \beta'J' \right\| \mathbf{d}
      \left\| \beta J \right\rangle \right|^2 }
    { \left( E_{\beta'J'}-E_{\beta J} \right)^2 - \hbar^2\omega^2 } 
  .
  \label{eq:ddpTens}
\end{align}
Note that algebraic expressions of those quantities can be found in Ref.~\cite{li2017}. Finally, the expression of the ac Stark shift is
\begin{align}
  \Delta E_{\beta JM}^{(2)} & = -\frac{I}{2\epsilon_0 c}
  \left[ \alpha_\mathrm{scal}(\omega)
    - i\left( \mathbf{e}^* \times \mathbf{e} \right) \cdot
      \mathbf{e}_z \frac{M}{2J} \, \alpha_\mathrm{vect}(\omega)
  \right. \nonumber \\
    & \left. \quad \phantom{-\frac{I}{2\epsilon_0 c}} 
    + \frac{ 3\left| \mathbf{e}\cdot\mathbf{e}_z \right|^2 - 1}{2}
      \times \frac{3M^2-J(J+1)}{J(2J-1)} 
    \, \alpha_\mathrm{tens}(\omega) \right]
  \label{eq:ddp2strk4}
\end{align}
where we have replaced the CG coefficients $C_{JMk0}^{JM}$ by their algebraic expressions, and introduced the field intensity $I = c\epsilon_0 \mathcal{E}^2 / 2$, with $c$ the speed of light and $\epsilon_0$ is the vacuum permitivity.

We consider two particular cases met in experiments: (i) a linearly-polarized electric field making an angle $\theta$ with the $z$ axis, and (ii) an elliptically-polarized field in the $xy$ plane propagating in the $z$ direction. In case (i), the cross product $\mathbf{e}^* \times \mathbf{e} = \mathbf{0}$ since $\mathbf{e}^* = \mathbf{e}$, and so the vector contribution vanishes. The Stark shift becomes
\begin{align}
  \Delta E_{\beta JM}^{(2),\mathrm{lin}} & 
  = -\frac{I}{2\epsilon_0 c}
  \left[ \alpha_\mathrm{scal}(\omega)
    + \frac{ 3\cos^2\theta - 1}{2}
      \times \frac{3M^2-J(J+1)}{J(2J-1)} 
    \, \alpha_\mathrm{tens}(\omega) \right]
  \label{eq:ddp2strkLin}
\end{align}
which gives $-I/2\epsilon_0 c \times [ \alpha_\mathrm{scal}(\omega) + \alpha_\mathrm{tens}(\omega)]$ for stretched Zeeman sublevels $M = \pm J$ in a $z$-polarized field. In case (ii), the field polarization can be written $\mathbf{e} = ( \cos\gamma \mathbf{e}_x + i\sin\gamma \mathbf{e}_y ) / \sqrt{2}$, so that $\mathbf{e}^* \times \mathbf{e} = i\sin(2\gamma) \mathbf{e}_z$. For example, $\gamma = \pm\pi/4$ corresponds to the $\sigma^\pm$ polarization. The light shift then reads
\begin{align}
  \Delta E_{\beta JM}^{(2),\mathrm{ell}} & 
  = -\frac{I}{2\epsilon_0 c}
  \left[ \alpha_\mathrm{scal}(\omega)
    + \sin(2\gamma) \frac{M}{2J} \, \alpha_\mathrm{vect}(\omega)
    - \frac{3M^2-J(J+1)}{2J(2J-1)} 
    \, \alpha_\mathrm{tens}(\omega) \right]
  \label{eq:ddp2strkEll}
\end{align}
which gives $-I/2\epsilon_0 c \times [ \alpha_\mathrm{scal}(\omega) \pm \sin(2\gamma) \alpha_\mathrm{vect}(\omega)/2 - \alpha_\mathrm{tens}(\omega)/2]$ for $M = \pm J$.

In cold-atom experiments, the atoms are placed in laser beams with inhomogeneous, for instance Gaussian, intensity profiles $I(\mathbf{r})$, which according to Eqs.~\eqref{eq:ddp2strk4}--\eqref{eq:ddp2strkEll}, result in energy shifts depending on the position of the atomic center of mass, and so in a mechanical potential for the atoms. If the term between the $[\,]$ braces is positive, the atoms are attracted towards the maximum of the intensity profile, and so they are trapped in the bright regions. If by contrast it is negative, the atoms are pushed towards the minimum of the intensity profile, in the dark regions.

\section{Calculation of dynamic dipole polarizabilities and comparison with experiments}
\label{sec:ddpCompar}

Among ultracold atoms in traps whose frequency is far detuned from atomic resonances \cite{grimm2000}, alkali metals only possess a scalar polarizability in their ground level, since their s valence orbital is spherically symmetric, and so insensitive to variations of the electric-field polarization. The situation is similar with alkaline-earth, whose orbital $L$, spin $S$ and total electronic $J$ angular momenta vanish. In this case, the CG coefficients of Eq.~\eqref{eq:ddpWEPola} are equal to zero whenever $k \neq 0$.
As for chromium, its ground level is [Ar]\,3d$^{5}$($^{6}$S)4s ${}^{7}$S$_3$. In spite of this S character, a yet very small tensor contribution comes into play \cite{chicireanu2007}, due to the presence of tiny higher-momentum components in the ground-level eigenvector.

Ytterbium excepted, lanthanide (Ln) atoms are characterized by an open 4f submerged subshell, closer to the nucleus than the closed 5s and 5p subsells. One can expect the non-spherical wave function of the unpaired 4f electrons to give rise to an anisotropic Stark shift, but due to the submerged nature of the 4f electrons, one can expect them to be weakly polarizable, and so the anisotropic Stark shift to be small. Moreover, due to the contraction of orbitals along the Ln series, one can expect the polarizability to decrease with increasing atomic number. This simple picture was confirmed by our calculations and several measurements.

\subsection{An example of joint theoretical and experimental study}

I was involved in several joint theoretical and experimental studies of Ln DDPs. Prior to that, I authored theoretical papers on erbium \cite{lepers2014}, dysprosium \cite{li2016} and holmium \cite{li2017}, whose results were compared with later experiments \cite{ravensbergen2018}. As an illustration, I have chosen here the first joint article to which I took part, with the Innsbruck group led by Francesca Ferlaino, see Ref.~\cite{becher2018} denoted as Paper III. The scalar and tensor polarizabilities are calculated and measured for ground-level erbium in three different trap wavelengths, as well as excited atoms in the level at 17147~cm$^{-1}$ in two different trap wavelengths.

When we started the work in 2012, dysprosium and erbium had just been Bose-condensed \cite{lu2011, aikawa2012}, and thulium had been laser-cooled \cite{sukachev2010}, following the pioneering work at NIST on erbium magneto-optical trapping \cite{mcclelland2006}. Later, holmium was also laser-cooled \cite{miao2014}, justifying our interest for that atom \cite{li2017}, and more recently, Bose-Einstein condensation of thulium and europium was also achieved \cite{davletov2020, miyazawa2022}. Regarding DDPs, there had been one measurement of the dysprosium scalar contribution at the common wavelength of 1064~nm, far from the broadest resonances, see Fig.~1 of Paper III below for an illustration. The result of that measurement was astonishingly low \cite{lu2011}, compared to calculations \cite{dzuba2011}.

Our calculations are performed using the sum-over-state formulas \eqref{eq:ddpScal}--\eqref{eq:ddpTens} in which transition energies and transition dipole moments (TDMs) are computed with the semi-empirical approach described in the previous chapter for Er$^+$. When available, experimental transition energies are incorporated in Eqs.~\eqref{eq:ddpScal}--\eqref{eq:ddpTens}. For the odd parity, the electronic configurations included in the calculations are \cite{lepers2014}: 4f$^{12}$6s6p, 4f$^{11}$5d6s$^2$, 4f$^{11}$5d$^2$6s and 4f$^{12}$5d6p. The even-parity configurations are split into three groups: 4f$^{12}$6s$^2$ + 4f$^{12}$5d6s + 4f$^{11}$6s$^2$6p, 4f$^{11}$5d6s6p, and 4f$^{12}$6s7s + 4f$^{12}$6s6d + 4f$^{12}$6p$^2$. Regarding the scaling factors of the monoelectronic TDMs, they are all equal to 0.807.

On the experimental side, the atoms are prepared in their lowest Zeeman sublevel $|J=6, M=-6 \rangle$ and placed in an optical trap in addition to the static magnetic field giving the quantization axis $z$. Three different wavelengths are tested: 1570, 1064 and 532~nm. The anisotropic contribution is probed changing the angle $\theta$ of Eq.~\eqref{eq:ddp2strkLin} between the trap electric field and the magnetic field. The ground-state DDP is measured by trap-frequency spectroscopy, \textit{i.e.}~shaking the atoms inside the trap, and monitoring their oscillations, whose frequency is proportional to the square root of the polarizability. Once the latter is known, the atoms are submitted to another beam at 583~nm driving the transition to the excited state 4f$^{12}$($^3$H$_6$)6s6p($^3$P$_1^\circ$) (6,1)$_7^\circ$ and $M=-7$. The Stark shift of the transition frequency due to the trapping light is measured for various intensities and polarization angles.

\includepdf[pages=-]{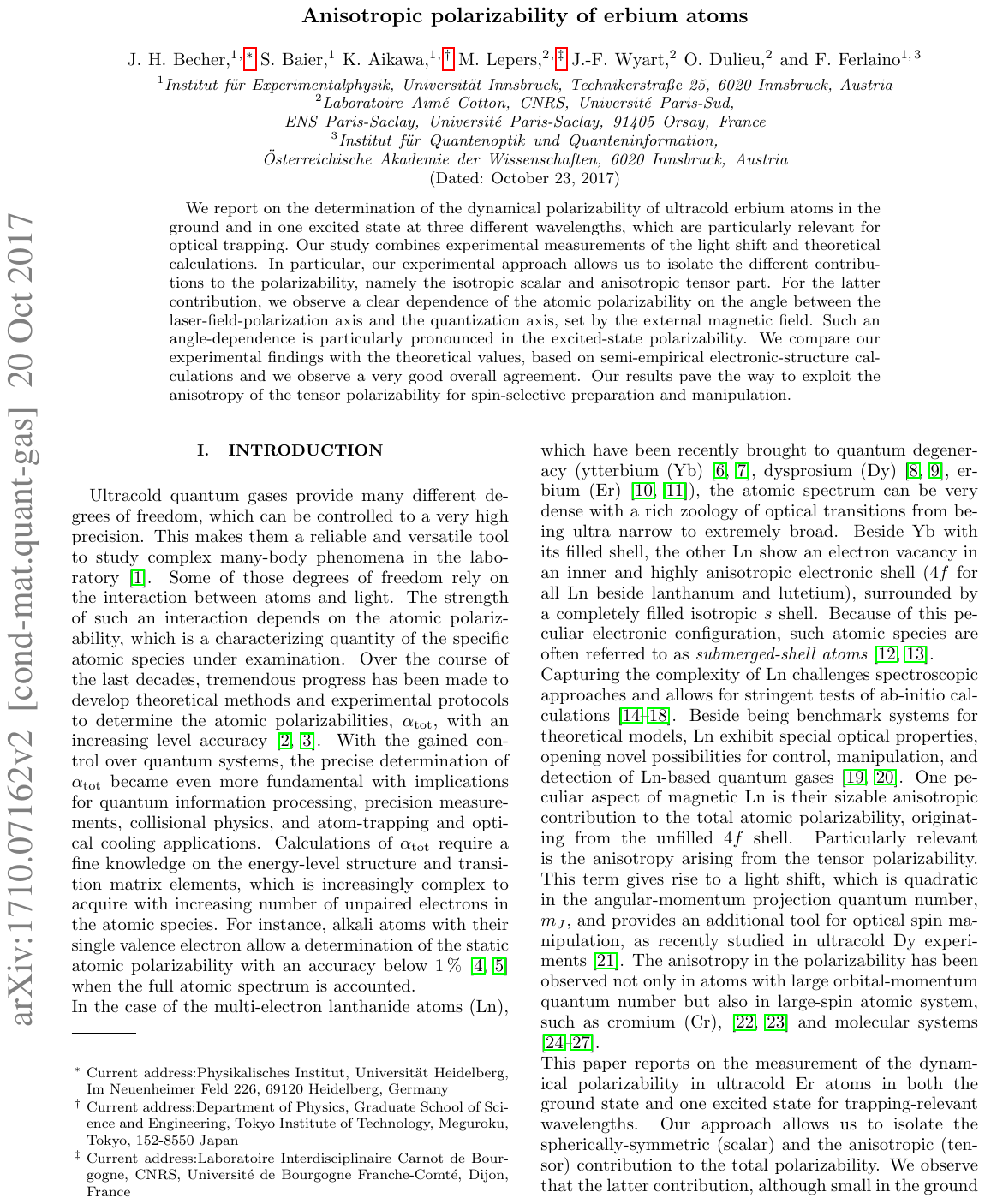}

The comparison shows a very satisfactory agreement between calculations and measurements. For the ground level in infrared trap, the agreement for the total and scalar polarizabilities is very good. As expected, the DDPs slightly increases for 1064~nm, a wavelength closer to the main Er resonances. The values around 170~a.u. supports the theoretical value of Ref.~\cite{dzuba2011} for Dy, and indeed, a later article measurement confirmed its validity, see Ref.~\cite{ravensbergen2018} and subsection \ref{sub:ddpSum}. Regarding the Er tensor contribution at 1064~nm, both theory and experiments find that it amounts to a few percents of the scalar DDP, with a negative sign. However, the calculated anisotropy is almost 3 times as large as the experimental one. The small calculated value ensues from quasi-cancellation of large terms coming from the main transitions of Er. The next subsection discusses this point as assessment of uncertainties.

At 532~nm, the calculated total DDP is 26~\% smaller than the measured one. This discrepancy affects the scalar contribution, since the tensor ones are in close agreement. At 532.26~nm (corresponding to a wave number of 18788~\cmi{}), the calculated DDPs are more sensitive to specific nearby transitions, compared to infrared wavelengths. For example, there is an odd-parity level of $J=7$ at 18774.123~\cmi{} \cite{NIST-ASD}. If one of those transitions, in particular its TDM, is not correctly described in our model, this can influence the accuracy of the computed DDPs.

As expected, the 17157~\cmi{} excited level shows a more pronounced anisotropy at both 1064 and 1570~nm, due to the 6p electron. The energy denominators of Eqs.~\eqref{eq:ddpScal}--\eqref{eq:ddpTens} are the smallest for even levels in the 23000--27000-\cmi{} energy range, which belong to the configurations 4f$^{12}$5d6s, 4f$^{12}$6s$^2$6p and 4f$^{11}$5d6s6p. As for the DDP at 532~nm, the proximity of such transitions can explain why the agreement is less good than for the ground level, even though it is globally satisfactory.

\subsection{Summary of results and uncertainty assessment}
\label{sub:ddpSum}

In this subsection, I summarize the available DDP measurements for Er and Dy, and I compare them to our calculations. I also give an estimate of the uncertainty of our results as in Ref.~\cite{bloch2024}. To do so, I assume that each term of Eq.~\eqref{eq:ddpAlphaK} brings a positive contribution to the uncertainty $\Delta \alpha_{(11)k}$, multiplied by the coefficient $\eta$ characterizing the uncertainty of computed TDMs (see previous chapter). Therefore 
\begin{align}
  \Delta \alpha_{(11)k} (\omega) 
   & = \eta \sqrt{2k+1} \sum_{\beta'J'}
    \left| \sixj{1}{1}{k}{J}{J}{J'} \right| \,
    \left| \left\langle \beta'J' \right\| \mathbf{d}
    \left\| \beta J \right\rangle \right|^2
  \nonumber \\
   & \phantom{\sqrt{2k+1}} \times \left|
     \frac{(-1)^{k}} {E_{\beta'J'}-E_{\beta J}-\hbar\omega}
   + \frac{1} {E_{\beta'J'}-E_{\beta J}+\hbar\omega} \right| .
  \label{eq:ddpDeltaAlpaK}
\end{align}
As an estimate, I take here $\eta = 0.124 = 12.4~\%$ \cite{bloch2024}. Our calculations have shown \cite{lepers2014, li2016, li2017} that the agreement between calculated and experimental TDMs is better than 10~\% for the strongest transitions, which contribute the most to the DDPs. By contrast, the agreement is less good for the weakest transitions, which contribute significantly less to the DDPs. Therefore a value of $\eta = 0.124$ seems like a good compromise to estimate $\Delta \alpha_{(11)k} (\omega)$. A particular value of $\eta$ for each transition would yield a thinner uncertainty calculation.

\begin{figure}
  \begin{center}
  \includegraphics[width=0.6\linewidth]
    {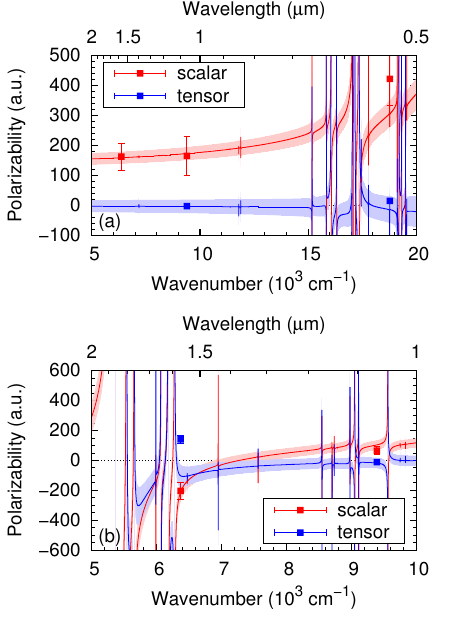}
  \caption{Examples of polarizability uncertainties (shaded regions) for the scalar and tensor components of erbium ground (a) and 17157-cm$^{-1}$ excited levels (b). The experimental values of Ref.~\cite{becher2018} are also presented with their uncertainties.
  \label{fig:ddpUncEr}}
  \end{center}
\end{figure}

An example of DDP curves including uncertainties is presented on Figure~\eqref{fig:ddpUncEr}, for the ground level of Er, as well as the 17157-\cmi{} excited one. The calculated and experimental scalar and tensor DDPs of Paper III are plotted versus the trapping wave number and wavelength. On panel (a), the peaks become more numerous above 15000~cm$^{-1}$. Because the uncertainties on scalar and tensor DDPs are similar, the uncertainty on the off-resonant tensor DDP is much larger than the value itself. Note that on panel (b), the wave number range is smaller, otherwise we would only see a succession of dense peaks. In contrast, the polarizability range is larger on panel (b), especially between 5000 and 6500 cm$^{-1}$, because of transitions towards levels of the 4f$^{12}$5d6s configuration. The measurement of 1570~nm falls very close to those transitions.

A comparison between available measurements and calculated values are given in Table \ref{tab:ddpCompar}, including uncertainties. The theoretical results of erbium slightly differ from Paper III, because here we include the experimental energies when they are known. The overall agreement is satisfactory. As observed in Paper III, it is all the better that the trapping wavelength falls far from absorption peaks. As in figure \ref{fig:ddpUncEr}, the uncertainty of the off-resonant tensor DDPs are significantly larger than their values. However, our computed values are close to the experimental ones, and they always have the same sign. Taking the experimental levels in the sum reduces the gap between calculated and measured tensor DDPs of Er at 1064~nm. By contrast, the DDPs at 532~nm show larger discrepancies, especially for the scalar and vector polarizabilities of Dy \cite{bloch2024}. To date, the latter is the only measurement of Ln DDPs made in an optical tweezer \cite{bloch2023}. Other measurements could reveal a systematic effect due to this specific environment, hence elucidating that yet unexplained strong discrepancy.

\begin{table}
  \begin{center}
  \caption{Comparison of our calculated Er and Dy polarizabilities with available experimental values.
  \label{tab:ddpCompar}}
  \begin{tabular}{|ccc|cc|cc|cc|}
   \hline
   \multirow{2}{*}{Atom} & Level & Wavel.
    & \multicolumn{2}{c|}{scalar} 
    & \multicolumn{2}{c|}{vector} & \multicolumn{2}{c|}{tensor} \\
   \cline{4-9}
    & (\cmi{}) & (nm) & cal. & exp. & cal. & exp. & cal. & exp. \\
   \hline
    Er & 0 & $\infty$ & 149$\pm$19 & 150$\pm$10$^\text{a}$ 
    & 0 & 0 & -1.9$\pm$20 & -2.9$\pm$0.2$^\text{b}$ \\
       &   & 1570 & 159$\pm$20 & 163$\pm$45$^\text{c}$ 
    & 0.4$\pm$10 & - & -2.4$\pm$21 & - \\
       &   & 1064 & 173$\pm$21 & 165$\pm$64$^\text{c}$ 
    & 0.7$\pm$16 & - & -3.2$\pm$22 & -1.9$\pm$2.0$^\text{c}$ \\
       &   &  532 & 304$\pm$43 & 422$\pm$88$^\text{c}$ 
    & -23$\pm$68 & - & -12$\pm$45 & -15$\pm$9$^\text{c}$ \\
    & 17157 & 1570 & -209$\pm$61 & -203$\pm$59$^\text{c}$
    & -25$\pm$107 & - & -102$\pm$72 & -141$\pm$28$^\text{c}$ \\
    &       & 1064 & 104$\pm$31 & 66$\pm$29$^\text{c}$ 
    & -222$\pm$56 & - & -17$\pm$35 & -11.3$\pm$2.5$^\text{c}$ \\
    & & & & & & & & \\
    Dy & 0 & $\infty$ & 163$\pm$20 & 163$\pm$15$^\text{a}$ 
    & 0 & 0 & 1.1$\pm$23 & 1.4$\pm$0.1$^\text{b}$ \\
       &   & 1064 & 193$\pm$24 & 184$\pm$2$^\text{d}$ 
    & 1.6$\pm$15 & - & 1.5$\pm$27 & 1.7$\pm$0.6$^\text{d}$ \\
       &   &  532 & 408$\pm$56 & 184$\pm$2$^\text{e}$ 
    & -57$\pm$93 & 4$\pm$15 & -18$\pm$61 & -25$\pm$12$^\text{d}$ \\
       & 15972 & 1070 & 161$\pm$19 & 188$\pm$12$^\text{f}$ 
    & -115$\pm$51 & - & 50$\pm$22 & 34$\pm$12$^\text{f}$ \\
       &   &  532 & 73$\pm$30 & 130$\pm$40$^\text{e}$ 
    & 125$\pm$60 & 260$\pm$80 & -44$\pm$38 & -68$\pm$18$^\text{d}$ \\
  \hline 
  \multicolumn{9}{l}{$^\text{a}$ Schwerdtfeger
    \textit{et al.}, Ref.~\cite{schwerdtfeger2019}} \\
  \multicolumn{9}{l}{$^\text{b}$ Rinkleff \textit{et al.},
    Ref.~\cite{rinkleff1994}} \\
  \multicolumn{9}{l}{$^\text{c}$ Becher \textit{et al.}, 
    Ref.~\cite{becher2018}} \\
  \multicolumn{9}{l}{$^\text{d}$ Ravensbergen \textit{et al.}, 
    Ref.~\cite{ravensbergen2018}} \\  
  \multicolumn{9}{l}{$^\text{e}$ Bloch \textit{et al.}, 
    Ref.~\cite{bloch2024}} \\  
  \multicolumn{9}{l}{$^\text{f}$ Chalopin \textit{et al.}, 
    Ref.~\cite{chalopin2018}}  
  \end{tabular}
  \end{center}
\end{table}

Regarding static scalar polarizabilities, Reference \cite{schwerdtfeger2019} is a critical compilation of experimental and theoretical literature results. The uncertainties given therein are estimated by the authors, taking into account the uncertainty associated with each compiled value, and the dispersion of the published polarizabilities for a particular element. Note that measured static scalar polarizabilities have been reported in Ref.~\cite{ma2015} for 35 metallic atoms; but I do not report them directly in Table \ref{tab:ddpCompar} since the value for erbium is unexpectedly large. The static tensor polarizabilities were measured with atomic beams submitted to optical pumping and radio-frequency detection in parallel electric and magnetic fields. In Ref.~\cite{rinkleff1994}, the polarizabilities are given in kHz/(kV/cm)$^2$, while in Table \ref{tab:ddpCompar}, we give them in atomic units $4\pi\epsilon_0 a_0^3$, using the relationship 1~a.u. = 0.248832~kHz/(kV/cm)$^2$.

Finally, in Refs.~\cite{patscheider2021, grun2024}, so-called magic trapping conditions are investigated. This corresponds to the situation where the ac Stark shifts of two levels are identical, or in other words, where the differential ac Stark shift vanishes. With Ln atoms, such conditions can be obtained by tuning not only the trapping wavelength as in spherically symmetric atoms, but also the light polarization. For example, in Ref.~\cite{patscheider2021}, magical conditions are found for a $\sigma^-$-polarized 532-nm trapping light, for the Er ground level and the long-lived excited level of $J=7$ at 7696.956~cm$^{-1}$, which is confirmed by our calculations. In Ref.~\cite{grun2024}, a magic elliptic polarization is used for the transition to the 17157-cm$^{-1}$ level of 488~nm, in order to load single atoms into an array of optical tweezers.

\section{Lanthanide atoms with an electric and a magnetic dipole moment}
\label{sec:atDblPol}

Among dipolar gases, a particular attention is paid on so-called doubly dipolar systems, namely possessing both an electric and a magnetic dipole moment. Such systems have the advantage of presenting more control opportunities via electric and magnetic fields, for example in quantum simulation \cite{micheli2006}, quantum computing \cite{karra2016} or ultracold chemistry \cite{son2022}.

Doubly dipolar gases are usually composed of paramagnetic polar molecules, \textit{i.e.}~open-shell heteronuclear diatomic molecules. A first family of such molecules are those composed of an alkali metal and an alkaline earth, such as rubidium-strontium (RbSr) \cite{zuchowski2014}. Experiments implying those systems are very challenging, since they require to laser-cool the two atoms separately \cite{pasquiou2013}, and then to assemble them, as done with heteronuclear bialkali molecules \cite{zuchowski2010a, barbe2018}. Along this direction, several studies have been published, dealing with molecules composed of an alkali and a more complex atom like chromium \cite{ciamei2022a, ciamei2022b, finelli2024} or a lanthanide \cite{khramov2014, tomza2014, frye2020, soave2023}, which brings its strong magnetic moment. Finally, LiNa molecules were produced in the lowest rovibrational level of their $a^3\Sigma^+$ triplet state, showing an electric and a magnetic dipole moment \cite{rvachov2017}.

A second family of paramagnetic polar molecules consists in alkaline-earth monofluorides, like CaF \cite{anderegg2018}, SrF \cite{barry2014} or BaF \cite{courageux2022, rockenhauser2024, zeng2024}, as well as yttrium monoxyde \cite{collopy2018}. Their peculiar electronic structure allow them to be cooled down from room temperature using the same laser-cooling and trapping techniques as for atoms (the so-called direct method). Recent improvements of those techniques have enabled to reach the microkelvin regime \cite{fitch2021}. Along these lines, alkaline earth hydroxile radicals were also investigated, paving the way toward laser-cooling of polyatomic molecules \cite{hallas2023, lasner2025}.

In Ref.~\cite{lepers2018a} (Paper IV), we proposed an alternative way to produce a doubly dipolar gas: using dysprosium atoms prepared in a superposition of quasi-degenerate opposite-parity energy levels. Indeed, in the quantum-mechanical point of view, an electric dipole moment is induced by coupling with an external electric field two energy levels of opposite parities and electronic angular momenta $J$ differing by at most one unity ($|\Delta J| \le 1$). Due to the dense spectrum of lanthanide atoms, several pairs of such levels can be identified, for example in Dy, Ho, Nd or Pr \cite{NIST-ASD}. Those levels are similar to Rydberg ones \cite{gallagher2005}, except that they are moderately excited.

In particular, the Dy levels at 19797.96~cm$^{-1}$ of angular momentum $J=10$ were used for various tests of fundamental physics, see for instance Refs.~\cite{budker1993, budker1994, cingoz2007, leefer2016}. However their reduced electric dipole moment, equal to 0.038 debye (D), is too low to induce a significant response to an electric field \cite{budker1994}. By contrast, the levels at 17313.33 and 17314.50~cm$^{-1}$, of configurations 4f$^{10}$6s6p and 4f$^{10}$5d6s, are more promising. Indeed, our electronic-structure calculations predict a reduced dipole moment of 8.16~D. Moreover, examining their possible decay channels by spontaneous emission suggests that they possess rather long radiative lifetimes. Therefore, in view of all these arguments, we investigated in Paper IV the response to static electric and magnetic fields of experimentally accessible amplitudes, and with an arbitrary relative (tilting) angle.

\includepdf[pages=-]{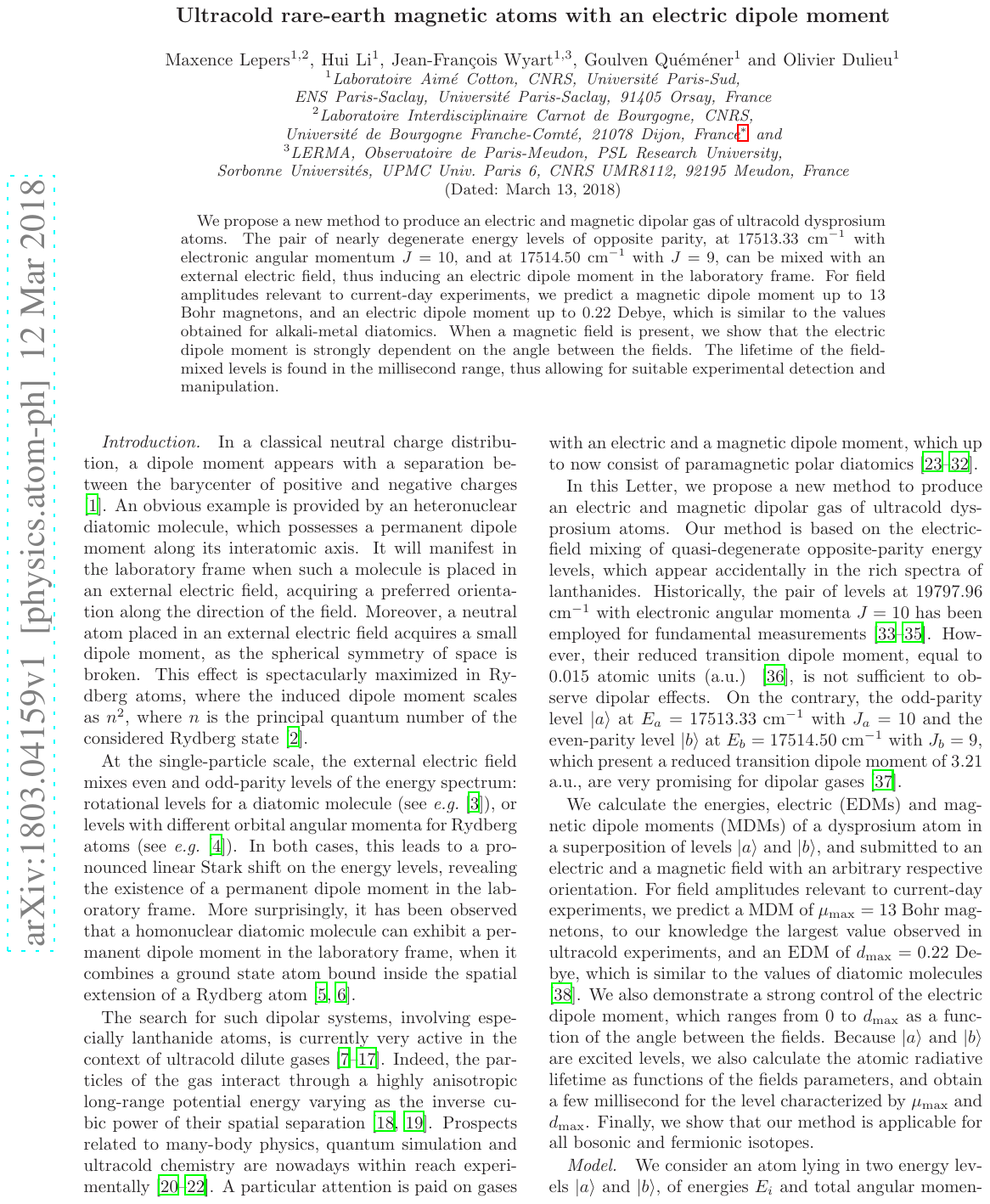}

In Paper IV, we have demonstrated the possibility to prepare a doubly dipolar gas of atoms possessing a magnetic moment up to $\mu^* = 13~\mu_B$ and an electric dipole moment up to $d^*=0.22$~D. The maximal values are obtained for perpendicular fields of amplitudes 100~G and 5~kV/cm. They are associated with the field-mixed state $|\bar{M}_a = -10 \rangle$ that converges to the lowest Zeeman sublevel of the $|a\rangle$ in the zero-electric-field limit. As Figs.~2(a) and (b) show, the electric dipole moment linearly increases with electric field, and strongly depends on the tilting angle $\theta$, from 0 in colinear fields to $d^*$ in perpendicular ones. Our conclusions are valid for bosonic and fermionic isotopes.

\begin{figure}
  \begin{center}
  \includegraphics[width=0.7\linewidth]
    {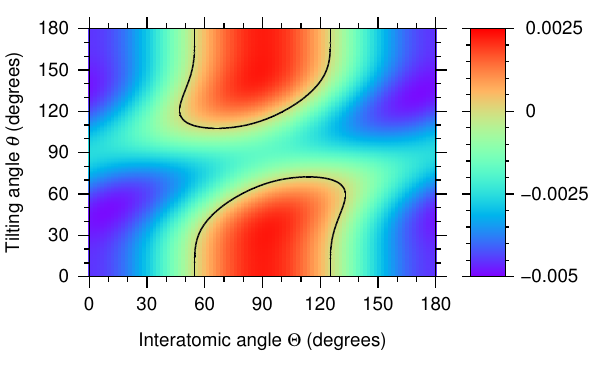}
  \caption{Color plot of the angular part of the electrostatic + magnetostatic potential energy \eqref{eq:ddpDPot} between two Dy atoms in the $|\bar{M}_a = -10 \rangle$ state in atomic units, as a function of the angle $\Theta$ between the $z$ axis defined by the 100-G magnetic field and the interatomic axis on the one hand, and the angle $\theta$ between the two fields on the other hand. The black lines gives the zero energy. The amplitude of the electric field is equal to 2.5~kV/cm.
  \label{fig:ddpDblPot}}
  \end{center}
\end{figure}

One may ask what is the potential energy between two such double dipoles, and to which extent it can be controlled. To answer this point which goes beyond the scope of Paper IV, I first point out that in our parameter range, the electric and magnetic dipole-dipole interactions (DDIs) are of the same order of magnitude, which opens the possibility to tailor the attractive or repulsive nature of the interaction. To stress that, I consider in this discussion an electric field of 2.5~kV/cm, inducing on the $|\bar{M}_a = -10 \rangle$ state an electric dipole moment $d(\theta)$ up to 0.113~D for $\theta = 90^\circ$. Its dependence is similar to Fig.~2(b) of Paper IV, and it can be very well fitted with the formula
\begin{equation}
  d(\theta) \approx a \sin^2\theta + b \sin^4\theta ,
\end{equation}
where $a=0.1311$~D and $b=-0.0183$~D. They can be expressed in atomic units ($ea_0$) using the conversion factor 1~D = 0.393456~$ea_0$. If I assume that the electric dipole is parallel to the electric field, thus in the direction given by $\theta$, the interaction energy between two dipoles depends on their distance $R$, the angle $\theta$, and the angle $\Theta$ defining the orientation of the interatomic axis \cite{quemener2015}
\begin{equation}
  V_e(R,\Theta,\theta) = -\frac{\left[ d(\theta) \right]^2}
    {4\pi\epsilon_0 R^3} \, \left(3\cos^2(\theta-\Theta)-1\right)
\end{equation}
where $\epsilon_0$ is the vacuum permitivity.
If I assume that in the $|\bar{M}_a = -10 \rangle$ state the magnetic moment $\mu$ is along the $z$ axis, the interaction energy is
\begin{equation}
  V_m(R,\Theta) = -\frac{\mu_0 \mu^2}{4\pi R^3}
    \, \left(3\cos^2\Theta-1\right) ,
\end{equation}
where $\mu_0$ is the vacuum permeability and $\mu = 13~\mu_B$.
Note that it is $\theta$-independent. The total potential energy reads
\begin{equation}
  V(R,\Theta,\theta) = V_e(R,\Theta,\theta) + V_m(R,\Theta)
    = \frac{C_3(\Theta,\theta)}{R^3} ,
  \label{eq:ddpDPot}
\end{equation}
where $C_3(\Theta,\theta)$, which gathers all the angular dependence, is plotted on Figure \ref{fig:ddpDblPot}. Note that in atomic units, $1/4\pi\epsilon_0 = 1$, $\mu_0/4\pi = 1/4\pi\epsilon_0 c^2 = \alpha^2$ ($c$ is the speed of light and $\alpha$ the fine-structure constant), and $\mu_B = 1/2$.

Figure \ref{fig:ddpDblPot} shows a rich landscape of attractive, repulsive or zero interactions. For $\theta = 0^\circ$, one sees the expected bare magnetic DDI, maximal in the side-by-side configuration $\Theta = 90^\circ$, and minimal in the head-to-tail configuration $\Theta = 0^\circ$. The influence of the electric DDI is mostly visible on a strip around $\theta = 90^\circ$. Unlike the magnetic one, it is maximal for $\Theta = 0^\circ$ and minimal for $\Theta = 90^\circ$. Because for the chosen electric field, the electric DDI is a little stronger than the magnetic one, and so the total energy is slightly negative in the center of the figure ($\theta = \Theta = 90^\circ$). Decreasing the field in order to obtain $d^* = 0.0852$~D would induce a vanishing interaction, increasing the size of the repulsive islands.

Let's note that Figure \ref{fig:ddpDblPot} is a first insight into the interaction landscape. It assumes that the fields' direction and the interatomic axis belong to the same plane, which could for instance correspond to a 2D optical lattice or tweezer array with coplanar fields. In any other situation, the potential energy \eqref{eq:ddpDPot} also depends on the azimuthal angles $\Phi$ and $\phi$. Moreover, Equation~\eqref{eq:ddpDPot} assumes {}``classical'' dipoles with well-defined orientations.  A calculation including all atomic sublevels is necessary in order to get a quantitative description of the interaction.

Going back to Paper IV, we also calculated the radiative lifetime of the field-mixed states as functions of the field parameters. According to our atomic-structure calculations, the lifetime of the bare $|a\rangle$ level is very large (28.1~s) since its main decay channels are magnetic-dipole transitions. The lifetime of $|b\rangle$ is much shorter (33.6~$\mu$s), though rather large for an atomic system. Therefore, the lifetime of the field-mixed states $|\bar{M}_a\rangle$ strongly decreases as the mixing caused by the electric field, and so the electric dipole moment, increase, hence a trade-off to find.

As corrolaries of their large lifetimes, levels $|a\rangle$ and $|b\rangle$ are not easily accessible by laser from the Dy ground level. In the conclusion of Paper IV, we discussed possible routes to access them, in particular by Stimulated Raman Adiabatic Passage (STIRAP) \cite{bergmann1998}. The situation is very different in other pairs of quasi-degenerate Ln levels. In holmium for example, the even level at 24360.81~\cmi{}, very close to the odd one at 24357.90~\cmi{}, possesses a very strong transition with the ground level, corresponding to a nanosecond-scale lifetime. This pair of levels was chosen in section \ref{sec:lriHo2} and Reference \cite{li2019}, in order to investigate how this doubly dipolar character can be transferred in the realm of ultracold molecules using photoassociation.

After its publication, our study generated discussions with Emil Kirilov at the University of Innsbruck. In Ref.~\cite{anich2024}, his team proposed our system as a platform to simulate so-called a $XYZ$ Heisenberg model \cite{pinheiro2013}, with experimentally tunable parameters in the Hamiltonian. Note in particular that Ref.~\cite{anich2024} suggests to use a microwave rather than a static electric field, which possesses two controllable parameters, its amplitude and frequency, instead of the sole amplitude of the static field. It allows for increasing the degree of admixture between the two quasi-degenerate levels.

\,

In this chapter, I have discussed various aspects of the interactions between ultracold lanthanide atoms and external electromagnetic fields. In section \ref{sec:atDblPol}, I focus on a dysprosium atom prepared in a superposition of quasi-degenerate energy levels of opposite parities and submitted to tilted static electric and magnetic fields. The atom then acquires an induced electric dipole moment comparable to molecular values, in addition to a large magnetic moment of 13~$\mu_B$. The angle between the two fields enables to control the value of the electric dipole moment, and also the radiative lifetime of the field-mixed states. This effect is observable with bosonic and fermionic isotopes of dysprosium. I also plot the total interaction energy between two such double dipoles, which is richer than the usual single-dipole case, since the magnetic and electric dipolar interactions are of the same order of magnitude. Those results increase the possibility of controlling dipolar gases.

The rest of the chapter is dedicated to the interaction with laser fields. After recalling in section \ref{sec:ddpDeriv} the derivations leading to the expression of the second-order AC Stark shift and of the atomic dipole polarizabilities, I summarize in section \ref{sec:ddpCompar} the comparisons between our calculated values and the measured ones in experimental groups. I also present the uncertainty evaluation for our calculated values. The overall agreement is satisfactory, even if it worsens for smaller wavelengths and excited levels. This corresponds to situations where the polarizabilities are sensitive to particular transitions for which the denominator of the sum terms is small. Moreover, we observe strong discrepancies for dysprosium placed in a tweezer array of 532-nm wavelength \cite{bloch2024}. Understanding the source of this difference is important, since optical tweezers are likely to play an important role in future quantum technologies with ultracold atoms \cite{bloch2023, grun2024, grun2025}. Possible explanations could be higher-order effects, either in amplitude (involving then hyperpolarizabilities due to the large intensity) or in multipole moments (quadrupole polarizabilities due to the large field gradient), or coupling between the internal and external atomic degrees of freedom.

To finish, I would like to mention a similar work in which I was involved \cite{vexiau2017}. It dealt with heteronuclear alkali-metal diatomic molecules, for which Romain Vexiau computed the DDPs in a large frequency window. To that end, he gathered the necessary data -- potential-energy curves and transition dipole moments -- and he computed all the vibrational levels. This work also came into play in the calculations of $C_6$ coefficients presented in Section \ref{sec:lriDiat}.

\chapter{Luminescent properties of trivalent lanthanide ions in solids}
\label{chap:ln3+}

Up to now, I have discussed the use of neutral lanthanide (Ln) atoms in ultracold gases. But actually, Ln elements are involved in many areas of current science and technology. One can think of lasers and optical-fiber telecommunications with the unmissable YAG-neodymium lasers or the erbium-doped amplifiers, but also of renewable energies, electric-car batteries, oil industry, or medical imaging. In Reference~\cite{kinos2021}, Ln elements are even discussed as possible systems for quantum computing. 
Unlike ultracold gases, those applications rely on trivalent ions \lntp{}, which most often corresponds to the most stable and natural ionization stage, embedded as impurities in solid materials. But just like ultracold gases, their appeal stems from their unpaired 4f electrons, resulting in their peculiar magnetic and optical properties. It is worthwhile mentioning that \lntp{}-doped solids can be laser-cooled, giving rise to so-called optical refrigeration, down to temperatures on the order of 100~K  \cite{seletskiy2016}.

Lanthanides are sometimes called lanthanoides, and together with scandium and yttrium, they form the group of rare-earth elements. The Ln series constitute a row of the Periodic Table, from lanthanum (atomic number 57) to lutetium (atomic number 71). In the trivalent form, their ground electronic configuration is [Xe]4f$^w$, where [Xe] is the ground configuration of xenon, omitted from now on, and $0 \le w \le 14$ from lanthanum to lutetium. For Nd and Er, $w=3$ and 11 respectively. The unpaired 4f electrons give rise to a large magnetic moment compared to other subshells, which at the macroscopic scale can give strong permanent magnets. Moreover, the 4f orbitals are said {}``submerged'', meaning that they are located closer to the nucleus than the outermost, filled 5s and 5p subshells. The 4f electrons are therefore shielded from the environment, and so when they are placed in a solid host, \lntp{} ions form weak covalent bonds. To some extent, one can say that the ions {}``keep their identity'', as in the gas phase. As a result, the energy levels of an ion in a crystal or a glass can be labeled with the quantum numbers of the corresponding free-ion ones. Understanding the free-ion structure, which we model with the semi-empirical method of Chapter \ref{chap:atoStr}, is a central step for understanding the spectra of rare-earth-doped solids.

The optical applications of \lntp{} ions involve transitions between energy levels of the ground configuration, see Fig.~\ref{fig:ln3+Lev}. Due to (Laporte) selection rules, the vast majority of those transitions are not observable in free space, but are activated by the environment around the ion. The most relevant tool to interpret those transitions is the celebrated Judd-Ofelt (JO) theory, named after its two independent founders, Brian Judd \cite{judd1962} and Georges Ofelt \cite{ofelt1962} in 1962. There are thousands of articles referring to the JO theory, see \textit{e.g.}~reviews \cite{walsh2006, hehlen2013, smentek2015}, as it allows to reproduce the observed transition intensities with a least-squares fit, but also to predict quantities difficult to measure, like spontaneous-emission branching ratios among excited levels. In spite of this great success, the theory gives less accurate results or cannot reproduce some transitions, in particular with europium (Eu$^{3+}$, $w=6$) \cite{tanner2013, binnemans2015}. To overcome these shortcomings, several improvements or extensions of the JO theory have been proposed 
\cite{tanaka1994, kushida2002, kushida2003, downer1988, burdick1989, florez1997, smentek1997, smentek2000, smentek2001, wybourne2002, ogasawara2005, dunina2008, wen2014, kornienko1990, burdick1993, burdick1995}, 
but even in the most recent studies, the standard version is still mostly used \cite{ciric2019a, ciric2019b, zanane2020, liu2021}.

In 2020, Gohar Hovhannesyan (at that time Master student) and I were contacted by the experimental group of Gérard Colas-des-Francs and Reinaldo Chac{\'o}n at Laboratoire ICB. They were looking for some theoretical insights into the light emission by Eu$^{3+}$-doped nanorods \cite{chacon2020, kim2021}, and in particular on the magnetic character of one transition \cite{taminiau2012}. Due to the conceptual similarity between the JO theory and the calculation of polarizabilities discussed in Chapter \ref{chap:ddp}, we tried to propose an extension of the JO model which would be based on the recent progress in the knowledge of the \lntp{} free-ion spectra, both with \textit{ab initio} \cite{radziute2015, freidzon2018, gaigalas2019, gaigalas2022} and semi-empirical methods \cite{meftah2007, wyart2007, meftah2016, arab2019, chikh2021}, involving spectroscopic measurements at Observatoire de Paris, Meudon. We benchmarked our model not only on Eu$^{3+}$, but also on Nd$^{3+}$ and Er$^{3+}$. Indeed, those two ions play a crucial role in rare-earth spectroscopy, due to the YAG laser and the fiber amplifier mentioned above, and they are at opposite places in the Ln row. Our work resulted in G.~Hovhannesyan's PhD thesis \cite{hovhannesyan2023} and in two articles in the Journal of Luminescence \cite{hovhannesyan2022, hovhannesyan2024}, the second of which will be presented in this chapter.

\section{Basics of the Judd-Ofelt theory}

When studying the spectroscopy of \lntp{} ions in solids, it is important to start with some orders of magnitude. The Coulombic interaction between pairs of electrons inside the ions is in the order of 10000~\cmi{}, the spin-orbit energy of the ion 4f electrons is in the order of 1000~\cmi{}, and the interaction energy between the ion and the host material is in the order of 100~\cmi{}. Therefore, the ion is poorly influenced by its environment, which justifies the use of quantum perturbation theory to calculate the spectroscopic properties.

A good knowledge of the free-ion spectra in the two lowest configurations is therefore necessary, as they represent the unperturbed (or zeroth-order) states in our theory. In principle, those configurations are mixed with higher ones due to configuration interaction (CI), especially 4f$^{w-1}$6p and 4f$^{w-1}$6s. However, Refs.~\cite{meftah2007, wyart2007, meftah2016, arab2019, chikh2021} have shown that this mixing is small, and so in our model, we consider one configuration in each parity, namely 4f$^w$ and 4f$^{w-1}$5d. This allows us to connect our extension to the usual tools of the JO theory like the unit-tensor operators $\mathrm{U}_\lambda$ of Eq.~\eqref{eq:ln3+SedJO}.

\begin{figure}
  \begin{center}
  \includegraphics[width=0.49\linewidth]
    {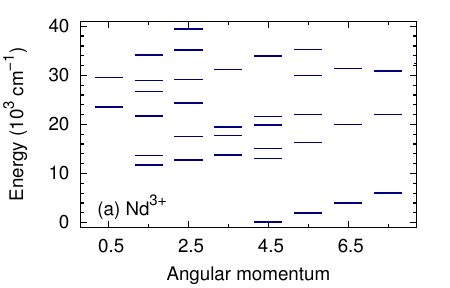}
  \includegraphics[width=0.49\linewidth]
    {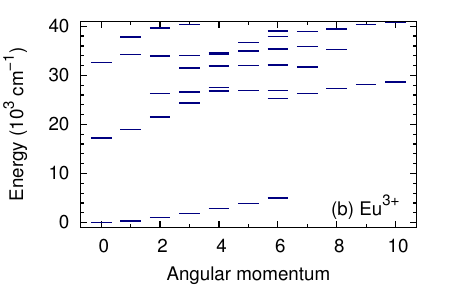}
  \includegraphics[width=0.49\linewidth]
    {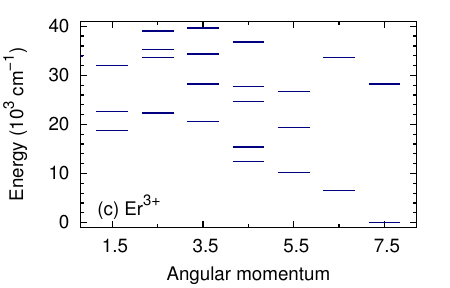}
  \caption{Calculated energy levels of the three free ions considered in this chapter sorted by atomic number: (a) Nd$^{3+}$, (b) Eu$^{3+}$, (c) Er$^{3+}$. The energy range is restricted between 0 and 40000~\cmi{}.
  \label{fig:ln3+Lev}}
  \end{center}
\end{figure}

Figure \ref{fig:ln3+Lev} presents the energy levels as functions of the electronic angular momentum $J$ of the three ions considered in Ref.~\cite{hovhannesyan2024}, called Paper V in what follows. Because the energy range is restricted to 40000~\cmi{}, the plotted levels all belong to the ground configuration 4f$^w$, with $w=3$, 6 and 11 for Nd$^{3+}$, Eu$^{3+}$ and Er$^{3+}$ respectively. At low energies, one can recognize the manifold to which the ground levels belong, namely $^4$I$^\circ$ for Nd$^{3+}$ and Er$^{3+}$, and $^7$F for Eu$^{3+}$. In the latter case, one can also mention the $^5$D manifold starting around 17300~\cmi{} for $J=0$.

The (unperturbed) eigenvectors associated with those energy levels are written in intermediate coupling as
\begin{equation}
  |\Psi_{i}^{(0)}\rangle = \sum_{\alpha_{i} L_{i} S_{i}} 
    c_{\alpha_{i} L_{i} S_{i}} \left| \text{4f}^{\,w}
    \alpha_{i} L_{i} S_{i} J_{i} M_{i} \right\rangle\,,
  \label{eq:ln3+0th}
\end{equation}
where $J_i$ represents the total electronic angular momentum, $M_i$ its $z$-projection, $L_i$ and $S_i$ the orbital and spin angular momenta, and $\alpha_i$ the seniority number \cite{cowan1981}. Usually, there exists a dominant term in Eq.~\eqref{eq:ln3+0th}, reflecting the appropriateness of Russel-Sanders (LS) coupling scheme. In this case, the level is labeled with this dominant term in spectroscopic notation ${}^{2S+1}L_J^p$, $p$ being the parity. As examples on figure \ref{fig:ln3+Lev}, the ground levels are $^4$I$_{9/2}^\circ$ (a), $^7$F$_0$ (b), $^4$I$_{15/2}^\circ$ (c), the superscript {}``$\circ$'' meaning odd parity (it will be omitted in the rest of the chapter). The eigenvalues and eigenvectors are calculated by diagonalizing the atomic Hamiltonian whose parameters are given in Ref.~\cite{hovhannesyan2023}.

Due to the host material, the \lntp{} ion is submitted to a so-called crystal-field (CF) potential,
\begin{equation}
  V_\mathrm{CF} = \sum_{kq} A_{kq} Q_{kq} ,
  \label{eq:ln3+vcf}
\end{equation}
expressed as a sum of tensor operators of ranks $k$ and components $q$. In this respect, $A_{kq}$ are called the crystal-field parameters and $Q_{kq}$ the multipole moments of the ion, see Eq.~\eqref{eq:lri-qlm-bf}. Note that they are denoted as $P^{(k)}_q$ in Paper V. The case $k=1$ corresponds to the electric dipole discussed in the previous chapters. The hermiticity of the CF potential imposes $A_{k,-q} = (-1)^q A_{kq}^*$, while the symmetry of the site where the ion sits can impose certain $A_{kq}$-values to be zero.

The first-order corrections of the perturbation theory yield the Stark energy splittings due to the CF interactions. For each free-ion level, the CF potential \eqref{eq:ln3+vcf} is diagonalized in the subspace of degeneracy spanned by the quantum numbers $M_i$ of Eq.~\eqref{eq:ln3+0th}. The matrix elements of the multipole operators of Eq.~\eqref{eq:ln3+vcf} between levels of the 4f$^w$ configuration are \cite{cowan1981}
\begin{equation}
  \left\langle JM \right| \mathrm{Q}_{kq} \left| JM' \right\rangle
    \propto C_{JM'kq}^{JM} C_{30k0}^{30}
    \left\langle 4f \right| r^k \left| 4f \right\rangle .
  \label{eq:ln3+qkev}
\end{equation}
The Clebsch-Gordan (CG) coefficient $C_{30k0}^{30}$ imposes $k = 2$, 4 and 6, while $C_{JM'kq}^{JM}$ imposes $0 \le k \le 2J$ and $M=M'+q$. The CF splittings are therefore due to the even-rank parameters $A_{kq}$. Note that the quantities $B_q^k = \langle 4f |r^k| 4f \rangle A_{kq}$ expressed in \cmi{}, are often used as fitting parameters in CF analysis. The number of Stark sublevels and their degeneracy depends on the site symmetry of the \lntp{} ion \cite{tanner2013}. In this chapter, we do not calculate Stark sublevels, because to calculate transition intensities, we make the hypothesis that all the sublevels are equally populated, corresponding to a large enough temperature.

In order to describe transitions between ground-configuration levels, we first point out that following Eq.~\eqref{eq:ln3+qkev}, electric-dipole (ED or E1) transitions are not allowed between such levels. By contrast, magnetic-dipole (MD or M1) and electric-quadrupole (EQ or E2) transitions are allowed. The latter being much weaker, they are often disregarded in rare-earth spectroscopy \cite{dodson2012}. The MD transition couple levels of the same $^{2S+1}L$ manifold, and such that $\Delta J = 0$ and $\pm 1$, $0 \nleftrightarrow 0$. However, transitions with $\Delta J$ up to 6 are observed in \lntp{}-doped material, which calls for another explanation.


Owing to the CF, the ground-configuration levels are slightly mixed with higher-configuration levels. The first excited configuration in \lntp{} ions is 4f$^{w-1}$5d, its lowest level lying at several ten thousands of \cmi{}, namely 70817.12 for Nd$^{3+}$ \cite{wyart2007} and 73426.17 for Er$^{3+}$ \cite{meftah2016}. Transitions between 4f$^{w}$ and 4f$^{w-1}$5d configurations are allowed at the ED approximation, since they correspond to the promotion of a 4f electron toward the 5d orbital. In consequence, ED transitions are induced by the admixture of a small 4f$^{w-1}$5d character into the 4f$^{w}$ levels. In terms of perturbation theory, this is captured by calculated the transition dipole moment $D_{12}$ between states $|\Psi_1^{(0)} \rangle$ and $|\Psi_2^{(0)} \rangle$, by expressing the first-order correction on eigenvectors \eqref{eq:ln3+0th} due to the CF perturbation operator \eqref{eq:ln3+vcf},
\begin{equation}
  D_{12} = \sum_{t} \left[
    \frac{ \langle\Psi_{1}^{(0)} |
    \mathrm{V}_\mathrm{CF} | \Psi_{t}^{(0)} \rangle
    \langle \Psi_{t}^{(0)} |
    \mathrm{Q}_{1p} | \Psi_{2}^{(0)} \rangle} {E_{1}-E_{t}}
  + \frac{ \langle\Psi_{1}^{(0)} | \mathrm{Q}_{1p}
    | \Psi_{t}^{(0)} \rangle\langle\Psi_{t}^{(0)}
    |\mathrm{V}_\mathrm{CF} | \Psi_{2}^{(0)} \rangle}
    {E_{2}-E_{t}} \right]
  \label{eq:ln3+d12}
\end{equation}
where $|\Psi_{t}^{(0)} \rangle$ represent the free-ion levels of the 4f$^{w-1}$5d configuration, and $E_{i,t} \equiv E_{i,t}^{(0)}$ are the free-ion energies. From $D_{12}$, one can calculate the ED transition line strength $S_{12}^\mathrm{ED} = \sum_{M_1M_2p} D_{12}^2$, $p$ denoting the light polarization, and then the usual quantities characterizing the transition intensities like the oscillator strength $f_{12}^\mathrm{ED}$ and transition probabilities of spontaneous emission $A_{21}^\mathrm{ED}$. It happens in \lntp{}-doped solids that those quantities are on the same order of magnitude as their MD counterparts.

As mentioned above, the levels $|\Psi_t^{(0)}\rangle$ of Eq.~\eqref{eq:ln3+d12} belong to the 4f$^{w-1}$5d configuration; they are expanded on LS-coupling basis states as
\begin{equation}
  |\Psi_t^{(0)} \rangle =
    \sum_{\overline{\alpha} \overline{L} \overline{S} LS}
    c_{\overline{\alpha} \overline{L} \overline{S} LS}
    |\text{4f}^{\,w-1} \overline{\alpha} \overline{L}
    \overline{S}, \, \text{5d} \, LSJM \rangle ,
  \label{eq:ln3+0thExc}
\end{equation}
where overlined quantum numbers describe the term of the 4f$^{w-1}$ subshell. The matrix element of the electric-multipole operator between states $|\Psi_i^{(0)}\rangle$ and $|\Psi_t^{(0)}\rangle$ reads
\begin{equation}
  \left\langle \Psi_i^{(0)} \right| \mathrm{Q}_{kq}
    \left| \Psi_t^{(0)} \right\rangle
    \propto C_{J_t M_t kq}^{J_i M_i} C_{20k0}^{30}
    \left\langle 4f \right| r^k \left| 5d \right\rangle .
  \label{eq:ln3+qkod}
\end{equation}
The CG coefficient $C_{20k0}^{30}$ imposes $k=1$, 3 and 5, and $C_{J_t M_t kq}^{J_i M_i}$ imposes $|J_i-J_t| \le k \le J_i+J_t$. Combined with the ED selection rule $|J_i-J_t| \le 1 \le J_i+J_t$, it explains the observed rule $0 \le |J_1-J_2| \le 6$, odd values being allowed.

In a similar way to scalar, vector and tensor polarizabilities, one can introduce coupled tensors of rank $\lambda$ and component $\mu$ built from the odd-rank multipoles of the CF and the electric dipole of the radiation, that is $T_{\lambda\mu} = {\{ Q_k \otimes Q_1 \}}_{\lambda\mu}$. For each $k$-value, there are three $\lambda$-values: $k$, $k\pm 1$. The smallest and largest $\lambda$-values are thus 0 and 6, associated with $k=1$ and $k=5$ respectively. The selection rules of ED-induced transitions become
\begin{equation}
  |J_1-J_2| \le \lambda \le J_1+J_2 ,
  \label{eq:ln3+JOSelRul}
\end{equation}
see Eq.~(5) of Paper V. 

In the standard version of the JO theory, a strong assumption is made on the energy difference $E_i-E_t$ of Eq.~\eqref{eq:ln3+d12}: it is replaced by a single value $\Delta E$, identical for all pairs of levels of the two lowest configurations. It can be justified because the 4f$^{w-1}$5d configuration is so high in energy that its detailed spectrum {}``is not visible'' from levels $|\Psi_1^{(0)}\rangle$ and $|\Psi_2^{(0)}\rangle$. Therefore, the sums on the quantum numbers of Eq.~\eqref{eq:ln3+0thExc} can be greatly simplified using Racah algebra (sometimes denoted as application of the closure relation in the literature). Finally, the ED line strength can be written
\begin{equation}
  S_{12}^\mathrm{ED} = \sum_{\lambda=2,4,6} \Omega_\lambda
    \left| \left\langle \Psi_1^{(0)} \right\| \mathrm{U}_{\lambda} 
    \left\| \Psi_2^{(0)} \right\rangle \right|^2
  \label{eq:ln3+SedJO}
\end{equation}
where $\Omega_\lambda$ are called JO parameters, and $\mathrm{U}_\lambda$ are the unit tensor operator of rank $\lambda$, discussed \textit{e.g.}~in Ch.~11 of Ref.~\cite{cowan1981} where they are written $\mathrm{U}^{(\lambda)}$. The latter can be calculated from the eigenvectors of the initial and final levels of the transition \eqref{eq:ln3+0th}, and with angular algebra. The three parameters $\Omega_{2,4,6}$ can be formally expressed as function of the crystal-field parameters (in particular $|A_{kq}|^2$), the transition integrals $\langle 4f |r^k| 5d\rangle$ and the energy difference $\Delta E$. But due to the difficulty to know those quantities, especially when the JO theory was formulated, the $\Omega_\lambda$'s are treated as adjustable parameters in a least-squares fit with experimental line strengths.

In Eq.~\eqref{eq:ln3+SedJO}, the terms with odd $\lambda$ values vanish because the two terms of Eq.~\eqref{eq:ln3+d12} exactly compensate each other with the approximation $E_i-E_t = \Delta E$, $\forall i,t$. The term with $\lambda=0$ is also equal to zero, because $\langle \Psi_1^{(0)} \| U^{(0)} \| \Psi_2^{(0)} \rangle$ is proportional to the identity matrix, and so it vanishes since $|\Psi_1^{(0)}\rangle \neq |\Psi_2^{(0)}\rangle$. According to the selection rules \eqref{eq:ln3+JOSelRul}, it results in strong restrictions in the predicted transitions, especially those involving Eu$^{3+}$ in its $J=0$ ground level: transitions $0 \leftrightarrow 0$, like $^7$F$_0 \leftrightarrow {}^5$D$_0$, and transitions $0 \leftrightarrow \text{odd }J$, like $^7$F$_0 \leftrightarrow {}^5$D$_{1,3}$ and $^5$D$_0 \leftrightarrow {}^7$F$_{1,3,5}$ are forbidden. But even though they are weak, those transitions are observed in experiments \cite{tanner2013, binnemans2015}.

\section{Extension of the Judd-Ofelt theory}

In order to overcome those limitations, many extensions have been proposed over the years, such as $J$-mixing \cite{tanaka1994, kushida2002, kushida2003}, spin-orbit interaction of the excited configuration \cite{downer1988, burdick1989}, odd-rank corrections \cite{florez1997}, velocity-gauge expression of the electric-dipole operator \cite{smentek1997}, relativistic or configuration-interaction effects \cite{smentek2000, smentek2001, wybourne2002, ogasawara2005, dunina2008}, purely \textit{ab initio} intensity calculations \cite{wen2014}, actual energies of the ground \cite{kornienko1990} and excited configurations \cite{burdick1993, burdick1995}. However, even the most recent experimental studies use the standard version of JO theory \cite{ciric2019a, ciric2019b, zanane2020, liu2021}.

When we started to work on the spectra of \lntp{}-doped solids, we were interested in Eu$^{3+}$, and in particular the transitions forbidden in the standard JO theory. Its strong selection rules can be relaxed by accounting for the actual free-ion energies of the ground configuration. Moreover, most Eu$^{3+}$ transitions are spin-changing ones, involving the lowest manifold $^7$F and the excited quintet ones $^5$D, $^5$G and $^5$L. Such transitions are determined by the spin-orbit interaction in both configurations. To characterize it accurately, we wanted to take advantage of the recent progress in the knowledge of the \lntp{} free-ion spectra \cite{meftah2007, wyart2007, meftah2016, arab2019, chikh2021}, which enabled us to fix the atomic properties and only take as adjustable the CF parameters. On the other hand, to conserve the simplicity of the JO theory, we kept its basic physical assumptions (perturbative treatment based on the CF potential induced by the 4f$^{w-1}$5d configuration). This resulted in line strengths of the form
\begin{equation}
  S_{12}^\mathrm{ED} = \sum_{k=1,3,5} C_k X_k \,, \quad
    X_k = \frac{1}{2k+1} \sum_{q=-k}^{+k} \left| A_{kq} \right|^2
  \label{eq:ln3+SedExt}
\end{equation}
where $C_k$, given in Eq.~(6) of Paper V, only depends on free-ion properties. From the rather unsignificant $X_k$ adjustable parameters, one can define the energies
\begin{equation}
  \bar{B}_k = \langle 4f|r^k|5d \rangle \sqrt{X_k}
  \label{eq:ln3+Bbar}
\end{equation}
that characterize the strength of the CF interaction (even though they are multiplied by angular factors). In our model, we also account for the spin-orbit mixing in the excited configuration (the so-called Wybourne-Downer mechanism), especially the one associated to the 5d electron. Compared to Eq.~\eqref{eq:ln3+d12}, $D_{12}$ contains additional terms of the kind $\langle \Psi_1^{(0)}| \mathrm{V}_\mathrm{CF} | \Psi_t^{(0)} \rangle \langle \Psi_t^{(0)} | \mathrm{H}_\mathrm{SO} | \Psi_u^{(0)} \rangle \langle \Psi_u^{(0)} | \mathrm{Q}_{1p} | \Psi_2^{(0)} \rangle$, stemming from the second-order correction on eigenvectors \cite{hovhannesyan2022}.

To calculate oscillator strengths or Einstein coefficients from line strength, there is a proportionality factor, given in Eq.~(8) of Paper V, that depends on the host-material refractive index $n_r$. In our work, we accounted for the wavelength-dependence of $n_r$ using the Sellmeier-Cauchy formula, see Eq.~(11) of Paper V. We checked the validity of our model with two experimental data sets for Eu$^{3+}$, Nd$^{3+}$ and Er$^{3+}$. The code and data sets studied can be found on GitLab \cite{joso-gitlab}.

\includepdf[pages=-]{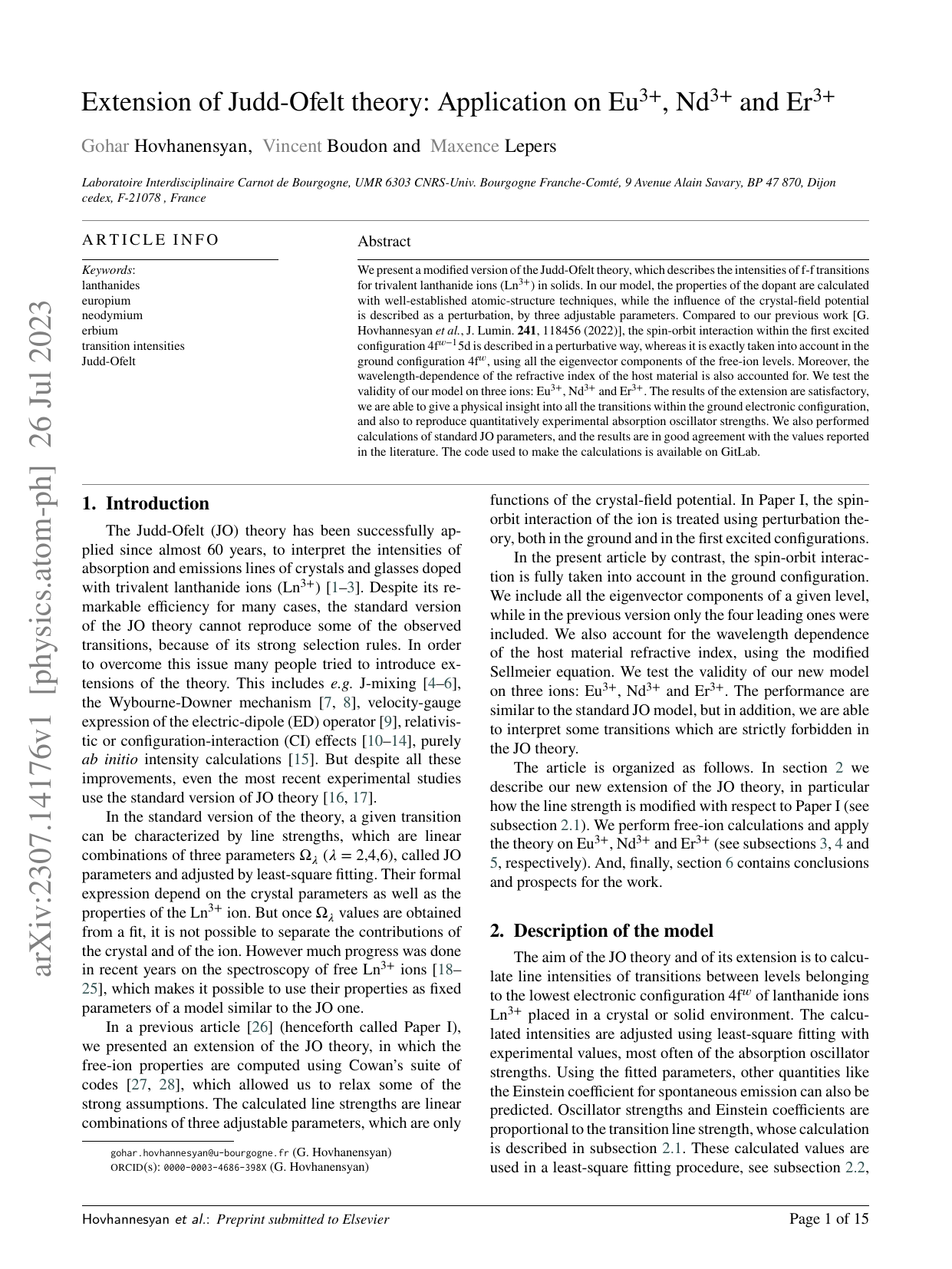}

The results obtained with Eu$^{3+}$ are very satisfactory: not only our model is able to describe transitions like ${}^7$F$_0 \leftrightarrow {}^5$D$_3$ forbidden in the standard JO model, but also it reproduces more accurately the other transitions. For one set in Nd$^{3+}$ \cite{zhang2010}, we obtained a large standard deviation on oscillator strengths, which we attributed to the overlapping transitions in the set, not accounted for in our model. In the case of Er$^{3+}{:}$Lu$_3$Ga$_5$O$_{12}$ \cite{liu2021}, we obtained negative values of $\Omega_4$ and $X_3$ fitting parameters in our standard and extended JO theory respectively. Our standard calculations significantly differ from those of the article, for which all $\Omega_\lambda$-values are positive. We attributed that discrepancies to the exclusion by us of two transitions from the fit. In the conclusion of Paper V, we proposed improvements of our model, the most important of which consists in accounting for overlapping transitions. This will be discussed in the next section.

\section{Impact of our work}

After the publication of Paper V and its presentation at the conference {}``Optique Normandie'' held in Rouen in July 2024, we were contacted by two researchers, Matias Velazquez from Grenoble and Richard Moncorg{\'e} from Caen, who wanted to test our model with other systems on which the standard JO theory is also challenged (work in progress). This convinced me (after G.~Hovhannesyan PhD defense) to implement in the code {}``jo{\_}so'' various improvements evoked above and in the conclusion of Paper V, namely:
\begin{itemize}
  \item Inclusion of overlapping transitions. When an absorption transition is characterized by several upper levels, we compare its measured line strength with the sums of calculated line strengths involving each upper level.
  \item Inclusion of magnetic contribution to the transitions. Each calculated oscillator strength is expressed as the sum of a magnetic and electric contribution, $f_{12} = f_{12}^\mathrm{MD} + f_{12}^\mathrm{ED}$, where $f_{12}^\mathrm{ED}$ is proportional to the line strength \eqref{eq:ln3+SedExt}. The magnetic part does not contain adjustable parameters, as it is assumed to solely depend on free-ion eigenvectors and the host refractive index.
  \item Better account for excited configuration energies. In Refs.~\cite{hovhannesyan2022, hovhannesyan2024}, the levels of the 4f$^{w-1}$5d configurations are supposed degenerate, and its energy is chosen as the one giving maximum ED transition strengths with ground configuration levels, see Figs.~3 and 6 of Paper V. Unlike those of the ground configuration, the energy levels of the excited configuration strongly depend on the host material (by thousands of \cmi{} \cite{dorenbos2000}), and the 5d orbital tend to expand under the effect of the surrounding ligands (nephelauxetic effect \cite{vanpieterson2002a, vanpieterson2002b}), decreasing the 4f-5d interaction parameter. To account empirically for those phenomena, it is now possible to choose 4f$^{w-1}$5d energy as an input parameter of the code.
\end{itemize}

\begin{figure}
  \begin{center}
  \includegraphics[width=0.6\linewidth]
    {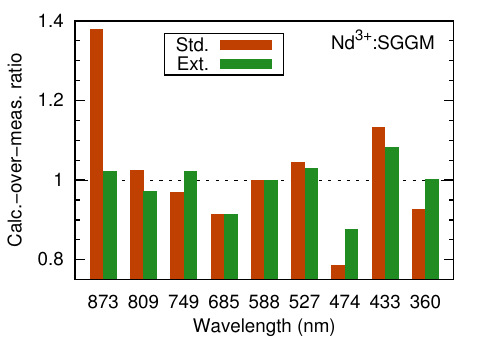}
  \caption{Ratio between calculated and measured oscillator strengths of Ref.~\cite{zhang2010} as a function of the experimental transition wavelength (not at scale). Calculations are performed with the standard ({}``Std.'') Judd-Ofelt model and with our extension ({}``Ext.'').
  \label{fig:ln3+NdFabs}}
  \end{center}
\end{figure}

To illustrate the impact of those modifications, I revisit the data set of Ref.~\cite{zhang2010}, in which there are several overlapping transitions. The result is spectacular, since the fit relative standard deviation shrinks from 23.8~\% (see Paper V, Sec.~4.2) to 1.22~\%, while for the standard JO theory I obtain 4.22~\%. The fitted parameters are $\Omega_2 = 2.04$, $\Omega_4 = 4.10$ and $\Omega_6 = 2.98 \times 10^{-20}$~cm$^2$ for the standard JO model, which are close to those of Ref.~\cite{zhang2010}, and $\bar{B}_1 = 1570$, $\bar{B}_3 = 2230$ and $\bar{B}_5 = 1850$~\cmi{} for the extended one, see Eq.~\eqref{eq:ln3+Bbar}. The hierarchy of parameters, namely $\Omega_4 > \Omega_6 > \Omega_2$ and $\bar{B}_3 > \bar{B}_5 > \bar{B}_1$, are identical in the two models.

Figure \ref{fig:ln3+NdFabs} presents the ratios between calculated and measured oscillator strengths as functions of the experimental transition wavelength, not at scale, for the standard and extended JO model. The trend suggested by the relative standard deviation is confirmed: the extended version systematically gives equivalent or better results than the standard one. The difference is pronounced for the 873-nm transition $^4$I$_{9/2} \to {}^4$F$_{3/2}$ with respective ratios of 1.02 and 1.38. Regarding the 474-nm transition, the ratios are 0.79 and 0.88: the extended model keeps underestimating the oscillator strength, even if I include four overlapping transitions with the 16-19th excited upper levels $^2$G$_{9/2}$, $^2$D$_{3/2}$, $^4$G$_{11/2}$ and $^2$K$_{15/2}$. 

I have also revisited the calculations with Er$^{3+}{:}\text{Lu}_3$Ga$_5$O$_{12}$ \cite{liu2021}, for which we obtained in Paper V negative $\Omega_4$ and $X_3$ fitted parameters. This problem is solved with the present version of our model; and the hierarchy of parameters $\Omega_6 > \Omega_2 > \Omega_4$ for the standard JO is the same as in Ref.~\cite{liu2021}, and the same as in the the present extension, \textit{i.e.}~$\bar{B}_5 > \bar{B}_1 > \bar{B}_3$.

\,

In conclusion, we have proposed a modified version of the Judd-Ofelt theory to describe transition intensities of lanthanide-ion-doped solids. Our initial motivation was to overcome the drawbacks of the original JO theory, especially in europium, mostly due to its too restrictive selection rules. In Refs.~\cite{hovhannesyan2022, hovhannesyan2023, hovhannesyan2024, joso-gitlab}, we presented various versions of our model, based on various perturbative treatments of the 4f$^{w-1}$5d configuration. Similarly to the original one, our model is based on a least-squares fit of experimental oscillator strengths or transition probabilities with three adjustable parameters. But unlike the original model, those parameters only depend on the crystal field, whereas the properties of the impurity are fixed parameters, computed with usual atomic-structure techniques. In this respect, we rely on the recent advances regarding the knowledge of spectra of free lanthanide ions. The obtained fitted parameters can be used to predicted unobserved properties, for instance oscillator strengths, transition probabilities or branching ratios characterizing transitions between pairs of excited levels. Not only our model allows for characterizing forbidden transitions in the standard version, but also it reproduces more accurately the experimental measurements in various cases, which opens the possibility to predict more accurate unobserved quantities.

The most natural prospect of this work is to apply our model to other lanthanide ions, which is currently in progress. In any case, the first step of such works consists in calculating accurate free-ion eigenvectors for levels of the lowest two electronic configurations 4f$^w$ and 4f$^{w-1}$5d, as well as transition integrals between them. The Cowan codes \cite{kramida2019} are particularly well suited for this purpose, since they present all the necessary information in a humane- and machine-readable format. However, other atomic-structure packages like GRASP \cite{froese-fischer2019} could also be employed. Another prospect is to describe transitions involving polarized light \cite{zhang2010, chacon2020} or between Stark sublevels \cite{burdick1994}, which would open the possibility to model low-temperature spectra. Currently, our fitted parameters $X_k$ are expressed as averages over different light polarizations, and initial and final Stark sublevels. One can go one step backward in the calculation of the transition strengths $\propto D_{12}^2$, in which case the fitting parameters would be the $A_{kq}$ themselves, and the fit would be nonlinear with terms of the form $A_{kq} A_{k'q'}$. The symmetry of the site occupied by the ion would come into play in order to determine the vanishing crystal-field parameters.

\part{Long-range interactions involving atoms and diatomic molecules}

\chapter{Basics of long-range interactions in ultracold gases}
\label{chap:lriPres}

In Chapters \ref{chap:atoStr} and \ref{chap:ddp}, I have focused on the interaction of one ultracold atom or ion, especially a lanthanide, with external electromagnetic fields. But even if they are strongly dilute, the properties of ultracold gases also depend on the interactions between pairs of their constituents. For example, Feshbach resonances, sometimes called Fano-Feshbach resonances, allow to control the stability of the gas, or to turn free atoms into weakly bound molecules \cite{kohler2006}. Due to the tiny kinetic energy of colliding ultracold particles, their relative motion is strongly sensitive to weak interactions taking place where the particles are far away from each other, well beyond the region of chemical bonding: so-called long-range interactions. This is \textit{a fortiori} the case when the individual particles carry a dipole moment, hence forming a dipolar gas. I have already discussed dipolar gases made up of paramagnetic atoms in the previous chapters \cite{norcia2021, chomaz2022}; in the rest of this manuscript, I will also discuss dipolar gases composed of molecules \cite{fitch2021, softley2023, langen2024}, which hold a lot of promises regarding ultracold chemistry \cite{bohn2023, karman2024} or quantum computation and quantum simulation \cite{cornish2024}, in which long-range interactions play a central role \cite{defenu2023}.

This research area recently experienced the lifting of a 20-year scientific lock, with the Bose-Einstein condensation of molecules in the lowest rovibrational and hyperfine level \cite{bigagli2024}, which followed the production of a Fermi degenerate gas \cite{demarco2019, schindewolf2022}. Those achievements required to engineer repulsive long-range intermolecular interactions, so as to inhibit their chemical reactivity. This shielding mechanism represents one of the most advanced examples in which a detailed understanding of the long-range molecule-molecule and molecule-field interactions is crucial. Such a detailed description is the subject of the second part of this manuscript, containing the present chapter and the next two ones.

After the pioneering work by F.~London \cite{london1937}, the general formalism of long-range interaction was mostly established in the 1950's and 60's, involving scientists like Hirschfelder \cite{buehler1951}, Dalgarno \cite{dalgarno1956}, Fontana \cite{fontana1961a, fontana1961b, fontana1962}, Buckingham \cite{buckingham1965}, Meath \cite{meath1966a, meath1966b}, Gray \cite{gray1968, gray1976a, gray1976b}, Langhoff \cite{langhoff1970, langhoff1971} or Tang \cite{tang1969}. Numerical calculations were also performed on rather simple systems \cite{bell1965, bell1966, langhoff1971}. Later, with the progress on quantum chemistry, calculations were performed on a wider range of species, see for example Refs.~\cite{graff1990, spelsberg1993, hettema1994, derevianko2000, porsev2002, chu2005, groenenboom2007, bussery-honvault2008, bussery-honvault2009, derevianko2010}. In particular, alkali-metal atoms \cite{bussery1987, merawa1994, marinescu1994, marinescu1995, marinescu1997a, marinescu1997b, merawa1997, merawa1998a, derevianko2002, porsev2003} and diatomic molecules \cite{merawa1998b, merawa2003, rerat2003, byrd2012a, byrd2012b, buchachenko2012, zuchowski2013} attracted a lot of interest, firstly because of their simple electronic structure, but also of the advent of laser-cooling and trapping techniques. Indeed, due to the weak relative kinetic energy in the cold and ultracold regimes, namely far below 1~\cmi{}, the colliding partners are very sensitive to the long-range interactions which are on the order of the \cmi{}.

Because ground-level alkali-metal and alkaline-earth atoms have an S symmetry, their mutual interactions are simply characterized by one potential-energy curves with long-range dispersion forces equal to $-C_6/R^6 - C_8/R^8 - C_{10}/R^{10} - \cdots$. When one atom is excited as in photoassociation experiments \cite{jones2006}, the picture gets more complex, with several $C_6$ coefficients and in homonuclear cases resonant dipole-dipole interactions, which result in a long-range well in one of the potential curves \cite{stwalley1978, wang1997, comparat2000}. With open-shell atoms, like chromium and lanthanides, and with heteronuclear diatomics, interactions within the ground level can display competition between permanent-dipole and induced-dipole terms. Moreover, those interactions can be tuned by external electromagnetic fields, inducing for example Feshbach resonances \cite{kohler2006}, or field-linked states \cite{avdeenkov2004, quemener2023, chen2024}. Furthermore, one can cite Rydberg atoms \cite{flannery2005, deiglmayr2016} and ion-neutral hybrid traps \cite{guillon2007, stoecklin2016, schaetz2017, tomza2019}, as systems for which a knowledge of long-range interactions is required.

In view of all these arguments, we can conclude that it is fully relevant today to investigate long-range interactions between ultracold atoms or molecules, which can be prepared in a well-defined ground or excited electronic, vibrational, rotational, fine or hyperfine level or sublevel, in the presence of electric, magnetic and laser fields. We will do so in this chapter and in the next two ones. Dealing with various systems (atoms, molecules, with or without fine and hyperfine structure), we will highlight the central role of tensor operators and angular momenta \cite{gray1976a, gray1976b, groenenboom2007}, to determine whether a given term of the long-range energy is positive, negative or zero. Once they are calculated, long-range energies can be used \textit{e.g.}~to estimate the density of rovibrational levels close to dissociation \cite{leroy1970, comparat2004}, or as in this manuscript, to prolong short-range potential-energy curves and surfaces, or to feed a scattering code. 

As discussed in Ref.~\cite{chomaz2022}, the expression {}``long-range" has different meanings depending on the discipline in which it is employed. In this work, it corresponds to relative distances for which the individual electronic clouds do not overlap. In other words, each partner conserves its identity as the exchange energy between them vanishes. R.~LeRoy has proposed an estimate of the lower bond of the long-range or asymptotic region, known as the LeRoy radius \cite{leroy1974}
\begin{align}
  R_\mathrm{LR} & = 2\left( \sqrt{\langle r_A^2 \rangle}
                         + \sqrt{\langle r_B^2 \rangle} \right)
  \label{eq:lriLRRad}
\end{align}
where $\langle r_{A,B}^2 \rangle$ is the mean squared radius of the individual electronic clouds. In ground or moderately excited atoms or diatomics, $R_\mathrm{LR}$ ranges from a few to a few tens of atomic units. Therefore, contrary to the terminology of ultracold dipolar gases \cite{chomaz2022}, to which most of this work is dedicated, we consider here that the van der Waals interaction belongs to long-range ones. Furthermore, the present developments are based on electro- and magnetostatic interactions, assumed to be instantaneous. Because no retardation effects are considered \cite{meath1966b, meath1968}, the upper bond of mutual distances is given by transition wavelengths of the partners, which are on the order of $10^4$ atomic units.

In this chapter, we recall the general framework of our calculations of long-range (LR) electrostatic and magnetostatic interactions. Detailed and pedagogical calculations can be found in our book chapter \cite{lepers2018}, and in other sources \cite{stone1996, kaplan2006}. In section \ref{sec:lriPot}, we recall the major steps to obtain the potential energy as a multipolar expansion, in terms of spherical tensors. The latter are essential in atomic and molecular physics, since they enable us to exploit the symmetries of the systems, and so to derive strong selection rules. In section \ref{sec:lriBFSF}, we discuss the different expressions obtained, whether the coordinates are taken in the frame of the complex (body-fixed, BF) or the frame of the laboratory (space-fixed, SF). Once the expressions of the potential energy are established, in section \ref{sec:lriPert}, we describe how to apply it with atomic or molecular systems, using time-independent degenerate quantum perturbation theory, up to the second-order correction. We also discuss the choice of angular-momentum uncoupled \textit{versus} coupled bases. Section \ref{sec:lriFld} gives the matrix elements of the Stark and Zeeman operators, characterizing the interaction with an electric and a magnetic field of arbitrary orientation. Finally, section \ref{sec:lriSym} contains developments not present in our book chapter \cite{lepers2018}: it discusses the symmetry properties of the complex, in particular the effect of inversion, reflection, permutation and parity operations on the basis states, in analogy to diatomic molecules.

\section{Interaction energy and irreducible spherical tensors}
\label{sec:lriPot}

\begin{figure}
  \begin{center}
  \includegraphics[width=0.6\linewidth]{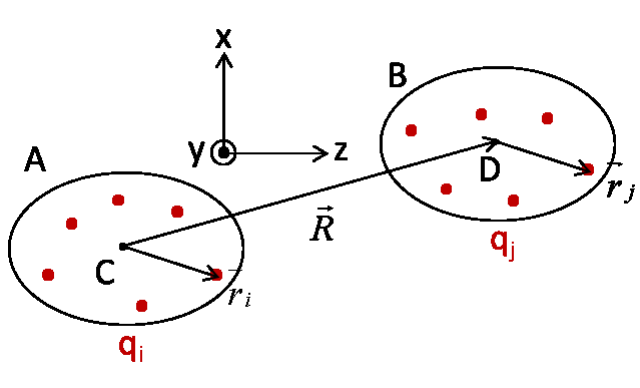}
  \caption{Schematics of the charge distributions $A$ and $B$, and of the coordinate systems $XYZ$ of the space-fixed frame and the $z$ axis of the body-fixed frame.
    \label{fig:chrg-distr}}
  \end{center}
\end{figure}

We consider two charge distributions $A$ and $B$ located in two different bounded regions of space. Their centers of mass, respectively C and D, are connected by the vector $\mathbf{R}$. We assume that each pair of point-like charges $(q_i,q_j)$ with $q_i\in A$ and $q_j\in B$ interact through Coulombic forces giving rise to an electrostatic potential energy. In the BF frame, the $z$ axis is the inter-partner axis with unit vector $\mathbf{u}_z \equiv \mathbf{u}$. In the SF frame, the direction of $\mathbf{u}$ is given by the polar angles $(\Theta,\Phi)$. The situation is depicted on Figure \ref{fig:chrg-distr}.

The Coulombic potential energy between $A$ and $B$ is given by
\begin{equation}
  V_{AB} = \frac{1}{4\pi\epsilon_{0}}
    \sum_{\substack{i\in A \\ j\in B}}
    \frac{q_{i}q_{j}}
      {\left|\mathbf{R}+\mathbf{r}_{j}-\mathbf{r}_{i}\right|}
  \label{eq:lri-vlr-1}
\end{equation}
where $\epsilon_0$ is the vacuum permitivity. The so-called long-range or asymptotic region is such that the two distributions are very far away from each other, namely
\begin{equation}
  |\mathbf{R}| \gg |\mathbf{r}_i|,|\mathbf{r}_j|, \, \forall \,i,j.
  \label{eq:lri-cond-rij}
\end{equation}
We can express the distance in Eq.~\eqref{eq:lri-vlr-1} as
\begin{equation}
  |\mathbf{R} + \mathbf{r}_j - \mathbf{r}_i|
    = R \sqrt{1 - 2\frac{\mathbf{u}\cdot\mathbf{r}_{ij}}{R}
                + \frac{\mathbf{r}_{ij}^2}{R^2}},
  \label{eq:lri-rij-1}
\end{equation}
where $\mathbf{r}_{ij} = \mathbf{r}_{i} - \mathbf{r}_{j}$ and $r_{ij} = |\mathbf{r}_{ij}|$. Calling $\theta_{ij}$ the angle between $\mathbf{u}$ and $\mathbf{r}_{ij}$, we have $\mathbf{u} \cdot \mathbf{r}_{ij} = r_{ij} \cos \theta_{ij}$. Using the generatrix series of Legendre polynomials $P_\ell (\cos\theta_{ij})$, we rewrite the inverse of Eq.~\eqref{eq:lri-rij-1} as
\begin{equation}
  \frac{1}{|\mathbf{R} + \mathbf{r}_j - \mathbf{r}_i|}
    = \frac{1} {R\sqrt{ 1 - \frac{2r_{ij}\cos\theta_{ij}}{R}
      + \frac{r_{ij}^{2}} {R^{2}}}}
    = \sum_{\ell=0}^{+\infty} \frac{r_{ij}^{\ell}} {R^{\ell+1}}
      P_{\ell}(\cos\theta_{ij}) .
  \label{eq:lri-ser-leg}
\end{equation}
Equation \eqref{eq:lri-ser-leg} is valid for $r_{ij}<R$, which is always satisfied in view of condition \eqref{eq:lri-cond-rij}. It allows us to separate the contribution of the inter-partner distance $R$. But to go one step further, we also separate the coordinates of $q_i$ and $q_j$, by expressing Legendre polynomials in terms of normalized spherical harmonics, $P_{\ell}(\cos\theta_{ij}) = \sqrt{4\pi / (2\ell+1)} \times Y_{\ell0}(\theta_{ij},\phi_{ij})$, and by using the property
\begin{align}
  r_{ij}^{\ell}Y_{\ell m}(\theta_{ij},\phi_{ij}) & =  
   \sqrt{4\pi\left(2\ell+1\right)!}
   \sum_{\ell_{A},\ell_{B}=0}^{+\infty} 
   \delta_{\ell_{A}+\ell_{B},\ell} 
   \frac{\left(-1\right)^{\ell_{B}} 
         r_{i}^{\ell_{A}}r_{j}^{\ell_{B}}}
   {\sqrt{\left(2\ell_{A}+1\right)!\left(2\ell_{B}+1\right)!}}
  \nonumber \\
   & \times \sum_{m_{A}=-\ell_{A}}^{+\ell_{A}} 
   \sum_{m_{B}=-\ell_{B}}^{+\ell_{B}} 
   C_{\ell_{A}m_{A}\ell_{B}m_{B}}^{\ell m}
   Y_{\ell_{A}m_{A}}(\theta_{i},\phi_{i})
   Y_{\ell_{B}m_{B}}(\theta_{j},\phi_{j})
  \label{eq:lri-rij-2}
\end{align}
where the Kronecker symbol $\delta_{\ell_{A}+\ell_{B},\ell}$ imposes $\ell_{A}+\ell_{B} = \ell$, and the quantity $C_{\ell_{A}m_{A}\ell_{B}m_{B}}^{\ell m} = \langle \ell_{A}m_{A} \ell_{B}m_{B} | \ell_{A}\ell_{B} \ell m \rangle$ is a Clebsch-Gordan (CG) coefficient. We use the notation of Ref.~\cite{varshalovich1988}, in which the subscripts are the uncoupled angular momenta and the superscripts are the coupled ones. Here the lowercase polar angles $(\theta_i,\phi_i)$ and $(\theta_j,\phi_j)$ give the orientation of vectors $\mathbf{r}_i$ and $\mathbf{r}_j$ in the BF frame. The CG coefficient of Eq.~\eqref{eq:lri-rij-2} imposes $m=m_{A}+m_{B}$ for the spherical harmonics $Y_{\ell m}(\theta_{ij},\phi_{ij})$. Since $m=0$ in the present case, on can set $m_A = -m_B = m$.
Plugging Eqs.~\eqref{eq:lri-rij-1}--\eqref{eq:lri-rij-2} into Eq.~\eqref{eq:lri-vlr-1}, and using the algebraic expression of $C_{\ell_{A}m_{A} \ell_{B}m_{B}}^{ \ell_{A}+\ell_{B}, m_{A}+m_{B}}$ \cite{varshalovich1988}, we then obtain the usual multipolar expansion
\begin{equation}
  V_{AB}^{\mathrm{BF}}(R) = \frac{1}{4\pi\epsilon_{0}}
    \sum_{\ell_{A},\ell_{B}=0}^{+\infty} \sum_{m=-\ell_{<}}^{+\ell_{<}}
    \frac{f_{\ell_{A}\ell_{B}m}}{R^{1+\ell_{A}+\ell_{B}}}
    Q_{\ell_{A}m}^{\mathrm{BF}}(A)
    Q_{\ell_{B},-m}^{\mathrm{BF}}(B)
  \label{eq:lri-vbf}
\end{equation}
where $\ell_{<} = \min(\ell_A,\ell_B)$,
\begin{equation}
  f_{\ell_{A}\ell_{B}m} = \frac{\left(-1\right)^{\ell_{B}}
    \left(\ell_{A}+\ell_{B}\right)!}
    {\sqrt{\left(\ell_{A}+m\right)! \left(\ell_{A}-m\right)!
    \left(\ell_{B}+m\right)! \left(\ell_{B}-m\right)! }}
  \label{eq:lri-flm}
\end{equation}
and the quantities
\begin{equation}
  Q_{\ell_{A}m}^{\mathrm{BF}}(A) 
    = \sqrt{\frac{4\pi}{2\ell_{A}+1}} \sum_{i\in A}
    q_{i}r_{i}^{\ell_{A}} Y_{\ell_{A}m}(\theta_{i},\phi_{i}),
  \label{eq:lri-qlm-bf}
\end{equation}
and similarly for $Q_{\ell_{B},-m}^{\mathrm{BF}}(B)$, are the electric multipole moments of charge distributions $A$ and $B$. They are expressed in the BF frame, which means that they depend on the angles $\theta_{i,j}$ and $\phi_{i,j}$, and in the form of irreducible tensors of ranks $\ell_{A,B}$. As examples, $\ell_{A,B} = 0, 1$ and 2 correspond to the total charge, dipole moment and quadrupole moment.

\section{Body-fixed and space-fixed frames}
\label{sec:lriBFSF}

The calculations of the previous section are performed in the BF frame attached to the complex. Because the interaction energy between atoms or molecules merely depend on their relative coordinates, the BF frame is more {}``natural'': it captures the essential features of the interactions (for instance their attractive or repulsive character), and it allows for extrapolating the potential-energy curves or surfaces calculated using quantum chemistry (see Section \ref{sec:lri-o3}). However, ultracold gases are always submitted to external electromagnetic fields which can for instance be used to tune the interparticle interactions. Because the orientation and polarization of those fields are defined in the laboratory, the SF frame is also relevant for calculations with ultracold particles.

Let us note however that the terms {}``body-fixed'' and {}``space-fixed'' come from scattering theory. They are based on the fact that the complex rotates in the laboratory, where the end-over-end rotation is described in terms of partial waves. However, in current experiments with arrays of optical tweezers \cite{kaufman2021}, it is possible to put one particle per site. In this case, all interparticle axes are fixed in the laboratory, and so the distinction between BF and SF frames is irrelevant.

In this section, we consider the coordinate system $XYZ$ attached to the laboratory frame, in which the direction of the inter-partner axis $\mathbf{u}$ is given by the polar angles $(\Theta,\Phi)$. To calculate the interaction energy in the SF frame, we start with Eq.~\eqref{eq:lri-ser-leg}, and we use the addition theorem of spherical harmonics,
\begin{equation}
  P_{\ell}(\cos\theta_{ij}) = \frac{4\pi}{2\ell+1}
    \sum_{m=-\ell}^{+\ell}Y_{\ell m}^{*}(\Theta,\Phi)
    Y_{\ell m}(\Theta_{ij},\Phi_{ij})\,.
\end{equation}
where $(\Theta_{ij},\Phi_{ij})$ designate the orientation of the vector $\mathbf{r}_{ij}$ in the SF frame. We can apply Eq.~\eqref{eq:lri-rij-2} to the angles $(\Theta_{ij},\Phi_{ij})$, except that $m$ is not necessarily equal to zero. We get to the final expression
\begin{align}
  V_{AB}^{\mathrm{SF}}(\mathbf{R}) & = \frac{1}{4\pi\epsilon_{0}}
    \sum_{\ell_{A}\ell_{B}\ell=0}^{+\infty} 
    \delta_{\ell_{A}+\ell_{B},\ell}
    \,\frac{\left(-1\right)^{\ell_{B}}}{R^{1+\ell}}
     \binom{2\ell}{2\ell_A}^{1/2}
    \sum_{m=-\ell}^{+\ell}\sqrt{\frac{4\pi}{2\ell+1}}
      Y_{\ell m}^{*}(\Theta,\Phi) \nonumber \\
  & \times \sum_{m_{A}=-\ell_{A}}^{+\ell_{A}}
    \sum_{m_{B}=-\ell_{B}}^{+\ell_{B}}
      C_{\ell_{A}m_{A}\ell_{B}m_{B}}^{\ell m} 
      Q_{\ell_{A}m_{A}}^{\textrm{SF}}(A)
      Q_{\ell_{B}m_{B}}^{\textrm{SF}}(B) ,
  \label{eq:lri-vsf-1}
\end{align}
where $\binom{n}{p}$ is a binomial coefficient, and the SF-frame multipole moments $Q_{\ell_{A}m_{A}}^{\textrm{SF}}(A) $ and $Q_{\ell_{B}m_{B}}^{\textrm{SF}}(B)$ are obtained by replacing lowercase angles by uppercase ones in Eq.~\eqref{eq:lri-qlm-bf}. Starting from Eq.~\eqref{eq:lri-vsf-1}, we can retrieve the BF-frame potential energy \eqref{eq:lri-vbf} by setting $\Theta = \Phi = 0$, which imposes $m=0$ since $Y_{\ell m} (0,0) = \sqrt{(2\ell+1)/4\pi} \times \delta_{m0}$.

We can rewrite Eq.~\eqref{eq:lri-vsf-1} in a more compact way, by noting that its last line is the tensor product $\left\{ Q_{\ell_{A}}^{\textrm{SF}} \otimes Q_{\ell_{B}}^{\textrm{SF}} \right\}_{\ell m}$ of the multipole moments, which can itself be combined with the spherical harmonics as a scalar product $(\cdot)$ of operators. Introducing the Racah spherical harmonics $C_{\ell m} (\Theta,\Phi) = \sqrt{4\pi/(2\ell+1)} \times Y_{\ell m} (\Theta,\Phi)$, we obtain
\begin{equation}
  V_{AB}^{\mathrm{SF}}(\mathbf{R}) = \frac{1}{4\pi\epsilon_{0}}
    \sum_{\ell_A\ell_B\ell=0}^{+\infty} \delta_{\ell_A+\ell_B,\ell}
    \,\frac{\left(-1\right)^{\ell_{B}}}{R^{1+\ell}}
    \binom{2\ell}{2\ell_A}^{1/2}
    \left( C_{\ell} (\Theta,\Phi) 
    \cdot \left\{ Q_{\ell_{A}}^{\textrm{SF}} 
      \otimes Q_{\ell_{B}}^{\textrm{SF}}  
      \right\}_{\ell} \right) ,
  \label{eq:lri-vsf-2}
\end{equation}
which is very convenient for practical calculations with atoms and molecules, as it enables to use the relations given in Chapter 13 of Ref.~\cite{varshalovich1988}.

\paragraph{Magnetostatic interactions.}

In addition to the electrostatic interactions, two partners can also interact through magnetostatic forces, created by stationary current distributions. They result in a potential energy involving magnetic multipole moments $\mathrm{M}_{\ell_k m_k}$. This potential energy is obtained by replacing $1/\epsilon_0$ by $\mu_0$ and $\mathrm{Q}_{\ell_k m_k}^{\mathrm{BF/SF}}$ by $\mathrm{M}_{\ell_k m_k}^{\mathrm{BF/SF}}$ in Eqs.~\eqref{eq:lri-vbf}, \eqref{eq:lri-vsf-1} and \eqref{eq:lri-vsf-2}. The detailed calculations are presented for example in Ref.~\cite{gray1976b}.

In practice, magnetostatic terms are roughly $(1/\alpha)^2 \approx 10^4$ smaller than electrostatic ones, $\alpha$ being the fine-structure constant. Only the magnetic dipole-dipole interaction is likely to compete with electrostatic (say, the van der Waals) ones. The magnetic (dipole) moment of an atom is given by
\begin{equation}
  \mathbf{M} = -\mu_B ( g_L \mathbf{L} + g_S \mathbf{S} )
\end{equation}
where $\mu_B$ is the unsigned Bohr's magneton, $g_L \approx 1$ and $g_S = 2.0023$ are the orbital and spin gyromagnetic ratios, $\mathbf{L}$ and $\mathbf{S}$ are the dimensionless total orbital and spin angular momenta. In what follows, atomic magnetic moments will be expressed as $\mathbf{M} = -\mu_B g_J \mathbf{J}$, with $g_J$ the energy-level dependent Land\'e factor and $\mathbf{J} = \mathbf{L} + \mathbf{S}$ the total electronic angular momentum. In open-shell diatomic molecules, only the electronic spin term comes into play (see Sec.~\ref{sec:lri-o3}). As vectors, $\mathbf{M}$, $\mathbf{L}$, $\mathbf{S}$ and $\mathbf{J}$ are expressed as rank-1 irreducible tensors. Note finally that $\mathbf{M}$ can also possess a term proportional to the nuclear spin $\mathbf{I}$, but it is much weaker than the electronic contribution.

\section{Perturbation theory}
\label{sec:lriPert}

Besides the LR potential energy, the other crucial ingredient for practical calculations is the time-independent quantum-mechanical perturbation theory, assuming that the interparticle interaction energy is much smaller than the intraparticle ones, and thus than the energies of individual partners. This is basically true for electronic and vibrational degrees of freedom in deeply bound atoms and molecules, but rarely true for rotational or hyperfine degrees of freedom. In this case, particular attention should be paid to the definition of the unperturbed Hamiltonian. In our calculations, we have used the first- and second-order energy corrections, for nondegenerate and degenerate unperturbed states.

From now on, we apply the correspondence principle to the LR potential energies \eqref{eq:lri-vbf}, \eqref{eq:lri-vsf-1}, and \eqref{eq:lri-vsf-2}. They are transformed into quantum operators, for which we use non-italicized letters, \textit{e.g.}~$\mathrm{V}_{AB}^{\mathrm{BF/SF}}$ or $\mathrm{Q}_{\ell_{A,B}m_{A,B}}^{\textrm{BF/SF}}$. We use the time-independent quantum perturbation theory in which the Hamiltonian $\mathrm{H}$ of the complex is split into two terms, namely $\mathrm{H} = \mathrm{H}_0+\mathrm{V}$, where $\mathrm{H}_0$ is the unperturbed or zeroth-order Hamiltonian, whose eigenvalues $E_n^{(0)}$ and eigenvectors $|\Psi_n^{(0)}\rangle$ are assumed to be known, and where $\mathrm{V}$ is the perturbation, whose matrix elements are much smaller than those of $\mathrm{H}_0$. Because Condition \eqref{eq:lri-cond-rij} implies that the interpartner interactions are much weaker than the intrapartner ones, it seems logical that the perturbation operator $\mathrm{V}$ is, or at least contains, the long-range one. In this case, the unperturbed Hamiltonian is the sum of individual Hamiltonians, $\mathrm{H}_0 = \mathrm{H}_A + \mathrm{H}_B$, and so the the zeroth-order energies are the sum of individual ones, $E_n^{(0)} = E_n^{(0)}(A) + E_n^{(0)}(B)$. The zeroth-order eigenvectors are tensor products of individual ones, $|\Psi_n^{(0)}\rangle = |\Psi_n^{(0)}(A)\rangle \otimes |\Psi_n^{(0)}(B)\rangle$ (or linear combinations of tensor products accounting for symmetries, see Sec.~\ref{sec:lriSym}). In subsection \ref{sub:lriPertBF}, we present the general principles of LR perturbation theory in the BF frame, and then in subsection \ref{sub:lriPertSF}, we give the specificities of the SF frame.

\subsection{Body-fixed frame}
\label{sub:lriPertBF}

Without loss of generality, we can write the partner eigenvectors as $|\Psi_n^{(0)}(k)\rangle = |\beta_k J_k M_k\rangle$, $k=A,B$, where $J_k$ and $M_k$ are the quantum numbers representing the total angular momentum and its $z$-projection, and $\beta_k$ stands for all the other quantum numbers of partner $k$. In the absence of external fields, the individual energy $E_n^{(0)}(k)$ is $(2J_k+1)$-fold degenerate, and so the unperturbed one $E_n^{(0)}$ is $(2J_A+1) \times (2J_B+1)$-fold degenerate. In this case, the first-order energy corrections $E_n^{(1)}$ are obtained by diagonalizing the perturbation operator restricted to the subspace of degeneracy. Its matrix elements are functions of the electric-multipole matrix elements, which can be expressed using the Wigner-Eckart theorem
\begin{align}
  \left\langle \qnpr{\beta_k} \qnpr{J_k} \qnpr{M_k} \right| 
    \mathrm{Q}_{\ell_k m_k}^{\mathrm{BF/SF}}
    \left| \beta'_k J'_k M'_k \right\rangle
  & = \frac{C_{J'_k M'_k \ell_k m_k}^{J_k M_k}}{\sqrt{2J_k+1}}
    \left\langle \qnpr{\beta_k} \qnpr{J_k} \right\| 
    \mathrm{Q}_{\ell_k} \left\| \beta'_k J'_k \right\rangle
  \nonumber \\
  & = \left(-1\right)^{J_k-M_k}
    \thrj{J_k}{\ell_k}{J'_k}{-M_k}{m_k}{M'_k}
    \left\langle \beta_k J_k \right\| \mathrm{Q}_{\ell_k}
    \left\| \beta'_k J'_k \right\rangle ,
  \label{eq:lriWE}
\end{align}
where the quantity $(:::)$ is a Wigner 3-j symbol \cite{varshalovich1988}, and $\left\langle \beta_k J_k \right\| \mathrm{Q}_{\ell_k} \left\| \beta'_k J'_k \right\rangle$ is the $(\qnpr{M_k},M'_k)$- and frame-independent reduced matrix element.

In order to treat second-order corrections, we introduce the effective operator
\begin{equation}
  \mathrm{W} = -\sum_{p\neq n} \frac{\mathrm{V}
    |\Psi_p^{(0)}\rangle \langle\Psi_p^{(0)}|
    \mathrm{V}} {E_p^{(0)}-E_n^{(0)}} \,.
  \label{eq:lriPert2nd}
\end{equation}
The second-order energy corrections $E_n^{(2)}$ are obtained by diagonalizing its restriction to the same subspace of degeneracy as in the first-order case \cite{landau2013}. Note that higher-order corrections are sometimes considered in three-body interactions (see \textit{e.g.}~\cite{marinescu1997} and references therein), but it will not be the case in this manuscript.

By plugging Eq.~\eqref{eq:lriWE} into \eqref{eq:lri-vbf} and setting $|\Psi_{p}^{(0)}\rangle = |\beta''_A J''_A M''_A \beta''_B J''_B M''_B \rangle$, we obtain terms of the kind $\langle \beta_k J_k \qnpr{M_k} | \mathrm{Q}_{\ell_k m_k}^{\mathrm{BF}} | \beta''_k J''_k M''_k \rangle \langle \beta''_k J''_k M''_k | \mathrm{Q}_{\ell'_k m'_k}^{\mathrm{BF}} | \beta'_k J'_k M'_k \rangle$. Applying the same method as in Chapter \ref{chap:ddp} on polarizabilities, we introduce coupled tensors of ranks $k_A$ and $k_B$ on which the Wigner-Eckart theorem can be applied. The details of the calculations are given in paragraph~4.3.4.3 of Ref.~\cite{lepers2018}; we give here their final result:
\begin{align}
  & \left\langle \qnpr{\beta_A} \qnpr{J_A} \qnpr{M_A}
    \qnpr{\beta_B} \qnpr{J_B} \qnpr{M_B} 
    \right| \mathrm{W}_{AB}^{\mathrm{BF}}(R) \left|
    \beta'_A J'_A M'_A \beta'_B J'_B M'_B \right\rangle 
  \nonumber \\
  = & -\frac{1}{16\pi^2\epsilon_0^2}
    \sum_{\ell_{A}\ell_{B}} \sum_{\ell'_{A}\ell'_{B}}
    \frac{\left(-1\right)^{\ell_B+\ell'_B+J_A+J'_A+J_B+J'_B}}
         {R^{2+\ell_{A}+\ell_{B}+\ell'_{A}+\ell'_{B}}}
    \sqrt{\frac{\left(2\ell_{A}+2\ell_{B}+1\right)!
                \left(2\ell'_{A}+2\ell'_{B}+1\right)!}
               {\left(2\ell_{A}\right)!\left(2\ell_{B}\right)!
                \left(2\ell'_{A}\right)!\left(2\ell'_{B}\right)!}}
  \nonumber \\
  \times & \sum_{k_{A}k_{B}kq} \left(-1\right)^{k_{A}+k_{B}}
    \left(2k_{A}+1\right)\left(2k_{B}+1\right)
    C_{\ell_{A}+\ell_{B},0,\ell'_{A}+\ell'_{B},0}^{k0}
    C_{k_{A}qk_{B},-q}^{k0}
    \ninej{\ell_A       }{\ell'_A        }{k_A}
          {\ell_B       }{\ell'_B        }{k_B}
          {\ell_A+\ell_B}{\ell'_A+\ell'_B}{k}
  \nonumber \\
  \times & \sum_{\beta''_A J''_A} \sum_{\beta''_B J''_B}
    \frac{\left\langle \qnpr{\beta_A} \qnpr{J_A} \right\|
      \mathrm{Q}_{\ell_{A}}
      \left\| \beta''_A J''_A \right\rangle
      \left\langle \beta''_A J''_A \right\|
      \mathrm{Q}_{\ell'_{A}}
      \left\| \beta'_A J'_A \right\rangle
      \left\langle \qnpr{\beta_B} \qnpr{J_B} \right\|
      \mathrm{Q}_{\ell_{B}}
      \left\| \beta''_B J''_B \right\rangle 
      \left\langle \beta''_B J''_B \right\| 
      \mathrm{Q}_{\ell'_{B}}
      \left\| \beta'_B J'_B \right\rangle }
    {E_{\beta''_{A}J''_{A}}+E_{\beta''_{B}J''_{B}}
    -E_{\beta_{A}J_{A}}    -E_{\beta_{B}J_{B}}}
  \nonumber \\
  \times & \sixj{\ell_A}{\ell'_A}{k_A}{J'_A}{J_A}{J''_A}
    \sixj{\ell_B}{\ell'_B}{k_B}{J'_B}{J_B}{J''_B}
    \frac{C_{J'_{A}M'_{A}k_{A}q}^{J_{A}M_{A}}
      C_{J'_{B}M'_{B}k_{B},-q}^{J_{B}M_{B}}}
    {\sqrt{\left(2J_{A}+1\right)\left(2J_{B}+1\right)}},
  \label{eq:lriPert2Tens}
\end{align}
where $\{:::\}$ and $\{ \vdots \vdots \vdots \}$ are respectively 6-j and 9-j Wigner symbols \cite{varshalovich1988}. For the sake of generality, we write Eq.~\eqref{eq:lriPert2Tens} with different states in the bra and the ket, for example two hyperfine or rotational levels, but we assume that they have the same unperturbed energies, namely $E_{\beta'_A J'_A} = E_{\beta_A J_A}$ and $E_{\beta'_B J'_B} = E_{\beta_B J_B}$. Note that we dropped the superscript {}``(0)''.

For a given set of $(\ell_A, \ell'_A, \ell_B, \ell'_B)$ values, there are several possibilities for $(k_A, k_B, k, q)$ imposed by the Wigner symbols and CG coefficients: $|\ell_A-\ell'_A| \le k_A \le \ell_A+\ell'_A$, $|\ell_B-\ell'_B| \le k_B \le \ell_B+\ell'_B$, $|k_A-k_B| \le k \le k_A+k_B$ and $-k \le q \le +k$. The radial dependence is given by the $\ell$'s indices, while the tensorial part of the interaction is governed by the $k$'s indices. The most common example of second-order correction is the van der Waals or induced-dipole interaction, scaling as $R^{-6}$, for which $(\ell_A, \ell'_A, \ell_B, \ell'_B) = (1,1,1,1)$. The possible sets of $(k_A, k_B, k)$ values are: $(0,0,0)$, $(0,2,2)$, $(2,0,2)$, $(1,1,0)$, $(1,1,2)$, $(2,2,0)$, $(2,2,2)$ and $(2,2,4)$.
Moreover, the CG coefficients of the last line impose additional conditions on $k_A$ and $k_B$: $|J_A-J'_A| \le k_A \le J_A+J'_A$ and $|J_B-J'_B| \le k_B \le J_B+J'_B$, see subsection \ref{sec:lriLn} for an illustration. They also impose that the $z$-projection of the total angular momentum is conserved, $M_A + M_B = M'_A + M'_B$.

\subsection{Space-fixed frame}
\label{sub:lriPertSF}

In the SF frame, the coordinates of the particles, see Eqs.~\eqref{eq:lri-vsf-1} and \eqref{eq:lri-vsf-2}, are expressed with respect to the laboratory, in which external electromagnetic fields are exerted. Moreover, collisional events are observed by a person or an apparatus located in the laboratory, in which the BF frame moves. Because in this manuscript, we study interactions and collisions without external fields or with homogeneous ones, our systems are invariant with respect to the translation of the complex center of mass. Therefore, we only account for the rotational motion of the BF frame in the SF one. This motion is associated with the polar angles $(\Theta,\Phi)$, and on a quantum-mechanical point of view, it is described by the partial-wave quantum numbers $|L M_L\rangle$, such that $\langle \Theta, \Phi |L M_L\rangle = Y_{LM_L} (\Theta,\Phi)$. The quantum number $L$ can \textit{a priori} go to infinity; in practice convergence of the calculated potential energies or scattering observables must be investigated. The unperturbed eigenvectors of our perturbation theory are thus $|\beta_A J_A M_A \beta_B J_B M_B L M_L\rangle$, where $(M_A, M_B, M_L)$ now represent the angular-momentum projections on the SF-frame $Z$ axis.

In this basis, the angular part of the relative kinetic energy $\hbar^2 \mathbf{L}^2/2\mu R^2$ only possesses diagonal matrix elements equal to $\hbar^2 L(L+1) /2\mu R^2$, $\mu$ being the reduced mass of the complex. The latter are often included in the LR potential curves in the laboratory frame, even though they are usually small compared to the contribution of $V_{AB}^{\mathrm{SF}}(\mathbf{R})$. Recalling the matrix elements of the Racah spherical harmonics of Eq.~\eqref{eq:lri-vsf-2} \cite{varshalovich1988}
\begin{align}
  \left\langle L \qnpr{M_L} \right| C_{\ell m}^* (\Theta,\Phi)
    \left| L' M'_L \right\rangle = \left(-1\right)^m
    \sqrt{\frac{2L'+1}{2L+1}}
    C_{L'0\ell 0}^{L0} C_{L'M'_L\ell,-m}^{LM_L} ,
\end{align}
we obtain the matrix element of the LR potential-energy operator
\begin{align}
  & \left\langle \qnpr{\beta_A} \qnpr{J_A} \qnpr{M_A} 
    \qnpr{\beta_B} \qnpr{J_B} \qnpr{M_B} L \qnpr{M_L} 
    \right| \mathrm{V}_{AB}^{\mathrm{SF}}(R)
    \left| \beta'_A J'_A M'_A \beta'_B J'_B M'_B L' M'_L \right\rangle
  \nonumber \\
  = & \frac{1}{4\pi\epsilon_{0}} 
    \sum_{\ell_{A}\ell_{B}\ell} \delta_{\ell_{A}+\ell_{B},\ell}
    \,\frac{\left(-1\right)^{\ell_{B}}}{R^{1+\ell}}
    \binom{2\ell}{2\ell_A}^{1/2}
    \frac{\left\langle \qnpr{\beta_A} \qnpr{J_A} \right\| 
      \mathrm{Q}_{\ell_A} \left\| \beta'_A J'_A \right\rangle
      \left\langle \qnpr{\beta_B} \qnpr{J_B} \right\| 
      \mathrm{Q}_{\ell_B} \left\| \beta'_B J'_B \right\rangle
    }{\sqrt{(2J_A+1)(2J_B+1)}}
  \nonumber \\
  \times & \sqrt{\frac{2L'+1}{2L+1}} C_{L'0\ell 0}^{L0}
    \sum_{m_A m_B m} \left(-1\right)^m 
    C_{L'M'_L\ell,-m}^{LM_L}
    C_{\ell_{A}m_{A}\ell_{B}m_{B}}^{\ell m}
    C_{J'_A M'_A \ell_A m_A}^{J_A M_A}
    C_{J'_B M'_B \ell_B m_B}^{J_B M_B} \,,
  \label{eq:lriVsf3}
\end{align}
where the boundaries of the sums are the same as in Eq.~\eqref{eq:lri-vsf-1}. The CG coefficients of the last two lines imply: (i) $M_L = M'_L-m$,  (ii) $m = m_A+m_B$, (iii) $M_A = M'_A+m_A$, (iv) $M_B = M'_B+m_B$. After eliminating the lower-case indices, conditions (i)--(iv) give $M_A + M_B + M_L = M'_A + M'_B + M'_L$, meaning that the $Z$-projection of the total angular momentum is conserved.

If the $AB$ complex is not submitted to any field, the modulus of its total angular momentum $\mathbf{J} = \mathbf{J}_A + \mathbf{J}_B + \mathbf{L} = \mathbf{J}_{AB} + \mathbf{L}$ is also conserved. We thus introduce the fully-coupled basis state,
\begin{align}
  \left| \beta_A \beta_B; ((J_A J_B) J_{AB} L) J M \right\rangle
  & = \sum_{M_{AB}M_L} C_{J_{AB} M_{AB} L M_L}^{J M}
  \nonumber \\
  & \times \sum_{M_A M_B} C_{J_A M_A J_B M_B}^{J_{AB} M_{AB}}
    \left| \beta_A J_A M_A \beta_B J_B M_B L M_L \right\rangle ,
  \label{eq:lriCplBas}
\end{align}
where the parentheses, which will be omitted in what follows, symbolize the successive angular-momentum coupling steps. In this basis, the LR potential is diagonal in $J$ and $M$,
\begin{align}
  & \left\langle \qnpr{\beta_A} \qnpr{J_A} \qnpr{\beta_B} \qnpr{J_B} 
    \qnpr{J_{AB}} L J M \right| 
    \mathrm{V}_{AB}^{\mathrm{SF}}(\mathbf{R})
    \left| \beta'_A J'_A \beta'_B J'_B J'_{AB} L' J' M' \right\rangle
    = \frac{\delta_{JJ'} \delta_{MM'}}{4\pi\epsilon_{0}}
  \nonumber \\
  & \times
    \sum_{\ell_{A}\ell_{B}\ell} \delta_{\ell_{A}+\ell_{B},\ell}
    \, \frac{(-1)^{\ell_B+J'_{AB}+L+J}}{R^{1+\ell}}
    \binom{2\ell}{2\ell_A}^{1/2}
    \sqrt{(2\ell+1)(2J_{AB}+1)(2J'_{AB}+1)(2L'+1)}
  \nonumber \\
  & \times C_{L'0\ell 0}^{L0} 
    \ninej{J_A   }{J_B   }{J_{AB}}
          {J'_A  }{J'_B  }{J'_{AB}}
          {\ell_A}{\ell_B}{\ell}
    \sixj{J_{AB}}{L      }{J}
         {L'    }{J'_{AB}}{\ell}
    \left\langle \beta_{A}J_{A} \right\| \mathrm{Q}_{\ell_A}
    \left\| \beta'_A J'_A \right\rangle
    \left\langle \beta_B J_B \right\| \mathrm{Q}_{\ell_B}
    \left\| \beta'_B J'_B \right\rangle .
  \label{eq:lriVsf4}
\end{align}
The details are given in Appendix~\ref{sec:AppLriVsfCpl}.

Regarding the second-order correction, we can introduce an effective operator similar to Eq.~\eqref{eq:lriPert2nd}, namely
\begin{equation}
  \mathrm{W}_{AB}^{\mathrm{SF}} = 
    -\sum_{\beta''_A J''_A} \sum_{\beta''_B J''_B} \sum_{C''}
    \frac{\mathrm{V}_{AB}^{\mathrm{SF}}
      \left| \beta''_A J''_A \beta''_B J''_B C'' \right\rangle
      \left\langle \beta''_A J''_A \beta''_B J''_B C'' \right|
      \mathrm{V}_{AB}^{\mathrm{SF}}}
    {E_{\beta''_A J''_A} + E_{\beta''_B J''_B}
    -E_{\beta_A J_A} - E_{\beta_B J_B}} \,.
  \label{eq:lriPert2SF}
\end{equation}
where $C''$ is a collective label representing the quantum numbers of the complex except $\beta''_A$, $J''_A$, $\beta''_B$ and $J''_B$. It is general enough to be applicable in the coupled and uncoupled bases. In both cases, the unperturbed energies at the denominator of Eq.~\eqref{eq:lriPert2SF} are those of the separated partners, which means that they do not depend on the $C''$ quantum numbers. In the uncoupled basis, the matrix elements of the effective operator, calculated in paragraph 4.3.4.3 of Ref.~\cite{lepers2018}, are equal to
\begin{align}
  & \left\langle \qnpr{\beta_A} \qnpr{J_A} \qnpr{M_A}
    \qnpr{\beta_B} \qnpr{J_B} \qnpr{M_B} L \qnpr{M_L}
    \right| \mathrm{W}_{AB}^{\mathrm{SF}}(R) \left|
    \beta'_A J'_A M'_A \beta'_B J'_B M'_B L' M'_L \right\rangle
  \nonumber \\
  = & -\frac{1}{16\pi^2\epsilon_0^2}
    \sum_{\ell_A\ell_B\ell} \sum_{\ell'_A\ell'_B\ell'}
    \delta_{\ell_A+\ell_B,\ell} \delta_{\ell'_A+\ell'_B,\ell'}\,
    \frac{\left(-1\right)^{\ell_B+\ell'_B+J_A+J'_A+J_B+J'_B}}
      {R^{2+\ell+\ell'}}
    \sqrt{\frac{\left(2\ell+1\right)! \left(2\ell'+1\right)!}
      {\left(2\ell_{A} \right)! \left(2\ell_{B} \right)!
       \left(2\ell'_{A}\right)! \left(2\ell'_{B}\right)!}}
   \nonumber \\
  \times & \sum_{k_{A}k_{B}k}
    \left(-1\right)^{k_{A}+k_{B}}
    \left(2k_{A}+1\right)\left(2k_{B}+1\right)
    \ninej{\ell_A}{\ell'_A}{k_A}
          {\ell_B}{\ell'_B}{k_B}
          {\ell  }{\ell'  }{k}
    C_{\ell,0,\ell',0}^{k0} \sqrt{\frac{2L'+1}{2L+1}} C_{L'0k0}^{L0} 
  \nonumber \\
  \times & \sum_{\beta''_{A}J''_{A}} \sum_{\beta''_{B}J''_{B}}
    \frac{\left\langle \beta_{A}J_{A} \right\|
      \mathrm{Q}_{\ell_{A}}
      \left\| \beta''_{A}J''_{A} \right\rangle
      \left\langle \beta''_{A}J''_{A} \right\|
      \mathrm{Q}_{\ell'_{A}}
      \left\| \beta'_{A}J'_{A} \right\rangle
      \left\langle \beta_{B}J_{B} \right\|
      \mathrm{Q}_{\ell_{B}}
      \left\| \beta''_{B}J''_{B} \right\rangle 
      \left\langle \beta''_{B}J''_{B} \right\| 
      \mathrm{Q}_{\ell'_{B}}
      \left\| \beta'_{B}J'_{B} \right\rangle }
    {E_{\beta''_{A}J''_{A}}+E_{\beta''_{B}J''_{B}}
    -E_{\beta_{A}J_{A}}    -E_{\beta_{B}J_{B}}}
  \nonumber \\
  \times & \sixj{\ell_A}{\ell'_A}{k_A}{J'_A}{J_A}{J''_A}
    \sixj{\ell_B}{\ell'_B}{k_B}{J'_B}{J_B}{J''_B}
    \sum_{q_{A}q_{B}q} \left(-1\right)^q
    C_{L'M'_Lk,-q}^{LM_L} C_{k_Aq_Ak_Bq_B}^{kq}
    \frac{C_{J'_{A}M'_{A}k_{A} q_A}^{J_{A}M_{A}}
          C_{J'_{B}M'_{B}k_{B} q_B}^{J_{B}M_{B}}}
    {\sqrt{\left(2J_{A}+1\right)\left(2J_{B}+1\right)}},
\end{align}
The detailed calculations, as well as the expression in the fully-coupled basis in given in Appendix~\ref{sec:AppLriVsf2Cpl}.

\section{Interaction with external fields}
\label{sec:lriFld}

This question was addressed in Refs.~\cite{lepers2018, li2019}. We consider first a static electric field of amplitude $\mathcal{E}$, polarized along a direction given by the polar angles $(\theta_E,\phi_E)$ with respect to the $z$- or $Z$ axes. The response of the complex to the field is the sum of individual responses. The correspondind Stark operator is $\mathrm{V}_\mathrm{S} = -[ \mathbf{Q}_1(A) + \mathbf{Q}_1(B) ] \cdot \mathbf{E} = \sum_q [ \mathrm{Q}_{1q}(A) + \mathrm{Q}_{1q}(B) ] \mathrm{E}^*_q$, with $\mathrm{E}_q = C_{1q} (\theta_E,\phi_E) \mathcal{E}$.

The matrix elements of $\mathrm{V}_\mathrm{S}$ in the uncoupled basis are composed of two terms: in the first one, the quantum numbers of partner $B$ are unaffected, while in the second one, the quantum numbers of partner $A$ are unaffected. In both terms, partial-wave quantum numbers are also spectators. Therefore,
\begin{align}
  & \left\langle \qnpr{\beta_A} \qnpr{J_A} \qnpr{M_A} 
    \qnpr{\beta_B} \qnpr{J_B} \qnpr{M_B} L \qnpr{M_L} 
    \right| \mathrm{V}_\mathrm{S}^\mathrm{SF} \left| 
    \beta'_A J'_A M'_A \beta'_B J'_B M'_B L' M'_L \right\rangle
  = - \mathcal{E} \delta_{LL'} \delta_{\qnpr{M_L} M'_L}
  \nonumber \\
  & \times \sum_q C_{1q}^* (\theta_E,\phi_E)
    \left[ \delta_{\qnpr{\beta_B} \beta'_B}
    \delta_{\qnpr{J_B} J'_B} \delta_{\qnpr{M_B} M'_B}
    \frac{C_{J'_A M'_A 1 q}^{J_A M_A}}{\sqrt{2J_A+1}}
    \left\langle \qnpr{\beta_A} \qnpr{J_A} \right\| 
    \mathrm{Q}_1 \left\| \beta'_A J'_A \right\rangle
  \right. \nonumber \\
  & \left. \phantom{\times \sum_q C_{1q}^* (\theta_E,\phi_E)}
  + \delta_{\qnpr{\beta_A} \beta'_A}
    \delta_{\qnpr{J_A} J'_A} \delta_{\qnpr{M_A} M'_A}
    \frac{C_{J'_B M'_B 1 q}^{J_B M_B}}{\sqrt{2J_B+1}}
    \left\langle \qnpr{\beta_B} \qnpr{J_B} \right\| 
    \mathrm{Q}_1 \left\| \beta'_B J'_B \right\rangle
  \right]
  \label{eq:lriEFldUnc}
\end{align}
where we have applied the Wigner-Eckart theorem \eqref{eq:lriWE}. In an electric field, the individual levels coupled by the electric dipole moment have different parities. In the BF frame, $\mathrm{V}_\mathrm{S}^\mathrm{BF}$ has the same expression but without $L$, $L'$, $M_L$ and $M'_L$ quantum numbers.

Even if the Stark operator acts on individual quantum numbers separately, its matrix elements can still be written in the fully coupled basis by applying Eq.~\eqref{eq:lriCplBas} in the bra and the ket,
\begin{align}
  & \left\langle \qnpr{\beta_A} \qnpr{J_A} \qnpr{\beta_B} \qnpr{J_B} 
    \qnpr{J_{AB}} L J M \right| 
    \mathrm{V}_{\mathrm{S}}^{\mathrm{SF}}
    \left| \beta'_A J'_A \beta'_B J'_B J'_{AB} L' J' M' \right\rangle
  \nonumber \\
  = & -\mathcal{E} \delta_{LL'}
    \sqrt{ (2J_{AB}+1) (2J'_{AB}+1) (2J'+1) }
    \sixj{J_{AB}}{L}{J}{J'}{1}{J'_{AB}}
    \sum_q C_{1q}^* (\theta_E,\phi_E) C_{J'M'1q}^{JM}
  \nonumber \\
  \times & \left[
    \delta_{\qnpr{\beta_B}\beta'_B} \delta_{\qnpr{J_B}J'_B}
    (-1)^{J_A + J_B + J_{AB} + J'_{AB} + L + J'}
    \sixj{J_A}{J_B}{J_{AB}}{J'_{AB}}{1}{J'_A}
    \left\langle \qnpr{\beta_A} \qnpr{J_A} \right\| 
    \mathrm{Q}_1 \left\| \beta'_A J'_A \right\rangle
  \right. \nonumber \\
   & + \left.
    \delta_{\qnpr{\beta_A}\beta'_A} \delta_{\qnpr{J_A}J'_A}
    (-1)^{J_A + J_B - L - J'}
    \sixj{J_B}{J_A}{J_{AB}}{J'_{AB}}{1}{J'_B}
    \left\langle \qnpr{\beta_B} \qnpr{J_B} \right\| 
    \mathrm{Q}_1 \left\| \beta'_B J'_B \right\rangle
  \right] .
  \label{eq:lriEFldCpl}
\end{align}
Unlike the long-range operator, the Stark one couples total angular momenta such that $|J'-J| \le 1$. It also couples the intermediate ones $J_{AB}$ according to the same selection rule. Therefore, in order to calculate long-range potential-energy curves in the presence of an external field in the fully-coupled basis, it is necessary to test the convergence with respect to $J$.

The situation is very similar for a static magnetic field of amplitude $\mathcal{B}$ and with an arbitrary orientation defined by the angles $(\theta_B,\phi_B)$. The Zeeman operator is $\mathrm{V}_\mathrm{Z} = -[ \mathbf{\upmu}_1(A) + \mathbf{\upmu}_1(B) ] \cdot \mathbf{B} = \mathcal{B} \sum_q [ \upmu_{1q}(A) + \upmu_{1q}(B) ] C_{1q}^* (\theta_B,\phi_B)$. Equations \eqref{eq:lriEFldUnc} and \eqref{eq:lriEFldCpl} may be used after replacing the electric-field parameters by the magnetic-field ones. For an atomic level, the reduced matrix element of the magnetic moment is $\langle \qnpr{\beta_k} \qnpr{J_k} \| \upmu_1 \| \beta_k J_k \rangle = -\mu_B g_k \sqrt{J_k (J_k+1) (2J_k+1)}$.

\section{Symmetrized basis states}
\label{sec:lriSym}

In the previous section, we have seen that the LR potential preserves the SF $z$- or BF $Z$-projection of the total complex angular momentum, and even its modulus in the SF frame. On a practical point of view, this allows to write a large Hamiltonian in a block-diagonal form, and so to diagonalize several smaller matrices instead of a large one, which is computationally less costly. In this section, we go further into this process, by discussing the effect of additional symmetries, with respect to inversion, reflection, parity and permutation operations, in a similar way to diatomic molecules \cite{brown2003, herzberg2013}. We do that firstly in the BF frame, and secondly in the SF frame.

\subsection{Body-fixed frame}

The symmetries in pairs of open-shell atoms are thoroughly discussed in Ref.~\cite{chang1967}. They give raise to the so-called Wigner-Witmer rules discussed for instance in Herzberg's book on diatomic molecules \cite{herzberg2013}. Here we extend the discussion to complexes of atoms or diatoms. To do so, in the general description of partners $k=A$ and $B$, we add the parity $p_k$ with respect to the inversion of all coordinates around the partner center of mass. The BF unperturbed states are thus $|\beta_A p_A J_A M_A \beta_B p_B J_B M_B \rangle$.

\paragraph{Reflection symmetry.}

We consider the operator $\upsigma_{xz}$ representing the reflection of all coordinates through the $xz$ plane. This operation can be decomposed into an inversion followed by a rotation of $\pi$ radians around the $y$ axis \cite{chang1967, varshalovich1988}. The action of the reflection operator can therefore be written as
\begin{align}
  \upsigma_{xz} |\beta_A p_A J_A M_A \beta_B p_B J_B M_B \rangle
    & = \left(-1\right)^{J_A-M_A+J_B-M_B} p_A p_B
    |\beta_A p_A J_A, -M_A \beta_B p_B J_B, -M_B \rangle ,
\end{align}
and so for $M_A + M_B = 0$, the states can be separated into even and odd ones ($\sigma = \pm 1$) according to
\begin{align}
  & \left|\beta_A p_A J_A M_A \beta_B p_B J_B, -M_A; 
    \sigma = \pm 1 \right\rangle
  \nonumber \\
  & = \frac{1}{\sqrt{2(1+\delta_{M_A,0})}}
    \left[ \left|\beta_Ap_A J_AM_A \beta_Bp_B J_B,-M_A \right\rangle
  \right. \nonumber \\
  & \phantom{=\sqrt{2(1+\delta_{M_A,0})}} \left.
    \pm (-1)^{J_A+J_B} p_A p_B
    \left| \beta_Ap_A J_A,-M_A \beta_Bp_B J_BM_A \right\rangle \right].
  \label{eq:lriSymBFRefl}
\end{align}
The normalization factor is $1/\sqrt{2}$ for $M_A \neq 0$; for $M_A=0$, the state $|\beta_A p_A J_A,0 \beta_B p_B J_B,0 \rangle$ is of sign $\sigma = (-1)^{J_A+J_B} p_A p_B$.

\paragraph{Inversion symmetry.}

For identical partners (but not necessarily in the same state), the inversion operator $\mathrm{i}$ acts as
\begin{align}
  \mathrm{i} |\beta_A p_A J_A M_A \beta_B p_B J_B M_B \rangle
    & = \eta p_A p_B
    |\beta_B p_B J_B M_B \beta_A p_A J_A M_A \rangle ,
\end{align}
where $\eta = \pm 1$ for identical bosons/fermions, and so the symmetrized states are similar to $g/u$ states in diatomic molecules
\begin{align}
  & \left|\beta_A p_A J_A M_A \beta_B p_B J_B M_B; 
    \epsilon = \pm \right\rangle
  \nonumber \\
  & = \frac{1}{\sqrt{2(1+\delta_{AB})}}
    \left[ \left|\beta_Ap_A J_AM_A \beta_Bp_B J_B M_B \right\rangle
  \right. \nonumber \\
  & \phantom{=\sqrt{2(1+\delta_{AB})}} \left.
    \pm \eta p_A p_B
    \left| \beta_Bp_B J_BM_B \beta_Ap_A J_A M_A \right\rangle \right],
  \label{eq:lriSymBFInv}
\end{align}
where the Kronecker symbol is $\delta_{AB} = 1$ if the two partners are in the same sublevel, in which case $\epsilon = \eta$ is the only possibility since $p_A = p_B$.

Finally, for identical partners with $M_A = -M_B$, one needs to combine the two symmetries. The fully symmetrized basis states are obtained as
\begin{align}
  \left|\beta_A p_A J_A M_A \beta_B p_B J_B, -M_A; 
    \epsilon \sigma \right\rangle
    \propto \left( \mathrm{I} + \epsilon \mathrm{i} \right)
    \left( \mathrm{I} + \sigma \upsigma_{xz} \right)
    \left|\beta_Ap_A J_AM_A \beta_Bp_B J_B, -M_A \right\rangle
\end{align}
where $\mathrm{I}$ is the identity matrix and the proportionality symbol means that the proper normalization factors of Eqs.~\eqref  {eq:lriSymBFRefl} and \eqref  {eq:lriSymBFInv} should be applied.

\subsection{Space-fixed frame}

\paragraph{Parity operation.}

We discussed in details symmetrized basis states in Ref.~\cite{li2019}. Firstly, the parity operator $\mathrm{E}^*$ is obtained by changing the sign of all electronic and nuclear coordinates. Similarly to $+/-$ parity states of diatomic molecules, this operator commutes with the long-range one for any complex, even with different partners. In the uncoupled and coupled bases, a given state possesses a given parity equal to $p_A p_B (-1)^L$.

\paragraph{Reflection symmetry.}

We consider the reflection operation $\sigma_{XZ}$ through the $XZ$ plane of the SF frame. This operation can be decomposed into an inversion (parity) operation discussed in the previous paragraph, followed by a rotation of $\pi$ radians around the $Y$ axis. In the uncoupled basis, the operation is applied on each partner and on the inter-partner axis separately,
\begin{align}
  & \upsigma_{XZ} |\beta_Ap_A J_A M_A \beta_Bp_B J_B M_B L M_L \rangle
  \nonumber \\
  & = p_A \left(-1\right)^{J_A-M_A} p_B
    \left(-1\right)^{J_B-M_B} (-1)^L (-1)^{L-M_L}
    |\beta_A p_A J_A, -M_A \beta_B p_B J_B, -M_B L, -M_L \rangle
  \nonumber \\
  & = p_A p_B \left(-1\right)^{J_A+J_B-M_A-M_B-M_L}
    |\beta_A p_A J_A, -M_A \beta_B p_B J_B, -M_B L, -M_L \rangle .
\end{align}
For $M_A + M_B + M_L = M = 0$, even and odd symmetrized states corresponding to $\sigma = \pm 1$ are constructed by applying $\mathrm{I} + \sigma \upsigma_{XZ}$ to the bare states,
\begin{align}
  & \left|\beta_A p_A J_A M_A \beta_B p_B J_B M_B L M_L; 
    \sigma = \pm 1 \right\rangle
  \nonumber \\
  & = \frac{1}{\sqrt{2(1+\delta_{M_A,0}\delta_{M_B,0})}}
    \left[ \left|\beta_Ap_A J_AM_A \beta_Bp_B J_BM_B LM_L \right\rangle
  \right. \nonumber \\
  & \phantom{=\sqrt{2(1+\delta_{M_A,0}\delta_{M_B,0})}} \left.
    \pm (-1)^{J_A+J_B} p_A p_B
    \left| \beta_Ap_A J_A,-M_A \beta_Bp_B J_B,-M_B
    L, -M_L \right\rangle \right] .
  \label{eq:lriSymSFRefl}
\end{align}
In the fully coupled basis, a given $M=0$ state has a well defined even or odd character corresponding to the value of $p_A p_B (-1)^{L+J}$, since the rotation around the $Y$ axis is applied to the total angular momentum of the complex.

\paragraph{Permutation symmetry.}

For identical partners (not necessarily in the same state), the complex states must be either symmetric ($\eta = 1$) or anti-symmetric ($\eta = -1$) with respect to permutation, whether the partners are bosons or fermions, respectively. In the uncoupled basis the action of the permutation operator $\mathrm{P}_{AB}$ is
\begin{align}
  & \mathrm{P}_{AB} |\beta_A p_A J_A M_A 
    \beta_B p_B J_B M_B L M_L \rangle
    = (-1)^L | \beta_B p_B J_B M_B
    \beta_A p_A J_A M_A L M_L \rangle .
\end{align}
The symmetrized basis states are therefore
\begin{align}
  & \left|\beta_A p_A J_A M_A \beta_B p_B J_B M_B; 
    \eta = \pm 1 \right\rangle
  \nonumber \\
  & = \frac{1}{\sqrt{2(1+\delta_{AB})}}
    \left[ \left|\beta_A p_A J_A M_A
    \beta_B p_B J_B M_B L M_L \right\rangle
  \right. \nonumber \\
  & \phantom{=\sqrt{2(1+\delta_{AB})}} \left.
    \pm (-1)^L \left| \beta_B p_B J_B M_B
    \beta_A p_A J_A M_A L M_L \right \rangle \right],
  \label{eq:lriSymSFPerm}
\end{align}
where the Kronecker symbol is $\delta_{AB} = 1$ if the two partners are in the same sublevel, and 0 otherwise. In the former case, one finds the well-known result that identical bosons (resp.~fermions) in the same sublevel only collide in even (resp.~odd) partial waves.

As demonstrated in Appendix A of Ref.~\cite{li2019}, in the fully-coupled basis, the action of the permutation operator is
\begin{align}
  & \mathrm{P}_{AB} |\beta_A p_A J_A  
    \beta_B p_B J_B J_{AB} L J M \rangle
   = (-1)^{J_A+J_B-J_{AB}+L} | \beta_B p_B J_B  
    \beta_A p_A J_A J_{AB} L J M \rangle ,
\end{align}
which gives the symmetrized basis states
\begin{align}
  & \left| \beta_A p_A J_A \beta_B p_B J_B
    J_{AB} L J M; \eta = \pm 1 \right\rangle
  \nonumber \\
  & = \frac{1}{\sqrt{2(1+\delta_{AB})}}
    \left[ \left| \beta_A p_A J_A  
    \beta_B p_B J_B J_{AB} L J M \right\rangle
  \right. \nonumber \\
  & \phantom{=\sqrt{2(1+\delta_{AB})}} \left.
    \pm (-1)^{J_A+J_B-J_{AB}+L} \left| \beta_B p_B J_B 
    \beta_A p_A J_A J_{AB} L J M \right \rangle \right].
  \label{eq:lriSymSFPerm2}
\end{align}
If the partners are in a stretched (or polarized) sublevel, $M_A = \pm J_A$ and $M_B = \pm J_B$, then necessarily $M_{AB} = \pm (J_A + J_B)$ and $J_{AB} = J_A + J_B$. The phase factor of Eq.~\eqref{eq:lriSymSFPerm2} is $(-1)^{J_A+J_B-J_{AB}+L} = (-1)^L$, and so we retrieve that identical bosons (resp.~fermions) then collide in even (resp.~odd) partial waves. For non-stretched sublevels, this rule does not come out so obviously.

\, 

In this chapter, I have recalled the tools that are employed in the next chapters, to compute long-range interactions with various atomic and molecular systems an in various conditions. In particular, I have made the distinction between the body-fixed and the space-fixed frames, as well as between uncoupled and fully coupled and symmetrized basis states. The matrix elements of the long-range potential, at the first and second orders of perturbation theory, are given in both frames. The matrix elements of the Stark and Zeeman Hamiltonians are also given.

Because ultracold gases are very dilute, the formalism of spherical tensor operators is very well adapted to long-range interactions inside them. The long-range energy comprises a radial part, scaling as inverse powers of the interpartner distance $R$, the powers depending on the multipole moments at play. The long-range energy also comprises an {}``angular'' part, responsible for anisotropic interactions coupling different angular-momentum states of the individual partners. In the first-order correction, the angular part is directly given by the ranks of the partners' multipole moments, whereas in the second-order correction, it is associated with coupled tensor operators, whose introduction is crucial to characterize the anisotropy of \textit{e.g.}~the van der Waals interaction. In both cases, it imposes strong selection rules, that can help us determine if a given term of the multipolar expansion or field-partner interaction is zero or not.

In consequence, a strong highlight is put on angular-momentum theory and Racah algebra. By contrast, little is said about the reduced matrix elements of the partners. We will see in the next chapters that they can be found in the literature, as experimental data (for instance permanent electric dipole moments) or as the results of our calculations or calculations from other groups of theoreticians.


\chapter{Long-range interactions in the body-fixed frame}
\label{chap:lriBF}

In the previous chapter, I have described various aspects of long-range interactions allowing for their practical calculations. I have discussed in particular their expression in the body- and in the space-fixed frames. The body-fixed frame seems more obvious or natural, since interpartner energy is function of their relative distances and angles. That is why, I have used this frame in my first investigations on long-range interactions.

The aim of this chapter is to trace the history of those investigations, and to highlight the progression of the ideas toward more a complete account for high-order terms, tensor operators, fine and hyperfine structures, symmetries and external fields. With the exception of ozone in Sec.~\ref{sec:lri-o3}, all the considered systems belong to the field of ultracold gases: atom-homonuclear molecule photoassociation in Sec.~\ref{sec:lri-pa}, heteronuclear molecules submitted to a static electric field in Sec.~\ref{sec:lriDiat}, and pairs of lanthanides and close shell atoms in Sec.~\ref{sec:lriLn}.

\section{Photoassociation of an atom and a diatomic molecule}
\label{sec:lri-pa}

Using the formalism of Chapter \ref{chap:lriPres}, we have investigated, in a series of five articles, the long-range interactions between a homonuclear diatomic molecule in its lowest electronic and vibrational level and an excited alkali-metal atom. The objective of those calculations was to determine the feasibility of ultracold trimer photoassociation (PA), which consists in binding a pair of colliding partners (here a molecule and an atom) into a bound level of the complex (here a triatomic molecule) after a photon absorption. PA is an interesting example of ultracold chemical reaction controlled by laser, and it opens the possibility to study ultracold few-body systems \cite{jones2006, ulmanis2012}. In 2009, it had been observed with homo- and heteronuclear alkali-metal diatomic molecules. Chronologically, it was the first route to produce samples of ground-state molecules \cite{deiglmayr2008}, in particular Cs$_2$ at Laboratoire Aim\'e Cotton (LAC) \cite{fioretti1998, viteau2008}. That is why, when I started my post-doc at LAC, I focused on the system Cs$_2$-Cs. More recently, atom-molecule, and even molecule-molecule PA has been numerically investigated with heteronuclear alkali diatoms \cite{gacesa2021, schnabel2021, elkamshishy2022, shammout2023}. It has also been observed experimentally \cite{cao2024}, as well as trimer formation via Feshbach resonances \cite{yang2022a, yang2022b}. Association of Efimov trimers \cite{ferlaino2011} using radiofrequency fields can also be considered as a PA process \cite{lompe2010, nakajima2011, tscherbul2011}.

%
%

\begin{figure}
  \centering
  \includegraphics[width=0.6\textwidth]
    {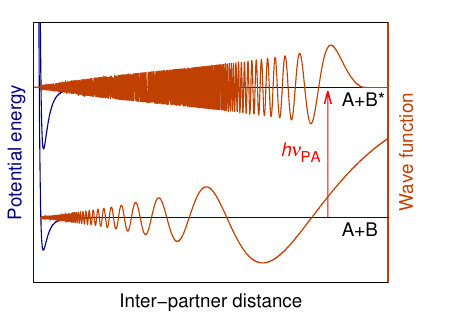}
  \caption{Schematic representation of photoassociation and the role of long-range interactions. Potential-energy curves correlated to the ground and excited dissociation limits are drawn in navy blue. For the lowest curve, a scattering wave function of low relative kinetic energy is drawn in dark orange, showing a large probability density far outside the potential well. For the excited curve, the wave function of a bound vibrational level is drawn. It is also associated with a large probability density at large distances, around the outer turning point of the potential curve. The photoassociation laser is represented, with a frequency red-detuned with respect to the dissociation limit, in order to reach the excited vibrational level.}
  \label{fig:lri-pa-schm}
\end{figure}

Because the wave function of ultracold colliding partners is mainly located in the long-range region of inter-partner distances, the wave function of the bound vibrational level should also be maximal in that region, due to the Franck-Condon principle, see Fig.~\ref{fig:lri-pa-schm}. Therefore, this vibrational level should be close to an excited dissociation threshold A+B*, where B* is reachable by electric-dipole transition from B ground state. In consequence, in our articles, we have computed in a step-by-step approach the long-range interactions between Cs$_2$ in the lowest vibrational level $v=0$ and a few rotational levels of the ground electronic state X$^1\Sigma_g^+$ and Cs in the first excited states $6p\,{}^2P_{1/2,3/2}^\circ$. Subsections \ref{sub:lri-cs3-quad} and \ref{sub:lri-cs3-dip} are respectively devoted to the calculation of the quadrupole-quadrupole and van der Waals interactions. They induce couplings between molecular rotational levels, which are investigated in Subsection \ref{sub:lri-cs3-nonpert}. Up to that point, the atom is considered without fine structure, namely $6p\,{}^2P^\circ$; the latter is also included in Subsection \ref{sub:lri-cs3-nonpert}. Finally in Subsection \ref{sub:lri-pa-rate}, the resulting LR curves are used to compute the PA rate in experimentally realistic conditions. Note that the hyperfine structures of Cs$_2$ and Cs have not been included in the model. Finally, even if many PA observations involved long-range vibrational levels, a so-called short-range PA was also reported, towards deeper levels of the excited electronic state \cite{gabbanini2011, fioretti2013, blasing2016, carollo2016, shimasaki2018}.

\subsection{First-order quadrupole-quadrupole interaction}
\label{sub:lri-cs3-quad}

In a first step \cite{lepers2010}, we calculated the quadrupole-quadrupole interaction energy given by $\ell_A = \ell_B = 2$ in Eq.~\eqref{eq:lri-vbf}, hence scaling as $R^{-5}$, using the degenerate first-order perturbation theory. Indeed, both Cs$_2$ and Cs* possess a large quadrupole moment. Namely, for $\text{A} = \text{Cs}_2$, one has
\begin{equation}
  \left\langle X, v_A=0, J_A, M'_{J,A} 
    \left| \mathrm{Q}_{2m}^\mathrm{BF} 
    \right| X, v_A=0, J_A, M_{J,A} \right\rangle 
    = C_{J_A M_{J,A} 2 m}^{J_A M'_{J,A}} 
     C_{J_A 0 2 0}^{J_A 0} \, q_{X,v_A=0}
  \label{eq:lri-cs2-quad}
\end{equation}
where $q_{X,v_A=0}$ is the $zz$ component of the traceless quadrupole tensor in the dimer coordinate system. We estimated it at 18.6 atomic units (a.u.) using the Gaussian software \cite{gaussian03}. For $\text{B} = \text{Cs}$,
\begin{equation}
  \left\langle 6p\,^2P^\circ, M'_{L,B} \left| 
    \mathrm{Q}_{2,-m}^\mathrm{BF} 
    \right| 6p\,^2P^\circ, M_{L,B} \right\rangle 
    = -C_{1 M_{L,B}, 2, -m}^{1 M'_{L,B}} C_{1 0 2 0}^{1 0} 
    \, e \langle r_{6p}^2 \rangle
  \label{eq:lri-cs-quad}
\end{equation}
where $\langle r_{6p}^2 \rangle = \int_0^{+\infty} dr r^2 [P_{6p}(r)]^2$ is the mean squared radius of the $6p$ orbital of cesium. We estimate it at 62.7~a.u. using the HFR method of Cowan \cite{cowan1981}. The unperturbed energies are given by the rotational structure of Cs$_2$, $E_n^{(0)} = B_0 J_A(J_A+1)$, with $B_0 = 1.173 \times 10^{-2}$~cm$^{-1}$ is the Cs$_2$ rotational constant in the lowest vibrational level \cite{amiot2002}.

Plugging Eqs.~\eqref{eq:lri-cs2-quad} and \eqref{eq:lri-cs-quad} into \eqref{eq:lri-vbf}, we see the central and similar roles played by the dimer rotational angular momentum $J_A$ and its projection $M_{J,A}$ on the inter-partner axis $\mathbf{u}$ on the one hand, and the atomic orbital angular momentum $L_B=1$ and its projection $M_{L,B}$. For each $J_A$ value, we diagonalize $V_{AB}^\text{BF}$ in the subspace spanned by $M_{J,A}$ and $M_{L,B}$. Because the quadrupole-moment operators couple different values of $M_{J,A}$ and $M_{L,B}$ such that $M_{J,A} + M_{L,B} = M'_{J,A} + M'_{L,B} = M$, we labeled the resulting eigenvectors using the diatomic-like symmetries $\Sigma^\pm$, $\Pi$, $\Delta$, etc. for $|M| = 0$, 1, 2 respectively. We obtained long-range potential energy curves (LR PECs) of the form $B_0 J_A(J_A+1) + C_5/R^5$, which are attractive or repulsive depending on the sign of the $C_5$ coefficient. For $R \le 100$~a.u., LR PECs correlated to different $J_A$ asymptotes cross each other, meaning that the quadrupolar and rotational energies are of similar magnitude, and thus that the formation of our perturbative calculation should be modified for $R_\text{LR} \le R \le 100$~a.u., where $R_\text{LR} \equiv 45$~a.u. according to our estimate.

\subsection{Second-order dipole-dipole interaction}
\label{sub:lri-cs3-dip}

The next term of the multipolar expansion scales as $R^{-6}$ \cite{lepers2011a}. It comes from the second-order dipole-dipole interaction, also called induced-dipole or van der Waals interaction. This term is always present in atoms and molecules which are by essence polarizable. For each first-order eigenvector of the previous subsection, written as
\begin{equation}
  \left| \Psi_n^{(0)} \right\rangle =
    \sum_{ \substack{ M_{J,A} M_{L,B} \\ M_{J,A}+M_{L,B}=M} } 
    c_{M_{J,A} M_{L,B}} \left| M_{J,A} M_{L,B} \right\rangle ,
\end{equation}
we calculated in Ref.~\cite{lepers2011a} the $C_6$ coefficient as
\begin{equation}
  C_6 = \sum_{M_{J,A} M_{L,B}} \sum_{M'_{J,A} M'_{L,B}}
    c_{M_{J,A} M_{L,B}} c_{M'_{J,A} M'_{L,B}}
    \left\langle \qnpr{M_{J,A}} \qnpr{M_{L,B}} \right| \text{C}_6
    \left| M'_{J,A} M'_{L,B} \right\rangle
\end{equation}
where
\begin{align}
  & \left\langle \qnpr{M_{J,A}} \qnpr{M_{L,B}} \right| \text{C}_6
    \left| M'_{J,A} M'_{L,B} \right\rangle
    \, = \, -4 \sum_{a,b} \frac{1}{\Delta E_a^{(0)} + \Delta E_b^{(0)}}
  \nonumber \\
  & \quad \times \sum_{m=-1}^{+1} 
    \frac{
      \left\langle X, v_A=0, J_A, M_{J,A}^{\phantom{(}} \right|
      \mathrm{Q}_{1m}^\mathrm{BF} \left| \Psi_a^{(0)} \right\rangle
      \left\langle 6p\,^2P^\circ, M_{L,B}^{\phantom{(0)}} \right| 
      \mathrm{Q}_{1,-m}^\mathrm{BF} \left| \Psi_b^{(0)} \right\rangle
    } {(1+m)!(1-m)!}
  \nonumber \\
  & \quad \times \sum_{m'=-1}^{+1}
    \frac{
      \left\langle \Psi_a^{(0)} \right| \mathrm{Q}_{1m'}^\mathrm{BF} 
      \left| X, v_A=0, J_A, M_{J,A}^{\prime\phantom{(}} \right\rangle
      \left\langle \Psi_b^{(0)} \right| \mathrm{Q}_{1,-m'}^\mathrm{BF}
      \left| 6p\,^2P^\circ, M_{L,B}'^{\phantom{(0)}} \right\rangle 
    } {(1+m')!(1-m')!}
  \label{eq:lri-c6cr}
\end{align}
where $\Delta E_{a,b}^{(0)}$ are the excitation energies of partners A and B.

In order to separate the contributions of the two partners, we rewrote the term $(\Delta E_{a}^{(0)} + \Delta E_{b}^{(0)})^{-1}$ using the identity
\begin{equation}
  \frac{1}{a+b} = \frac{2}{\pi} \int_{0}^{+\infty} du
    \frac{ab}{\left(a^{2}+u^{2}\right)\left(b^{2}+u^{2}\right)},
    \quad a,b > 0
  \label{eq:lri-resid}
\end{equation}
that we applied to $a = \Delta E_{a}^{(0)}$ and $b = \Delta E_{b}^{(0)}$. Because the atom $B$ is in its first-excited state, Eq.~\eqref{eq:lri-resid} cannot be used for the transition leading to the ground state ($6s\,^2S$ for Cs). In this case, setting $b<0$, we write
\begin{equation}
  \frac{1}{a+b} = \frac{1}{\left|a\right|-\left|b\right|} 
   = -\frac{ \left|a\right| + \left|b\right| }
      { \left(\left|a\right|+\left|b\right|\right)
        \left(\left|b\right|-\left|a\right|\right)}
   = - \frac{1}{\left|a\right|+\left|b\right|}
     - \frac{2a}{b^{2}-a^{2}} ,\,a>0,b<0 .
  \label{eq:lri-resid-2}
\end{equation}
The first term of the right-hand side can be rewritten using equation (\ref{eq:lri-resid}), with the numerator of the integrand $-|a||b| = ab$. Plugging Eqs.~\eqref{eq:lri-resid} and \eqref{eq:lri-resid-2} into Eq.~\eqref{eq:lri-c6cr}, we obtained expressions of the dynamic dipole polarizabilities $\upalpha_{1m1m'}$ as in Chapter~\ref{chap:ddp}, but as if they were calculated at so-called imaginary frequencies $iu$. 
As for the last term of Eq.~\eqref{eq:lri-resid-2}, $-2a/(b^2-a^2) = 2a/(a^2-b^2)$, it results in the dynamic polarizability of partner $A$, taken at the real frequency of downwards transitions of partner $B$ (like $6p\,^2P \rightarrow 6s\,^2S$).
Finally, Eq.~\eqref{eq:lri-c6cr} reads
\begin{align}
  & \left\langle M_{J,A} M_{L,B} \right| \text{C}_6
    \left| M'_{J,A} M'_{L,B} \right\rangle
    \, = \, -4 \sum_{m,m'=-1}^{+1} \frac{1}{(1+m)!(1-m)!(1+m')!(1-m')!}
  \nonumber \\
  & \quad \times \left[ \frac{1}{2\pi} \int_{0}^{+\infty} du
    \left\langle X, v_A=0, J_A, \qnpr{M_{J,A}} \right|
    \upalpha_{1m1m'}^\mathrm{BF}(iu)
    \left| X, v_A=0, J_A, M'_{J,A} \right\rangle
      \right.
  \nonumber \\
  & \left. \quad \quad \quad \times
    \left\langle 6p\,^2P^\circ, M_{L,B}^{\phantom{(0)}} \right| 
    \upalpha_{1,-m,1,-m'}^\mathrm{BF}(iu)
    \left| 6p\,^2P^\circ, M_{L,B}'^{\phantom{(0)}} \right\rangle 
      \right.
  \nonumber \\
  & \left. \quad \quad + \sum_{b<0}
    \left\langle X, v_A=0, J_A, \qnpr{M_{J,A}} \right|
    \upalpha_{1m1m'}^\mathrm{BF}(\Delta E_b^{(0)})
    \left| X, v_A=0, J_A, M'_{J,A} \right\rangle
      \right.
  \nonumber \\
  & \left. \phantom{\int} \quad \quad \times
    \left\langle 6p\,^2P^\circ, M_{L,B}^{\phantom{(0)}} \right| 
    \mathrm{Q}_{1,-m}^\mathrm{BF} \left| \Psi_b^{(0)} \right\rangle
    \left\langle \Psi_b^{(0)} \right| \mathrm{Q}_{1,-m'}^\mathrm{BF}
    \left| 6p\,^2P^\circ, M_{L,B}'^{\phantom{(0)}} \right\rangle 
    \right]
  \label{eq:lri-c6cr-2}
\end{align}
where the symbol $\sum_{b<0}$ means a sum on transitions with $\Delta E_b^{(0)} < 0$, and where the polarizabilities are calculated with transition dipole moments (TDMs) expressed in the BF frame.

Equation \eqref{eq:lri-c6cr-2} requires a significant set of transition energies and TDMs of both partners. For the Cs atom, we took experimental energies from the NIST database \cite{NIST-ASD} and computed TDMs from Ref.~\cite{iskrenova2007}, that we averaged over fine-structure manifolds. For Cs$_2$, the vibrational transition energies and TDMs were calculated at LAC by R.~Vexiau, using experimental electronic PECs, as well as PECs and TDMs computed in the LAC team \cite{aymar2005}. The transitions towards levels belonging to $\Sigma$ excited electronic states give rise to the so-called parallel polarizability $\upalpha_\parallel$, whereas those towards levels of the $\Pi$ excited electronic states give rise to the perpendicular one $\upalpha_\perp$. After some angular algebra, the expression of $\upalpha_{1m1m'}^\mathrm{BF} (\omega)$ is then
\begin{align}
  & \left\langle X, v_A=0, J_A, \qnpr{M_{J,A}} \right|
    \upalpha_{1m1m'}^\mathrm{BF} (z)
    \left| X, v_A=0, J_A, M'_{J,A} \right\rangle
    = \sum_{J''_A M''_A} \frac{2J_A+1}{2J''_A+1} 
    C_{J_AM_A1,-m}^{J''_AM''_A} C_{J_AM'_A1m'}^{J''_AM''_A}
  \nonumber \\
  & \times
    \left[ \left( C_{J_A010}^{J''_A0} \right)^2
    \left\langle X, v_A=0 \right|
    \upalpha_{\parallel} (z)
    \left| X, v_A=0 \right\rangle
    + 2 \left( C_{J_A011}^{J''_A1} \right)^2
    \left\langle X, v_A=0 \right|
    \upalpha_{\perp} (z) 
    \left| X, v_A=0 \right\rangle \right]
  \label{eq:lri-cs2-pola}
\end{align}
where $z$ can be equal to $\omega$ or $iu$. Note that the parallel and perpendicular polarizabilities do not depend on rotational quantum numbers, since rotational energies are much smaller than electronic and vibrational ones, and so can be neglected in the calculation of $\Delta E_a^{(0)}$.

Most of the resulting $C_6$ coefficients between Cs$_2$ in the lowest rotational levels $J_A=0$--4 and excited Cs are strongly negative (below -10000~a.u.), and so trigger attractive PECs, even when the quadrupolar interaction is repulsive. Note that in Ref.~\cite{lepers2011a}, we also published $C_6$ coefficients between Cs$_2(J_A=0-4)$ and ground-state Cs, which are all negative and approximately twice as large as the $C_6$ coefficients between two ground-state Cs atoms. Finally, we observed that the crossing between PECs correlated to different $J_A$-value are still around 100~a.u., as with the quadrupolar interaction. In this region, rotational, first-order quadrupolar and second-order dipolar energies are comparable, which appealed for an alternative formulation of the perturbation theory.

\subsection{New formulation of the perturbation theory}
\label{sub:lri-cs3-nonpert}

This was done in Ref.~\cite{lepers2011b}. Since the long-range energy was not a perturbation compared to the rotational energy, the latter was included as part of the perturbation operator $\mathrm{V}$. Moreover, since the first-order quadrupolar and second-order dipolar energies was of the same order of magnitude, they were both included in the operator $\mathrm{V}$. It may seem surprising to treat equally terms coming from different orders of perturbation, but one can keep in mind that they come from different terms of the multipolar expansion. In consequence the perturbation operator is now $R$-dependent,
\begin{equation}
  \mathrm{V}(R) = B_0 \mathbf{J}_A^2 
                + \mathrm{V}_\text{qq}^\text{BF}(R)
                + \mathrm{V}_\text{vdW}^\text{BF}(R) ,
  \label{eq:lri-v-non-prt}
\end{equation}
and it should be diagonalized for each $R$. Each matrix element of $\mathrm{V}(R)$ can be expressed as a sum on inverse powers of $R$, namely $\sum_n C_n/R^n$, but the PECs after diagonalization cannot.

Because the dimer rotational energy is part of the perturbation, the unperturbed energies only account for the electronic and vibrational ones. The subspace in which $\mathrm{V}$ is diagonalized is thus spanned by $M_{J,A}$, $M_{L,B}$ as previously, but also by $J_A$. It is \textit{a priori} of infinite size since $J_A = 0$, 1, ..., $+\infty$, but in practice it has an upper bond $J_{A,\mathrm{max}}$, on which convergence of the computed PECs should be checked. The later is facilitated by the selection rules associated with Eq.~\eqref{eq:lri-v-non-prt}: the rotational levels coupled by $\mathrm{V}_\text{qq}^\text{BF}(R)$ and $\mathrm{V}_\text{vdW}^\text{BF}(R)$ are such that $J'_A-J_A = 0, \pm 2$.

A similar reasoning can be applied when considering the fine structure of the $np\,^2P^{\circ}$ of the alkali-atom first excited state, as was done in Ref.~\cite{lepers2011c}. The fine-structure Hamiltonian, $A_B\mathbf{L}_B \cdot \mathbf{S}_B$, may be incorporated in the perturbation operator \eqref{eq:lri-v-non-prt}, depending on the value of the fine-structure constant $A_B$. Because the latter strongly varies along the alkali column, from 0.335~cm$^{-1}$ for Li to 554~cm$^{-1}$ for Cs, the fine-structure and long-range energies are only comparable for Li. For all the other alkali atoms, the fine-structure Hamiltonian belong to the unperturbed one, and so the subspace in which $\mathrm{V}(R)$ is diagonalized is spanned by $J_A$, $M_{J,A}$ and $M_{J,B}$, representing the projection of the electronic angular momentum of the atom $J_B$ on the interpartner axis. In Ref.~\cite{lepers2011c}, we discussed these various physical situations using the extended Hund's cases introduced in Ref.~\cite{dubernet1994} for weakly-bound molecules.

\begin{figure}
  \centering
  \includegraphics[width=0.48\textwidth]
    {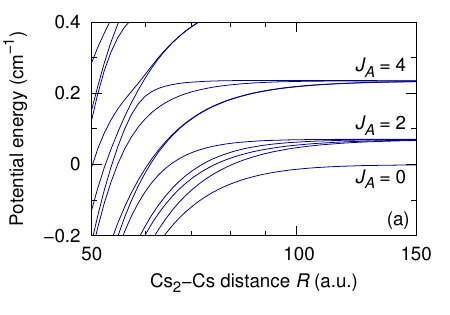}
  \includegraphics[width=0.48\textwidth]
    {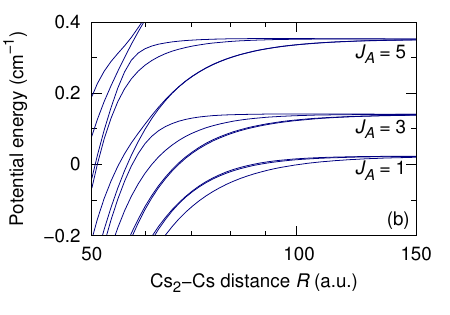}
  \caption{Long-range potential-energy curves between Cs$_2(X^1\Sigma_g^+, v_A=0, J_A)$ and Cs$(6p\,{}^2P_{3/2}^{\circ})$ for $M_{J,A} + M_{J,B} = 1/2$, and (a) even, (b) odd values of $J_A$.}
  \label{fig:lri-cs3-pecs}
\end{figure}

As an example, Figure \ref{fig:lri-cs3-pecs} presents LR PECs between Cs$_2$ in its lowest rotational levels and Cs in its $6p\,^2P_{3/2}^{\circ}$ fine-structure level, and for $M_{J,A} + M_{J,B} = 1/2$. Those PECs ensue from the diagonalization of Eq.~\eqref{eq:lri-v-non-prt} on a grid in interpartner distances from 40 to 150 a.u.. All the curves are attractive, and we observe avoided crossings between PECs correlated to different asymptotes. Moreover, due to the coupling between $J_A=0$ and $J_A=2$, the curve correlated to $J_A=0$ acquires a $R^{-5}$ character and so becomes more attractive.

\subsection{Photoassociation rate}
\label{sub:lri-pa-rate}

As stated above, we make the hypothesis that the PA between Cs$_2$ and Cs takes place in the long-range region, as observed with diatomic systems. From the point of view of calculations, it means that we prolong the LR PECs of the previous subsection towards smaller distances with a repulsive $C_{12}/R^{12}$ term, but we do not expect the latter to have a significant influence on the computed scattering and bound trimer wave functions, which are, to a very large extent, located in the LR region, as illustrated on Fig.~\ref{fig:lri-pa-schm}. In principle, a full quantum study of the PA process requires a global three-dimensional potential-energy surface (PES) of ground-state and excited Cs$_3$, which were not available in literature, and which anyway would not have a sufficient precision for the ultracold regime.
In consequence, based on the LR PECs described above, we have estimated the PA rate of Cs$_2$ and Cs in realistic experimental conditions \cite{perez-rios2015}. 

Applying a PA laser of intensity $I_\mathrm{PA}$ and frequency $\nu_{\mathrm{PA}}$, red-detuned by $\delta_\mathrm{PA}$ with respect to the Cs $6s\,^2S \rightarrow 6p\,^2P_{3/2}$  transition, the PA rate is \cite{pillet1997}
\begin{equation}
  R_{\mathrm{PA}} \propto (k_B T)^{-3/2} \, n_{\mathrm{mol}} \,
    I_{\mathrm{PA}} \, d_{sp}^2 \, | \langle\Psi_i| \Psi_f \rangle |^2,
  \label{eq:lri-pa-rate}
\end{equation}
where $k_B$ is Boltzmann's constant, $T$ the temperature, $n_\mathrm{mol}$ the density of Cs$_2$ molecules, $d_{sp}$ is the TDM of the atomic transition and $\langle\Psi_i| \Psi_f \rangle$ is the Franck-Condon factor between the initial and the final wave function. The initial wave function $\Psi_i(R)$ is a scattering one of the complex Cs$_2\,(X,v_A=0,J_A=0)$--Cs$(6s\,^2S_{1/2})$, with relative kinetic energy $k_B T$. Because of its ultralow value, the collision is assumed to take place in the $s$-wave regime.

The wave function $\Psi_i(R)$ was calculated by integrating using the Numerov method the time-independent Schr\"odinger equation on the interpartner distance $R$ and for a relative particle on reduced mass $\mu = m_{\mathrm{Cs}_2} m_{\mathrm{Cs}} (m_{\mathrm{Cs}_2} + m_{\mathrm{Cs}})^{-1}$ submitted to a Lennard-Jones potential characterized by the $C_6$ coefficient discussed in subsection \ref{sub:lri-cs3-dip} and an arbitrary, positive $C_{12}$ coefficient. As for the wave function $\Psi_f(R)$, it represents a bound level of the trimer, whose discrete energy is just below the Cs$_2\,(X,v_A=0,J_A=0)$ + Cs$(6p\,^2P_{3/2})$ dissociation limit. It was calculated using the mapped Fourier-grid method \cite{kokoouline1999}, which is particularly suited for wave functions located in the long-range region. The potential was defined by Eq.~\eqref{eq:lri-v-non-prt}, and each diagonal term was matched to a repulsive $C_{12}/R^{12}$ term in the short-range region. These terms influences the overall position of the trimer bound level, but not the general conclusions of the study.

\begin{figure}
  \centering
  \includegraphics[width=0.48\textwidth]
    {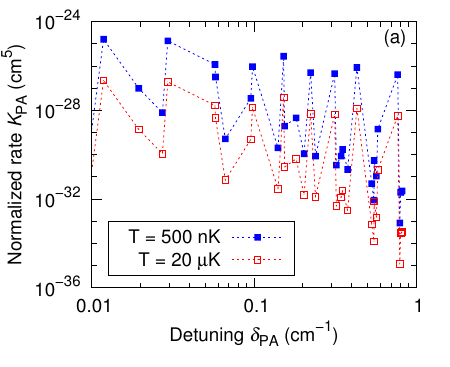}
  \includegraphics[width=0.48\textwidth]
    {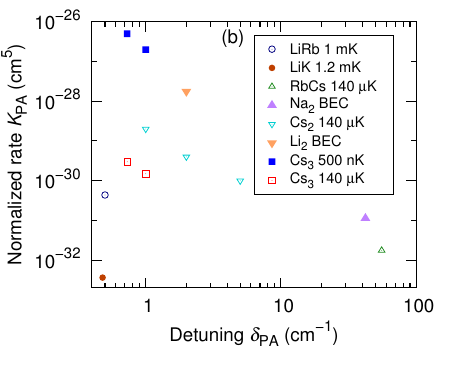}
  \caption{(a) Normalized photoassociation rate $K_{\mathrm{PA}}$ as a function of the detuning $\delta_{\mathrm{PA}}$ with respect to the Cs$_2\,(X,v_A=0,J_A=0)$ + Cs$(6p\;^2P_{3/2}^\circ)$ dissociation limit. The points correspond to vibrational levels of the Cs$_3^*$ complex. The rates are calculated at temperatures of 500~nK (closed blue squares) and 20~$\mu$K (open red squares). (b) Comparison of our results at 140~$\mu$K and 500~nK for two Cs$_3^*$ vibrational levels at 0.770 and 1.008~cm$^{-1}$ with the literature values for Cs$_2$ \cite{drag2000}, LiK \cite{ridinger2011}, LiRb \cite{dutta2014}, RbCs \cite{kerman2004}, Na$_2$ \cite{mckenzie2002}, and Li$_2$ \cite{prodan2003}.}
  \label{fig:lri-pa-rate}
\end{figure}

To allow for the comparison with other systems, figure \ref{fig:lri-pa-rate} shows, as a function of the detuning $\delta_{\mathrm{PA}}$, the PA rate $K_{\mathrm{PA}} = R_{\mathrm{PA}}/(n_{\mathrm{mol}}\phi_{\mathrm{PA}})$, normalized with respect to the molecular density and the photon flux $\phi_{\mathrm{PA}} = I_{\mathrm{PA}}/h\nu_{\mathrm{PA}}$, and expressed in cm$^5$ \cite{cote1998}. Panel (a) shows our results for two different temperatures, 20~$\mu$K typical of magneto-optical traps (MOTs), and 500~nK typical of Bose-Einstein condensates (BECs). The points correspond to energies of the trimer vibrational levels relative to the dissociation limit. To guide the eye, the points are related with a dotted line. The detuning is below 1~cm$^{-1}$, corresponding to the energy window of long-range interactions. Even if that window is narrow, it is likely to contain quite a lot of bound levels. The oscillations from one point to the next are due to the variation of the Franck-Condon overlap between the initial and the final wave functions. We observe larger rates when temperature decreases, as Eq.~\eqref{eq:lri-pa-rate} shows. 
Figure \ref{fig:lri-pa-rate} (b) presents experimental results obtained with homo- and heteronuclear diatomic systems in a broad range of temperatures. The normalized rates are in the same orders of magnitude as ours, which makes the Cs$_2$-Cs PA likely to observe. Again, the rates obtained in lower temperatures are larger. The largest rates, obtained for Cs$_2$, are due to peculiar structure of the electronically-excited state, which contains a long-range potential well.

\section{Formation of atmospheric ozone}
\label{sec:lri-o3}

While working on the Cs$_2$-Cs system, we became aware of the controversy regarding the electronic structure of the ozone molecule O$_3$. To extrapolate the computed PESs in the asymptotic region, it was thus relevant to characterize the long-range interactions between O$_2$ and O in their ground states, namely $X^3\Sigma_g^-$ and $1s^2 2s^2 2p^4 \;{}^3P$, which is formally similar to the Cs$_2$-Cs case. This was achieved in 2012 \cite{lepers2012}.

\subsection{Scientific context}

The ozone molecule plays a crucial role in the physics and chemistry of the Earth atmosphere. However, its mechanism of formation is not yet fully understood, especially its isotopic dependence. It is thought to take place in two steps (see the review articles \cite{mauersberger2005, schinke2006} and references therein): firstly, an oxygen atom O and an oxygen molecule O$_2$ collide to give a rovibrationally excited ozone complex O$_3^*$, which then stabilizes by inelastic collision with one surrounding atom or molecule, corresponding to the so-called deactivation process. However, this second step takes place provided that the excited complex O$_3^*$ does not dissociate into O$+$O$_2$ before colliding with the surrounding gas. In this respect, ozone formation is characterized by unconventional isotopic effects, such as mass-independent fractionation, which are particularly important in the isotope exchange reaction ${}^x\mathrm{O} + {}^y\mathrm{O} {}^z\mathrm{O} \rightarrow {}^y\mathrm{O} + {}^x\mathrm{O} {}^z\mathrm{O}$.

Those unconventional isotopic effects were very well understood in the beginning of the 2000's within the framework of the statistical RKRM (Rice-Kassel-Ramsperger-Marcus) theory \cite{gao2001,gao2002}. However, an adjustable parameter $\eta$ had to be added to that theory, in order to account for deviation from the energy equipartition theorem, after the formation of the O$_3^*$ molecule. Then, the need for a first-principle understanding of the ozone formation, based on quantum mechanics, became obvious. Since a full quantum treatment of the two-step process was beyond the computational resources in 2012, researchers had to focus on specific aspects of the process, like highly-excited vibrational levels of O$_3$ \cite{grebenshchikov2003, babikov2003b, lee2004}, the influence of resonances \cite{babikov2003a, grebenshchikov2009}, or to use less demanding computational techniques \cite{fleurat2003}.

Such studies require a reliable potential energy surface, at least for the electronic ground state. Since the formation of stable O$_3$ involves a wide variety of geometries, from almost separated O and O$_2$, to tightly bound O$_3$, one actually needs a global PES. When they are cut along the minimum-energy path, they all show a change in character between the inner and the asymptotic regions, due to an avoided crossing with an excited electronic state. However, in 2012, the consequence of this avoided crossing was still controversial: a potential barrier above \cite{siebert2001} or below the dissociation threshold (a so-called reef) \cite{babikov2003a, rosmus2002, holka2010}, or on the contrary, a monotonic evolution of the potential energy as suggested by the latest \textit{ab initio} calculations at that time \cite{dawes2011}. Even in the asymptotic region, those studies were based on quantum-chemical calculations on O$_3$, and not on the multipolar expansion on the O$_2$-O complex, hence our interest.

\subsection{Calculations at fixed geometries}

\begin{figure}
  \centering
  \includegraphics[scale=0.7]{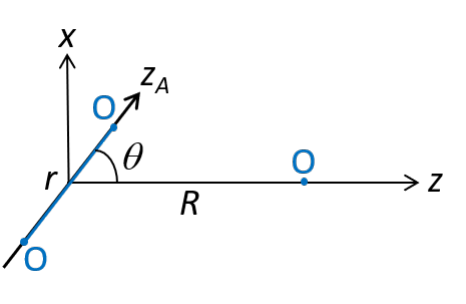}
  \caption{System of Jacobi coordinates for the O$_2$-O complex. The $y$ axis, not shown, points toward the reader.}
  \label{fig:o3-jacob}
\end{figure}

In order to connect a LR PES to a short-range one, we first need to choose a coordinate system. Here we choose the Jacobi coordinates often used to describe atom-diatom complexes and shown on Figure \ref{fig:o3-jacob}. Partner $A$ is the O$_2$ molecule, whose atoms define the axis $z_A$ and which are separated by the distance $r$. The unit vector $\mathbf{u}$ defines the $z$ direction of the BF frame. It joins the center of mass of O$_2$ and the O atom (partner $B$), separated by the distance $R$. The angle between the axes $z$ and $z_A$ is $\theta$. Unlike the Cs$_2$-Cs case of Sec.~\ref{sec:lri-pa}, we firstly derive expressions at fixed geometries, namely as functions of $(R,r,\theta)$. Afterwards, we will consider the rovibration of O$_2$ by integrating on $r$ and $\theta$.

In its electronic ground state, O$_2$ belongs to the $^3\Sigma_g^-$: it has an electronic spin $S_A=1$, with projections $\Sigma_A = 0, \pm 1$ on the internuclear axis $z_A$. The O$_2$ spectrum is characterized by a spin-spin interaction, equal to $\lambda(r) [3\Sigma_A^2 - S_A(S_A+1)] / 2$, where $\lambda(r)$ is the $r$-dependent spin-spin constant; $\lambda(r_e) = 1.980$~\cmi, with $r_e = 2.282$~a.u. the equilibrium distance. Because the interactions between electronic multipole moments only depend on spatial coordinates, the different $\Sigma_A$-values are not coupled by LR interactions, and so each of them provokes a global shift of the PESs calculated below. The fine-structure splitting of the O$(^3P_{J_B})$ level is $A_{\mathrm{O}}[J_B(J_B+1)-6]/2$ with $A_{\mathrm{O}} = -79.1$~\cmi. Therefore, the situation is similar to cesium in Subsec.~\ref{sub:lri-cs3-nonpert}: the different $J_B$-values define different subspaces of degeneracy, and so they also trigger a global shift of the PESs calculated below. In the next paragraphs, we drop those unperturbed shifts, and we focus on the perturbation-operator matrix elements.

The leading term of the multipolar expansion is the quadrupole-quadrupole interaction. The quadrupole moment of O$({}^3P)$ is expressed by applying Wigner-Eckaert theorem
\begin{equation}
  \left\langle {}^3P, M_{L,B} \right| \mathrm{Q}_{2,-m}^\mathrm{BF} 
    \left| {}^3P, M'_{L,B} \right\rangle 
    = C_{1 M'_{L,B}, 2, -m}^{1 M_{L,B}} 
    \left\langle {}^3P \right\| \mathrm{Q}_{2} 
    \left\| {}^3P \right\rangle
  \label{eq:lri-o-quad}
\end{equation}
where $\langle {}^3P \| \mathrm{Q}_{2} \| {}^3P \rangle$ is the reduced matrix element of the quadrupole moment, independent from the referential frame. Equation \eqref{eq:lri-o-quad} is very similar to \eqref{eq:lri-cs-quad}, except that the reduced quadrupole moment is more involved for a many-electron atom. During our study, the quadrupole moment of the $|{}^3P, M_{L,B}=1 \rangle$ stretched sublevel was calculated at -0.95~a.u. by B.~Bussery-Honvault \cite{lepers2012}. As for O$_2\,(X^3\Sigma_g^-)$, it possesses a nonzero $r$-dependent $z_Az_A$ component of the traceless quadrupole tensor in its own frame, denoted $\langle X | \mathrm{Q}_{20}^{\mathrm{O}_2} (r) | X \rangle$. Because it is an irreducible tensor of rank 2, the quadrupole moment in the BF frame can be expressed as a function of its value in the dimer frame as \cite{bussery-honvault2008}
\begin{equation}
  \left\langle X \right| \mathrm{Q}_{2m}^\mathrm{BF} 
    (r,\theta) \left| X \right\rangle
     = D_{m0}^{2*} (0,\theta,0) \, \left\langle X \right|
    \mathrm{Q}_{20}^{\mathrm{O}_2} (r) \left| X \right\rangle
  \label{eq:lri-o2-quad}
\end{equation}
where $D_{m0}^{2} (0,\theta,0)$ is a Wigner $D$ matrix describing the rotation form one frame to the other \cite{varshalovich1988}. The value $\langle X | \mathrm{Q}_{20}^{\mathrm{O}_2} (r_e) | X \rangle = q_{X}(r_e) = -0.253$~a.u. was taken from Ref.~\cite{lawson1997}.

The next term of the multipolar expansion is due to the van der Waals interaction. Just like in subsection \ref{sub:lri-cs3-dip}, it is calculated with the formalism of polarizabilities at imaginary frequencies. For atomic oxygen, the components $\upalpha_{zz} = \upalpha_{1010}$ of the sublevels $M_{L,B} = 0$ and 1 were calculated by B.~Bussery-Honvault using the methods of Pad{\' e} approximants \cite{langhoff1970}. It consists in expanding a polarizability as $\alpha(iu) = \sum_{k=0}^{N-1} a_k(iu)^{2k} / (1+\sum_{k=1}^{N} b_k(iu)^{2k})$. For O$_2$, the parallel and perpendicular polarizabilities at imaginary frequencies were calculated at $r=r_e$ using the method of pseudo dipole-oscillator-strength distributions (DOSDs) \cite{kumar1996}. Pad{\' e} approximants and pseudo-DOSDs are methods which allow for reducing the sum over excited atomic or molecular levels to a limited number of terms (around 10).

Even if the polarizabilities $\upalpha_{1m1m'}$ are insightful to directly characterize the response to an electric field, their main drawback for calculations is that they are not irreducible tensors, contrary to the so-called coupled polarizabilities 
\begin{equation}
  \upalpha_{(11)kq} = \sum_{m,m'=-1}^{+1}
    C_{1m1m'}^{kq} \upalpha_{1m1m'}
  \label{eq:lri-pola-cpl}
\end{equation}
which are defined for $0 \le k \le 2$ and $-k \le q \le k$. Note that Equation \eqref{eq:lri-pola-cpl} is valid in any referential frame. It can be inverted as $\upalpha_{1m1m'} = \sum_{k,q} C_{1m1m'}^{kq} \upalpha_{(11)kq}$. Regarding O$_2$, since $\upalpha_{(11)kq}$ is an irreducible tensor, we can apply a similar relationship as \eqref{eq:lri-o2-quad},
\begin{equation}
  \left\langle X \right| \upalpha_{(11)kq}^\mathrm{BF}
    (iu;r,\theta) \left| X \right\rangle
     = D_{q0}^{k*} (0,\theta,0) \times \left\langle X \right|
    \upalpha_{(11)k0}^{\mathrm{O}_2} (iu;r) \left| X \right\rangle .
  \label{eq:lri-o2-pola}
\end{equation}
Because $D_{00}^{0}(0,\theta,0) = 1$, the term in $k=0$ does not depend on $\theta$ and thus is said isotropic. The coupled polarizabilities of ranks 0 and 2 are respectively proportional to the isotropic polarizability $\overline{\alpha}$ and the anisotropic one $\Delta\alpha$.

In conclusion, to calculate LR PESs between O$_2$ and O, we diagonalized, for $r=r_e$ and different values of $R$ and $\theta$, the perturbation operator
\begin{align}
  \brapr{M_{L,B}} \mathrm{V}(R,r,\theta)
    \left| M'_{L,B} \right\rangle 
    & = \frac{D_{(M'_{L,B}-\qnpr{M_{L,B}}),0}
          ^{2*}(0,\theta,0)}{R^5}
    \left\langle M_{L,B} \right| \mathrm{C}_5(r)
    \left| M'_{L,B} \right\rangle
  \nonumber \\
   & + \frac{\delta_{\qnpr{M_{L,B}},M'_{L,B}}}{R^6}
    \left\langle M_{L,B} \right| \mathrm{C}_{6,0}(r)
    \left| M_{L,B} \right\rangle
  \nonumber \\
   & + \frac{D_{(M'_{L,B}-\qnpr{M_{L,B}}),0}^{2*}
         (0,\theta,0)}{R^6}
    \left\langle M_{L,B} \right| \mathrm{C}_{6,2}(r)
    \left| M'_{L,B} \right\rangle
  \label{eq:lri-o3-v1}
\end{align}
in the basis spanned by the atomic sublevels $M_{L,B}$ and $M'_{L,B}$, where
\begin{align}
  \brapr{M_{L,B}} \mathrm{C}_5(r)
    \left| M'_{L,B} \right\rangle
  & = \frac{24 q_{X}(r_e) \left\langle {}^3P \right\| 
        \mathrm{Q}_{2} \left\| {}^3P \right\rangle
      }
      {(2+M'_{L,B}-\qnpr{M_{L,B}})!
       (2+\qnpr{M_{L,B}}-M'_{L,B})!}
    C_{1 M'_{L,B}, 2, \qnpr{M_{L,B}}-M'_{L,B}}^{1 M_{L,B}} 
  \nonumber \\
  \left\langle M_{L,B} \right| \mathrm{C}_{6,0}(r)
    \left| M_{L,B} \right\rangle
  & = -\frac{1}{2\pi} \sum_{m=-1}^{+1} (-1)^m
    \left( f_{11m} \right)^2 \sum_{k_B=0,2} C_{1,-m1m}^{k_B0}
  \nonumber \\
  & \times \int_0^{+\infty} du \, \overline{\alpha}(iu;r)
    \left\langle M_{L,B} \right| \upalpha_{(11)k_B0} (iu)
    \left| M_{L,B} \right\rangle
  \nonumber \\
  \brapr{M_{L,B}} \mathrm{C}_{6,2}(r)
    \left| M'_{L,B} \right\rangle
  & = -\frac{1}{2\pi} \sqrt{\frac{2}{3}} 
    \sum_{m,m'=-1}^{+1} f_{11m} f_{11m'}
    \sum_{k_B=0,2} \sum_{q=-k_B}^{+k_B} 
    C_{1m1m'}^{2q} C_{1,-m1,-m'}^{k_B,-q}
  \nonumber \\
  & \times \int_0^{+\infty} du \, \Delta\alpha(iu;r)
    \left\langle \qnpr{M_{L,B}} \right| \upalpha_{(11)k_B,-q} (iu)
    \left| M'_{L,B} \right\rangle .
  \label{eq:lri-o3-cn}
\end{align}
The matrix elements of $\mathrm{C}_{6,0}(r)$ and $\mathrm{C}_{6,2}(r)$ contain one term proportional to the atomic scalar polarizability ($k_B=0$) and another to the tensor polarizability ($k_B=2$).

Finally, in order to account for the spin-orbit of the oxygen atom, we use the same method as in subsection \ref{sub:lri-cs3-nonpert} for cesium. We expand the fine-structure levels in LS coupling, $|L_B S_B J_B M_{J,B} \rangle = \sum_{M_{L,B} M_{S,B}} C_{L_B M_{L,B} S_B M_{S,B}}^{J_B M_{J,B}} |L_B M_{L,B} S_B M_{S,B} \rangle$, and we recall that the electric multipole moments do not act on the spin quantum numbers. In the basis spanned by the quantum number $M_{J,B}$, the perturbation operator becomes
\begin{align}
  \left\langle \qnpr{M_{J,B}} \right| \mathrm{V}(R,r,\theta) 
    \left| M'_{J,B} \right\rangle 
    = & \sum_{M_{L,B} M'_{L,B} M_{S,B}} 
    C_{L_B M_{L,B} S_B M_{S,B}}^{J_B M_{J,B}}
  \nonumber \\
    & \times C_{L_B M_{L,B}' S_B M_{S,B}}^{J_B M_{J,B}'}
    \left\langle \qnpr{M_{L,B}} \right| \mathrm{V}(R,r,\theta) 
    \left| M'_{L,B} \right\rangle .
  \label{eq:lri-o3-v2}
\end{align}
Separate perturbative calculations are performed for each fine-structure level, with unperturbed energies $A_{\mathrm{O}} (J_B(J_B+1) - L_B(L_B+1) - S_B(S_B+1)) / 2$.


\begin{figure}
  \centering
  \includegraphics[width=0.48\textwidth]
    {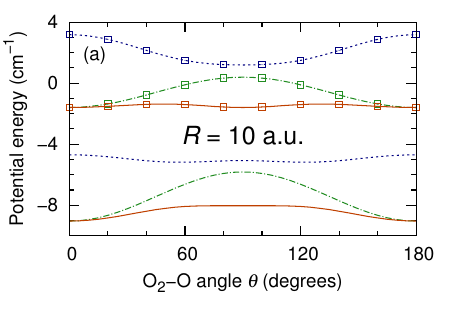}
  \includegraphics[width=0.48\textwidth]
    {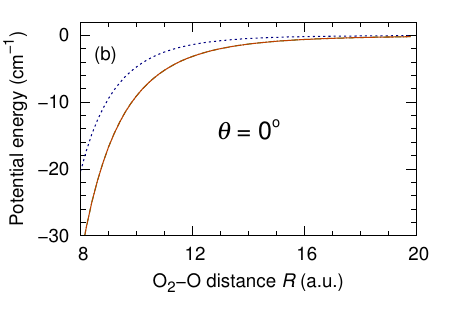}
  \caption{Cuts of the long-range potential-energy surfaces of the states $1A'$ (solid lines), $1A''$ (dash-dotted lines) and $2A'$ (dotted lines) correlated to O$(^3P)+$O$_2(X^3\Sigma_g^-)$ for $r = r_e = 2.282$~a.u. (a) at $R = 10$~a.u., (b) at $\theta = 0^{\circ}$. On panel (a), the curves with open squares result from the quadrupole-quadrupole interaction only.
  }
  \label{fig:lri-o3-pes}
\end{figure}

Figure \ref{fig:lri-o3-pes} presents cuts of the 3 PESs without atomic spin-orbit: (a) at a fixed distance $R = 10$~a.u. and (b) at a fixed angle $\theta=0^{\circ}$. The range of interpartner distances is smaller than in Cs$_2$-Cs, since we estimated Leroy's radius \eqref{eq:lriLRRad} at 8 a.u.. Figure \ref{fig:lri-o3-pes} (a) also shows PESs resulting from the quadrupole-quadrupole interaction only. The latter generates two attractive and one repulsive PES, which strongly depend on $\theta$, revealing the anisotropic character of the quadrupolar interaction. The lowest PES is of $A'$ symmetry. The vdW term tends to shift all the PESs towards lower energies, making them all attractive, whereas their $\theta$-dependence is not significantly modified. This illustrates that the isotropic vdW term, \textit{i.e.}~the second term of Eq.~\eqref{eq:lri-o3-v1}, is the largest one. The corresponding $\langle \qnpr{M_{L,B}} | \mathrm{C}_{6,0} | M'_{L,B} \rangle$ coefficients are of the order of -30~a.u., much weaker than in the Cs$_2$-Cs complex. In the colinear geometry $\theta=0$, the two PESs of $A'$ geometry are degenerate. The above results are not significantly changed by the atomic fine structure \cite{lepers2012}.

\subsection{Calculations for vibrating and rotating O$_2$}

At present, I show the same kind of PECs as in the Cs$_2$-Cs case, assuming O$_2$ in the vibrational ground level $v_A=0$ and in some of the lowest rotational levels. Complexes containing O$_2$ have been theoretically investigated \textit{e.g.}~in Refs.~\cite{vdavoird1983, vdavoird1987, avdeenkov2001}. Due to its nonzero electronic spin, the rotational structure of O$_2(X^3\Sigma_g^-)$ is more complex than that of Cs$_2(X^1\Sigma_g^+)$. Written in Hund's case b basis $|N_A S_A J_A M_{J,A} \rangle$, its effective Hamiltonian is
\begin{equation}
  \mathrm{H}_A = B_0 \mathbf{N}_A^2 
               + \mu_0 \mathbf{N}_A \cdot \mathbf{S}_A
               + \frac{\lambda_0}{2} \left( 3S_{z_A}^2
               - \mathbf{S}_A^2 \right) ,
   \label{eq:lri-o2-rot}
\end{equation}
where $\mathbf{N}_A$ is the (electronic + nuclear) orbital angular momentum, $\mathbf{J}_A = \mathbf{N}_A + \mathbf{S}_A$ is the total angular momentum, with its projection on the interpartner axis $z$ characterized by the quantum number $M_A$. The quantity $B_0 = 1.438$~\cmi{} is the rotational constant of the $v_A = 0$ level. The third term of Eq.~\eqref{eq:lri-o2-rot} is the spin-spin term, already discussed above, proportional to the constant $\lambda_0 = 1.983$~\cmi{} in the lowest vibrational level. Its BF-frame expression is given in Refs.~\cite{vdavoird1983, lepers2012}. The second term is the spin-rotation interaction, which is much smaller ($\mu_0 = 8.43\times 10^{-3}$~\cmi), and so it will be neglected in what follows. Regarding the O atom, it is in a fine-structure level $|J_B M_{J,B} \rangle$ of the ground term ${}^3P$, and as previously, the different $J_B$-values are assumed not coupled by the LR terms. Finally, the perturbation operator is
\begin{equation}
  \mathrm{V}(R) = \mathrm{H}_A + \mathrm{V}_\text{qq}^\text{BF}(R)
                + \mathrm{V}_\text{vdW}^\text{BF}(R)
\end{equation}
and for various $R$-values, it will be diagonalized in the subspace spanned by $N_A$, $J_A$, $M_{J,A}$ and $M_{J,B}$ quantum numbers, keeping $M_{J,A} + M_{J,B}$ constant.

In order to calculate $\mathrm{V}_\text{qq}^\text{BF}(R)$ and $\mathrm{V}_\text{vdW}^\text{BF}(R)$, we average the matrix elements of $\mathrm{V}(R,r,\theta)$ over the rovibrational wave functions of O$_2$. Regarding vibration, the $\mathrm{C}_n(r)$ matrix elements of Eq.~\eqref{eq:lri-o3-cn} are averaged over the wave function $\psi_{v_A}(r)$ (assumed independent from $J_A$),
\begin{align}
  \left\langle \qnpr{v_A,M_{J,B}} \right| \mathrm{C}_n 
    \left| v_A,M'_{J,B} \right\rangle
    & = \sum_{\qnpr{M_{L,B}} M'_{L,B} \qnpr{M_{S,B}}} 
    C_{L_B M_{L,B}  S_B M_{S,B}}^{J_B M_{J,B} }
    C_{L_B M'_{L,B} S_B M_{S,B}}^{J_B M'_{J,B}}
  \nonumber \\
    & \quad \times \int_{0}^{+\infty} dr [\psi_{v_A}(r)]^2 
    \brapr{M_{L,B}} \mathrm{C}_n(r) \left| M'_{L,B} \right\rangle
\end{align}
which implies to average the molecular quadrupole moment for $\mathrm{C}_5$ and polarizabilities $\mathrm{C}_{6,k}$. In the latter case, this point was discussed in details in Ref.~\cite{lepers2012}. As for the rotational part, it acts on the Wigner $D$-matrices of Eq.~\eqref{eq:lri-o3-v2}. In Hund's case b, the rotational wave function is
\begin{equation}
  \psi_{N_A J_A M_{J,A}} (\theta) = \sqrt{\frac{2N_A+1}{8\pi^2}}
    \sum_{M_{N,A}M_{S,A}} C_{N_{A}M_{N,A}S_{A}M_{S,A}}^{J_{A}M_{J,A}}
    D_{M_{N,A},0}^{N_{A}*} (0,\theta,0) .
\end{equation}
Expressing the integrals of three Wigner $D$-matrices as in Eqs.~\eqref{eq:appInt3D}--\eqref{eq:appSum3CG}, we get to the final results
\begin{align}
  & \brapr{N_A J_A M_{J,A} M_{J,B}} \mathrm{V}(R)
    \left| N'_A J'_A M'_{J,A} M'_{J,B} \right\rangle
  \nonumber \\
  & = \frac{\delta_{\qnpr{N_A},N'_A} 
            \delta_{\qnpr{M_{J,A}},M'_{J,A}} 
            \delta_{\qnpr{M_{J,B}},M'_{J,B}}} {R^6}
    \left\langle \qnpr{v_A, M_{J,B}} \right| \mathrm{C}_{6,0} 
    \left| v_A, M_{J,B} \right\rangle
  \nonumber \\
  & + (-1)^{S_A+J'_A+N_A}
    \sqrt{(2N'_A+1)(2J'_A+1)}
    C_{N'_A020}^{N_A0}
    \sixj{N_A}{S_A}{J_A}{J'_A}{2}{N'_A}
    \nonumber \\
  & \quad \times C_{J'_A M'_{J,A}2,M_{J,A}-M'_{J,A}}^{J_A M_{J,A}}
    \left[
      \left\langle \qnpr{v_A,M_{J,B}} \right| \mathrm{C}_{5} 
      \left| v_A,M'_{J,B} \right\rangle
    + \left\langle \qnpr{v_A,M_{J,B}} \right| \mathrm{C}_{6,2} 
      \left| v_A,M'_{J,B} \right\rangle \right]
  \label{eq:lri-o3-v3}
\end{align}
where the symbol between curly brackets is a Wigner 6-j symbol. The isotropic vdW term of Eq.~\eqref{eq:lri-o3-v2} results in diagonal terms in Eq.~\eqref{eq:lri-o3-v3}, while the anisotropic ones results in off-diagonal terms obeying the selection rules $N_A - N'_A = 0, \pm 2$, $J_A - J'_A = 0, \pm 1, \pm 2$ and $M_{J,A} + M_{J,B} = M'_{J,A} + M'_{J,B}$.


\begin{figure}
  \centering
  \includegraphics[width=0.48\textwidth]
    {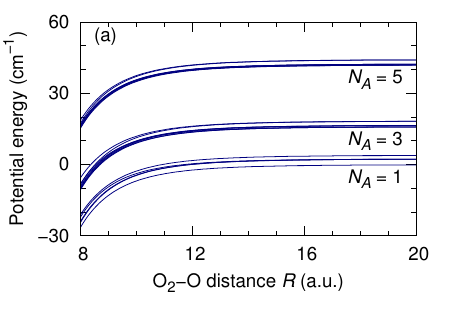}
  \includegraphics[width=0.48\textwidth]
    {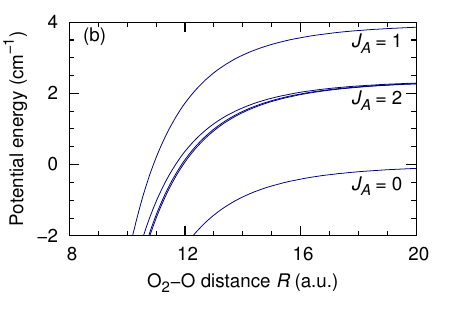}
  \caption{Examples of long-range potential energy curves of interaction between $^{16}$O$_2(X^3\Sigma_g^- ,v_A = 0, N_A = 1, 3, 5)$ and $^{16}$O$({}^3P_2)$, for $M_{J,A} + M_{J,B} = 0$ and the $(+)$ reflection symmetry.}
  \label{fig:lri-o3-pec}
\end{figure}

Figure \eqref{fig:lri-o3-pec} contains examples of LR PECs characterizing the interaction between O in its ground level $^3P_2$ and $^{16}$O$_2$ in the three lowest rotational levels $N_A = 1$, 3 and 5, see panel (a). The mere presence of odd rotational levels comes from the nuclear-spin symmetry in $^{16}$O$_2$. Each rotational level is split into three fine-structure levels with $J_A = N_A,\, N_A \pm 1$. This is visible on panel (b), which is a zoom on the $(N_A = 1)$-manifold. The PECs of figure \ref{fig:lri-o3-pec} are such that $M_{J,A} + M_{J,B} = 0$, and they belong to the $(+)$ reflection symmetry. All the curves are attractive and almost parallel to each other. Indeed they are dominated by the isotropic vdW interaction. This suggests that, in the LR region, the rotation of O$_2$ is not hindered by the presence of the atom.

\paragraph{Impact of the work.}

After the publication of Ref.~\cite{lepers2012}, our fixed-geometry multipolar expansion was employed to extrapolate PESs or to check the correct behavior of \textit{ab initio} points in the asymptotic region \cite{dawes2013, tyuterev2013, ayouz2013, egorov2023}. Some of those PESs possess a reef while other predict a shoulder along the minimum energy path. Because the LeRoy radius is estimated around 8~a.u., namely further than those peculair structures, the LR expansion could not directly put an end to the controversy. However, it allowed for constructing global PESs, for the ground and for the excited electronic states of O$_3$, which were then the basis of isotope-exchange scattering calculations \cite{rao2015a, xie2015, guillon2020}, or spectroscopic calculations close to dissociation \cite{ndengue2016, barbe2022}. Comparisons with experimental results are better for PESs without a reef.

\section{Polar alkali diatomic molecules and external electric fields}
\label{sec:lriDiat}

At present, we come back to ultracold gases, by considering two heteronuclear alkali-metal diatomic molecules, possibly submitted to an external static field. This corresponds to References \cite{lepers2013} and \cite{vexiau2015}. At that time, LiCs and KRb were the only molecules which had been produced in the lowest rovibrational and hyperfine level \cite{deiglmayr2008, ni2008, ospelkaus2010}. But there were several ongoing experiments following the same objective with other molecules. The great advantage of heteronuclear molecules is their permanent electric dipole moment (PEDM) in their own frame, which give rise to long-range and anisotropic dipole-dipole interactions. Namely, for two polar particles separated by a distance $R$, the interaction energy scales as $R^{-3}$, whereas it scales as $R^{-6}$ for non-polar ones (like ground-state atoms). The interaction energy also depends on the relative orientation of the dipoles; it can be attractive (head-to-tail configuration) or repulsive (side-by-side configuration). This effect can in particular be highlighted in confined geometries \cite{micheli2010, demiranda2011, quemener2015, vexiau2019}.

However, molecules prepared in a well-defined rotational level, for instance the lowest one, have no PEDM, and so interact via an $R^{-6}$  term. To induce an electric dipole moment and an $R^{-3}$ interaction, it is necessary to apply an external electric field which, at least partially, polarizes the molecules along its direction. Consequently, in 2013, a thorough study of the long-range interactions between polar bi-alkali molecules including an external electric field was needed, in order to sort out in which conditions the dipole-dipole $R^{-3}$ or vdW $R^{-6}$ terms were dominant. In the LAC team, we based our study on a combination of experimental and computed PECs and TDMs \cite{aymar2005}, and vibrational wave functions computed with our mapped Fourier-grid method \cite{kokoouline1999}. We did so for a bosonic isotopologue of the ten heteronuclear molecules composed of Li, Na, K, Rb and Cs. Our work complemented previous ones, performed with different methodologies \cite{byrd2012b, zuchowski2013}.

\subsection{Giant $C_6$ coefficients between molecules in their lowest rotational level}

We consider two molecules $A$ and $B$ in their ground rovibrational level $X^1\Sigma^+, v_A=v_B=0, J_A=J_B=0$. We ignore the hyperfine structure as the nuclear spin is not affected by electric-multipole operators. In the same spirit as Eq.~\eqref{eq:lri-cs2-quad}, we express the dipole-moment operator between different rotational levels as
\begin{equation}
  \left\langle X, v_A, J'_A, M'_{J,A} 
    \left| \mathrm{Q}_{1m}^\mathrm{BF} 
    \right| X, v_A, J_A, M_{J,A} \right\rangle 
     = \sqrt{\frac{2J_A+1}{2J'_A+1}}
    C_{J_A M_{J,A} 1 m}^{J'_A M'_{J,A}} 
    C_{J_A 0 1 0}^{J'_A 0} \, d_{X,v_A=0}
  \label{eq:lriDiatDip}
\end{equation}
where $d_{X,v_A}$ is the PEDM of the molecule along its internuclear axis $z_A$; we set $d_{X,v_A=0} \equiv d_0$ in what follows. Because the CG coefficient $C_{J_A 0 1 0}^{J'_A 0}$ is nonzero if $J_A+J'_A+1$ is even, Eq.~\eqref{eq:lriDiatDip} is zero for $J_A = J_B$, and so the PEDM of a given rotational level vanishes, as well as the first-order dipole-dipole correction.

The leading term of the multipolar expansion is therefore the vdW one $-C_6/R^6$ (with $C_6>0$). Because the unperturbed state is non-degenerate, the vdW term is characterized by a single, isotropic $C_6$ coefficients. Starting from Eq.~\eqref{eq:lri-c6cr} and adapting it to the case of two molecules, we note that the excited levels $|\Psi_{a,b}^{(0)} \rangle$ have necessarily rotational quantum numbers $J''_A=J''_B=1$. Noting that $m=m'$ and replacing the CG coefficients by their values (in particular $C_{001m}^{1m}=1$), we obtain
\begin{equation}
  C_6 = \frac{2}{3} \sum_{e''_A v''_A e''_B v''_B} 
    \frac{\left| d_{X\qnpr{v_A} ,\, e''_Av''_A} \,
          d_{X\qnpr{v_B} ,\, e''_Bv''_B}\right|^2}
         {E_{e''_Av''_A}^{(0)} + E_{e''_Bv''_B}^{(0)}}
\end{equation}
where $e''_i$ and $v''_i$ denote the electronic and vibrational parts of the excited level of $i = A, B$. The states $e''_i$ are characterized by the angular-momentum projection $\Lambda''_i$ on the internuclear axis $z_i$ ($\Lambda''_i=0$ for $\Sigma$ states, and $\Lambda''_i=\pm 1$ for $\Pi$ states). The quantity $d_{X\qnpr{v_i} ,\, e''_iv''_i} = \langle e''_i v''_i | \mathrm{Q}_{1,\Lambda''_i}^\mathrm{mol} | X, v_i \rangle$ is the vibrationally-averaged TDM in the molecule-$i$ frame.
As in Cs$_2$, $e''_i$ can be an electronically excited state, in which case the excitation energy $E_{e''_iv''_i}^{(0)}$ is on the order of $10^4$~\cmi. But because molecule $i$ is polar, $e''_i$ can also be the ground state $X$. In this case, $v''_i$ is either a vibrationally excited level -- but the TDM $d_{X\qnpr{0} ,\, X,v''_i \neq 0}$ is vanishingly small -- or the vibrational ground level, for which $d_{X0 ,\, X0} = d_0$ and $E_{X,0,j''_i=1}^{(0)} = E_{X,0,j_i=0}^{(0)} = 2B_0$ is a fraction of \cmi. The vdW coefficient $C_6 = C_6^{\mathrm{g}} + C_6^{\mathrm{g-e}} + C_6^{\mathrm{e}}$ comprises three contributions: $C_6^{\mathrm{g}} = d_0^4/6B_0$ when both molecules are in the ground vibrational level, $C_6^{\mathrm{e}}$ when both molecules are in an excited electronic state, and $C_6^{\mathrm{g-e}}$ when one molecule is in the ground vibrational level and the other in an excited electronic state.

\begin{figure}
  \centering
  \includegraphics[width=0.48\textwidth]
    {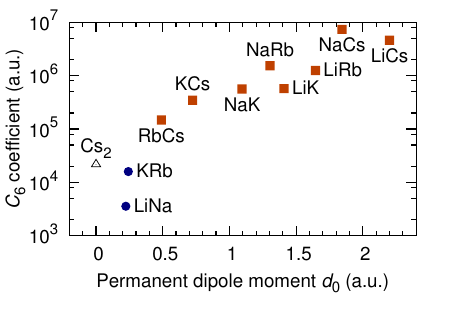}
  \caption{$C_6$ coefficients between two ground-level molecules as functions of the permanent electric-dipole moment in their own frame. The molecules are sorted in three groups: the strongly polar ones (full squares), weakly polar ones (full circles) and Cs$_2$ ($d_0=0$, open triangles).
  }
  \label{fig:lriDiatC6Dip}
\end{figure}

In Ref.~\cite{lepers2013}, we computed those three contributions for the ten heteronuclear bi-alkali molecules in their ground level. This part of the work was performed by R.~Vexiau. Similarly to alkali atom-atom interactions, the $C_6^{\mathrm{e}}$ coefficients equal a few thousands of atomic units, ranging from 3321 for the least polarizable pair LiNa to 17707 for the most polarizable pair RbCs. Due to its $d_0^4$ dependence, the variation of $C_6^{\mathrm{g}}$ is more spectacular, ranging from 241~a.u. for the least polar molecule LiNa, to $7.311 \times 10^6$~a.u. for the second most polar one NaCs. For all molecules except LiNa and KRb, the $C_6^{\mathrm{g}}$ contribution is at least one order of magnitude larger than $C_6^{\mathrm{e}}$, yielding giant values in comparison with atom-atom interactions. As for $C_6^{\mathrm{g-e}}$, it is always very small. On figure \ref{fig:lriDiatC6Dip}, the $C_6$ coefficients are plotted as functions of the PEDM. Our computed values are in rather good agreement with published values, even though our $C_6^{\mathrm{g}}$ are almost systematically larger, and $C_6^{\mathrm{e}}$ smaller, by up to 20~\%. For the $C_6^{\mathrm{g}}$, this may come from the overestimation of the PEDM observed in Ref.~\cite{aymar2005}, when compared to the most accurate measurements on that time, or with later measurements like in RbCs \cite{takekoshi2014}.

In Ref.~\cite{vexiau2015} and its supplementary material, we calculated $C_6(v)$ coefficients between molecules in the same vibrationally-excited level $v_A=v_B=v \neq 0$ of the electronic ground state. Regarding the three contributions, the same hierarchy is observable in a wide range of vibrational levels, even if $C_6^{\mathrm{g}}(v)$ decreases with $v$, just like the PEDM $d_v$. Regarding molecules in different vibrational levels, we discussed the validity of Tang's combination rule \cite{tang1969},
\begin{equation}
  \left( \frac{C_6(v_A,v_B)}{2} \right)^{-1} \approx
    \left( \frac{\overline{\alpha}(0;v_B)}
                {\overline{\alpha}(0;v_A)} C_6(v_A) \right)^{-1}
  + \left( \frac{\overline{\alpha}(0;v_A)}
                {\overline{\alpha}(0;v_B)} C_6(v_B) \right)^{-1}
  \label{eq:lriDiatMH}
\end{equation}
where $\overline{\alpha}(0;v)$ is the static (isotropic) polarizability in the vibrational level $v$. We demonstrated with analytical arguments and numerical examples that it is more accurate to apply Eq.~\eqref{eq:lriDiatMH} to each contribution $C_6^{\mathrm{g}}$,  $C_6^{\mathrm{e}}$,  $C_6^{\mathrm{g-e}}$ and $C_6^{\mathrm{e-g}}$ separately, rather than to the total $C_6$ coefficient. This requires to expand the polarizabilities as $\overline{\alpha} (0;v) = \overline{\alpha}_{\mathrm{g}} (0;v) + \overline{\alpha}_{\mathrm{e}} (0;v)$, distinguishing the contributions of the purely rotational transition $\overline{\alpha}_{\mathrm{g}} (0;v)$ and the contributions of electronic transitions $\overline{\alpha}_{\mathrm{e}} (0;v)$.

\subsection{Coupled rotational levels in free space}

As discussed in subsection \ref{sub:lri-cs3-nonpert}, the LR energy of the previous subsection are comparable to the rotational splittings for distances around 200~a.u. \cite{lepers2013}, that is above the LeRoy radius estimated around 40~a.u.. Again, the perturbation formalism must be reformulated so as to allow couplings between rotational levels under the effect of the dipole-dipole interaction. Similarly to Eq.~\eqref{eq:lri-v-non-prt}, the perturbation operator becomes
\begin{equation}
  \mathrm{V}(R) = B_0 \left[ \mathbf{J}_A^2 + \mathbf{J}_B^2 \right] 
    + \mathrm{V}_\mathrm{dd}^\mathrm{BF}(R) 
    + \mathrm{V}_\mathrm{vdW}^\mathrm{BF}(R) ,
  \label{eq:lriDiatNonPrt}
\end{equation}
where $\mathrm{V}_\mathrm{vdW}^\mathrm{BF}(R)$ accounts for the contributions of electronically-excited states, \textit{i.e.}~$\langle \mathrm{V}_\mathrm{vdW}^\mathrm{BF}(R) \rangle = -C_6^{\mathrm{e}} / R^6$. In what follows, we consider the 8 molecules for which $C_6^{\mathrm{g}} \gg C_6^{\mathrm{e}}$, so that we can safely neglect the term $\mathrm{V}_\mathrm{vdW}^\mathrm{BF}(R)$ in Eq.~\eqref{eq:lriDiatNonPrt}. 
The dipole-dipole interaction (DDI) $\mathrm{V}_\mathrm{dd}^\mathrm{BF}(R)$ couples rotational levels such that $J'_A - J_A = \pm 1$, $J'_B - J_B = \pm 1$ and $M_{J_A} + M_{J_B} = M'_{J_A} + M'_{J_B} = M$ is conserved. 
The eigenvectors of Eq.~\eqref{eq:lriDiatNonPrt} can be labeled $|M|_{g/u}^{\pm}$ in analogy to diatomic-molecule PECs. The symbols $g(u)$ correspond to eigenvectors $\propto (|J_A M_{J,A} J_B M_{J,B} \rangle \pm (-1)^p |J_B M_{J,B} J_A M_{J,A} \rangle)$ with the parity $p = (-1)^{J_A+J_B}$. For $M=0$, the reflection symmetry is $\pm 1$ for eigenvectors $\propto (|J_A M_{J,A} J_B, -M_{J,A} \rangle \pm |J_A, -M_{J,A} J_B M_{J,A} \rangle)$. For example, the eigenvector $(|1000 \rangle + |0010 \rangle) / \sqrt{2}$ belongs to the $0_u^+$ symmetry.


\begin{figure}
  \centering
  \includegraphics[width=0.64\textwidth]
    {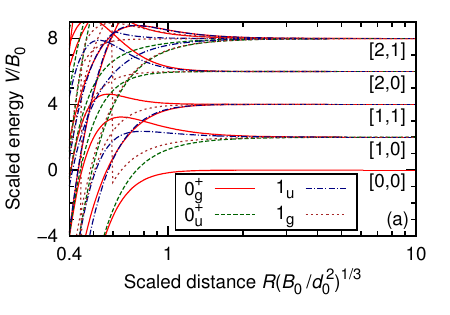}
  \includegraphics[width=0.32\textwidth]
    {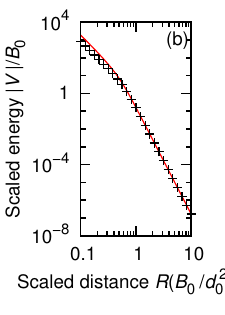}
  \caption{(a) Long-range potential-energy curves (in scaled units) of $0_g^+$ (solid lines), $0_u^+$ (dashed lines), $1_g$ (dotted lines) and $1_u$ (dash-dotted lines) symmetries of two identical $v=0$ ground state molecules. The asymptotes are labeled $[J_A,J_B]$. (b) The lowest $0_g^+$ curve in log scale (solid line: numerical; crosses: Eq.~(\ref{eq:lriDiat2chan})).
  }
  \label{fig:lriDiatPecs1}
\end{figure}

Figure \ref{fig:lriDiatPecs1} (a) shows the resulting PECs correlated to the lowest dissociation channels $[J_A,J_B]$, and sorted by diatomic-like symmetries. The curves are displayed in scaled units of distances and energy
\begin{equation}
  \bar{R} = R (B_0/d_0^2)^{1/3} ,\quad \bar{V} = V/B_0
\end{equation}
where the characteristic length $R^* = (d_0^2/B_0)^{1/3}$ determines the crossing region as in subsection \ref{sub:lri-cs3-nonpert}. Values up to $J_i=6$ for $\bar{R}>10$, $J_i=10$ for $0.25<\bar{R}<10$ and $J_i=15$ for $0.1<\bar{R}<0.25$ ($i=A,B$) have been included in the basis $\{| J_A M_{J,A} J_B M_{J,B} \rangle\}$. Panel (b) shows a comparison in log-log scale of the ground PEC $0_g^+$ calculated numerically and estimated analytically. This estimate is done by diagonalizing Eq.~\eqref{eq:lriDiatNonPrt} in the two-channel approximation $[0,0]$ and $[1,1]$, since $[1,1]$ is the closest channel to which $[0,0]$ is coupled by DDI. This calculation gives
\begin{equation}
    \bar{V}_{0}(\bar{R}) \approx 2-2\sqrt{1+\frac{1}{6\bar{R}^{6}}} \,.
  \label{eq:lriDiat2chan}
\end{equation}

It predicts a sudden change in $R$-dependence of the PEC around $\bar{R} = 1$ ($R = R^*$). For $R \gg 1$, we retrieve the $R^{-6}$-character described in the previous subsection, with the coefficient $C_6 = d_0^4/6B_0$. By contrast for $R \ll 1$, Eq.~\eqref{eq:lriDiat2chan} becomes $V_0(R) \approx 2 - 2 / \sqrt{6} \bar{R}^3$, which means the usual $R^{-3}$-dependence of the DDI. This change in $R$-dependence is confirmed by the numerical diagonalization, see Fig.~\ref{fig:lriDiatPecs1} (b), but the prefactor of the $R^{-3}$ part is not the one of Eq.~\eqref{eq:lriDiat2chan}, since higher channels like $[2,0]$, $[2,2]$, $[3,1]$, etc. come into play.
The PECs correlated to higher dissociation channels present a similar behavior: for $R \gg 1$, a one-channel $R^{-6}$ dependence with attractive or repulsive interactions, and for $R \ll 1$, an attractive DDI due to avoided crossings with curves coming from higher dissociation limits. This can give rise to potential barriers, interesting in the framework of shielding.

The channel $[0,1]$ is special. The PECs of $0_g^+$ and $1_u$ symmetries also possess a potential barrier. But because the states $|001M\rangle$ and $|1M00\rangle$ are degenerate and coupled by DDI, a resonant interaction or excitation exchange takes place, and the corresponding PECs have a $R^{-3}$ character even when $\bar{R} \gg 1$. Namely $V(R) = 2 \pm 2/3\bar{R}^3$ for $0_g^-$ ($0_u^+$) and $V(R) = 2 \pm 1/3\bar{R}^3$ for $1_u$ ($1_g$) respectively. A similar phenomenon takes place close to asymptotes of the kind $(J_A,J_A \pm 1)$. Note that this resonant DDI is also observed with the LR PECs involving identical atoms close to $S+P$ dissociation limits, see for instance Ref.~\cite{jones2006}. Finally, it is worthwhile to note that the $0_u^+$ curve is degenerate with the lowest $0_g^+$ one for $\bar{R} < 0.5$.

\subsection{Application of an external electric field}

At present, we consider that the two molecules are submitted to a static homogeneous electric field $\mathbf{E}$, whose amplitude $\mathcal{E}$ is sufficient to couple molecular rotational levels. Therefore, the perturbation operator becomes
\begin{equation}
  \mathrm{V}(R) = B_0 \left[ \mathbf{J}_A^2 + \mathbf{J}_B^2 \right] 
    + \mathrm{V}_\mathrm{dd}^\mathrm{BF}(R) 
    + \mathrm{V}_\mathrm{S}^\mathrm{BF} ,
  \label{eq:lriDiatNonPrt2}
\end{equation}
$\mathrm{V}_\mathrm{S}^\mathrm{BF}$ is the ($R$-independent) Stark operator given by $\mathrm{V}_\mathrm{S}^\mathrm{BF} = -( \mathbf{Q}_1^\mathrm{BF}(A) + \mathbf{Q}_1^\mathrm{BF}(B) ) \cdot \mathbf{E}$. The matrix elements of dipole-vector operator $\mathbf{Q}_1^\mathrm{BF}(A)$, given in Eq.~\eqref{eq:lriDiatDip}, couples $J_A$ with $J'_A = J_A \pm 1$ leaving $J_B$ unaffected. Conversely, the matrix elements of $\mathbf{Q}_1^\mathrm{BF}(B)$ couples $J_B$ with $J'_B = J_B \pm 1$ leaving $J_A$ unaffected. The Stark operator thus couples basis states of different parities, $g \leftrightarrow u$. As an example, $|0000 \rangle$, of $g$ symmetry, is coupled to the symmetric superposition $(|1M00\rangle + |001M\rangle) / \sqrt{2}$, of $u$ symmetry.
The field is taken either parallel (along $z$) or perpendicular (along $x$) to the interpartner axis. In the former case, the Stark operator is $-( \mathrm{Q}_{10}^\mathrm{BF}(A) + \mathrm{Q}_{10}^\mathrm{BF}(B) ) \mathcal{E}$ and it conserves $M$, and the reflection symmetry $(\pm)$ for $M=0$. In the latter case, it is $-( \mathrm{Q}_{1,-1}^\mathrm{BF}(A) - \mathrm{Q}_{1,1}^\mathrm{BF}(A) + \mathrm{Q}_{1,-1}^\mathrm{BF}(B) - \mathrm{Q}_{1,1}^\mathrm{BF}(B) ) \mathcal{E} / \sqrt{2}$, and it couples $M$ to $M'=M\pm 1$.


\begin{figure}
  \centering
  \includegraphics[width=0.48\textwidth]
    {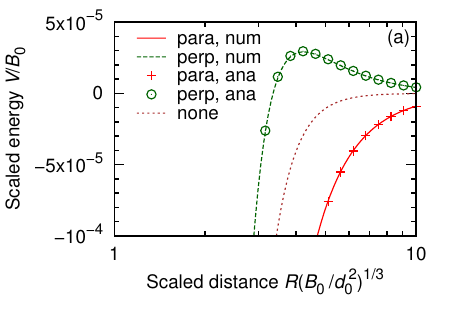}
  \includegraphics[width=0.48\textwidth]
    {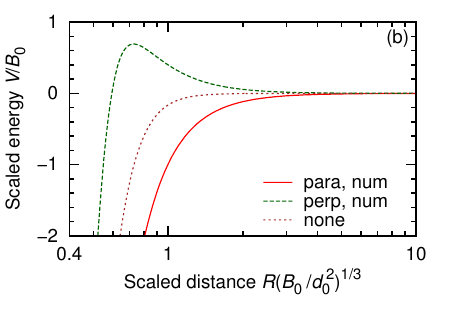}
  \caption{Long-range potential-energy curves (in scaled units) for two identical $v=0$ ground state polar diatoms submitted to an external electric field: (a) $\mathcal{E} = B_0/5d_0$; and (b) $\mathcal{E} = 5B_0/d_0$. The solid (dashed) lines correspond to the numerical results in a parallel (perpendicular) field. The dotted line is the lowest $0_g^+$ field-free curve of Fig.~\ref{fig:lriDiatPecs1}. On panel (a), the plus signs (open circles) correspond to the analytical approximations given in Eq.~\eqref{eq:lriDiatEFld}.
  }
  \label{fig:lriDiatPecs2}
\end{figure}

Figure \ref{fig:lriDiatPecs2} shows the lowest PEC with a parallel and a perpendicular electric field of amplitude (a) $\mathcal{E} = \mathcal{E}^{*}/5$ and (b) $\mathcal{E} = 5\mathcal{E}^*$, with $\mathcal{E}^* = B_0/d_0$. Panel (a) shows a very good agreement between the numerical PECs and perturbative calculations assuming $R \gg R^*$ and $\mathcal{E} \ll \mathcal{E}^*$, giving
\begin{equation}
  \bar{V}_{0,\parallel} (\bar{R},\bar{\mathcal{E}})
    \approx -\frac{2\bar{\mathcal{E}}^2}{9\bar{R}^3}
    - \frac{1}{6\bar{R}^6}
  \: \text{and} \;
  \bar{V}_{0;\bot} (\bar{R},\bar{\mathcal{E}})
    \approx \frac{\bar{\mathcal{E}}^2}{9\bar{R}^3}
    - \frac{1}{6\bar{R}^6},
  \label{eq:lriDiatEFld}
\end{equation}
where $\bar{\mathcal{E}} = \mathcal{E} / \mathcal{E}^*$ is the scaled electric field. Note that the PECs are shifted to have a zero dissociation energy. The electric fields brings a $R^{-3}$-character to the lowest PECs at very large distances, $\bar{R} \gtrsim \bar{\mathcal{E}}^{-2/3}$ according to Eq.~\eqref{eq:lriDiatEFld}. Parallel fields strengthen the intermolecular attraction as they favor the head-to-tail configuration. By contrast, perpendicular fields which provoke a repulsive interaction (side-by-side configuration), which, due to the competition with the huge vdW interaction, creates a potential barrier of height $\bar{\mathcal{E}}^4 / 54$, see Eq.~\eqref{eq:lriDiatEFld} and Fig.~\ref{fig:lriDiatPecs2} (a). With increasing field amplitudes, the barrier moves towards smaller interpartner distances and gets higher.


\begin{figure}
  \centering
  \includegraphics[width=0.96\textwidth]
    {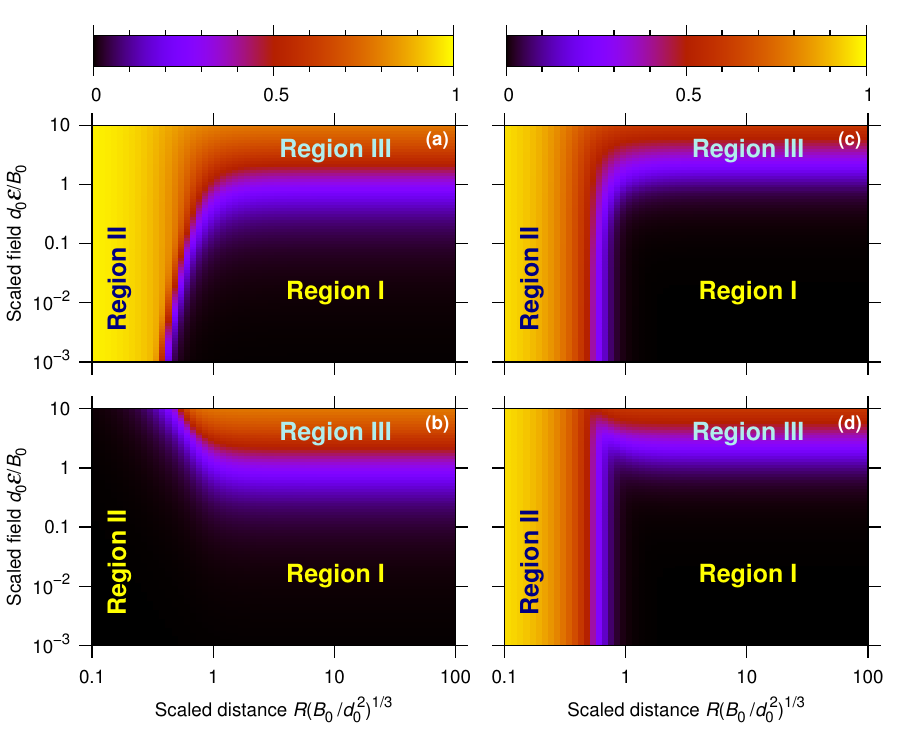}
  \caption{Molecular orientation in the lowest PEC in parallel (upper row) and perpendicular (lower row) electric field as a function of the scaled distance and electric field. The color scale ranges from black (minimal values) to white (maximal values). (a)-(b) Induced dipole moment along the field axis, see Eq.~\eqref{eq:lriDiatDip1}; (c)-(d) scalar product of the two dipole moments, see Eq.~\eqref{eq:lriDiatDip2}. The Roman numbers correspond to regions of the $(\bar{R}, \bar{\mathcal{E}})$ plane characterized by different types of interactions (see text).
  }
  \label{fig:lriDiatDip}
\end{figure}

In order to confirm our interpretation in terms of head-to-tail or side-by-side configurations, we calculated for the lowest PECs two quantities characterizing the molecular orientation: (i) the scaled induced dipole moment along the field direction
\begin{equation}
  d(\bar{R},\bar{\mathcal{E}}) = \frac{1}{2d_0\mathcal{E}}
    \left\langle \left[
      \mathbf{\mathbf{Q}_{1}^{\mathrm{BF}}} (A)
    + \mathbf{\mathbf{Q}_{1}^{\mathrm{BF}}} (B)
    \right] \cdot \mathbf{E} \right\rangle
  \label{eq:lriDiatDip1}
\end{equation}
and the reduced scalar product
\begin{equation}
  s(\bar{R},\bar{\mathcal{E}}) = \frac{1}{d_0^2}
    \left\langle \mathbf{\mathbf{Q}_{1}^{\mathrm{BF}}} (A)
    \cdot \mathbf{\mathbf{Q}_{1}^{\mathrm{BF}}} (B) \right\rangle
  \label{eq:lriDiatDip2}
\end{equation}
Both quantities are defined so as to vary between $-1$ and $+1$. On Figure \ref{fig:lriDiatDip}, they are plotted as functions of $\bar{R}$ and $\bar{\mathcal{E}}$ for parallel and perpendicular fields. The four panels are divided in three regions in Roman numbers.

In Region I (large distances, low field), both $d(\bar{R},\bar{\mathcal{E}})$ and $s(\bar{R},\bar{\mathcal{E}})$ are zero. The molecules have no preferential orientation, as they mostly occupy the (isotropic) rotational ground level. This is even the case in the $R^{-3}$ region visible on Fig.~\ref{fig:lriDiatPecs2} (a). By contrast, in Region III (strong distance and field), the field is strong enough to significantly orient the two molecules, namely $d(\bar{R},\bar{\mathcal{E}}) > 0.5$. \textit{A fortiori}, the mutual orientation also increases but slower. In Region III, the results are identical for a field parallel or perpendicular to the intermolecular axis. In the first case, the molecules are preferentially in a head-to-tail configuration, hence the attractive curves of Figure \ref{fig:lriDiatPecs2}. In the second case, the molecules are preferentially in a side-by-side configuration, hence the repulsive curves of Figure \ref{fig:lriDiatPecs2}.

Region II (low distance) is characterized by a strong mutual orientation $s(\bar{R}, \bar{\mathcal{E}}) \to 1$ in both geometries. Moreover, the border between Regions I and II is field-independent. It corresponds to the left part of Figures \ref{fig:lriDiatPecs1} (a) and (b), where the two molecules lock on each other. They are in a head-to-tail configuration, but unlike Region III, without preferential orientation along the electric field. This is at last the case in perpendicular and vanishing parallel field. However, in nonzero parallel fields, see Fig.~\ref{fig:lriDiatDip} (a), the molecular orientation $d(\bar{R},\bar{\mathcal{E}})$ is close to unity. This is due to the degeneracy between the lowest $0_g^+$ and $0_u^+$ curves of Fig.~\ref{fig:lriDiatPecs1} (a). Even a small $\bar{\mathcal{E}}$-value is sufficient to lift that degeneracy and induce a strong dipole moment. However, there exists a second, close field-mixed state for which $d(\bar{R}, \bar{\mathcal{E}}) \to -1$. In Ref.~\cite{lepers2013}, we mentioned the possibility to use that large induced dipole moment to perform radiative association of tetramers by microwave stimulated emission. However, the strong losses observed in molecular samples seem to prevent that process. Anyway, Ref.~\cite{lepers2013} triggered our deep understanding of the molecular interactions which revealed crucial for shielding of ultracold collisions.

\section{Alkali-metal and lanthanide atoms}
\label{sec:lriLn}

Along with polar molecules, lanthanide (Ln) atoms represent other prime systems to observe dipolar effects with ultracold gases \cite{norcia2021, chomaz2022}. The dipole-dipole interactions are triggered by the strong magnetic moments of Ln atoms, up to 10 Bohr magnetons ($\mu_B$) for dysprosium (Dy). However, the vdW term also bring an important contribution to the interaction energy, and is even dominant for $R \lesssim 100$~a.u. \cite{petrov2012}. The anisotropic vdW interaction is thought to be responsible of the quantum chaos observed in ultracold collisions of Ln atoms \cite{frisch2014, maier2015}. Moreover, mixtures of two Ln atoms, or of one Ln (or chromium) with one alkali have also been produced and investigated \cite{trautmann2018, ravensbergen2020, schaefer2022, ciamei2022a}.

In consequence, with the sets of transition energies and TDMs employed in Chapter~\ref{chap:ddp} to compute dynamical dipole polarizabilities (DDPs), we also computed isotropic and isotropic coefficients for Er, Dy and Ho \cite{lepers2014, li2016, li2017}. More recently, in a collaboration with M.~Tomza's group in Warsaw, we calculated the $C_6$ coefficients for systems composed of Er or Dy in the one hand, and a closed-shell alkali or alkaline-earth atom on the other hand \cite{zaremba-kopczyk2024}.

Generally speaking, the matrix elements of the BF-frame vdW operator between sublevels $|\beta_i J_i M_{J,i} \rangle$ ($i=A,B)$ of the same atomic level read
\begin{equation}
  \left\langle \qnpr{M_{J,A}} \qnpr{M_{J,B}} \right|
    \mathrm{V}_\mathrm{vdW}^\mathrm{BF}(R)
    \left| M'_{J,A} M'_{J,B} \right\rangle
    = -\frac{1}{R^6} \sum_{
                     \substack{k_A,k_B=0 \\ k_A+k_B\text{ even}} }^{2}
      A_{\qnpr{k}_A \qnpr{M_{J,A}} M'_{J,A} 
      \qnpr{k}_B \qnpr{M_{J,B}} M'_{J,B} } \times C_{6,k_A,k_B}
\end{equation}
where $A_x$ is an angular factor containing CG coefficients and Wigner symbols and $C_{6,k_A,k_B}$, which depend on the atomic properties, are frame-independent. For $k_A = k_B = 0$, the angular factors is $\delta_{\qnpr{M_{J,A}},M'_{J,A}} \, \delta_{\qnpr{M_{J,B}},M'_{J,B}}$, and so the vdW operator only contains equal diagonal terms $-C_{6,00} / R^6$, with $C_{6,00}$ the isotropic vdW coefficient. All the terms with $(k_A,k_B) \neq (0,0)$ are anisotropic. The isotropic coefficient has the well known expression, as a function of scalar polarizabilities at imaginary frequencies $iu$
\begin{equation}
  C_{6,00} = \frac{3}{\pi} \int_{0}^{+\infty} du
    \alpha_\mathrm{scal}(iu;A) \alpha_\mathrm{scal}(iu;B) .
  \label{eq:lriLnVdw}
\end{equation}
As for the anisotropic ones, the expression can change from one author or one article to the other. But of course, any change in $C_{6,k_A,k_B}$ coefficient (say a factor of $1/2$) is compensated in the $A_x$ factor (say a factor of 2), so that the matrix elements \eqref{eq:lriLnVdw}, and \textit{in fine} the $C_6$ coefficients after diagonalization are the same. As an example, for $(k_A,k_B) = (2,0)$, one has $C_{6,20} \propto \int_{0}^{+\infty} du \alpha_\mathrm{tens}(iu;A) \alpha_\mathrm{scal}(iu;B)$; in Ref.~\cite{zaremba-kopczyk2024}, we took a prefactor of $3/\pi$ while in Ref.~\cite{chu2005} they took $3(2J_A+3) / 2\pi J_A$. Note that in Ref.~\cite{zaremba-kopczyk2024}, we took the polarizabilities of close-shell atoms from Ref.~\cite{derevianko2010}.

The key result is that, for Ln-Ln pairs and for Ln--closed-shell-atom pairs, the isotropic vdW term is strongly dominant, at least two orders of magnitude larger than the anisotropic ones. Among the latter, the $C_{6,20}$ and $C_{6,02}$ are dominant in Ln-Ln pairs. In Ln--closed-shell-atom pairs, $C_{6,20}$ is the only nonzero anisotropic coefficient, since the zero orbital angular momentum ($L_B=0$) of closed-shell atoms imposes $k_B = 0$. The consequence of that weak anisotropy is that, after diagonalization of the operator \eqref{eq:lriLnVdw}, the $C_6$ coefficients are spread over a small range of values around the $C_{6,00}$ coefficient.

\begin{figure}
  \centering
  \includegraphics[width=0.5\textwidth]{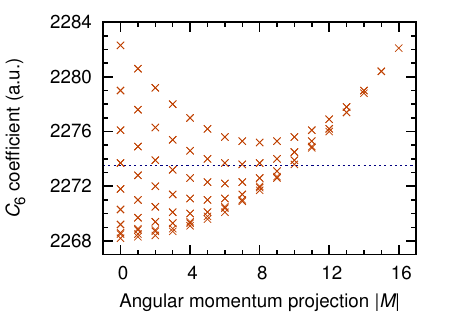}
  \caption{$C_6$ coefficients as functions of $|M|$ for two Dy atoms in their ground level. The horizontal dotted line is the isotropic vdW coefficient $C_{6,00} = 2273.5$~a.u..}
  \label{fig:lriLnC6}
\end{figure}

This is illustrated on Figure \ref{fig:lriLnC6} with the Dy-Dy pair with both atoms in their ground level [Xe]$4f^{10}6s^{2}~{}^5I_8$. The $C_6$ coefficients are sorted by values of $M = M_A + M_B$, with $|M| = 0$ to 16, which means that they may be used to prolong Dy$_2$ PECs in Hund's case (c) (where our $M$ is called $\Omega$). Note that the coefficients may further be sorted according to the inversion symmetry: eigenvectors containing $( |M_{J,A} M_{J,B} \rangle + |M_{J,B} M_{J,A} \rangle ) /\sqrt{2}$ and $|M_{J,A} M_{J,A} \rangle$ states are \textit{gerade} ($g$), while eigenvectors containing $( |M_{J,A} M_{J,B} \rangle - |M_{J,B} M_{J,A} \rangle ) /\sqrt{2}$ states are \textit{ungerade} ($u$). The $C_6$ coefficients are only spread by  16~a.u. around the isotropic coefficient $C_{6,00} = 2273.5$~a.u.. Our coefficients are larger than those of Ref.~\cite{kotochigova2011}, but our spread is smaller, which means a less pronounced anisotropy.

Alkali-metal as well as fermionic Er and Dy atoms possess a nonzero nuclear spin, and so a hyperfine structure, which must be accounted for in ultracold collisions. Therefore, in Ref.~\cite{zaremba-kopczyk2024}, we also gave the expression of the vdW operator in the hyperfine-structure basis $| F_A M_{F,A} F_B M_{F,B} \rangle$, as functions of the $C_{6,k_A,k_B}$ discussed above. Again, due to their zero orbital angular momentum, alkali metals merely give rise to diagonal terms in $F_B$ and $M_{F,B}$. Then, since $M_{F,A} + M_{F,B}$ is conserved by the vdW operator, $M_{F,A}$ is also conserved. By contrast, different $F_A$-values can be coupled by the anisotropic term proportional to $C_{6,20}$, according to the selection rules $|F'_A-F_A| \le 2$.  

\,

In this chapter, I have presented in chronological order a first group of studies applying the formalism of long-range interactions, described in the body-fixed frame. The systems under investigation range from atom pairs with erbium or dysprosium, and alkali or alkaline-earth metals including hyperfine structure, to pairs of heteronuclear bialkali molecules in the presence of a static electric field. A strong focus is also set on atom-molecule pairs like O$_2$-O and Cs$_2$-Cs. In the first case, our computed energies were added to short-range potential-energy surfaces calculated with quantum-chemistry codes, used to study the ozone formation. In the second case, we employed our computed potential-energy curves to compute long-range rovibrational and continuum states of Cs$_3$, in order to model the photoassociation of that molecule. Finally, I mention that I computed long-range potential curve of the B$^+$ + F system, that were used to study the dissociative recombination of the boron monofluoride ion in the context of cold plasmas \cite{mezei2016}.

\chapter{Long-range interactions in the space-fixed frame}
\label{chap:lriSF}

After studying long-range interactions with various systems and in the body-fixed frame in the previous chapter, I turn here to the space-fixed frame. The space-fixed frame is well suited since I will consider two situations involving external fields. It also enables to account for the rotation of the interpartner axis through the partial-wave quantum numbers.

Section \ref{sec:lri1OS} is dedicated to the optical shielding of ultracold collisions between heteronuclear bialkali molecules in order to suppress their reactive collisions. This implies the presence of one and two laser fields. The calculated curves and couplings serve as inputs of a scattering code. Then in section \ref{sec:lriHo2}, I discuss the formation of long-range molecules of Ho$_2$ (holmium dimer), possessing both a magnetic and an electric dipole moment in the laboratory frame. Section \ref{sec:lriHo2} is closely related to section \ref{sec:atDblPol} which deals with doubly dipolar gases of lanthanide atoms.

In both cases, we use the symmetrized and fully coupled basis, in order to reduce the computational cost. In Sec.~\ref{sec:lriSym}, those basis states are written using the general letter $J_{A,B}$ to designate the individual angular momenta. When dealing with examples, we need to decide in particular if the total angular momentum accounts or not for the nuclear spin. In section \ref{sec:lri1OS} that is dedicated to ultracold polar molecules, nuclear spin will not be included in the basis, which will be justified, while it will be in section \ref{sec:lriHo2} dedicated to lanthanide atoms.

\section{Optical shielding of destructive collisions between ultracold polar molecules}
\label{sec:lri1OS}

\subsection{Scientific context}

In section \ref{sec:lriDiat}, I evoke the context of ultracold polar (\textit{i.e.}~heteronuclear) bialkali molecules in 2012 when I started to work on that topic. At that time, LiCs and KRb had been obtained in their rovibrational and hyperfine ground level \cite{deiglmayr2008, ni2008, ospelkaus2010}, but those gases suffered from limited lifetime (typically a fraction of second). This loss mechanism was attributed to the barrierless chemical reaction $2\,\mathrm{KRb} \to \mathrm{K}_2 + \mathrm{Rb}_2$ which according to Ref.~\cite{zuchowski2010} is exothermic, along with reactions involving lithium compounds. Unfortunately, later experiments involving presumably stable molecules also reported similar two-body losses \cite{takekoshi2014, park2015}, including NaRb \cite{guo2016}, RbCs \cite{gregory2019} and NaK \cite{voges2020}. The origin of this loss mechanism was attributed to so-called {}``sticky'' collisions \cite{mayle2013, croft2014}, in which two molecules form a long-lived complex until they are hit by a third one and are expelled from the trap. Still, this phenomenon is not yet fully understood \cite{christianen2019a, jachynski2022}, and photoinduced processes can also cause trap losses \cite{christianen2019a, liu2020, bause2021, gersema2021, gregory2021}.

\begin{figure}
  \begin{center}
  \includegraphics[width=0.49\linewidth]
    {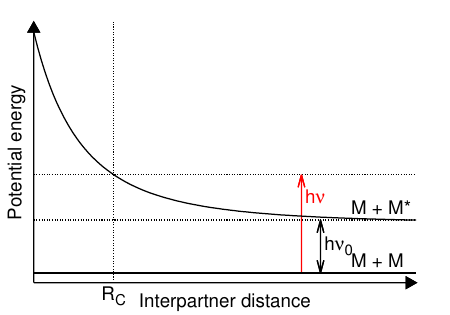}
  \includegraphics[width=0.49\linewidth]
    {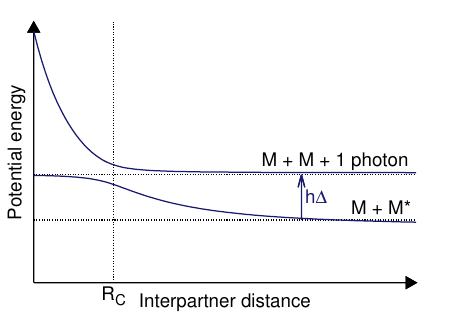}
  \caption{Schematics of the shielding mechanism illustrated with two potential-energy curves: a flat one dissociating into two ground-state molecules, hence enabling a barrierless reaction, and a repulsive one dissociating into an excited molecule. Panel (a): a laser with a frequency $\nu$ larger than the frequency $\nu_0$ of the molecular transition is shined. Panel (b): in a dressed-state picture, the lowest curve is raised by an energy of $h\nu$, so that its dissociation limit is higher by $h\Delta$ than the dissociation limit of the excited curve. The two curves cross at a distance $R_C$ called the Condon point. The electric field of the laser beam induces an avoided crossing between the curves, whose width is equal to $h\Omega$. $\Delta$ and $\Omega$ are respectively called the detuning and Rabi frequency.
  \label{fig:lriShldSchm}}
  \end{center}
\end{figure}

Because the loss mechanism implies collisions, it appeared necessary to keep the molecules far away from each other, in order to prevent those destructive collisions, hence the idea of ``{}shielding'' presented on Figure \ref{fig:lriShldSchm}. It can be achieved by tailoring repulsive long-range interactions between molecules by means of well designed electromagnetic fields. This shielding mechanism was reported in the 1990's in cold alkali-metal atomic gases submitted to a blue-detuned laser field with respect to the alkali-atom $D_2$ line \cite{marcassa1994, bali1994, zilio1996}. However,  spontaneous emission from the short-lived atomic excited states was proven to deteriorate the shielding efficiency \cite{suominen1995}. 

Due to their permanent electric dipole moment (PEDM), heteronuclear diatomic molecules possess purely rotational transitions, say $J=0 \to J=1$, inside the ground electronic and vibrational level, usually in the microwave (MW) region. Because the $J=1$ excited levels has an extremely large radiative lifetime, MW shielding is not hindered by spontaneous emission. Several theoretical proposals were dedicated to the control of molecular collisions with MW fields \cite{lassabliere2018, karman2018}, followed by experimental observations of larger sample lifetimes \cite{anderegg2021, bigagli2023, lin2023}, finally resulting in the quantum degeneracy of fermionic \cite{schindewolf2022} and bosonic molecules \cite{bigagli2024}. In a related work, a relatively strong static electric field was employed to reach Fermi degeneracy with KRb molecules \cite{matsuda2020, li2021}.

In the meantime, the Theomol team investigated optical shielding between pairs of unlike atoms \cite{orban2019, xie2022}, and between molecules \cite{xie2020}, trying to eliminate the problem of spontaneous emission. The idea was to shine a laser close to a weak transition, whose excited level has a large radiative lifetime. In polar bialkali molecules, this is the case of the $X^1\Sigma^+ \to b^3\Pi$ intercombination transition with a radiative lifetime in the microsecond range, see Ref.~\cite{vexiau2017} and references therein. In addition, optical fields also present easily controllable polarizations compared to MW ones, whereas circular polarization is required for MW shielding \cite{lassabliere2018, karman2018}.

\subsection{Description of basis states}

\subsubsection{One-molecule ground and excited states}

As mentioned above, our one-photon optical shielding (1-OS) mechanism is based on a transition between the ground state and the long-lived excited state $b^3\Pi$. The rovibrational levels of the ground state are presented in Subsection \ref{sec:lriDiat}: they are labeled $|X, v, p, j, M \rangle$ with the parity $p=(-1)^J$. Due to spin-orbit interaction, the $b^3\Pi$ excited state is mixed with the $A^1\Sigma^+$ one, and so a vibrational level $|i\rangle$ of the coupled $(A,b)$ system can be written
\begin{align}
  \left| i \right\rangle & =
   c_{i,A} \left| \Lambda=0, S=0, \Sigma=0, \Omega=0^+ \right\rangle
  \nonumber \\
    & + \frac{c_{i,b}}{\sqrt{2}} \left[ 
    \left| \Lambda=1, S=1, \Sigma=-1, \Omega=0^+ \right\rangle
     \pm \left| \Lambda=-1, S=1, \Sigma=1, \Omega=0^+ \right\rangle 
    \right]
\end{align}
where $\Lambda, \Sigma, \Omega$ denote the projections of the internuclear axis on the molecule of the orbital, spin and total angular momenta respectively. The superscript $+$ corresponds to the reflection symmetry through a plane containing that axis. Here, $A$ and $b$, which refer to the electronic state of one molecule, should not be mixed with the labels $A$ and $B$ of the molecules in two-body states. The coefficients are such that $|c_{i,A}|^2 + |c_{i,b}|^2 = 1$. For the lowest levels, because spin-orbit mixing is modest compared to the energy gap between $b$ and $A$ states, one coefficient is large and the other one is small. Those levels strongly inherits from the vibrational levels without spin-orbit, and can be labeled accordingly. For example, the lowest $|i=0 \rangle$ will be labeled $|b, v=0 \rangle$, even if it contains a few-percent component $|c_{i,A}|^2$ from the $A$ state. Due to electric-dipole selection rules, the latter will be responsible for the transition dipole moment (TDM) from the ground rovibronic level $|X,v=0 \rangle$, see Ref.~\cite{vexiau2017} and references therein.

At present, we consider rotation and parity. Being a $^1\Sigma$ state, a rovibrational level of the $A$ state has a parity $p = (-1)^J$. In contrast, a given $J$ value in the $b$ state possesses an even and an odd-parity level. The general expression of the rovibrational levels of $b$ including spin-orbit is therefore
\begin{align}
  \left| b, v, p=\pm 1, J, M \right\rangle & =
   \frac{1+p(-1)^J}{2} c_{v,A}
     \left| \Lambda=0, S=0, \Sigma=0, \Omega=0^+, J, M \right\rangle
  \nonumber \\
    & + \frac{c_{v,b}}{\sqrt{2}} \left[ 
    \left| \Lambda=1, S=1, \Sigma=-1, \Omega=0^+, J, M \right\rangle
  \right. \nonumber \\
  & \quad \left. + p(-1)^J 
     \left| \Lambda=-1, S=1, \Sigma=1, \Omega=0^+, J, M \right\rangle 
    \right] .
\end{align}
When $p = -(-1)^J$, the contribution of the $A$ state vanishes, leaving a pure $f$ rovibrational level of the $b$ electronic state. The latter cannot undergo transitions with rovibrational levels of the ground electronic state $X$, of $e$ character \cite{brown2003, lefebvre-brion2004}.

\subsubsection{Two ground-state molecules}

We apply the ideas of Section \ref{sec:lriSym} to the situation of identical ultracold bosonic molecules described in the space-fixed (SF) frame.
The basis states are symmetric with respect to the permutation of the molecules, given by $\eta=+1$ in Eqs.~\eqref{eq:lriSymSFPerm} and \eqref{eq:lriSymSFPerm2}. For the sake of pedagogy, we assume that the molecules are cold enough to collide in the $s$-wave regime $L=0$. Since they are initially in their lowest rovibrational level, $J_A=J_B=0$, the parity in the initial scattering state is $p = 1$ and its reflection symmetry is $\sigma=1$, see Eq.~\eqref{eq:lriSymSFRefl}. In the fully-coupled basis, states with $J_A=J_B=L=0$ can only give rise to $J_{AB} = J = M = 0$, owing to the triangle inequalities of Eq.~\eqref{eq:lriCplBas}.

\begin{table*}
  \begin{center}
  \caption{Selection rules of the electric dipole-dipole interaction, the one-photon and two-photon electric-dipole optical transitions, written for the fully-coupled and symmetrized basis states.
  }
  \label{tab:lriSFSlctRlDiat}
  \begin{tabular}{|c|ccc|}
    \hline
    Quantum & Dipole-dipole & One-photon & Two-photon \\
    numbers &  interaction  & transition & transition \\
    \hline
      \multirow{2}{*}{$[p_A,p_B]$} &
        $[\pm,\pm] \leftrightarrow [\mp,\mp]$ & 
        \multirow{2}{*}{$[\pm,\pm] \leftrightarrow [\pm,\mp]$} & 
        \multirow{2}{*}{unchanged} \\
      & $[\pm,\mp] \leftrightarrow [\mp,\pm]$ & & \\
      $[\Delta J_A,\Delta J_B]$ & $[\pm 1,\pm 1]$ or $[\pm 1,\mp 1]$ &
        $[0, \pm 1]$ & $[0, \pm 2]$ \\
      $\Delta J_{AB}$ & $0^a$ or $\pm 1$ or $\pm 2$ &
        $0^a$ or $\pm 1$ & $0^b$ or $\pm 2$ \\
      $\Delta L$ & $0^a$ or $\pm 2$ & 0 & 0 \\
      $\Delta J$ & 0 & $0^{a,c}$ or $\pm 1$ & $0^b$ or $1$ or $\pm 2$
      \\
      $\Delta M$ & 0 & $q$ & $q_1-q_2$ \\
      Parity & $\pm \leftrightarrow \pm$ & 
        $\pm \leftrightarrow \mp$ & $\pm \leftrightarrow \pm$ \\
      Reflection & $\pm \leftrightarrow \pm$ & 
        $\pm \leftrightarrow \pm$ & $\pm \leftrightarrow \pm$ \\
      Permutation & $\pm \leftrightarrow \pm$ & 
        $\pm \leftrightarrow \pm$ & $\pm \leftrightarrow \pm$ \\
    \hline
    \multicolumn{4}{l}{$^a$ $\Delta X=0$ except $0 \leftrightarrow 0$}
    \\
    \multicolumn{4}{l}{$^b$ $\Delta X=0$ except $0 \leftrightarrow 0$ 
      and $1/2 \leftrightarrow 1/2$} \\
  \end{tabular}
  \end{center}
\end{table*}

The molecules interact through the dipole-dipole term $\mathrm{V}_{\mathrm{dd}}^{\mathrm{SF}}$, for which $\ell_A = \ell_B = 1$ and $\ell = 2$ in Eq.~\eqref{eq:lriVsf4}. The reduced dipole matrix elements are $\langle X, \qnpr{v_k=0}, \qnpr{J_k} \| \mathrm{Q}_{1} \| X, \qnpr{v_k=0}, J'_k \rangle = \sqrt{2J_k+1} \times C_{J'_k010}^{J_k0} d_0$, with $d_0$ the permanent electric dipole moment of the vibrational ground level ($d_0 = 1.304$~a.u.~for \narb{} \cite{lepers2013}). The selection rules associated with the $\mathrm{V}_{\mathrm{dd}}^{\mathrm{SF}}$ operator are given in Table~\ref{tab:lriSFSlctRlDiat}. Both rotational quantum numbers vary by one unit at the same time: $[J_A,J_B] = [0,0]$ states are coupled with $[1,1]$ ones, themselves with $[0,2]$, and then $[1,3]$, and so on. The initial $s$-wave state is coupled to $d$-wave and then $g$-wave ones, etc. In the absence of external field, the complex is invariant upon rotation, parity and reflection. The entrance channel is therefore coupled with $|[ J_A=1, J_B=1], J_{AB}=2, L=2, J=0, M=0 \rangle$, itself coupled with $|[ J_A=0, J_B=2], J_{AB}=2, L=2, J=0, M=0 \rangle$, and so on. Note that we dropped the quantum numbers $X$, $v_B=0$ and $\eta=1$ and indicated symmetrized states with the braces $[\,]$. In principle, the number of coupled states is infinite, but in practice the coupling decreases with increasing $[J_A,J_B]$, so that the convergence of the lowest LR PECs can be quickly reached. Typically, we took $\, 0\le J_A,J_B \le 4\,$ and $\,0 \le L \le 4\,$ in our calculations.

\begin{figure}
  \begin{center}
  \includegraphics[width=0.6\textwidth]
    {fig-lri/Diat_PECs_E0}
  \includegraphics[width=0.6\textwidth]
    {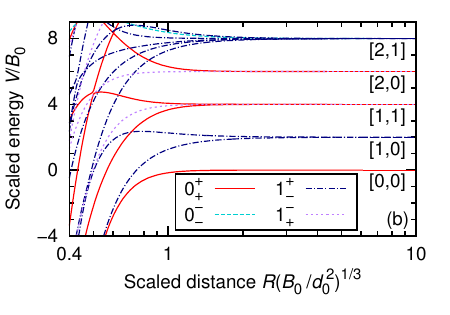}
  \caption{(a) Long-range potential-energy curves between two \narb{} molecules in the ground electronic and vibrational level: (a) in the body-fixed frame and (b) in the space-fixed frame. The curves are plotted in scaled units and their asymptotes are labeled $[J_A,J_B]$. The curves are sorted by symmetries: on panel (a), $0_g^+$ (solid lines), $0_u^+$ (dashed lines), $1_g$ (dotted lines) and $1_u$ (dash-dotted lines); on panel (b), $0_+^+$ (solid lines), $0_-^-$ (dashed lines), $1_+^-$ (dotted lines) and $1_-^+$ (dash-dotted lines), see text for explanation of labeling.
  }
  \label{fig:lriSFDiatPecs1}
  \end{center}
\end{figure}

This is illustrated on Fig.~\ref{fig:lriSFDiatPecs1}, where the lowest PECs are shown in the body-fixed (BF) frame on panel (a), and in the SF frame on panel (b). The left panel is reproduced from Figure \ref{fig:lriDiatPecs1} (a). The characteristic distance and energy are respectively equal to 175~a.u. and 0.0697~\cmi{}. In the SF frame, the potential operator
\begin{equation}
  \mathrm{V}(R) = \frac{\hbar^2 \mathbf{L}^2}{2\mu R^2}
    + B_0 \left[ \mathbf{J}_A^2 + \mathbf{J}_B^2 \right] 
    + \mathrm{V}_\mathrm{dd}^\mathrm{SF}(R) 
    + \mathrm{V}_\mathrm{vdW}^\mathrm{SF}(R) ,
  \label{eq:lriSFDiatV}
\end{equation}
is diagonalized at various $R$-values. In addition to the DDI and the individual rotational energies, $\mathrm{V}(R)$ also accounts from the isotropic electronic van der Waals interaction $\mathrm{V}_{\mathrm{vdW}}^{\mathrm{SF}} = -C_6^{e} / R^6$, where $C_6^{e} = 7731$~a.u., see Sec.~\ref{sec:lriDiat} and Ref.~\cite{lepers2013}. The SF curves also comprise the angular part of the relative kinetic energy, equal to $\hbar^2 L(L+1) / 2\mu R^2$, with $\mu$ the reduced mass of the complex. Those two terms only bring diagonal terms to the potential operator $\mathrm{V}(R)$. Their influence is not visible at the scale of the two graphs, whose curves look identical. The curves of panel (b) are characterized by even partial waves, $M=0$ and $\eta=1$; they are labeled $J_{p}^{\sigma}$, with $p$ and $\sigma$ corresponding to the parity and reflection symmetries. There are fewer curves on panel (b) because the SF-frame symmetries are more restrictive.

\subsubsection{One electronically-excited molecule}

At present, I describe the effect of a laser with a frequency close to the $|X, v_k=0, p_k=1, J_k=0 \rangle \to |b, v_k=0, p_k=-1, J_k=1 \rangle$ molecular transition ($k=A,\,B$). Regarding the states of the complex, the one-photon selection rules are given in Table \ref{tab:lriSFSlctRlDiat}. They can be deduced from Eq.~\eqref{eq:lriEFldCpl}, giving the matrix element of the Stark operator in the fully-coupled basis. The entrance channel $|[J_X=0, J_X=0], J_{AB}=0, L=0, J=0, M=0 \rangle$ is coupled to the electronically-excited one $|[J_X=0, J_b=1^-], J_{AB}=1, L=0, J=1, M=q \rangle$ with $p=-1$ and $\sigma = \eta = 1$, where $J_b^{p_b}$ stands for the rotational level and parity of the $b$ states, and $q=0$ (resp.~$\pm 1$) for $\pi$ (resp.~$\sigma^\pm$) light polarization. In what follows, we consider the case $M=q=0$. We recall that the parity of a $J_X$ level is $(-1)^{p_X}$.

\begin{figure}
  \begin{center}
  \includegraphics[width=0.6\textwidth]
    {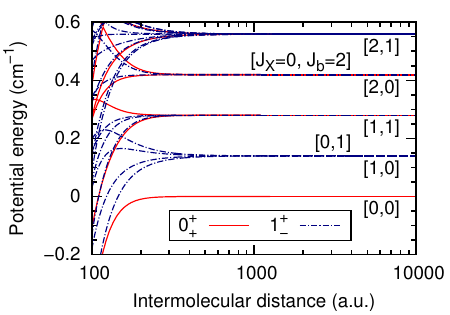}
  \caption{Long-range potential-energy curves between two \narb{} molecules, one being in the ground electronic and vibrational level and the other being in the lowest vibrational level of the excited electronic state $b^3\Pi_{0^+}$. The asymptotes are labeled $[J_X,J_b]$ using the rotational numbers in both states. The curves are sorted by symmetries: $0_+^+$ (solid lines) and $1_-^+$ (dash-dotted lines). The labeling is the same as in Figure \ref{fig:lriSFDiatPecs1}.
  }
  \label{fig:lriSFDiatPecs2}
  \end{center}
\end{figure}

The states of the family $[J_X=0, J_b=1^-]$ are coupled among themselves, in particular $L=0$ and 2, due to the resonant DDI (or the so-called excitation exchange). In unsymmetrized basis, it corresponds to states of type $J_X=0, J_b=1^-$ and $J_b=1^-, J_X=0$. This term is proportional to $d_{X0,b0}^2$, with $d_{X0,b0} = 0.1918$~a.u.~the TDM between the lowest vibrational levels of the $X$ and $b$ states. But the states $[J_X=0, J_b=1^-]$ are also coupled to the $[J_X=1, J_b=0^+]$ ones under the effect of the direct DDI proportional to the product $d_{X0}d_{b0}$ of PEDMs ($d_{b0} = 1.735$~a.u.). For strongly polar molecules, the direct term is significantly larger than the resonant one. In polar bialkali molecules, the asymptotes $[J_X=0, J_b=1^-]$ and $[J_X=1, J_b=0^+]$ are almost degenerate since the rotational constants of the $X$ and $b$ are almost equal \cite{xie2020}. The direct term then arises at the first order of degenerate perturbation theory, scaling as $R^{-3}$, which is visible on Figure \ref{fig:lriSFDiatPecs2}. In particular, we see repulsive PECs which are promising for optical shielding, as they are similar to those on which MW shielding relies.

\subsubsection{Role of the hyperfine structure}

In ultracold experiments, atoms or molecules are often prepared in a well defined hyperfine sublevel (\textit{e.g.}~the lowest one), so that hyperfine structure (HFS) can in principle not be ignored in the models. However, in molecular scattering calculations, HFS strongly increases the size of the basis and so the computational time. It also makes more complex the physical understanding of the processes at play. Various strategies have been proposed to treat HFS in ultracold molecular collisions, see for instance Ref.~\cite{simoni2006, gonzalez2011, tscherbul2023}. In our study of optical shielding, we ignore it, and in what follows, we discuss in which conditions it can be done regarding our symmetrized basis states.

The hyperfine levels of a bialkali molecule is often described with the nuclear-spin quantum numbers of its composing atoms, namely $I_{A1}$, $M_{I,A1}$, $I_{A2}$ and $M_{I,A2}$, and similarly for $B$. Because the two molecules are identical, then $A1=B1$ (say Na) and $A2=B2$ (say Rb). Starting back from the fully uncoupled basis, the complex states are $| e_A, v_A, J_A, M_{J,A},$  $I_{A1}, M_{I,A1}, I_{A2}, M_{I,A2}, $  $e_B, v_B, J_B, M_{J,B},$  $I_{B1},$ $M_{I,B1}, I_{B2}, M_{I,B2},$  $L, M_L \rangle$, with $e_k$ and $v_k$ denote the electronic and vibrational levels of $k=A,B$. Since the parity and reflection operations do not act on nuclear spins, the character of the complex states is the same with or without HFS. To express the action of permutation, we drop electronic and vibrational quantum numbers and we gather all the HFS ones,
\begin{align}
  & \mathrm{P}_{AB} \left| J_A, M_{J,A}, I_{A1}, M_{I,A1}, 
    I_{A2}, M_{I,A2}, J_B, M_{J,B}, I_{B1}, M_{I,B1}, I_{B2}, M_{I,B2}, 
    L, M_L \right\rangle
  \nonumber \\
  =\; & \mathrm{P}_{AB} \left| J_A, M_{J,A}, J_B, M_{J,B}, L, M_L
    \right\rangle \left| I_{A1}, M_{I,A1}, I_{A2}, M_{I,A2}, 
    I_{B1}, M_{I,B1}, I_{B2}, M_{I,B2} \right\rangle
  \nonumber \\
  =\; & (-1)^L \left| J_B, M_{J,B}, J_A, M_{J,A}, L, M_L
    \right\rangle \left| I_{B1}, M_{I,B1}, I_{B2}, M_{I,B2},
    I_{A1}, M_{I,A1}, I_{A2}, M_{I,A2} \right\rangle
\end{align}
If both molecules are in the same HFS sublevel, the permutation of the HFS quantum numbers (second half of the last line) leaves the complex state unchanged, and so the angular-momentum coupling scheme of Eq.~\eqref{eq:lriCplBas} can be applied on the sole rotational quantum numbers. On a physical point of view, it corresponds to the situation where a significantly strong magnetic field decouples the HFS from other degrees of freedom (hyperfine Paschen-Bach regime). Furthermore, since the molecules interact via electrostatic forces, we also assume that the HFS states are not modified during the collision.

\subsection{One-photon shielding}
\label{sub:lri1OS}

\begin{figure}
  \begin{center}
  \includegraphics[width=0.6\textwidth]
    {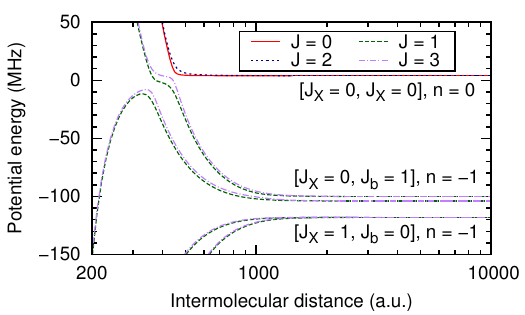}
  \caption{Dressed long-range potential-energy curves between two \narb{} molecules submitted to a linearly-polarized optical field of Rabi frequency $\Omega = 10$~MHz and blue-detuned by $\Delta = 100$~MHz with respect to the transition $X^1\Sigma^+, v_X=0, J_X=0 \to b^3\Pi_{0^+}, v_b=0, J_b=1$.
  }
  \label{fig:lriSFDiatPecs3}
  \end{center}
\end{figure}

The results presented in this section were obtained by Ting Xie and published in Ref.~\cite{xie2020}. In order to give a first answer about the 1-OS feasibility, we plot on Figure \ref{fig:lriSFDiatPecs3} dressed LR PECs for two molecules submitted to an optical field whose frequency is blue-detuned by $\Delta = 100$~MHz with respect to the transition $X^1\Sigma^+,v_X=0 \to b^3\Pi_{0^+},v_b=0$ of energy $h \times 338.960$~THz. The curves are obtained by diagonalizing at each $R$ the potential operator \eqref{eq:lriSFDiatV} in the two blocks of electronic states $[X,X]$ and $[X,b]$, plus a molecule-field interaction taken as the Stark operator of Eq.~\eqref{eq:lriEFldCpl}, coupling states of the families $[J_X=0,J_X=0]$ and $[J_X=0,J_b=1^-]$, and parametrized by the Rabi frequency $\Omega = d_{X0,b0}\mathcal{E} / 2\pi\hbar$. On Figure \ref{fig:lriSFDiatPecs3}, one Floquet block, characterized by the photon numbers $n=0$ and -1, is sufficient to obtain converged PECs. According to the one-photon selection rules of Table \ref{tab:lriSFSlctRlDiat}, states of the kind $|[X,X],J=0 \rangle$ interact with $|[X,b],J=1 \rangle$, themselves interacting with $|[X,X],J=2 \rangle$ and even to $|[X,X],J=1 \rangle$ in circular polarization, and so on. States up to $|[X,b],J=3 \rangle$ ensures convergence of the dressed PECs.


Along with the photon number, the rotational quantum numbers of the dissociation limits are given in Figure \ref{fig:lriSFDiatPecs3}. But unlike Figs.~\ref{fig:lriSFDiatPecs1} and \ref{fig:lriSFDiatPecs2}, this labeling is approximate, since the molecule-field coupling is $R$-independent, and thus the eigenstates of the potential operator are mixed states of the $[X,X]$ and $[X,b]$ blocks.
Due to the field dressing, the $[J_X=0,J_b=1]$ asymptote lies below the entrance channel $[J_X=0,J_X=0]$, which corresponds to the situation where one photon is taken from the field to perform an absorption. The spacing between the asymptotes is equal to the detuning $\Delta$. The curves correlated to $[J_X=0,J_X=0]$ cross the repulsive ones correlated to $[J_X=0,J_b=1]$, resulting in two avoided crossings whose widths are proportional to the Rabi frequency. Therefore, two colliding ground-state molecules are likely to follow adiabatically one of the two crossing, and therefore to turn back to large distances, under the effect of the repulsive curves. The curves look similar in linear, panel (a), or circular, panel (b), polarization, even though in the latter case, there is an additional avoided crossing, between $|[J_X=0,J_b=1],J=1 \rangle$ and $|[J_X=0,J_X=0],J=1 \rangle$.


To confirm the efficiency of 1-OS, T.~Xie also calculated the rate coefficients characterizing the three types of collisions
\begin{eqnarray}
  \mathrm{elastic \;}k_\mathrm{el} : \quad 
   2\mathrm{\:NaRb\:}(J_X=0) & \to 
   & 2 \mathrm{\:NaRb\:}(J_X=0) \\
  \mathrm{inelastic \;}k_\mathrm{in} : \quad 
   2\mathrm{\:NaRb\:}(J_X=0) & \to 
   & \mathrm{NaRb\:}(J_X=0) + \mathrm{NaRb\:}(J_b=1) \\
  \mathrm{reactive \;}k_\mathrm{re} : \quad 
   2\mathrm{\:NaRb\:}(J_X=0) & \to
   & \mathrm{Na}_2 + \mathrm{Rb}_2
\end{eqnarray}
for various temperatures, detunings and Rabi frequencies. Shielding is all the more efficient that the elastic, so-called good, collisions dominate over the inelastic and reactive, so-called bad, ones. This is quantified by the ratio $\gamma = k_\mathrm{el}/(k_\mathrm{in} + k_\mathrm{re})$. A value larger than 1000 indicates the feasibility of evaporative cooling. 

The rates are obtained by calculating the reactance $\mathrm{K}$ and scattering $\mathrm{S}$ matrices using the time-independent Schr\"odinger equation. Following Johnson \cite{johnson1973} and Manolopoulos \cite{manolopoulos1986}, the log-derivative of the multi-channel $R$-dependent wave function is propagated from small to larger intermolecular distances. The interaction potential $\mathrm{V}(R)$ is the LR + molecule-field one described in the previous subsection. Moreover, to simulate the short-range losses due to reactive collisions, the $\mathrm{K}$ matrix is taken purely imaginary at the minimal $R$-value of the propagation. This procedure, described in details in Ref.~\cite{karam2024} and references therein, was used with success to calculate the reactive rate coefficients without any field. In the case of \narb{}, it gives $k_\mathrm{re} = 4.0 \times 10^{-10}$~cm$^3$.s$^{-1}$, which is consistent with the experimental one $4.5(2) \times 10^{-10}$~cm$^3$.s$^{-1}$ \cite{ye2018}.


\begin{figure}
  \begin{center}
  \includegraphics[width=0.48\textwidth]
    {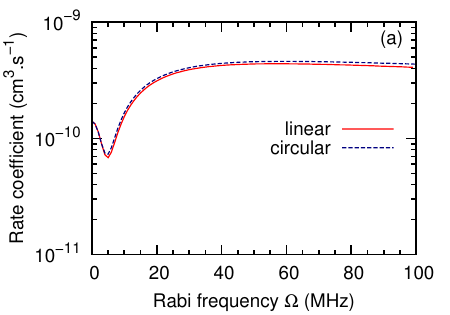}
  \includegraphics[width=0.48\textwidth]
    {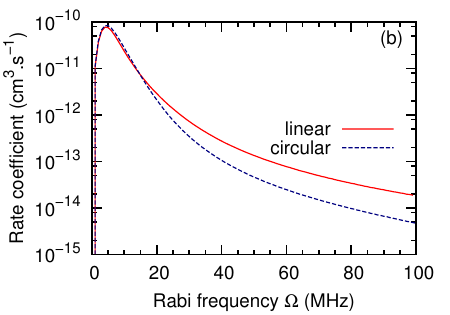}
  \includegraphics[width=0.48\textwidth]
    {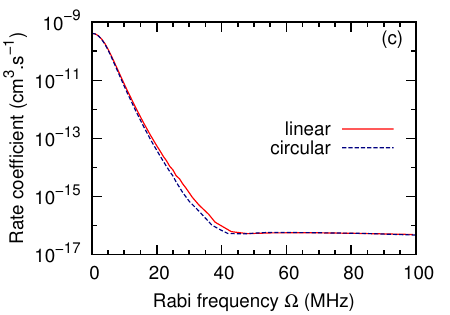}
  \includegraphics[width=0.48\textwidth]
    {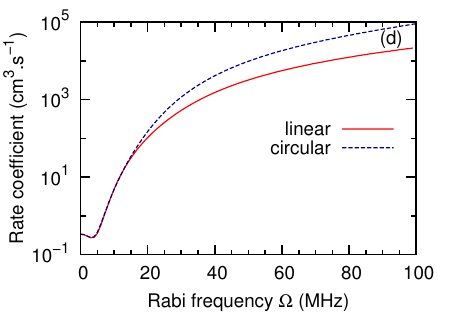}
  \caption{Rate coefficients of elastic (a), inelastic (b) and reactive collisions (c), as well as the ratio $\gamma$ of good-over-bad collisions for two ground-level \narb{} molecules, as functions of the Rabi frequency, for a fixed detuning of $\Delta = 100$~MHz, and for both linear and circular polarizations.
  }
  \label{fig:lriSF1OSRate}
  \end{center}
\end{figure}

Figure \ref{fig:lriSF1OSRate} displays the rates characterizing the three types of collisions, as well as the ratio $\gamma$ of good-over-bad collisions, as functions of the Rabi frequency for a fixed detuning $\Delta = 100$~MHz and a fixed temperature $T = 400$~nK. The 1-OS efficiency is confirmed in those simulations, since the reactive rates drops by 7 orders of magnitude, reaching a minimum around $\Omega = 40$~MHz. The inelastic collisions of panel (b) also strongly decrease from 0 to 100~MHz, but they pass through a maximum around 5~MHz, which can be interpreted using the PECs of Figure \ref{fig:lriSFDiatPecs3}. When they approach each other, the molecules are sensitive to the avoided crossing and follow the repulsive branch. But on their way back, because the avoided crossing is not very large, a significant fraction cross diabatically and end up in the $[J_X = 0, J_b = 1]$ channel, giving rise to inelastic collisions. This phenomenon vanishes as $\Omega$ increases, since the avoided crossing enlarges. Compared to inelastic collisions, the opposite evolution is observed for elastic ones, see panel (a). Finally, the ratio $\gamma$ globally increases with Rabi frequency, reaching 1000 at 30 -- 40 MHz. The increase is stronger for circular polarization, which may be due to the existence of an additional avoided crossing evoked in the previous subsection.

In Ref.~\cite{xie2020}, we deduce the Rabi frequency and the corresponding laser intensity $I$ that are necessary to reached $\gamma=1000$, not only for NaRb, but for all the other polar bialkali molecules except LiNa and KRb. For the two latter molecules, the PEDM is too weak to ensure a good shielding effect. Apart from them, a ratio of 1000 is reachable for experimentally relevant intensities: for example 6.3~W.cm$^{-2}$ in \narb{}. That intensity tends to shrink with the molecular PEDM.

\paragraph{Spontaneous emission and photon scattering.}

We recall here that the motivation of choosing a 1-OS intercombination transition was to get rid of the spontaneous emission observed in pairs of cold atoms. To quantify the effect of spontaneous emission, we used a semi-classical picture, which proved to be reliable with atoms, to calculate the time $\tau$ spent by the complex on an excited, \textit{i.e.}~curve of Fig.~\ref{fig:lriSFDiatPecs3}, in other words the time spent to go from the crossing to the turning point. This duration depends on the initial velocity, hence the temperature, and on the detuning. For \narb{}, $\tau$ varies from 0.79~ns for $\Delta = 10$~MHz down to 1.2~ps for 500~MHz. Since it is way smaller than the radiative lifetime of the $b$ state $\tau_b = 6.97~\mu$s, we concluded that spontaneous emission during the collision was negligible. 

The publication of our article \cite{xie2020} resulted in discussions with experimentalists who pointed out the harmful role of one-molecule photon scattering for the 1-OS scheme. Indeed, molecules spend most of their time very far away from each other. During those moments, interacting with the shielding laser, they undergo photon scattering whose rate is \cite{grimm2000} (assuming two non-degenerate levels)
\begin{equation}
  \Gamma_\mathrm{sc} = \frac{3\pi c^2}{2\hbar\omega_0^2} 
   \left( \frac{\Gamma_b}{\Delta'} \right)^2 I
   = \left( \frac{\Omega}{\Delta} \right)^2 
   \frac{\Gamma_b}{4}
\end{equation}
where $\Delta' = 2\pi\Delta$, $\omega_0 = 2\pi\nu_0$ and $\Gamma_b = \tau_b^{-1} = 1.43 \times 10^5$~s$^{-1}$ is the radiative relaxation rate of the $b$ state. If we assume after Fig.~\ref{fig:lriSF1OSRate} that 1-OS becomes efficient for $\Omega / \Delta \approx 0.4$, we obtain a photon-scattering rate of $\Gamma_\mathrm{sc} \approx 0.04 \times \Gamma_b = 5.72 \times 10^3$~s$^{-1}$. This is problematic because, whenever it happens, the molecules kinetic energy increases by the recoil energy, which tends to heat up the sample. This does probably not destroy the 1-OS process, but it makes it counteract against evaporative cooling, which is not suitable for reaching Bose-Einstein condensation.

\subsection{Two-photon shielding}
\label{sub:lri2OS}

Then, the idea came out from Silke Ospelkaus's group in Hannover to use a two-photon transition, hence realizing two-photon optical shielding (2-OS). Indeed, if the frequency of the two lasers are chosen at the two-photon resonance, there exists a so-called dark state in which the molecules do not feel the presence of the laser beams, and so do not undergo photon scattering. This is equivalent to electromagnetically-induced transparency (EIT) \cite{fleischhauer2005}.

\subsubsection{Position of the problem}


\begin{figure}
  \begin{center}
  \includegraphics[width=0.48\textwidth]
    {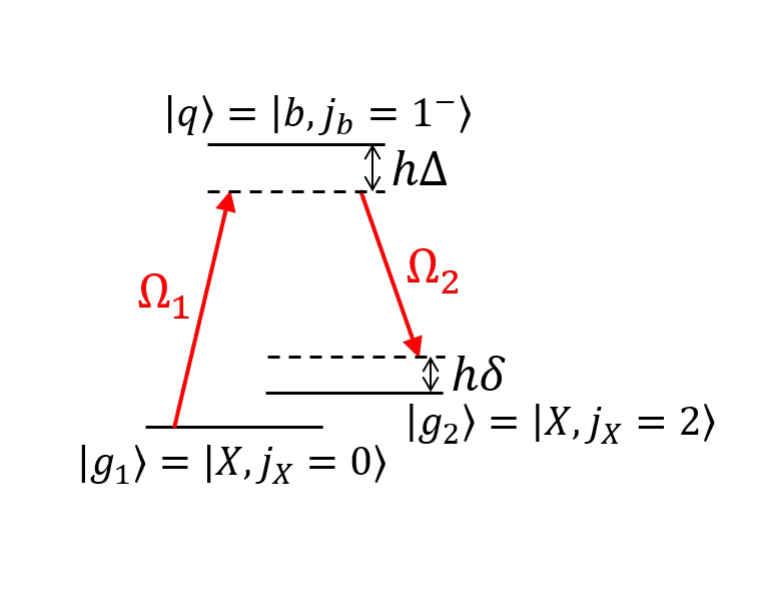}
  \caption{Scheme of $\Lambda$ two-photon transition with one molecule.
  }
  \label{fig:lriSF2OSSchm}
  \end{center}
\end{figure}

Figure \ref{fig:lriSF2OSSchm} presents a schematics of the $\Lambda$ three-level system for one molecule. It consists of two close ground states $|g_1\rangle$ and $|g_2\rangle$, each of which is coupled to an excited states $|e\rangle$ by lasers of Rabi frequencies $\Omega_1$ and $\Omega_2$, respectively. The two-photon resonance condition is achieved when $\delta = 0$. Under the rotating-wave approximation, the Hamiltonian of the three-level system can be written in the field-dressed basis $\{ |\bar{g}_1\rangle, |\bar{g}_2\rangle, |\bar{e}\rangle \}$
\begin{equation}
  \mathrm{H} = 2\pi\hbar \left( \begin{array}{ccc}
    0 & 0 & \Omega_1/2 \\
    0 & \delta & \Omega_2/2 \\
    \Omega_1/2 & \Omega_2/2 & \Delta
  \end{array} \right) .
  \label{eq:lriSFH3Lev}
\end{equation}
At the two-photon resonance $\delta=0$, one of the eigenvectors is proportional to $\Omega_2 |\bar{g}_1\rangle - \Omega_1 |\bar{g}_2\rangle$. Having no $|\bar{e}\rangle$-component, it is as if this {}``dark" state were insensitive to the presence of the fields. Note that for $\Omega_2 \gg \Omega_1$, this eigenvector is close to $|\bar{g}_1\rangle$. One can go one step further by applying adiabatic elimination \cite{brion2007}, namely assume that the excited state is not populated. In this condition, Eq.~\eqref{eq:lriSFH3Lev} can be reduced to a $2\times 2$ effective Hamiltonian consisting of an effective detuning $\Delta_\mathrm{eff} = \delta + (\Omega_2^2 - \Omega_1^2) / 2\Delta$ and an effective Rabi frequency $\Omega_\mathrm{eff} = -\Omega_1 \Omega_2 / 2\Delta$. It is valid for $\Delta \gg \delta, \Omega_1, \Omega_2$. This is another condition to minimize spontaneous emission.

The choice of the molecular levels composing the basis of the Hamiltonian \eqref{eq:lriSFH3Lev} is of course crucial. Taking $|g_1\rangle = |X, v_X=0, p_X=1, J_X=0 \rangle$ seems obvious since molecules are prepared in their rovibrational ground level. Because $|e\rangle$ must satisfy electric-dipole selection rules, see Table \ref{tab:lriSFSlctRlDiat}, it must be an odd level $p_k=-1$ in the first excited rotational level $J_k=1$. But it can be any vibronic level with a sizable TDM with the ground one. Here, in coherence with the previous subsection, we take the $|b, v_b=0 \rangle$ level. Therefore, the level $|g_2\rangle$, which must be coupled to $|e\rangle$ is naturally $|X, v_X=0, p_X=1, J_X=2 \rangle$. Combining the selection rules of the two one-photon transitions, we can derive those of the two-photon transitions, given in the last column of Table \ref{tab:lriSFSlctRlDiat}.

\subsubsection{Relevant potential-energy curves}

In a first study, presented in Ref.~\cite{karam2023}, we selected and computed the LR PECs which, according to two-photon selection rules, are likely to play an important role in 2-OS, and we made a comparison with the PECs of the successful microwave shielding.

Because we have two interacting molecules, the parameter of Hamiltonian \eqref{eq:lriSFH3Lev} are now $R$-dependent. If we assume that we can associate each state of \eqref{eq:lriSFH3Lev} with one LR PEC correlated to the asymptotes $[g_1,g_1]$, $[g_1,g_2]$ and $[g_1,e]$ described in the previous paragraph, the detunings $\delta(R)$ and $\Delta(R)$ are the differences between the PECs. Then the question arises at which distance $R$ the two-photon resonance condition should be applied. Since molecules spend most of the time far away from each other, we decided to apply it at $R\to +\infty$, \textit{i.e.}~in the one-molecule situation described above. In return, the two-photon resonance is not any more achieved in the crossing region. However, selecting a large detuning $\Delta (R\to +\infty)$ with the excited states ensures that, even in the crossing region, the detuning is still large enough to validate adiabatic elimination, and so minimize spontaneous emission. The relevant electronically-excited PECs are those of Figure~\ref{fig:lriSFDiatPecs2}, except that we can take a red-detuned frequency.


\begin{figure}
  \begin{center}
  \includegraphics[width=0.48\textwidth]
    {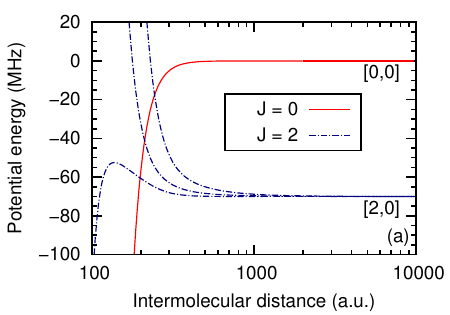}
  \includegraphics[width=0.48\textwidth]
    {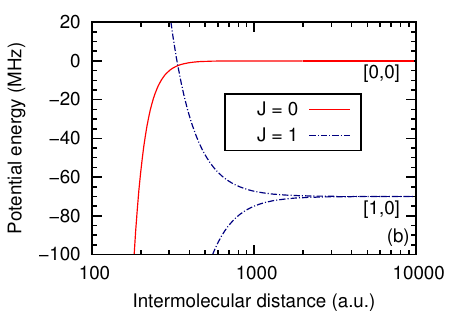}
  \caption{Shifted long-range potential-energy curves between two \nak{} molecules mimicking a blue detuning of 70~MHz. On panel (a), the even-parity curves dissociating into $[0,2]$ with $J=2$ are lowered by $6B_0+h \times 70$~MHz; on panel (b), the odd-parity curves dissociating into $[0,1]$ with $J=1$ are lowered by $2B_0+h \times 70$~MHz. In both cases, the unshifted lowest PEC is shown.}
  \label{fig:lriSFDiatPecs4}
  \end{center}
\end{figure}

As for the ground PEC, its asymptote $[g_1,g_1]$ is naturally the entrance channel $|[J_X=0,J_X=0], J_{AB}=0, L=0, J=0, M=0 \rangle$, while the second asymptote $[g_1,g_2]$ is $|[J_X=0,J_X=2], J_{AB}=2, L=0, J=2, M=0 \rangle$, following two-photon selection rules. At $R\to +\infty$, the latter state is degenerate with $|[J_X=0,J_X=2], J_{AB}=2, L=2\text{\,and\,}4, J=2, M=0 \rangle$. Those three basis states give rise to three LR PECs visible on Figs.~\ref{fig:lriSFDiatPecs3} and \ref{fig:lriSFDiatPecs4}. Their repulsive nature can be explained as follows: due to DDI, states of the family $[J_X=0,J_X=2]$ are coupled to the $[J_X=1,J_X=1]$ and $[J_X=1,J_X=3]$ ones, the first one being located $2B_0$ below $[J_X=0,J_X=2]$, and the second one $8B_0$ above. The family $[J_X=1,J_X=1]$ has thus the strongest influence, repelling $[J_X=0,J_X=2]$ towards higher energies. This results in strong and repulsive vdW interaction $C_6/R^6$, with $C_6$ of the same order of magnitude as in section \ref{sec:lriDiat}.

As an illustration, on Fig.~\ref{fig:lriSFDiatPecs4} (a), the PECs dissociating to $[J_X=0,J_X=2]$ have been lowered to mimic an effective blue detuning of 70~MHz \cite{karam2023}. On panel (b), the PECs dissociating to $[J_X=0,J_X=1]$, at play in MW shielding, have been lowered in a similar way. On both cases, the entrance channel crosses repulsive curves, for slightly smaller distance on panel (a), but that distance can be modified by changing the detuning. The Rabi frequencies are set to zero, but we checked that the eigenvectors of the repulsive curves of panel (a) contain a sizable component of $|[J_X=0,J_X=2], J_{AB}=2, L=0, J=2, M=0 \rangle$, which suggests a good two-photon coupling with the entrance channel.

\subsubsection{Collision rates}


\begin{figure}
  \begin{center}
  \includegraphics[width=0.98\textwidth]
    {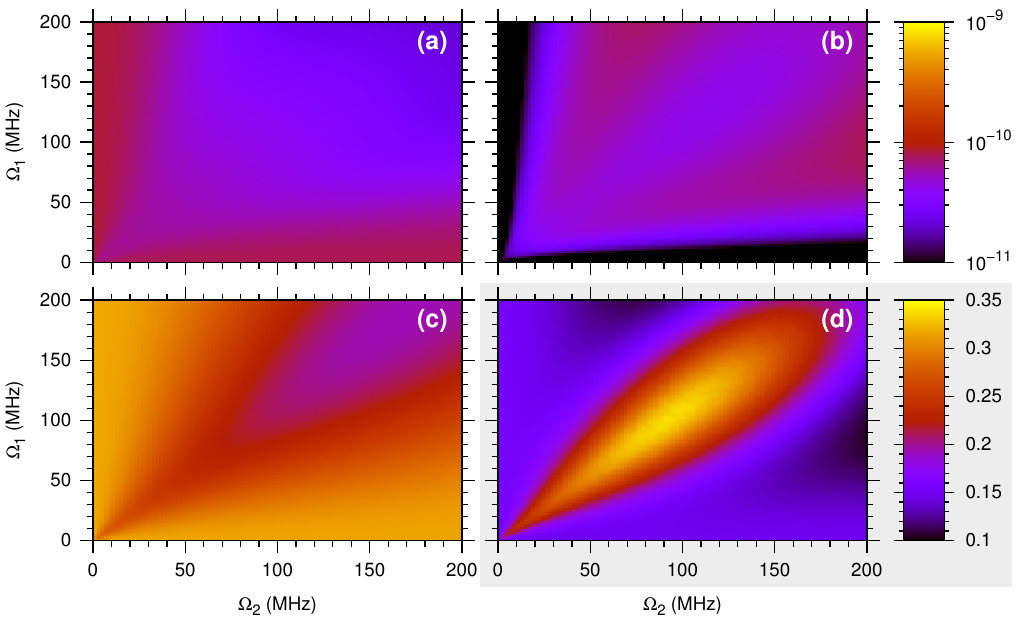}
  \caption{Rate coefficients of elastic (a), inelastic (b) and reactive collisions (c), as well as the ratio $\gamma$ of good-over-bad collisions (d) for two ground-level \nak{} molecules, as functions of the two Rabi frequencies $\Omega_1$ and $\Omega_2$, at the two-photon resonance $\delta=0$, the detuning of $\Delta = 1$~GHz and the temperature $T=300$~nK. The color scale of panels (a)-(c) is given on the top-right color box, while that of panel (d) is on the bottom-right color box.
  }
  \label{fig:lriSF2OSRate}
  \end{center}
\end{figure}

In the same spirit as 1-OS, we have calculated the rates of elastic, inelastic and reactive collisions, as well as the ratio of good-over-bad collisions. This work was done by Charbel Karam in his Ph.~D \cite{karam2024}. Results are presented on Fig.~\ref{fig:lriSF2OSRate} as function of the Rabi frequencies $\Omega_1$ and $\Omega_2$, and for fixed detunings $\delta=0$ and $\Delta=1$~GHz. The collision rates are all plotted on the same color scale, whereas the ratio $\gamma$ has its own scale. The most pronounced variations are observed for the reactive rate, which decreases by one order of magnitude in the region $\Omega_1 \approx \Omega_2$. However, this decrease is probably not sufficient to significantly increase the lifetime of molecular samples, all the more since the inelastic rate grows up with Rabi frequencies. A closer look at the scattering matrix shows that the main exit channel of inelastic collisions is $[J_X=0, J_X=2]$ \cite{karam2024}. The elastic rate slightly shrinks with the Rabi frequencies, which makes all in all that the largest $\gamma$-value is around 0.33, far below the expected value of 1000.
	
This somewhat disappointing result is in contradiction with the aspect of LR PECs of Figure \ref{fig:lriSFDiatPecs4}. This is hard to interpret since, as the two-photon resonances, the dressed PECs dissociating to $[J_X=0, J_X=0]$, $[J_X=0, J_X=2]$, and also $[J_X=2, J_X=2]$ are almost degenerate. This creates a congested landscape of curves, in which the avoided crossings between the initial and the repulsive curves do not seem broad enough. Because the latter are due to the DDI, it is likely that a static electric field will enlarge the crossings by orienting the molecules, and so increase their interaction energy. Furthermore, the DC field is inescapable to induce the molecular electric dipole moment in the SF frame on which all the dipolar effects that are envisioned are based. In consequence, modeling 2-OS in presence of a DC electric field is the main prospect of this work.

\section{Long-range doubly polar homonuclear molecules of lanthanides}
\label{sec:lriHo2}

To finish this part on long-range interactions, I describe Hui Li's post-doctoral work published in Ref.~\cite{li2019}. On a methodological point of view, it is my most elaborate investigation on LR interactions, including direct and resonant terms, hyperfine structure of lanthanide atoms, and external electric and magnetic fields. It was also the first time that I used a symmetrized basis, as described in Section \ref{sec:lriSym}. This work is related to the prediction of a  doubly polar gas of dysprosium atoms presented in Section \ref{sec:atDblPol} and published in Ref.~\cite{lepers2018}.

In section \ref{sec:atDblPol}, I have already presented the interest in doubly dipolar gases, namely composed of particles carrying both an electric and a magnetic dipole. I reported on the possibility to produce a doubly dipolar gas of dysprosium atom prepared in a superposition of opposite-parity quasi-degenerate ($1.3$~\cmi{} away) energy levels, submitted to tilted electric and magnetic fields. The levels have a rather long radiative lifetime (in the $\mu$s range), which in return make them inaccessible by one-photon transition from the ground state.

On the other hand, so-called purely long-range homonuclear molecules were produced by photoassociation of ultracold atoms \cite{movre1977, stwalley1978, leonard2003, enomoto2008, hollerith2023}. Due to the competition between the resonant DDI and the atomic fine or hyperfine interaction, a shallow potential well exists in an excited dimer electronic state, which can contain a few vibrational levels accessible by laser from the continuum of the ground electronic state. The minimum of this well is located at interatomic distances much larger than the Leroy radius, hence its long-range nature.

Considering these elements, we wondered if a purely LR molecule could be produced by photoassociation (PA) in a gas of lanthanide atoms, due to the interplay between interatomic interactions and the small energy splitting between close energy levels. Using a pair of opposite-parity levels in an electric field could give to this LR molecule an electric dipole moment, in addition to the strong magnetic one present in lanthanide atoms. We identified several candidates, including the pair evoked above, and finally chose the pair of levels presented in Table \ref{tab:lriSFHoLev} in the spectrum of holmium (Ho).


\begin{table}
  \begin{center}
  \caption{Characteristics of the three holmium energy levels proposed to support the long-range molecule. The two last columns give reduced matrix elements of the electric transition multipole moments between pairs of levels, namely $|\langle g\| \mathrm{Q}_2 \|a \rangle|$, $|\langle a\| \mathrm{Q}_1 \|b \rangle|$ and $|\langle g\| \mathrm{Q}_1 \|b \rangle|$. Those quantities, as well as the magnetic moments of the excited levels are calculated with the Cowan codes \cite{cowan1981} using the parameters given in Ref.~\cite{li2017}. The other data are extracted from the NIST ASD database \cite{NIST-ASD}.
  }
  \label{tab:lriSFHoLev}
  \begin{tabular}{|cccc|cc|}
    \hline
    Level & Configuration, term, &  Energy  & Mag.
     & \multicolumn{2}{c|}{Elect.~trans.} \\
    name  &    parity and $J$    & (\cmi{}) & mom. ($\mu_B$)
     & \multicolumn{2}{c|}{mult.~mom. (a.u.)} \\
    \hline
    $|g\rangle$ & $4f^{11}\,^4I_{15/2}^{\circ\phantom{I^I}}$ & 0 & 8.96
    & \multirow{2}{*}{35.3} & \multirow{3}{*}
     {$\left\} \begin{array}{c}
       \, \\ \textrm{11.6} \\ \,
     \end{array} \right.$} \\
    $|a\rangle$ & $4f^{11}(^4I^\circ) 5d6s(^1D) \,^4I_{15/2}^\circ$ & 24357.90 & 8.86 
    & \multirow{2}{*}{2.56} & \\
    $|b\rangle$ & $4f^{11}(^4I_{15/2}^\circ) 6s6p(^1P_1^\circ) \,(\nicefrac{15}{2},1)_{17/2}$ & 24360.81 & 10.0 
    & & \\
    \hline
  \end{tabular}
  \end{center}
\end{table}

The three states are characterized by an identical term in their $4f$ subshell, $4f^{11} \,^4I^\circ$, which can thus be considered as a spectator in multipole transitions. Regarding valence electrons, the ground level is closed-shell $6s^2 \,^1S$, while level $|b\rangle$ corresponds to an electric-dipole excitation toward the $6p$ orbital, \textit{i.e.}~$6s6p\,^1P^\circ$. Another electric-dipole excitation brings the $6p$ orbital to the $5d$ one, giving the valence term $5d6s\,^1D$ for $|a\rangle$, itself related to $|g\rangle$ by a $6s$-$5d$ electric-quadrupole coupling. Because all the valence terms are singlet, the transition multipole moments between the three levels are large. The corresponding reduced transition dipole and quadrupole moments are also given in Table \ref{tab:lriSFHoLev}. They have been calculated with the Cowan codes \cite{cowan1981}, using the parameters given in our paper dedicated to Ho spectroscopy \cite{li2017}.


\begin{table}
  \begin{center}
  \caption{Schematic representation of the atom-atom and atom-field interactions in the unsymmetrized basis with one atom in its ground level $|g\rangle$ and the other atom in one of the two quasi-degenerate excited levels $|a\rangle$ and $|b\rangle$. The abbreviations mean: {}``dir" = direct, {}``res" = resonant, {}``d" = dipole, {}``q" = quadrupole, {}``$\mu$" = magnetic dipole, {}``Z" = Zeeman for the two atoms, {}``S\,A" = Stark for atom A, {}``S\,B" = Stark for atom B. For example, {}``S\,B + res qd" in row $\langle ga|$ and column $|gb\rangle$ stands for: Stark interaction for the second atom (B) plus resonant quadrupole-dipole interaction.
  }
  \label{tab:lriSFHoInter}
  \begin{tabular}{|c|cccc|}
    \hline
     & $|ga\rangle$ & $|ag\rangle$ 
     & $|gb\rangle$ & $|bg\rangle$ \\
    \hline
    $\langle ga|$ & Z + dir qq + dir $\mu\mu$ 
     & res qq & S\,B + res qd & res dq \\
    $\langle ag|$ & res qq & Z + dir qq + dir $\mu\mu$ 
     & res qd & S\,A + res dq \\
    $\langle gb|$ & S\,B + res qd & res qd
     & Z + dir qq + dir $\mu\mu$ & res dd \\
    $\langle bg|$ & res dq & S\,A + res dq
     & res dd & Z + dir qq + dir $\mu\mu$ \\
    \hline
  \end{tabular}
  \end{center}
\end{table}

Those transition dipole and quadrupole moments give rise to resonant interactions, represented schematically on Table~\ref{tab:lriSFHoInter}. In the uncoupled Ho-Ho$^*$ basis, there are three resonant interaction types:
\begin{enumerate}
  \item dipole-dipole between states of the blocks ($|gb\rangle$, $|bg\rangle$), proportional to $|\langle g\| \mathrm{Q}_1 \|b \rangle|^2$ on the one hand, and ($|ab\rangle$, $|ba\rangle$) proportional to $|\langle a\| \mathrm{Q}_1 \|b \rangle|^2$ on the other hand; 
  \item quadrupole-quadrupole between $|ga\rangle$ and $|ag\rangle$, proportional to $|\langle g\| \mathrm{Q}_2 \|a \rangle|^2$;
  \item dipole-quadrupole between ($|ga\rangle$, $|bg\rangle$), proportional to $\langle g\| \mathrm{Q}_1 \|b \rangle \times \langle g\| \mathrm{Q}_2 \|a \rangle$ on the one hand, and ($|ag\rangle$, $|bg\rangle$), proportional to $\langle a\| \mathrm{Q}_1 \|b \rangle \times \langle g\| \mathrm{Q}_2 \|g \rangle$ on the other hand.
\end{enumerate}
There are also direct interactions, which appear in the diagonal of Table~\ref{tab:lriSFHoInter}: between (permanent) magnetic dipoles and between (permanent) electric quadrupoles for all pairs of states. Note that because the quadrupole moment of the ground level $\langle g\| \mathrm{Q}_2 \|g \rangle$ is estimated smaller than 1~a.u., the quadrupolar terms proportional to it are ignored. The diagonals also contain the Zeeman interaction, proportional to the sum of the magnetic (dipole) moments of the two levels present in the basis state, see Table~\ref{tab:lriSFHoLev}. As for the Stark interaction, it couples states for which one atom remains in the $|g\rangle$, and the other goes from $|a\rangle$ to $|b\rangle$.

The calculations are performed in the lab-frame, symmetrized and fully-coupled basis including the atomic hyperfine structure (HFS)
\begin{align}
  & \left| [\beta_A p_A J_A I F_A, \beta_B p_B J_B I F_B],
    F_{AB} L F M \right\rangle
  \nonumber \\
  & = \frac{1}{\sqrt{2}}
    \left[ \left| \beta_A p_A J_A I F_A,  
    \beta_B p_B J_B I F_B, F_{AB} L F M \right\rangle
  \right. \nonumber \\
  & \phantom{=\sqrt{2}} \left.
    + (-1)^{F_A+F_B-F_{AB}+L} \left| \beta_B p_B J_B I F_B, 
    \beta_A p_A J_A I F_A, F_{AB} L F M \right \rangle 
  \right].
  \label{eq:lriSFHoBas}
\end{align}
where the braces $[\,]$ indicate that we consider two identical bosons, which corresponds to $\eta = 1$ in Eq.~\eqref{eq:lriSymSFPerm2}. Indeed, holmium possesses one stable isotope, $^{165}$Ho, with a nuclear spin $I=7/2$. The HFS splittings are proportional to the HFS $A$ and $B$ constants which, for the three states considered here, are in the GHz range \cite{dankwort1974, wyart1978, miao2014}. Since, this range is comparable to that of LR interactions, HFS cannot be ignored in the present study. In Eq.~\eqref{eq:lriSFHoBas}, the letters $(A,B)$ stand for the possible couples of energy levels, namely $(A,B) = (g,a)$ and $(g,b)$. In the first case, the parity of the state is equal to $p=(-1)^L$, in the second case, it is equal to $p=-(-1)^L$.

Even if we do not make calculations for two ground-level atoms, we make a few assumptions on their collisions: (i) the ground-level atoms are prepared in their stretched HFS sublevel $|F_g=J_g+I=11, M_{F,g}=11 \rangle$, as in the experiment \cite{miao2014}; (ii) they are cold enough to collide in the $s$-wave regime. In consequence, the only coupled and symmetrized state for two ground-level atom is $|[g,g], F_{AB}=22, L=0, F=M=22 \rangle$. If a linearly-polarized ($q=0$) or circularly-polarized ($q=1$) PA laser is applied red-detuned with respect to the $|g\rangle \to |b\rangle$ transition (wavelength of 410.5~nm), the excited complex states that are reached have a projection $M = 22 + q$. Therefore, in this study, we are interested in the LR PECs close to the $(|g\rangle + |a\rangle)$ and $(|g\rangle + |b\rangle)$ asymptotes with $M=22$ and 23, in the presence of colinear static electric and magnetic fields in the $z$ direction. Those fields mixes states with $F=22$ with states with higher $F$-values.

The DC fields interact with the atomic dipole moments, whose reduced matrix elements including HFS are
\begin{align}
  \left\langle \qnpr{\beta_k} \qnpr{J_k} I \qnpr{F_k} 
    \right\| \mathrm{X}_{\ell_k} \left\| 
    \beta'_k J'_k I F'_k \right\rangle
  & = (-1)^{J_k+I+F'_k+\ell_k} \sqrt{(2F_k+1)(2F'_k+1)}
  \nonumber \\
  & \times \sixj{J_k}{I}{F_k}{F'_k}{\ell_k}{J'_k}
    \left\langle \qnpr{\beta_k} \qnpr{J_k} \right\| 
    \mathrm{X}_{\ell_k} \left\| \beta'_k J'_k \right\rangle ,
  \label{eq:lriSFHoMultMom}
\end{align}
where $\ell_k=1$ and $\mathrm{X}_{\ell_k}$ is a general notation covering electric multipoles $\mathrm{Q}_{\ell_k}$ or magnetic ones $\mathrm{M}_{\ell_k}$. In the magnetic case, $\langle \beta_k J_k \| \mathrm{M}_1 \| \beta'_k J'_k \rangle = -\delta_{\beta_k\beta'_k} \delta_{J_kJ'_k} \mu_B g_J \sqrt{J_K(J_k+1)(2J_k+1)}$, with $g_J$ the Land\'e $g$-factor of the level, obtained by dividing the magnetic moment of Table~\ref{tab:lriSFHoLev} by $J_k$. In the electric case, $\langle \beta_k J_k \| \mathrm{Q}_1 \| \beta'_k J'_k \rangle$, between ($|g\rangle$, $|b\rangle$) and ($|a\rangle$, $|b\rangle$) is also given on Table~\ref{tab:lriSFHoLev}. The matrix element of the quadrupole moment, coupling $|g\rangle$ and $|a\rangle$, are obtained with $\ell_k=2$ in Eq.~\eqref{eq:lriSFHoMultMom}. All those multipolar matrix elements can be plugged in Eqs.~\eqref{eq:lriVsf4} and \eqref{eq:lriEFldCpl}, after replacing their $J$ quantum numbers by $F$ ones.


\begin{figure}
  \begin{center}
  \includegraphics[width=0.49\textwidth]
    {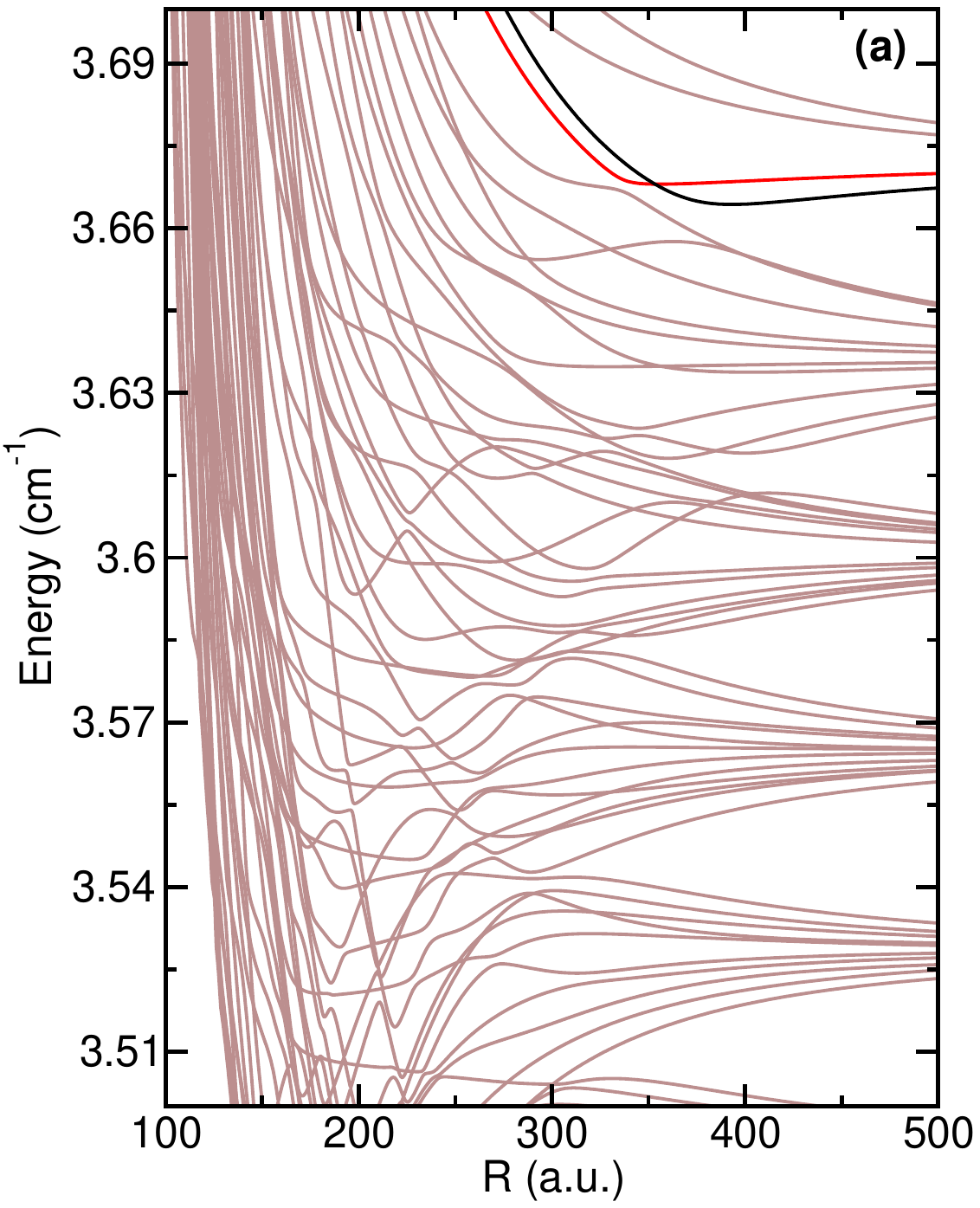}
  \includegraphics[width=0.49\textwidth]
    {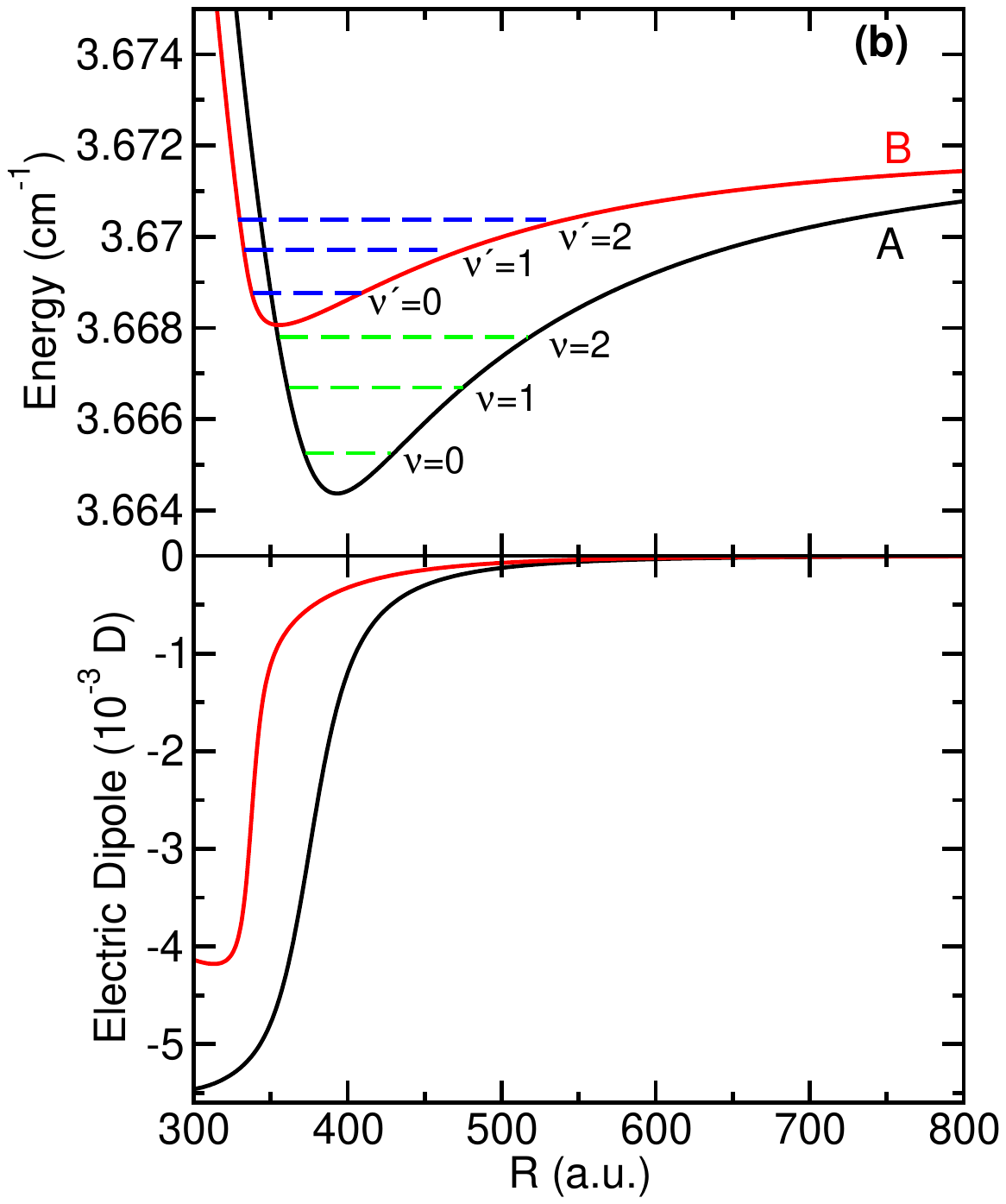}
  \caption{Panel (a): examples of long-range potential-energy curves correlated to various hyperfine sublevels of $(|g\rangle + |b\rangle)$ asymptote, in the presence of a static electric field $\mathcal{E} = 5$~kV/cm and magnetic field $B=1000$~G, both in the $z$-direction. Panel (b): zoom on the two thicker black (A) and red (B) curves of panel (a) with their three lowest vibrational levels. Panel (c): their respective average electric dipole moment.
  }
  \label{fig:lriSFHoPecs}
  \end{center}
\end{figure}

Calculations were carried out for an electric field $\mathcal{E} = 5$~kV/cm and magnetic fields $B$ up to 1000~G. With such amplitudes, values of $M \le F \le 27$ and $0 \le L \le 4$ are included to obtain converged PECs. Due to the triangle inequality, the smallest $F_{12}$ value is $M-4 = 18$ or 19, and the largest is $\max(F_g)+\max(F_b) = 23$. Figure~\ref{fig:lriSFHoPecs} (a) displays PECs close to the $(|g\rangle + |b\rangle)$ asymptote for $M=23$ (circular PA laser). The zero of energy is the average of field- and HFS-free energies of $|a\rangle$ and $|b\rangle$. Groups of curves converge to various $M_{F,b}$ values, the highest of which interests us. In particular, the two thick curves are indeed long-range wells, which are zoomed in on panel (b). Their minimum is located at 300--400~a.u., and their depth is a few thousands of \cmi{} (a few tens of MHz). To check the existence of vibrational levels, H.~Li used the Mapped-Fourier-grid method \cite{kokoouline1999} on each PEC separately. He found that curve A (resp.~B) could contain 16 (resp.~11) levels, the three lowest of which are represented on Fig.~\ref{fig:lriSFHoPecs} (b).

To each point of the PECs can be associated an eigenvector expressed in our basis. In the same spirit as in Eq.~\eqref{eq:lriDiatDip1} for diatomic molecules, we calculate the $R$-dependent average electric dipole moment $\langle \mathrm{Q}_{10}^\mathrm{SF} \rangle$. The results are shown on Fig.~\ref{fig:lriSFHoPecs} (c) for curves A and B. They amount to a few thousands of Debye, which is small. They tend to 0 for $R \to \infty$, because the asymptotic sublevel $M_{F,b}=13$ is not field-coupled to any $M_{F,a}$ counterpart, of maximum 12. In the region of the well minimum, several $M_{F,b}$ sublevels are coupled by the LR potential, which turns on the field coupling with $M_{F,a}$ counterparts. Larger dipole moments could be reached with larger DC-field amplitudes or with a MW field.

Note finally that shallower LR wells containing one bound level were found close to the highest $(|g\rangle + |a\rangle)$ asymptote with $M=22$. The dipole moment was also of a thousands of Debye \cite{li2019}. We can expect this level to have a larger radiative lifetime than those of Fig.~\ref{fig:lriSFHoPecs}, since the lifetime of level $|a\rangle$ is 650 times larger than that of $|b\rangle$ \cite{denhartog1999}. Still, in both cases, the levels have a limited lifetime due to predissociation. This can be understood on Fig.~\ref{fig:lriSFHoPecs} (a), where PECs A and B exhibit avoided crossings with dissociative curves dissociating to lower asymptotes. The widths of the resulting resonances could be estimated with a scattering code as in Sec.~\ref{sec:lri1OS}.

\,

In this chapter, I have described examples of long-range interactions described in the space-fixed frame. In section \ref{sec:lri1OS}, I have presented their application in the framework of the shielding of ultracold reactive collisions with one- and two-photon transitions. Combining with purely absorbing short-range conditions, the computed long-range matrices have allowed for calculating scattering observables like elastic, inelastic and reactive collision rates using a close-coupling code. One-photon shielding is efficient according to those calculations; but the photon scattering induced by the blue-detuned laser seems prohibitive, even near a forbidden transition, since it results in a significant heating of the molecular samples. 
To overcome this problem, we have proposed two-photon shielding, based on a $\Lambda$ scheme that generates a dark states immune to photon scattering. For the moment, the scattering calculations have not proved the efficiency of two-photon shielding. Investigations continue in yet unexplored detuning and Rabi-frequency regions. Addition of a static field is also envisioned in order to induce larger avoided crossing, hence to increase the shielding efficiency.

In section \ref{sec:lriHo2}, we have explored the possiblity of creating doubly dipolar diatomic holmium molecules via photoassociation. To that end, one atom is in the ground level and the other in a superposition of close opposite-parity excited levels. We have demonstrated the existence of shallow long-range wells that can accommodate a few bound levels. However, their electric field is found disappointingly small. We attribute this to the large energy difference between the excited levels compared to the Stark energy at play. A possibility to increase this small electric dipole moment would be to use a microwave rather than a static field, as it allows for reaching larger field amplitudes. Still, this last description is certainly the most elaborate long-range problem addressed in this manuscript, with an excited atom in a superposition of levels including hyperfine structure, two external fields, and the use of the symmetrized and fully coupled basis. It is a milestone in our methodology.

A possible prospect of all those developments could be the study of few-body or many-body interactions in such a detailed manner. Indeed, long-range many-body quantum interacting systems can be realized with various experimental platforms \cite{defenu2023}. A first step in this direction was done in Ref.~\cite{lepers2015}, where we studied the interactions between two weakly-bound Er$_2$ molecules described  as four-body atom-atom interactions. The general formalism was given in the first and the second orders of perturbation theory.

\chapter*{Conclusion}  \addcontentsline{toc}{chapter}{Conclusion}\markboth{CONCLUSION}{ }

In this manuscript, I have discussed the structure and interactions of rather complex quantum systems, mainly in gas phase. Most of the presented results concern the realm of ultracold gases, especially the atom- and molecule-light interactions. Some studies were also performed in the framework of atmospheric or plasma physics. Furthermore, I have discussed the luminescent properties of crystals or glasses doped with lanthanide trivalent ions. This topic is \textit{a priori} far from gas-phase physics; but due to the weak interaction between the dopant and its environment, the situation is similar to a single ion submitted to external electromagnetic fields.

More specifically, my research activities follow two main directions: the electronic structure of lanthanide atoms and ions submitted to external electromagnetic fields, and the long-range interactions between atoms and/or diatomic molecules also submitted to external fields. In both cases, I have highlighted the central role of angular algebra, which allows for deriving selection rules resulting from symmetry properties. As for the atomic and molecular properties necessary for our calculations are either taken from experimental measurements of calculated with quantum chemistry, and in particular semi-empirical methods.

In addition to the work presented here, I have studied in collaboration with P.~Honvault and G.~Guillon from ICB, reactive collisions between atoms and diatomics using the time-independent quantum method, similar to the one used in the shielding investigations, and hyperspherical coordinates \cite{honvault2004}. I have focused on the reaction H$^+$ + HD$\,\to\,$D$^+$ + H$_2$ \cite{lepers2019, lepers2021}, relevant for the chemistry of the primordial universe \cite{galli2013}. Moreover, with Etienne Brion from Laboratoire des Collisions, Agrégats, Réactivité at Toulouse, we have investigated the interaction (energy and spontaneous emission rates) of one and two alkali-metal Rydberg atoms at the vicinity of an optical nanofiber \cite{stourm2019, stourm2020, stourm2023}. Such calculations rely on atomic transition energies, dipole and quadrupole moments, hence my participation.

I have currently two major prospects of my research work. The first one, based on both electronic-structure and long-range interactions of dysprosium atom, consists in modeling two- and three-body collisions between such atoms, their Feshbach resonances and weakly-bound molecular levels. This system is believed to exhibit quantum chaos, preventing the predictive character of collisional calculations \cite{frisch2014, maier2015, augustovivcova2018, mccann2021}. But together with Charbel Karam, now in post-doc, we want to tackle that problem with a different point of view. This work is performed in collaboration with the experimental team of Jean Dalibard and Raphael Lopes at Coll{\` e}ge de France, and the theoretical team of Olivier Dulieu at Laboratoire Aimé Cotton, in the framework of the ANR project {}``FewBoDyK''. 

The second direction is a continuation of Chapter \ref{chap:ln3+} on trivalent ions. After bringing to it some improvements, I am currently testing our model with additional ions for which the usual Judd-Ofelt theory is not fully satisfactory. This work is performed in collaboration with Matias Velazquez, Richard Moncorgé and Yannick Guyot. The fact that the adjustable parameters of our model merely depend on the crystal-field parameters opens the possibility to model spectra at low temperatures, between individual Stark sublevels, and in polarized light. Finally, if the fitting process gives more accurate results than the standard Judd-Ofelt, it may also serve to predict quantities difficult to measure, like branching ratios among excited levels.

\part* {Appendices}
\addcontentsline{toc}{part}{Appendices}

\appendix
\addtocontents{toc}{\protect\setcounter{tocdepth}{0}}

\chapter{Useful relations}

\section{Miscellaneous}

I present here various relations useful in the main text, and that are extracted from Ref.~\cite{varshalovich1988}. The integral on Euler angles involving three Wigner $D$-matrices are
\begin{align}
  & \int_0^{2\pi} d\alpha \int_0^{\pi} d\beta \int_0^{2\pi} d\gamma
    D_{m_1 n_1}^{j_1  } (\alpha, \beta, \gamma)
    D_{m_2 n_2}^{j_2  } (\alpha, \beta, \gamma)
    D_{m_3 n_3}^{j_3,*} (\alpha, \beta, \gamma)
  \nonumber \\
   & = \frac{8\pi^2} {2j_3+1}
     C_{j_1 m_1 j_2 m_2}^{j_3 m_3}
     C_{j_1 n_1 j_2 n_2}^{j_3 n_3} .
  \label{eq:appInt3D}
\end{align}
The link between $D$-matrices, Racah and normalized spherical harmonics is
\begin{equation}
  C_{\ell m} (\beta,\alpha)
   = \sqrt{\frac{4\pi}{2\ell+1}} Y_{\ell m} (\beta,\alpha)
   = D_{m 0}^{\ell,*} (\alpha, \beta, \gamma) .
  \label{eq:appCYD}
\end{equation}
The sum of products of three Clebsch-Gordan coefficients is
\begin{align}
  & \sum_{\alpha \beta \delta}
     C_{a\alpha b\beta}^{c\gamma}
     C_{d\delta b\beta}^{e\epsilon}
     C_{a\alpha f\phi }^{d\delta  }
   \nonumber \\
   & = (-1)^{b+c+d+f} \sqrt{(2c+1)(2d+1)}
     C_{c\gamma f\phi }^{e\epsilon}
     \sixj{a}{b}{c}{e}{f}{d} .
  \label{eq:appSum3CG}
\end{align}

\section{Long-range interactions and irreducible tensors}

In this appendix, we extensively use the relations given of Chapter 13 of Ref.~\cite{varshalovich1988} We start with the space-fixed long-range operator expressed in terms of scalar and tensor products of operators, see Eq.~\eqref{eq:lri-vsf-2}
\begin{equation}
  \mathrm{V}_{AB}^{\mathrm{SF}}(\mathbf{R})
    = \frac{1}{4\pi\epsilon_{0}}
    \sum_{\ell_A\ell_B\ell=0}^{+\infty} \delta_{\ell_A+\ell_B,\ell}
    \,\frac{\left(-1\right)^{\ell_{B}}}{R^{1+\ell}}
    \binom{2\ell}{2\ell_A}^{1/2}
    \left( \mathrm{C}_{\ell} (\Theta,\Phi) 
    \cdot \left\{ \mathrm{Q}_{\ell_{A}}^{\textrm{SF}} 
      \otimes \mathrm{Q}_{\ell_{B}}^{\textrm{SF}}  
      \right\}_{\ell} \right) .
  \label{eq:appLriVsf}
\end{equation}

\subsection{First-order correction in fully coupled basis}
\label{sec:AppLriVsfCpl}

We seek to evaluate the matrix elements of operator \eqref{eq:appLriVsf} in the fully coupled basis \\ $\{| \beta_A J_A \beta_B J_B J_{AB} L J M \rangle\}$, see Eq.~\eqref{eq:lriCplBas}. To that end, we use some relationships on tensor operators, for which we consider to following general notations: two angular momenta are coupled as $\mathbf{j} = \mathbf{j}_1 + \mathbf{j}_2$; the operator $\mathrm{T}_a$ acts on $\mathbf{j}_1$, while $\mathrm{U}_a$ and $\mathrm{U}_b$ act on $\mathbf{j}_2$. We apply
\begin{align}
  & \left\langle \qnpr{j_1} \qnpr{j_2} j \right\|
    \left( \mathrm{T}_a \cdot \mathrm{U}_a \right)
    \left\| j'_1 j'_2 j' \right\rangle
  \nonumber \\
  & = \delta_{jj'} \delta_{mm'} (-1)^{j'_1+j_2+j}
    \sixj{j_1 }{j_2 }{j}
         {j'_2}{j'_1}{a}
    \left\langle \qnpr{j_1} \right\| \mathrm{T}_a
    \left\| j'_1 \right\rangle
    \left\langle \qnpr{j_2} \right\| \mathrm{U}_a
    \left\| j'_2 \right\rangle
\end{align}
with $\mathbf{j}_1 = \mathbf{J}_{AB}$, $\mathbf{j}_2 = \mathbf{L}$, $\mathbf{j} = \mathbf{J}$ and $a=\ell$. Then we apply
\begin{align}
  & \left\langle \qnpr{j_1} \qnpr{j_2} j \right\|
    \left\{ \mathrm{T}_a \otimes \mathrm{U}_b \right\}_c
    \left\| j'_1 j'_2 j' \right\rangle
  \nonumber \\
  & = (-1)^{2c} \sqrt{(2c+1)(2j+1)(2j'+1)}
    \ninej{a   }{b   }{c }
          {j_1 }{j_2 }{j }
          {j'_1}{j'_2}{j'}
    \left\langle \qnpr{j_1} \right\| \mathrm{T}_a
    \left\| j'_1 \right\rangle
    \left\langle \qnpr{j_2} \right\| \mathrm{U}_b
    \left\| j'_2 \right\rangle
\end{align}
with $\mathbf{j}_1 = \mathbf{J}_A$, $\mathbf{j}_2 = \mathbf{J}_B$, $\mathbf{j} = \mathbf{J}_{AB}$, $a=\ell_A$, $b=\ell_B$ and $c=\ell$. We get to Eq.~\eqref{eq:lriVsf4}.

\subsection{Second-order correction in space-fixed frame}
\label{sec:AppLriVsf2Cpl}

To account for second-order corrections, we introduce in Eq.~\eqref{eq:lriPert2nd} the effective operator
\begin{align}
  \mathrm{W}_{AB}^{\mathrm{SF}} 
  & = -\sum_{A''B''} \mathrm{V}_{AB}^{\mathrm{SF}} 
    \frac{|A''B''\rangle \langle A''B''|}
      {\Delta E''_A+\Delta E''_B} \mathrm{V}_{AB}^{\mathrm{SF}}
  \nonumber \\
   & = -\frac{1}{16\pi^2 \epsilon_0^2} 
    \sum_{\ell_A \ell_B \ell} \sum_{\ell'_A \ell'_B \ell'}
    \delta_{\ell_A+\ell_B,\ell} \, \delta_{\ell'_A+\ell'_B,\ell'}
    \frac{(-1)^{\ell_B+\ell'_B}}{R^{2+\ell+\ell'}}
    \binom{2\ell }{2\ell_A }^{\frac{1}{2}}
    \binom{2\ell'}{2\ell'_A}^{\frac{1}{2}}
  \nonumber \\
   & \times \left( \mathrm{C}_{\ell} \cdot \left\{ \mathrm{Q}_{\ell_A}
    \otimes \mathrm{Q}_{\ell_B} \right\}_{\ell} \right)
    \sum_{A''B''} \frac{|A''B''\rangle \langle A''B''|}
    {\Delta E''_A+\Delta E''_B}
    \left( \mathrm{C}_{\ell'} \cdot \left\{ \mathrm{Q}_{\ell'_A}
    \otimes \mathrm{Q}_{\ell'_B} \right\}_{\ell'} \right)
  \label{eq:AppLriWsf1}
\end{align}
where $| A''B'' \rangle$ is a condensed notation of the complex {}``excited'' states (namely $|A'' \rangle = |\beta''_A J''_A \rangle$, $|B'' \rangle = |\beta''_B J''_B \rangle$
), and $\Delta E''_k = E_{\beta''_k J''_k} - E_{\beta_k J_k}$ are the excitation energies of individual partners. We want to work out Eq.~\eqref{eq:AppLriWsf1} in order to gather in three distinct groups the spherical harmonics, the operators of partner $A$ and the operators of partner $B$. To that end, we use the relationships
\begin{equation}
  \left( \mathrm{T}_a \cdot \mathrm{U}_a \right)
     \left( \mathrm{T}_b \cdot \mathrm{U}_b \right)
   = \sum_{k=|a-b|}^{a+b} (-1)^{a+b-k} \left( 
     \left\{ \mathrm{T}_a \otimes \mathrm{T}_b \right\}_k \cdot
     \left\{ \mathrm{U}_a \otimes \mathrm{U}_b \right\}_k
     \right)
  \label{eq:AppLriRecpl1}
\end{equation}
with $\mathrm{T}_{a,b} = \mathrm{C}_{\ell,\ell'}$, $\mathrm{U}_a = \left\{ \mathrm{Q}_{\ell_A} \otimes \mathrm{Q}_{\ell_B} \right\}_{\ell}$ and $\mathrm{U}_b = \left\{ \mathrm{Q}_{\ell'_A} \otimes \mathrm{Q}_{\ell'_B} \right\}_{\ell'}$,
and also
\begin{align}
   & \left\{ \left\{ \mathrm{T}_a \otimes \mathrm{U}_b \right\}_c 
     \otimes \left\{ \mathrm{T}_d \otimes \mathrm{U}_e \right\}_f
     \right\}_k
  \nonumber \\
   = & \sum_{k_1=|a-d|}^{a+d} \sum_{k_2=|b-e|}^{b+e}
     \sqrt{(2c+1)(2f+1)(2k_1+1)(2k_2+1)}
     \ninej{a}{d}{k_1}{b}{e}{k_2}{c}{f}{k}  
  \nonumber \\
   & \times
     \left\{ \left\{ \mathrm{T}_a \otimes \mathrm{T}_d \right\}_{k_1} 
     \otimes \left\{ \mathrm{U}_b \otimes \mathrm{U}_e \right\}_{k_2}
     \right\}_k
\end{align}
with $\mathrm{T}_{a,d} = \mathrm{Q}_{\qnpr{\ell_A},\ell'_A}$ and $\mathrm{U}_{b,e} = \mathrm{Q}_{\qnpr{\ell_B},\ell'_B}$. Finally, the effective operator reads
\begin{align}
  \mathrm{W}_{AB}^{\mathrm{SF}} & = -\frac{1}{16\pi^2 \epsilon_0^2}
    \sum_{\ell_A \ell_B \ell} \sum_{\ell'_A \ell'_B \ell'}
    \delta_{\ell_A+\ell_B,\ell} \, \delta_{\ell'_A+\ell'_B,\ell'}
    \frac{(-1)^{\ell_B+\ell'_B}}{R^{2+\ell+\ell'}}
    \binom{2\ell }{2\ell_A }^{\frac{1}{2}}
    \binom{2\ell'}{2\ell'_A}^{\frac{1}{2}}
  \nonumber \\
   & \times \sum_{A''B''} \frac{1}{\Delta E''_A+\Delta E''_B}
    \sum_{k_A k_B k} \sqrt{(2\ell+1)(2\ell'+1)(2k_A+1)(2k_B+1)}   
    \ninej{\ell_A}{\ell'_A}{k_A}
          {\ell_B}{\ell'_B}{k_B}
          {\ell  }{\ell'  }{k  }
  \nonumber \\
   & \times \left( \left\{ \mathrm{C}_{\ell} 
    \otimes \mathrm{C}_{\ell'} \right\}_k \cdot \left\{ 
    \left\{ \mathrm{Q}_{\ell_A} \otimes \left\| A'' \right\rangle
    \left\langle A'' \right\| \mathrm{Q}_{\ell'_A} \right\}_{k_A}
    \otimes
    \left\{ \mathrm{Q}_{\ell_B} \otimes \left\| B'' \right\rangle
    \left\langle B'' \right\| \mathrm{Q}_{\ell'_B} \right\}_{k_B}
    \right\}_k \right).
  \label{eq:AppLriWsf2}
\end{align}
The tensor product of spherical harmonics can be written
\begin{equation}
  \left\{ \mathrm{C}_{\ell} \otimes \mathrm{C}_{\ell'} \right\}_{kq}
   = \sum_{m m'} C_{\ell m \ell' m'}^{kq}
    C_{\ell m} (\Theta,\Phi) C_{\ell' m'} (\Theta,\Phi)
   = C_{\ell 0 \ell' 0}^{k0} C_{kq} (\Theta,\Phi) ,
\end{equation}
where the CG coefficient of the right-hand side imposes $\ell+\ell'+k$ even. Note that the BF frame expression can be retrieved by setting $\Theta = \Phi = 0$, which imposes $q = 0$ in the scalar product of Eq.~\eqref{eq:AppLriWsf2}.

Note that alternatively, we can introduce the dynamic dipole polarizabilities at imaginary frequencies as in sections \ref{sec:lri-pa} and \ref{sec:lri-o3},
\begin{align}
  & \sum_{A''B''} \frac{1}{\Delta E''_A+\Delta E''_B} \left\{
    \left\{ \mathrm{Q}_{\ell_A} \otimes \left\| A'' \right\rangle
    \left\langle A'' \right\| \mathrm{Q}_{\ell'_A} \right\}_{k_A}
    \otimes
    \left\{ \mathrm{Q}_{\ell_B} \otimes \left\| B'' \right\rangle
    \left\langle B'' \right\| \mathrm{Q}_{\ell'_B} \right\}_{k_B}
    \right\}_k 
  \nonumber \\
  & = \frac{1}{2\pi} \int_0^{+\infty} du \left\{
    \upalpha_{(\ell_A\ell'_A)k_A} (iu) \otimes
    \upalpha_{(\ell_B\ell'_B)k_B} (iu) \right\}_k 
\end{align}
where $\upalpha_{(\ell_{A,B} \ell'_{A,B}) k_{A,B}}$ are irreducible tensor operators of ranks $k_{A,B}$, discussed in Chapter \ref{chap:ddp}.

To calculate the matrix elements in the fully coupled basis, we use the relationship
\begin{align}
  & \left\langle j \right\| \left\{ \mathrm{T}_a 
    \otimes \mathrm{T}_b \right\}_c
    \left\| j' \right\rangle
  \nonumber \\
  & = (-1)^{j+j'-c} \sqrt{2c+1} \sum_{j''}
    \sixj{a }{b}{c  }
         {j'}{j}{j''}
    \left\langle j \right\| \mathrm{T}_a \left\| j'' \right\rangle
    \left\langle j'' \right\| \mathrm{T}_b \left\| j' \right\rangle
  \label{eq:AppLriProdTens}
\end{align}
with $\mathrm{T}_a = \mathrm{Q}_{\ell_{A,B}}$, $\mathrm{T}_b = \mathrm{Q}_{\ell'_{A,B}}$, $c=k_{A,B}$, $j=J_{A,B}$ (and the corresponding primed and double-primed quantum numbers). Finally, we obtain
\begin{align}
  & \left\langle \qnpr{\beta_A} \qnpr{J_A}
    \qnpr{\beta_B} \qnpr{J_B} \qnpr{J_{AB}} L J M
    \right| \mathrm{W}_{AB}^{\mathrm{SF}}(R) \left|
    \beta'_A J'_A \beta'_B J'_B J'_{AB} L' J' M' \right\rangle
  \nonumber \\
  = & -\frac{\delta_{JJ'}\delta_{MM'}}{16\pi^2\epsilon_0^2}
    \sum_{\ell_A\ell_B\ell} \sum_{\ell'_A\ell'_B\ell'}
    \delta_{\ell_A+\ell_B,\ell} \delta_{\ell'_A+\ell'_B,\ell'}\,
    \frac{(-1)^{\ell_B+\ell'_B+J_A+J'_A+J_B+J'_B+J'_{AB}+L+J}}
      {R^{2+\ell+\ell'}}
    \binom{2\ell }{2\ell_A }^{\frac{1}{2}}
    \binom{2\ell'}{2\ell'_A}^{\frac{1}{2}}
   \nonumber \\
  \times & \sum_{k_{A}k_{B}k}
    \left(-1\right)^{k_{A}+k_{B}} \left[k_A k_B\right] 
    \left[\ell \ell' k J_{AB} J'_{AB} L' \right]^{1/2}
    C_{\ell,0,\ell',0}^{k0} C_{L'0k0}^{L0} 
    \ninej{\ell_A}{\ell'_A}{k_A}
          {\ell_B}{\ell'_B}{k_B}
          {\ell  }{\ell'  }{k}
  \nonumber \\
  \times & \sixj{J_{AB}}{L}{J}{L'}{J'_{AB}}{k}
    \ninej{k_A }{k_B }{k      }
          {J_A }{J_B }{J_{AB} }
          {J'_A}{J'_B}{J'_{AB}}
    \sum_{\beta''_{A}J''_{A}} \sum_{\beta''_{B}J''_{B}}
    \sixj{\ell_A}{\ell'_A}{k_A}{J'_A}{J_A}{J''_A}
    \sixj{\ell_B}{\ell'_B}{k_B}{J'_B}{J_B}{J''_B}
  \nonumber \\
  \times &
    \frac{\left\langle \beta_{A}J_{A} \right\|
      \mathrm{Q}_{\ell_{A}}
      \left\| \beta''_{A}J''_{A} \right\rangle
      \left\langle \beta''_{A}J''_{A} \right\|
      \mathrm{Q}_{\ell'_{A}}
      \left\| \beta'_{A}J'_{A} \right\rangle
      \left\langle \beta_{B}J_{B} \right\|
      \mathrm{Q}_{\ell_{B}}
      \left\| \beta''_{B}J''_{B} \right\rangle 
      \left\langle \beta''_{B}J''_{B} \right\| 
      \mathrm{Q}_{\ell'_{B}}
      \left\| \beta'_{B}J'_{B} \right\rangle }
    {E_{\beta''_{A}J''_{A}}+E_{\beta''_{B}J''_{B}}
    -E_{\beta_{A}J_{A}}    -E_{\beta_{B}J_{B}}} \,,
  \label{eq:AppLriWsf3}
\end{align}
where $[ab \cdots c] = (2a+1) \times (2b+1) \times \cdots \times (2c+1)$.



\printbibliography

\end{document}